\documentclass[12pt]{article}
\pdfoutput=1

\usepackage{amsmath,graphicx,wrapfig,mathtools,latexsym}
\usepackage{epstopdf} 
\usepackage{multirow}
\usepackage{amsfonts}
\usepackage{amssymb}
\usepackage{amscd}
\usepackage{enumitem}
\usepackage{subcaption}
\usepackage{comment}

\usepackage{amsthm,stmaryrd,fancyhdr,latexsym,color,float,fullpage,bm,amscd,tikz-cd,lscape,epsfig,authblk,array,tikz}

\usetikzlibrary{decorations.markings}

\newcommand{\dynkinradiusold}{.08cm}
\newcommand{\dynkinstepold}{.35cm}
\newcommand{\dynkindotold}[2]{\fill (\dynkinstepold*#1,\dynkinstepold*#2) circle (\dynkinradiusold);}
\newcommand{\dynkinlineold}[4]{\draw[thin] (\dynkinstepold*#1,\dynkinstepold*#2) -- (\dynkinstepold*#3,\dynkinstepold*#4);}

\newcommand{\dynkinradius}{0.108cm}
\newcommand{\dynkinstep}{.4725cm}
\newcommand{\dynkindot}[2]{\fill (\dynkinstep*#1,\dynkinstep*#2) circle (\dynkinradius);}
\newcommand{\dynkinXsize}{2.025}
\newcommand{\dynkincross}[2]{
\draw[thick] (#1*\dynkinstep-\dynkinXsize,#2*\dynkinstep-\dynkinXsize) -- (#1*\dynkinstep+\dynkinXsize,#2*\dynkinstep+\dynkinXsize);
\draw[thick] (#1*\dynkinstep-\dynkinXsize,#2*\dynkinstep+\dynkinXsize) -- (#1*\dynkinstep+\dynkinXsize,#2*\dynkinstep-\dynkinXsize);
}

\newcommand{\dynkinlabel}[3]{
\node at (\dynkinstep*#1,\dynkinstep*#2) {{\footnotesize #3}};}

\newcommand{\dynkinline}[4]{\draw[thin] (\dynkinstep*#1,\dynkinstep*#2) -- (\dynkinstep*#3,\dynkinstep*#4);}
\newcommand{\dynkindots}[4]{\draw[dotted] (\dynkinstep*#1,\dynkinstep*#2) -- (\dynkinstep*#3,\dynkinstep*#4);}

\newenvironment{dynkin}{\begin{tikzpicture}[decoration={markings,mark=at position 0.7 with {\arrow{>}}}]}
{\end{tikzpicture}}

\newcommand{\imagen}{\includegraphics[width=.72\linewidth]}

\def\hybrid{\topmargin -20pt    \oddsidemargin 0pt
        \headheight 0pt \headsep 0pt
        \textwidth 6.25in       
        \textheight 9.5in       
        \marginparwidth .875in
        \parskip 5pt plus 1pt   \jot = 1.5ex}

\hybrid

\numberwithin{equation}{section}
\numberwithin{table}{section}

\def\nn{\nonumber}

\usepackage{physics}
\usepackage{cancel} 

\usepackage[utf8]{inputenc}
\usepackage{amsthm,stmaryrd,fancyhdr,latexsym,color,float,fullpage,bm,amscd,tikz-cd,lscape,epsfig,authblk,array,tikz}
\usepackage{simpler-wick}
\usetikzlibrary{decorations.markings}

\def\hybrid{\topmargin -20pt    \oddsidemargin 0pt
        \headheight 0pt \headsep 0pt
        \textwidth 6.25in       
        \textheight 9.5in       
        \marginparwidth .875in
        \parskip 5pt plus 1pt   \jot = 1.5ex}

\hybrid

\numberwithin{equation}{section}
\numberwithin{table}{section}

\newcommand{\beq}{\begin{equation}\begin{aligned}}
\newcommand{\eeq}{\end{aligned}\end{equation}}
\newcommand{\bse}{\begin{subequations}}
\newcommand{\ese}{\end{subequations}}
\newcommand{\bi}{\begin{itemize}}
\newcommand{\ei}{\end{itemize}}
\newcommand{\bea}{\begin{eqnarray}}
\newcommand{\eea}{\end{eqnarray}}
\newcommand{\ba}{\begin{array}}
\newcommand{\ea}{\end{array}}
\newcommand{\bt}{\begin{tabular}}
\newcommand{\et}{\end{tabular}}
\newcommand{\bc}{\begin{center}}
\newcommand{\ec}{\end{center}}

\newcommand{\ax}{\alpha}
\newcommand{\bx}{\beta}

\newcommand{\dx}{\delta}
\newcommand{\ex}{\epsilon}

\def\nn{\nonumber}

\newcommand{\munu}{{\mu\nu}}

\newcommand{\Goo}{\Gamma_{8+8}}

\newcommand{\cref}{{\bf [check ref]}}

\newcommand{\Rsd}{R_{\rm{fp}}}

\definecolor{verde}{HTML}{66CD4B}
\definecolor{rojo}{HTML}{D01F0E}
\definecolor{azul}{HTML}{0000FE}
\definecolor{amarillo}{HTML}{CBCB00}
\definecolor{violeta}{HTML}{9819FE}
\definecolor{negro}{HTML}{050505}
\definecolor{naranja}{HTML}{FC7E00}
\definecolor{cyan}{HTML}{00FEFE}
\definecolor{rosa}{HTML}{FD9EEA}

\definecolor{magenta}{HTML}{FF00FF}
\definecolor{celeste}{HTML}{9FC5E8}
\definecolor{verde2}{HTML}{93C47D}
\definecolor{gris}{HTML}{CCCCCC}

\newcommand{\cua}[1]{{\fcolorbox{black}{#1}{\rule{0pt}{3pt}\rule{3pt}{0pt}}}}
\newcommand{\cuad}[2]{{\fcolorbox{black}{#1}{\rule{0pt}{3pt}\rule{-2pt}{0pt}}\fcolorbox{black}{#2}{\rule{0pt}{3pt}\rule{-2pt}{0pt}}}}

\newcommand{\bleyenda}{\begin{footnotesize}\begin{equation}\begin{aligned}}

\newcommand{\eleyenda}{\nn\end{aligned}\end{equation}\end{footnotesize}}

\newcommand{\dynkindotc}[3]{\fill[#3] (\dynkinstep*#1,\dynkinstep*#2) circle  (\dynkinradius);}

\begin{document}

\pagestyle{empty}
\begin{titlepage}
\begin{center}

\vskip .20cm

\LARGE{\LARGE\bf A new twist on heterotic string compactifications 
}

\vskip 0.3cm

\large{Bernardo Fraiman$^{\dag\, *}$, Mariana Gra\~na$^{\#}$ and Carmen A. Nu\~nez$^{\dag\, *}$
 \\[6mm]}
{\small\it  $^\dag$ Instituto de Astronom\'ia y F\'isica del Espacio (IAFE-CONICET-UBA) \\
Ciudad Universitaria, Pabell\'on IAFE, 1428 Buenos Aires, Argentina\\ [.3 cm]}
{\small\it  $^*$ Departamento de F\'isica, FCEyN, Universidad de Buenos Aires (UBA)\\ [.3 cm]}
{\small\it  $^{\#}$ Institut de Physique Th\'eorique,
CEA/ Saclay \\
91191 Gif-sur-Yvette Cedex, France}

{\footnotesize  E-mail: \quad bfraiman@iafe.uba.ar, \quad mariana.grana@ipht.fr, \quad carmen@iafe.uba.ar}

\vskip 1cm

\end{center}

{\bf Abstract}:  A rich pattern of gauge symmetries is found  in
the moduli space of heterotic string toroidal compactifications, at fixed points of the T-duality transformations. We analyze this pattern for generic tori, and scrutinize in full detail compactifications on a circle, where we find all  the maximal gauge symmetry groups and the points where they arise. We present figures of two-dimensional slices of the 17-dimensional moduli space of Wilson lines and circle radii, showing the rich pattern of points and curves of symmetry enhancement.  We then study the target space realization of  the duality symmetry.   Although the global continuous duality symmetries of dimensionally reduced heterotic supergravity  are completely broken by the structure constants of the maximally enhanced gauge groups,  the low energy effective  action   can be written in a manifestly  duality covariant  form using heterotic double field theory.  As a byproduct,   we show that a unique deformation of the generalized diffeomorphisms accounts for both $SO(32)$ and $E_8\times E_8$  heterotic effective field theories, which can thus be considered  two different backgrounds of the same  double field theory even before compactification. Finally we discuss the spontaneous gauge symmetry breaking 
and Higgs mechanism that occurs when  slightly perturbing the background fields, both from the string and the field theory perspectives.

\end{titlepage}

\newpage

\pagestyle{plain}

\tableofcontents

\newpage


\section{Introduction}

 The distinct backgrounds of heterotic string theory on a $k$ dimensional torus with constant  metric, antisymmetric tensor field and Wilson lines are characterized by the points of the $\frac{O(k, k+16; \mathbb R)}{O(k;\mathbb R)\times O(k+16;\mathbb R)\times  O(k,k+16;\mathbb Z)}$  coset manifold,
where $O(k,k+16;\mathbb Z)$ is the T-duality group  \cite{narain,witten}. 
At  self-dual points of this manifold, some massive modes become massless and the $U(1)^{2k+16}$ gauge symmetry  becomes non-abelian. 
In particular, for zero Wilson lines, the massless fields give rise to $SO(32) \times U(1)^{2k}$ or $E_8 \times E_8 \times U(1)^{2k}$ at generic values of the metric and B-field. By introducing  Wilson lines, not only is it possible to totally or partially break the non-abelian  gauge symmetry of the uncompactified theory, but it is also possible to enhance these groups.   The construction of \cite{witten} further allowed to continuously interpolate between the $SO(32)$ and $E_8 \times E_8$ heterotic theories after compactification \cite{ginsparg}, and even  suggested
 that these  superstrings  are two different vacuum states in the same theory {\it before} compactification.

Enhancement  of the gauge symmetry occurs at fixed points of the T-duality transformations \cite{giveon}. 
Massless fields become massive at the neighborhood of such points and the T-duality group mixes massless modes with massive ones \cite{GPR}.    
 Moreover, by identifying different string backgrounds
that provide identical theories, T-duality gives rise to  stringy features that are rather surprising from the viewpoint of  particle field theories. Nevertheless, some of these ingredients  have a correspondence in toroidal compactifications of heterotic supergravity.  In particular, although the field theoretical  reduction of heterotic supergravity cannot describe the non-abelian fields that give rise to maximally enhanced gauge symmetry\footnote{\label{foot:maximal}``Maximal'' stands here for an enhanced semi-simple and simply-laced symmetry group.},  being a gauged supergravity, the reduced theory  is completely determined by the gauge group, which can  be chosen to be one of maximal enhancement.   Likewise, the global  symmetries of heterotic supergravities are linked to T-duality. While the theory with the full set of $SO(32)$ or $E_8 \times E_8$ gauge fields has a global continuous
$O(k,k;\mathbb R)$ symmetry, when introducing  Wilson lines,  the symmetry enlarges to $O(k,k+16;{\mathbb R})$
 \cite{ms,km,hsz}, which is related to  the discrete T-duality symmetry of the parent string theory.

The global duality  symmetries  are not manifest in heterotic supergravity. 
To  manifestly display these symmetries,  as well as to account for the maximally enhanced gauge groups in a field theoretical setting, one appeals to the double field theory/generalized geometric reformulation of the string effective actions \cite{siegel,Arg} (for reviews and more references  see \cite{reviews}). Specifically, these frameworks  not only  describe  the  enhancement of gauge symmetry  \cite{agimnr}-\cite{aamp}, but  also give a geometric description of the non-geometric backgrounds that are obtained from T-duality \cite{non-geom}  and provide a gauge principle that requires and fixes the $\alpha'$-corrections  of the string effective actions \cite{mn}. Dependence of the fields on double internal coordinates and an extension of the tangent space are some of the elements that allow to go beyond the standard dimensional reductions of supergravity.

Motivated by  deepening our understanding of  heterotic string toroidal compactifications,  in section 2 we  review the  main features
 of heterotic string propagation on a $(10-k)$-dimensional Minkowski space-time times an internal $k$-torus with constant background metric, antisymmetric tensor field and Wilson lines, and recall their $O(k, k+16)$ covariant formulation. We focus on the phenomenon of symmetry enhancement arising at special points in moduli space.

In section 3, we concentrate on the simplest case, namely circle compactifications ($k=1$). 
We first find all the possible maximal enhancement groups, and the point in the fundamental region of moduli space where they arise, using the generalized Dynkin diagram of the lattice $\Gamma^{1,17}$ \cite{go,cv}.  
To explore the whole moduli space, we split the discussion into the situations in which the Wilson line $A$  preserves the $E_8\times E_8$ or $SO(32)$ gauge symmetry, and those where it breaks it. 
In the former case, the circle direction can give a further enhancement of symmetry to $E_8\times E_8\times SU(2)$ at radius $R=1$, and either to $SO(32)\times SU(2)$  at  $R=1$ or to $SO(34)$ at $R=\frac1{\sqrt2}$. 
When the Wilson line breaks the $E_8\times E_8$ or $SO(32)$ gauge symmetry, the pattern of gauge symmetries is very interesting. Not only is it possible to restore the original $E_8\times E_8$ or $SO(32)$ gauge symmetry  for specific values of $R$ and $A$, but also  larger groups of rank 17 can be obtained.  We explicitly work out enhancements of the $SO(32)$ theory to
$SO(34)$ at $R^2=\tfrac12$;
$SU(18)$ at $R^2=\tfrac14$;
$E_{p+1}\times SO(32-2p)$ at $R^2= 1 - \tfrac{p}{8}$;
$E_{p+1}\times SU(16-p)$ at $R^2= 1 - \tfrac{8}{16-p}$, and in the $E_8 \times E_8$ to 
$SO(34)$ at $R^2 = \tfrac{1}{18}$;
$SU(18)$ at $R^2=\tfrac19$;
$SO(18) \times E_8$ at $R^2=\frac12$;
$SU(2) \times E_8 \times E_8$ at $R^2 = \tfrac14$.
We depict  slices of the moduli space for different values of 
$R$ and Wilson lines in several figures, which clarify the analysis and neatly  exhibit the curves and points with special properties.

Examining the action of T-duality, we can see  that  all points in moduli space where there is maximal symmetry enhancement, namely enhancement to groups that do not have $U(1)$ factors, are fixed points of T-duality, or more general $O(1,17,{\mathbb Z})$ dualities that involve some exchange of momentum and winding number on the circle. In the simplest cases, such as those listed above, the enhanced symmetry arises at the self-dual radius given by $R^2_{sd}=1-\frac12|A|^2$. We explore the action of T-duality and its fixed points in section \ref{sec:T-duality}. One can have  other points of symmetry enhancement, which are fixed points of duality symmetries that involve shifts of Wilson lines on top of the exchange of momentum and winding. This is studied in detail in section \ref{sec:36}, where we obtain the most general duality symmetries that change the sign of the right-moving momenta and rotate the left-moving momenta, leaving the circle direction invariant. Concentrating on the case where the Wilson lines have only one non-zero component, we find a rich pattern of fixed points that correspond to $SU(2) \times SO(32)$ or $SU(2) \times E_8 \times E_8$ enhanced gauge symmetry, arising at $R_{\rm sd}^{-1}=C$, with $C$
 an integer number with prime divisors congruent to 1 or 3 (mod 8), and $SO(34)$ or $SO(18)\times E_8$ at $R_{\rm sd}^{-1}=\sqrt{2}\, C$ with $C$ a Pythagorean prime number or a product of them.

We then turn to the target space realization of the theory. In section 4, we construct the low energy effective actions of  (toroidally compactified) heterotic strings from the three and four point functions of string states. We first consider only the massless states and  compare the effective action obtained from the  string amplitudes with the dimensional reduction of heterotic supergravity performed in \cite{hsz}. As expected, we get a gauged supergravity which only differs from the effective action of  \cite{hsz}  in the cases of maximal enhancement, in which all the (left-moving) $U(1)^k$  Kaluza-Klein (KK) gauge fields of the compactification  become part of the Cartan subgroup of the enhanced gauge symmetry.  

The higher dimensional origin of the  low energy theory with maximally enhanced gauge symmetry cannot be found in supergravity, and one has to refer to DFT. 
Although the structure constants of the gauge group completely break the global duality symmetry of dimensionally reduced supergravity, the action can still be written in terms of $O(k,N)$ multiplets, with $N$ the dimension of the gauge group.
 We show in section 5 that the low energy effective action  of the toroidally compactified heterotic string at self-dual points  of the moduli space can be reproduced through a generalized  Scherk-Schwarz reduction of heterotic DFT. Furthermore, extending the construction of \cite{cgin}, we find  the generalized  vielbein that reproduces the structure constants of the enhanced gauge groups through a deformation of the generalized diffeomorphisms. An important output of the construction is that a unique   deformation is required for the $SO(32)$ and $E_8\times E_8$ groups, and hence the $SO(32)$ and $E_8\times E_8$ theories can be considered  two different solutions of the same heterotic DFT,  even {\it before} compactification. 

   When perturbing the background fields away from the enhancement points, some massless string states  become massive. The vertex operators of the massive vector bosons develop a cubic pole in their OPE with the energy-momentum tensor, and it is necessary to combine them with the vertex operators of the massive scalars in order to cancel the anomaly. This fact had been already noticed in \cite{agimnr}, but unlike the case of the  bosonic string, in the heterotic string all the massive scalars are ``eaten'' by the massive vectors. We compute the three point functions involving massless and  slightly massive states\footnote{For consistency, we consider only small perturbations because we are not including other massive states from the string spectrum.} and construct the corresponding effective massive gauge theory coupled to gravity.  Comparing the string theory results with the spontaneous  gauge symmetry breaking and Higgs mechanism in DFT, we see that the masses acquired by the sligthly massive string states fully agree with those of the DFT fields, provided there is a specific relation between the vacuum expectation value of the scalars along the Cartan directions of the gauge group and the deviation of the metric, B-field and Wilson lines from the point of enhancement. 

We have included  seven appendices. 
Appendix A collects some known facts  about lattices that are used in the main body of the paper. Details of the procedures leading to find the maximal enhancement points from Dynkin diagrams, to construct the curves of enhancement, more slices of the moduli space   and   the fixed points of the duality transformations  are contained in Appendices \ref{app:Dyndia}, \ref{app:enhancement},  \ref{app:extrafigures} and \ref{app:B6}, respectively. The  three and four point amplitudes of the massless and slightly massive string states are reviewed in Appendix \ref{app:3point}. Finally we count the number of non-vanishing structure constants of $SO(32)$ and $E_8\times E_8$ in Appendix \ref{app:X}.

\section{Toroidal compactification of the heterotic string}
\label{sec:torusstring}

In this section we recall the main features of heterotic string compactifications on $T^k$. We first discuss the generic $k$ case and then we concentrate on the $k=1$  example. For a more complete review see \cite{GPR}.

\subsection{Compactifications on $T^k$}

Consider the heterotic string propagating on a background manifold that is a product of a $d=10-k$ dimensional flat space-time times an internal torus $T^k$ with constant background metric 
$G=e^t e \ \left( \Rightarrow G_{mn}=e^a{}_m \delta_{ab} e^b{}_n\right.$),
 antisymmetric two-form field $B_{mn}$ and $U(1)^{16}$ gauge field $A_m^A$, where $m,n, a, b=1,...,k$ and $A=1, \dots , 16$. For simplicity we take the background dilaton to be zero.  The set of vectors $e_m$ define a basis in the compactification lattice $\Lambda^k$ such that the internal part of the  target space is the $k$-dimensional torus $T^k=\mathbb R^k/\pi\Lambda^k$. The vectors $\hat e_a$ constitute the canonical basis for the dual lattice $\Lambda^{k*}$, {\em i.e.} $\hat e_a{}^{m} e^a{}_n = \delta^{m}{}_n$, and thus they obey
$\hat e^t \hat e=G^{-1} \  \left( \Rightarrow \hat e_a{}^{m} \delta^{ab} \hat e_b{}^{n} = G^{mn} \right)$.

The contribution from the internal sector to the world-sheet action (we consider only the bosonic sector here) is 
\bea
\label{action}
S&=&\frac{1}{4\pi} \int_M d\tau d\sigma \left( \delta^{\alpha\beta}G_{mn} -i\epsilon^{\alpha\beta} B_{mn} \right) \partial_\alpha Y^{m}\partial_{\beta}Y^{n} 
\nn\\
&&  +\frac{1}{8\pi} \int_M d\tau d\sigma \left(\delta^{\alpha \beta}\partial_{\alpha}Y^{A}\partial_{\beta}Y^{A}-2 i\epsilon^{\alpha \beta} A_m^A \partial_{\alpha}Y^{m}\partial_{\beta}Y^{A}\right)\, ,
\eea
where we take $\alpha'=1$, $Y^A$ are chiral bosons  and the currents $\partial Y^A$  form a maximal commuting set of the  $SO(32)$ or $E_8\times E_8$ current algebra.
The world-sheet metric has been gauge fixed to $\delta^{\alpha\beta}$ ($\alpha,\beta=\tau,\sigma$) and $\epsilon^{01}=1$. The internal string coordinate fields satisfy
\beq \label{periodicityY}
Y^m(\tau,\sigma+2\pi)\simeq Y^m(\tau,\sigma) + 2 \pi w^m\ ,  
\eeq  
where  $w^m\in \mathbb Z$ are the winding numbers. It is convenient to define holomorphic $Y^m_L(z)$ and antiholomorphic $Y^m_R(\bar z)$ fields as 
\begin{equation}
Y^m(z, \bar z)=\left(\frac{1}2\right)^{1/2}\left[Y^m_L(z)+Y^m_R(\bar z) \right]\ , \quad z=\exp(\tau + i\sigma) \ , \ \bar z=\exp(\tau - i\sigma) \ , \label{y}
\end{equation}
with Laurent expansion
\bea
Y^m_L(z)&=& y_{L}^m-ip^m_L\,  lnz+\cdots, \qquad Y^A_L(z)= y_{L}^A-ip^A\,  lnz+\cdots,
\label{ya}\\
Y^m_R(\bar z)&=& y_{R}^m-ip^m_R \,  ln\bar z+\cdots,
\eea
 the dots standing for the oscillators contribution. Then the periodicity condition is
\beq
Y^m(\tau,\sigma+2\pi)-Y^m(\tau,\sigma)=2\pi\left(\frac{1}2\right)^{1/2}(p^m_L-p_R^m)=2\pi w^m \ .
\eeq

The canonical momentum has components\footnote{The unusual $i$ factors are due to the use of Euclidean world-sheet metric.}
\bea
\Pi_m & = & i \frac{\delta S}{\delta \partial_\tau Y^m}=\frac1{2\pi}\left[iG_{mn}\partial_{\tau} Y^{n}+B_{mn}\partial_{\sigma}Y^n+\frac{1}2A_m^A\partial_\sigma Y^A \right]\, ,\nn \\
&=&\frac1{2\pi}\left(\frac{1}{2}\right)^{1/2}\left[G_{mn}(p_L^n+p_R^n)+B_{mn}(p_L^n-p_R^n)\right]+\frac1{4\pi}A_m^Ap^A\nn\\
\Pi^A& = & i \frac{\delta S}{\delta \partial_\tau Y^A}= \frac{1}{4\pi}\left(i\partial_{\tau} Y^{A}-A_m^A\partial_\sigma Y^m\right)=\frac{1}{2\pi}\left[p^A-\left(\frac{1}{2}\right)^{1/2}A_m^A(p_L^m-p_R^m)\right]\, .\nn
\eea
The chirality constraint on $Y^A$ and the condition of vanishing Dirac brackets between momentum components require the redefinitions $\Pi_A\rightarrow \tilde\Pi_A=2\Pi_A$ and $\Pi_m\rightarrow \tilde\Pi_m=\Pi_m + \frac12 A_m^A \tilde\Pi_A$.
Integrating over $\sigma$, we get the center of mass momenta
\begin{subequations}
\bea
\pi_m&=&\int d\sigma\tilde \Pi_m= 2\pi\left(\Pi_m+\frac12A_m^A\tilde\Pi_A\right)={n_m} \in {\mathbb Z}\, ,\\
\pi^A&=&\int d\sigma\tilde \Pi^A=p^A-A_m^Aw^m\, ,
\eea 
\end{subequations}
where we used univaluedness of the wave function in the first line. Modular invariance requires $\pi^A\in \Gamma_{16}$ or $\Gamma_8\times\Gamma_8$,  corresponding to the $SO(32)$ or $E_8 \times E_8$ heterotic theory, respectively. 
 In Appendix \ref{app:lattices} we give all the relevant explanations and details about these lattices.

From these equations we get
\begin{subequations}
\bea
p_{Ra}&=&\left(\frac{1}{2}\right)^{1/2}\hat e_a{}^m\left[{n_m}-( G_{mn}+B_{mn}) w^n -\pi^AA_m^A
-\frac{1}2A_n^AA_m^A w^n\right] \, , \\
p_{La}&=&\left(\frac{1}{2}\right)^{1/2}\hat e_a{}^m\left[{n_m}+(G_{mn}-B_{mn}) w^n -\pi^A A_m^A
-\frac{1}2 A_n^AA_m^Aw^n\right], \label{momenta}  \\
p^A&=& \pi^A+w^m A^A_{m}\, .
\eea
\end{subequations}

The momentum $\bf{p}=(\bf{p_R,p_L})$, with ${\bf{p_R}}=p_{Ra}, \ {\bf{p_L}}=(p_{La}, p^{A})$, transforms as a vector under $O(k,k+16;\mathbb R)$. It expands the $2k+16$-dimensional momentum lattice $\Gamma^{(k,k+16)} \subset {\mathbb R}^{2k+16}$, satisfying
\beq
{\bf p}\cdot {\bf p}={\bf p_L}^2-{\bf p_R}^2=2w^m n_m+\pi^A\pi^A
 \in 2 \mathbb Z\, , \label{esd}
\eeq
because $\pi^A$ is on an even lattice, and  therefore ${\bf p}$ forms an even $(k, k+16)$ Lorentzian lattice. In addition, self-duality  $\Gamma^{(k,k+16)}=\Gamma^{(k,k+16)*}$ follows from modular invariance \cite{narain, polchi}. Note that $\bf{p_L,p_R}$ depend on $2k+16$ integer parameters  $n_m, w^m$ and $\pi^A$, and on the background fields $G$, $B$ and $A$.

The space of inequivalent lattices and inequivalent backgrounds reduces to
\begin{equation}
\frac{O(k,k+16;\mathbb R)}{O(k+16;\mathbb{R})\times O(k;\mathbb R) \times O(k,k+16;\mathbb Z) }\, ,
\end{equation}
where $O(k,k+16;\mathbb Z)$ is the T-duality group (we give more details about it in the next section).

The mass of the states and the level matching condition are respectively given by
\begin{subequations}
\begin{equation}
\label{Ham}
m^2={\bf p_L}^2+{\bf p_R}^2+2 \left({\cal N}+\overline{\cal N}-\left\{\begin{matrix}1&{\rm R \ sector}\\
\frac32&{\rm NS \ sector}\end{matrix}\right.\right)   ,
\end{equation}
\bea
\label{constraint}
0={\bf p_L}^2 -{\bf p_R}^2+2\left({\cal N}-\overline{\cal N} -\left\{\begin{matrix}1& {\rm  R \ sector}\\ \frac12&{\rm NS \ sector}\end{matrix} \right.\right)\, .
\eea
\end{subequations}

\subsection{$O(k,k+16)$ covariant formulation}
\label{sec:Okktoro}

The $O(k,k+16)$  invariant metric   $\eta$ is
\bea
\eta_{MN}=\begin{pmatrix} 0 & 1_{k\times k} &0\\ 1_{k\times k} & 0 &0\\0&0& \kappa_{IJ} \end{pmatrix} \ , \qquad 
\eea
where $\kappa$ is the Killing metric for the Cartan subgroup of $SO(32)$ or $E_8 \times E_8$, and the ``generalized metric'' of  the $k$-dimensional torus, given by the $(2k+16) \times (2k+16)$ scalar matrix, is
\beq
\label{G}
{\cal M}_{MN}=\begin{pmatrix} G_{mn}+C_{lm}G^{lk}C_{kn} +A_{m}{}^I A_{nI} &  -G^{nk}C_{km}& C_{km}G^{kl}A_{lJ}+A_{mJ} \\  -G^{mk}C_{kn}&G^{mn}&-G^{mk}A_{kJ}\\
C_{kn}G^{kl}A_{lI}+A_{nI}&-G^{nk}A_{kI}&\kappa_{IJ}+A_{kI} G^{kl}A_{lJ} \end{pmatrix} \  \in  O(k,k+16;\mathbb{R})\, ,
\eeq
 where
 \beq \label{C}
 C_{mn}=B_{mn}+\frac12A_{mI} \kappa^{IJ} A_{nJ} \ .
 \eeq
 This is a symmetric element of $O(k,k+16)$,
 accounting for the degrees of freedom of the $\frac{O(k,k+16)}{O(k)\times O(k+16)}$ coset. 

Combining the momentum and winding numbers  in an $O(k,k+16)$-vector 
\beq \label{Zeta}
Z^M= \begin{pmatrix} w^m \\ n_m \\ \pi^I\end{pmatrix}  \ , \qquad 
\pi^I\equiv \pi^A \hat{\tilde e}_A{}^I \ , \qquad  {\rm with} \quad   \hat{\tilde e}_A{}^I \hat{\tilde e}_A{}^J= \kappa^{IJ} \ ,
\eeq
the mass formula \eqref{Ham} and  level matching condition \eqref{constraint} read
\beq \label{MZhZ}
m^2=2 \left({\cal N}+\overline{\cal N}-\left\{\begin{matrix}1&{\rm R \ sector}\\
\frac32&{\rm NS \ sector}\end{matrix}\right.\right) +  Z^t {\cal M} Z \ ,
\eeq
\beq \label{LM}
0= 2 \left({\cal N}-\overline{\cal N}-\left\{\begin{matrix}1&{\rm R \ sector}\\
\frac12&{\rm NS \ sector}\end{matrix}\right.\right)  +  Z^t \eta Z \ ,
\eeq
respectively.
Note that these equations are invariant under the T-duality group $O(k,k+16;\mathbb Z) $
acting as
\beq
Z \to  \eta^{-1} O \eta Z \ , \quad {\cal M}\to O {\cal M} O^t \ , \quad \eta \to O \eta O^t=\eta \ , \quad O\in O(k,k+16,\mathbb Z) \ .
\eeq
The group $O(k,k+16;\mathbb Z)$ 
is generated by:
\begin{itemize}
\item[-] Integer $\Theta$-parameter shifts, associated with the addition of an antisymmetric integer matrix $\Theta_{mn}$ to the antisymmetric $B$-field, 
\beq \label{theta}
O_{\Theta}=\begin{pmatrix} 1  & \Theta&0\\ 0 & 1 &0\\
0&0&1_{16\times 16}\end{pmatrix} \ , \quad \Theta_{mn} \in {\mathbb Z}\, ,
\eeq
\item[-] Lattice basis changes 
\beq \label{GLk}
O_{M}=\begin{pmatrix} M & 0 &0\\ 0 & (M^t)^{-1}& 0\\
0&0&1_{16\times 16}\end{pmatrix} \ , \quad M \in GL(k;{\mathbb Z})\, ,
\eeq
\item[-] $\Lambda$-parameter shifts associated to the addition of 
vectors $\Lambda_m{}^A$ to  the Wilson lines\footnote{\label{foot:lambdashiftB}Note that this adds a shift to $B$ of the form $B\to B+\frac12 (A \Lambda^t - \Lambda A^t)$.}\\
\bea \label{OLambda}
O_\Lambda=\left(\begin{matrix}1&-\frac12\Lambda\Lambda^t&\Lambda\\
0&1&0\\
0&-\Lambda^t&1_{16\times 16}
\end{matrix}\right)\, , \quad \Lambda_m \in\Gamma_{16} \ \ {\rm or}\ \ \Gamma_8\otimes \Gamma_8\, ,
\eea
\item[-] Factorized dualities, which are generalizations of the $R\rightarrow 1/R$ circle duality, of the form
\beq \label{Di}
O_{D_i}=\begin{pmatrix} 1-D_i & D_i &0\\ D_i & 1-D_i &0\\0&0&1_{16\times 16}\end{pmatrix} \ , 
\eeq
where $D_i$ is a $k \times k$ matrix with all zeros except for a one at the $ii$ component. 
\end{itemize}
The first three generators comprise the so-called geometric dualities, transforming the background fields parameterizing the generalized metric (\ref{G}). The $O(k,k+16)$ group contains in addition 
\begin{itemize}
\item[-] Orthogonal rotations of the Wilson lines 
\beq \label{ON}
O_{N}=\begin{pmatrix} 1 & 0 & 0\\ 0 & 1 & 0 \\ 0 & 0 & N \end{pmatrix} \ , \quad N \in O(16;{\mathbb Z})\, ,
\eeq
\item[-] Transformations of the dual Wilson lines 
\beq \label{OGamma}
O_{\Gamma}=\begin{pmatrix} 1 & 0 & 0 \\ -\frac12\Gamma \Gamma^t & 1 & -\Gamma^t\\ \Gamma & 0 & 1\end{pmatrix} \ , \quad \Gamma^m \in \Gamma_{16} \ \ \text{ or } \Gamma_8 \times \Gamma_8\, ,
\eeq
\item[-] Shifts by a bivector
\beq \label{Obeta}
O_{\beta}=\begin{pmatrix} 1 & 0 & 0 \\ \beta & 1 & 0\\ 0 & 0 & 1\end{pmatrix} \ , \quad \beta^{mn} \in {\mathbb Z}\, ,\quad \beta^{mn}=-\beta^{nm}.
\eeq
\end{itemize}
The transformation of the charges under the action of $O_{\Theta} O_{\Lambda}$,  which will be useful later, is
\beq \label{transformZ}
 \begin{pmatrix} w \\ n \\ \pi \end{pmatrix} \to  \begin{pmatrix} w  \\ n + (\Theta-\frac12 \Lambda \Lambda^t) w +\Lambda \pi \\ \pi - \Lambda^t w \end{pmatrix}  \ . 
\eeq

Notice the particular role played by the element $\eta$ viewed as a sequence of factorized dualities in all tori directions, {\em i.e.} 
\beq
\eta^{-1}=O_D\equiv \prod_{i=1}^k O_{D_i} \begin{pmatrix} 1 & 0 & 0\\ 0 & 1 & 0 \\ 0 & 0 & \kappa^{-1} \end{pmatrix}  \ .
\eeq
Its action on the generalized metric is 
\beq \label{Tall} 
{\cal M} \to O_D {\cal M} O_D^t = \begin{pmatrix} G^{-1} & -G^{-1}C&-G^{-1}A  \\  -C^tG^{-1} & G+C^tG^{-1}C+AA^t&(1+C^tG^{-1})A  \\- A^tG^{-1} & A^t(1+G^{-1}C)& \kappa^{-1} +  A^tG^{-1}A \end{pmatrix} = {\cal M}^{-1} \ ,
\eeq
where $A\equiv A_m{}^I$ and, together with the transformation $Z \to \eta^{-1} O_D \eta Z $ which accounts for the exchange $w^m\leftrightarrow n_m$, it generalizes the $R\leftrightarrow 1/R$ duality of the circle compactification. These transformations determine the dual coordinate fields\footnote{The transformations also determine a {\it dual coordinate} $\tilde Y^A =Y^A+\frac1{\sqrt2}A_m^A(Y^m_L+Y^m_R)$, but this is not actually  independent of $Y^m(z,\bar z)$ and $Y^A(z)$.}
\bea
\tilde Y_m(z,\bar z)
&=& \frac1{\sqrt2}G_{mn}(Y^n_L-Y^n_R) + \frac1{\sqrt2}C_{mn}(Y^n_L+Y^n_R)+A_m^AY^A\, .  \label{tildeY}
\eea

A vielbein $E$ for the generalized metric 
\bea \label{E}
{\cal M}_{MN}=E^{\bf a} {}_M{}\delta_{\bf ab} E^{\bf b}{}_N, \ \ \ \ \ \ \ \ 
\eea
with $M,N={\bf a, b}=1,\dots,2k+16$, can be constructed from the vielbein for the internal metric  and inverse internal metric  as follows 
\beq \label{genframe}
E^{\bf a}{}_M \equiv E=\begin{pmatrix}-\hat e_a{}^n C_{nm} & \hat e_a{}^m & -\hat e_a{}^nA_n^I\kappa_{IJ}&\\  e^a{}_m & 0 &0\\ 
{\tilde e}^A{}_I A^I{}_{m}&0&{\tilde e}^A{}_J
\end{pmatrix}\, ,
\eeq
where ${\tilde e}$ is the vielbein for $\kappa$. In the basis of right and  left movers, that we denote ``RL", where the $O(k,k+16;\mathbb R)$ metric $\eta$ takes the diagonal form 
\beq \label{etaC+C-}
\eta_{RL}=(R \eta R^{T})=\begin{pmatrix} -\delta_{ab} & 0 &0\\ 0 & \delta_{ab} &0\\0&0&\delta_{AB}\end{pmatrix}\, ,  \quad R=\frac{1}{\sqrt2}\left(\begin{matrix}
\delta_a{}^b &-\delta_{ab}&0
\\\delta_a{}^b&\delta_{ab} &0
\\
0& 0&\sqrt2\delta^A{}_B\end{matrix}\right)\ ,
\eeq
the vielbein is 
\beq \label{genframeLR}
E_{RL}\equiv R E\equiv \begin{pmatrix} E_{aR} \\ E_{aL} \\ E_A\end{pmatrix}=\frac{1}{\sqrt2}\begin{pmatrix} 
-e_{am} -\hat e_a{}^n C_{nm}& \hat e_a{}^m & -\hat e_a{}^nA_n^I\kappa_{IJ}\\ 
e_{am} - \hat e_a{}^n C_{nm}&\hat e_a{}^m & -\hat e_a{}^n A_n^I\kappa_{IJ} \\
\sqrt2{\tilde e}^A{}_IA^I{}_m&0&\sqrt2{\tilde e}^A{}_J
\end{pmatrix} \ .
\eeq

Then the momenta $( p_{aR},p_{aL},p^A)$ in \eqref{momenta} are 
\beq \label{pE}
 \begin{pmatrix} p_{ aR} \\ p_{aL}\\p^{A}\end{pmatrix} = E_{RL} \, Z \ .
\eeq

\subsection{Massless spectrum}
\label{sec:masslessspectrum}

The massless  bosonic spectrum of the heterotic string in ten external dimensions is given, in terms of bosonic and fermionic creation operators $\alpha_{-1}^\mu,\bar\psi_{-1/2}^\mu$, respectively, by 
\begin{enumerate}
\item  ${\cal N}=1$, $\overline{\cal N}=\frac{1}{2}$, ${p_A}=0$ :
\begin{itemize}  
\item Gravitational sector: 
\bea
\alpha_{-1}^{\mu} \bar \psi^{\nu}_{-\frac{1}{2}} \ket{0,k}_{NS} \nn 
\eea
where the symmetric traceless, antisymmetric and trace pieces are respectively the graviton, antisymmetric tensor  and dilaton.
\item Cartan gauge sector: 
\bea
\alpha_{-1}^{I} \bar \psi^{\mu}_{-\frac{1}{2}} \ket{0,k}_{NS} \nn
\eea
containing 16 vectors $A_\mu^I$ in the Cartan subgroup of SO(32) or $E_8 \times E_8$. 
\end{itemize}

\item  ${\cal N}=0, \overline{\cal N}=\frac12$, ${p_A}^2=2$: 
\begin{itemize}
\item Roots gauge sector:  
\bea
\bar{\psi}^{\mu}_{-\frac{1}{2}} \ket{0,k,\pi_\alpha}_{NS}\nn
\eea
with ${\pi}_{\alpha}$ denoting one of the $480$ roots of $SO(32)$ or $E_8 \times E_8$.
 \end{itemize}
\end{enumerate}

In compactifications on $T^k$, the spectrum depends on the background fields.  In sector 1  there are the same number of massless states at any point in moduli space.  In sector 2, we see from \eqref{momenta} that  there are no massless states for generic values of  the metric, $B$-field and Wilson lines $A_m^I$,  while for certain values of these fields  the momenta can lie in the weight lattice of a rank $2k+16$ group $G_L \times G_R$. In this case, there is a subgroup with $|({\bf p_R},{\bf p_L})|^2=2$ which can give rise to massless states. Subtracting (\ref{Ham}) and (\ref{constraint}) we see that massless states have $\bf{p_R}=0$, and thus (unlike in the bosonic string theory), the non-abelian gauge symmetry comes from the left sector only. The group $G_L\times U(1)_R^k$ in which the massless states transform defines the gauge group of the theory, with  $G_L$ a simply-laced group of rank $16+k$ and dimension $N$, that depends on the point in moduli space (which is spanned by $G_{mn}, B_{mn}, A^I_n$). 
Specifically, the $10-k$ dimensional  massless bosonic spectrum and the corresponding vertex operators (in the $-1$ and $0$ pictures) are given by ($\mu, \nu=0,\dots , 9-k$; $m,n=1,\dots,k$; $I=1,\dots ,16$):

\begin{enumerate}
 \item ${\cal N}=1, \overline{\cal N}=\frac12$, ${\bf p_L}={\bf p_R}=0$: 
\begin{itemize}
\item Common  gravitational  sector: $g_{\mu\nu}, b_{\mu\nu}, D$
\bea
\alpha_{-1}^{\mu} \bar \psi^{\nu}_{-\frac{1}{2}} \ket{0,k}_{NS} \rightarrow 
\left\{ \begin{matrix}
\sqrt{2}\epsilon_{\munu}i\partial X^{\mu}(z) e^{-\phi} \bar\psi^{\nu}(\bar z)e^{i k \cdot X(z,\bar z)}\, \\
\sqrt{2}\epsilon_{\munu}i\partial X^{\mu}(z) \bar\Upsilon^\nu(\bar z) e^{i k \cdot X(z,\bar z)}\, 
\end{matrix}
\right. \label{gravity}
\eea
with $\phi$   the scalar from the bosonization of the superconformal ghost system,
\bea
\bar\Upsilon^\mu = \sqrt{2}i\bar\partial X^\mu + \frac1{\sqrt{2}} k\cdot\bar\psi\bar\psi^\mu  \, ,
\eea
and $k^\mu\epsilon_{\munu}=\epsilon_{\munu}k^\nu=0$.

\item $k$ KK  left abelian gauge vectors:  $ g_{m\mu}+ b_{m\mu}\equiv a_{m\mu}$  and 16 Cartan generators of $SO(32)$ or $ E_8\times E_8$: $a_\mu^I$ 
\bea
\alpha_{-1}^{\hat I} \bar \psi^{\mu}_{-\frac{1}{2}} \ket{0,k}_{NS}\rightarrow \left\{ \begin{matrix}
A_{\hat I\mu}i\partial Y^{\hat I}(z) e^{-\phi} \bar\psi^{\mu}(\bar z)e^{i k \cdot X(z,\bar z)}\\
A_{\hat I\mu}i\partial Y^{\hat I}(z) \bar\Upsilon^\mu(\bar z) e^{i k \cdot X(z,\bar z)}
\end{matrix}
\right. \, ,\label{vertexcartan}
\eea
where the index $\hat I=(I,m)$  includes  both the chiral ``heterotic'' directions  and the compact toroidal ones, labeling the Cartan sector of the gauge group $G_L$. 
\item $k$ KK  right abelian  gauge  vectors:
$g_{m\mu}- b_{m\mu}\equiv \bar a_{m\mu}$
\bea\label{vertexcartanR}
\alpha_{-1}^{\mu}  \bar \chi^m_{-\frac{1}{2}} \ket{0,k}_{NS}\rightarrow\left\{ \begin{matrix}
\sqrt{2}\bar A_{\mu m}i\partial X^{\mu}(z) e^{-\phi} \bar\chi^m(\bar z)e^{i k \cdot X(z,\bar z)}\\
\sqrt{2}\bar A_{\mu m}i\partial X^{\mu}(z)\bar\Upsilon^m(\bar z) e^{i k \cdot X(z,\bar z)}
\end{matrix}
\right. \, ,
\eea
with 
\bea
\bar\Upsilon^m=i\bar\partial Y^m + \frac{1}{\sqrt{2}} k\cdot\bar\psi\bar\chi^m\, .
\eea

\item $k(k+16)$ scalars: $g_{mn}, b_{mn}, a^I_m$ 
\bea
\alpha_{-1}^{\hat I }  \bar \chi^m_{-\frac{1}{2}} \ket{0,k}_{NS}\rightarrow \left\{ \begin{matrix}
S_{\hat I m}i\partial Y^{\hat I}(z) e^{-\phi} \bar\chi^m(\bar z)e^{i k \cdot X(z,\bar z)}\\
S_{\hat Im}i\partial Y^{\hat I}(z) \bar \Upsilon^m(\bar z) e^{i k \cdot X(z,\bar z)}
\end{matrix}
\right. \label{vertexcartanscalars}
\eea

\end{itemize}

\item ${\cal N}=0$, $\overline{\cal N}=\frac{1}{2}$, $\boldsymbol{p_L}^2=2$,  $\boldsymbol{p_R}=0$:
\begin{itemize}
\item $(N-k-16)$  root  vectors: $a_\mu^\alpha$
\bea
\bar{\psi}^{\mu}_{-\frac{1}{2}} \ket{0,k,\pi_\alpha}_{NS}\rightarrow \left\{ \begin{matrix} A_{\alpha\mu}J^\alpha(z) e^{-\phi} \bar\psi^{\mu}(\bar z)e^{i k \cdot X(z,\bar z)}\\
A_{\alpha\mu}J^\alpha(z) \bar\Upsilon^\mu(\bar z) e^{i k \cdot X(z,\bar z)} \end{matrix}
\right. \label{vertexroots}\, ,
\eea
with   $k^\mu A_\mu=0$ and  currents 
\bea
J^{\alpha}(z)=c_\alpha e^{i {\bf\alpha}\cdot  Y(z)}\, ,\label{cocycles}
\eea
where   $\bf\alpha$ are the roots of  $G_L$  (or equivalently the left momenta) and the cocycles $c_\alpha$ verify $c_\alpha c_\beta =\varepsilon(\alpha,\beta)c_{\alpha+\beta}$, with $\varepsilon(\alpha,\beta)=\pm 1$  the structure constants of $G_L$  in the Cartan-Weyl basis. 
 
\item   $(N-k-16) \times k$ scalars: $a_{\alpha n}$
\bea
\bar{\chi}^m_{-\frac{1}{2}} \ket{0,k,\pi_\alpha}_{NS}\rightarrow \left\{ \begin{matrix}
S_{\alpha m }J^\alpha(z) e^{-\phi} \bar\chi^m(\bar z)e^{i k \cdot X(z,\bar z)}\\
S_{\alpha m}J^\alpha(z)  \bar\Upsilon^m(\bar z) e^{i k \cdot X(z,\bar z)}
\end{matrix}
\right. \label{morescalars}
\eea

\end{itemize}
\end{enumerate}

It is convenient to define the index $\Omega = (\hat I, \alpha)=1,...,N$ and condense the vertex operators for left vectors and scalars as
\beq
A_{(-1)}&=A_{\Omega\mu}J^{\Omega}(z) e^{-\phi} \bar\psi^{\mu}(\bar z)e^{i k \cdot X(z,\bar z)}
\eeq
\beq
S_{(-1)}&=S_{\Omega m}J^{\Omega}(z) e^{-\phi} \bar\chi^m(\bar z)e^{i k \cdot X(z,\bar z)}
\eeq
where $J^{\hat I}=i\partial Y^{\hat I}$.

The massive states are obtained increasing the oscillation numbers ${\cal N}$ and $\overline{\cal N}$ or choosing $|(p_R,p_L)|^2 \ge 4$.

Due to the uniqueness of Lorentzian self-dual lattices \cite{go} both heterotic theories on $T^k$ can be connected continuously \cite{narain,witten}, i.e., they belong to the same moduli space. The possible enhanced non-abelian gauge symmetry groups are those with root lattices  admitting an embedding into $\Gamma^{k,16+k}$.  Although some theorems on lattice embeddings are known \cite{niku},  it is a non-trivial problem to determine which groups admit an embedding\footnote{A preliminary attempt can be found in \cite{mo}. }. Here we present a general discussion.

Using that ${\bf p_R}=0$, we get from \eqref{momenta} that the massless states have left-moving momentum
\beq
{\bf p_L}=\left(\sqrt{2} \, \hat e_{am} w^m,\pi^A+w^m A^A_m\right)\, ,
\eeq
while their momentum number on the torus is given by
\bea
{n_m}=\left( G_{mn}+B_{mn}\right) w^n +\pi^AA_m^A+\frac12 A_n^AA_m^Aw^n\,  . \label{cond}
\eea
Note that quantization of momentum number on the torus is a further condition to be imposed on top of ${\bf p_L}^2=2$.  

In the absence of Wilson lines $A_m^A=0$, the $k$ torus directions decouple from the 16 chiral ``heterotic directions" $Y^A$; $p^A=\pi^A$ is a vector of the weight lattice of $SO(32)$ or $E_8\times E_8$ and then $|p^A|^2\in 2{\mathbb N}$. The only possible massless states then have either momenta ${\bf p_L}=(0,\pi^A)$ with $|\pi|^2= 2$, or ${\bf p_L}=(\sqrt{2} \, e_{an} w^n,0)$ with $w^m g_{mn} w^n=1$ (and additionally $n_m w^m=1$). The former are the root vectors of $SO(32)$ or $E_8\times E_8$, while the latter have solutions only for certain values of the metric and $B$-field on the torus and lead to the same groups as in the (left sector of) bosonic string theory, namely all simply-laced groups $H$ of rank $k$. The total gauge group is then $SO(32) \times H \times U(1)_R^k$ or $E_8 \times E_8 \times H \times U(1)_R^k$. For $k=1$, i.e. a circle compactification, $H$ is $SU(2)$ at $g_{11}=R^2=1$, and $U(1)$ for any other value of the radius.     
For compactifications on $T^2$, the possible groups of maximal  enhancement (see footnote \ref{foot:maximal})  are 
$SO(32)\times SU(2)_L^2\times U(1)^2_R$ (for a diagonal metric with both circles at the self-dual radius and no $B$-field) or $SO(32)\times SU(3)_L\times U(1)^2_R$ (equivalently $SO(32)\rightarrow E_8\times E_8$). See \cite{cgin} for details. 


Turning on Wilson lines, the pattern of gauge symmetries is more complicated, and also richer. In the sector with zero winding numbers, $w^m=0$, we have  $p^A=\pi^A$ as before, but now requiring a quantized momentum number imposes $\pi^A A^A_m \in \mathbb Z$ (see \eqref{cond}) which, for a generic Wilson line breaks all the gauge symmetry leaving only $\pi^A=0$, which corresponds to the $U(1)^{16}$ Cartan subgroup. The opposite situation corresponds to $A^A_m \in \Gamma^*_g$ \footnote{We denote $\Gamma^*_{g}$ the dual of the root lattice, and one has $
\Gamma^*_g=\Gamma_8\times\Gamma_8$ for $E_8\times E_8$ and $ \Gamma^*_g=\Gamma_w=\Gamma_{16}+\Gamma_v+\Gamma_c$ for $SO(32)$ (see Appendix \ref{app:lattices} for more details).}. For $E_8 \times E_8$, since $\Gamma_g^*=\Gamma_8 \times \Gamma_8$,  $A_m^A$ can be eliminated through a $\Lambda$-shift of the form $O_{\Lambda}$ in \eqref{OLambda} and thus the pattern of gauge symmetries is as for no Wilson line\footnote{The only difference is that the massless states have  shifted momenta $\pi^A$ and a shifted momentum number along the circle compared to the ones without Wilson lines, see Eq.\eqref{transformZ}.}. In the $SO(32)$ theory, the same conclusions hold if $A\in \Gamma_{16}$, but one has the more interesting possibility $A \in \Gamma_v$ or $A \in \Gamma_c$, where the $SO(32)$ symmetry is not broken, and the 16 chiral heterotic directions can be combined with the torus ones, giving larger groups which are not products. 

Let us discuss the different groups that can arise in points of moduli space where the enhancement is maximal. In that case, the matrices  that embed the internal sector of the heterotic theory on $T^k$ into a $16+k$-dimensional bosonic theory are related to the Cartan matrix ${\cal C}$ by  \cite{GPR} 
\begin{align}
\begin{split}\label{Eenhanced}
\begin{pmatrix} (G + \frac12 A^I A^I)_{mn} & \frac 12 A_m{}^I \\ \frac12 A^I{}_n & {G}_{IJ} \end{pmatrix} &= \frac12 {\cal C}_{\hat I \hat J} \ ,  \\
\begin{pmatrix} B_{mn} &  \frac 12 A_m{}^I \\  -\frac 12 A^I{}_n  & B_{IJ} \end{pmatrix} 
&=\left\{ \begin{matrix} \ \ \frac12 {\cal C}_{\hat I \hat J} \ & {\rm for} \ \hat I< \hat J \\ -\frac12 {\cal C}_{\hat I \hat J} \ & {\rm for} \ \hat I> \hat J\\ 0 \ & {\rm for} \ \hat I= \hat J \end{matrix} \right. 
\end{split}
\end{align}   
One can then view the possible maximal enhancements from Dynkin diagrams. Let us first consider Wilson lines that do not break the original gauge group, i.e $A\in \Gamma_g^*$. We start with the SO(32) heterotic theory.
The Dynkin diagram of $SO(32)$ is
\bea
\begin{dynkin} \dynkinlineold{1}{0}{2}{0}; \dynkinlineold{2}{0}{3}{0}; \dynkinlineold{2}{0}{2}{1}; \dynkinlineold{3}{0}{4}{0}; \dynkinlineold{4}{0}{5}{0};\dynkinlineold{5}{0}{6}{0}; 
\dynkinlineold{6}{0}{7}{0}; \dynkinlineold{7}{0}{8}{0};\dynkinlineold{8}{0}{9}{0};\dynkinlineold{9}{0}{10}{0};\dynkinlineold{10}{0}{11}{0};\dynkinlineold{11}{0}{12}{0};\dynkinlineold{12}{0}{13}{0};\dynkinlineold{13}{0}{14}{0};\dynkinlineold{14}{0}{15}{0};\dynkindotold{2}{1} \foreach \x in {1,...,15} {\dynkindotold{\x}{0}} \nn \end{dynkin}
\eea
The Dynkin diagrams of the gauge symmetry groups arising at points of maximal enhancement in the compactification of the SO(32) theory on $T^k$ have $k$ extra nodes, with or without lines in between. Since the resulting groups have to be in the ADE class (they are all simply laced), one cannot add nodes with lines on the left side.
Therefore, the nodes should be added on the right side, and linked or not linked to the last node or not, and additionally add lines linking them to ech other, or not. For one dimensional compactifications ($k=1$), the only possibilities are
\bea
\begin{dynkin} \dynkinlineold{1}{0}{2}{0}; \dynkinlineold{2}{0}{3}{0}; \dynkinlineold{2}{0}{2}{1}; \dynkinlineold{3}{0}{4}{0}; \dynkinlineold{4}{0}{5}{0};\dynkinlineold{5}{0}{6}{0}; 
\dynkinlineold{6}{0}{7}{0}; \dynkinlineold{7}{0}{8}{0};\dynkinlineold{8}{0}{9}{0};\dynkinlineold{9}{0}{10}{0};\dynkinlineold{10}{0}{11}{0};\dynkinlineold{11}{0}{12}{0};\dynkinlineold{12}{0}{13}{0};\dynkinlineold{13}{0}{14}{0};\dynkinlineold{14}{0}{15}{0};\dynkindotold{2}{1} \foreach \x in {1,...,16} {\dynkindotold{\x}{0}} \end{dynkin}\qquad
\begin{dynkin} \dynkinlineold{1}{0}{2}{0}; \dynkinlineold{2}{0}{3}{0}; \dynkinlineold{2}{0}{2}{1}; \dynkinlineold{3}{0}{4}{0}; \dynkinlineold{4}{0}{5}{0};\dynkinlineold{5}{0}{6}{0}; 
\dynkinlineold{6}{0}{7}{0}; \dynkinlineold{7}{0}{8}{0};\dynkinlineold{8}{0}{9}{0};\dynkinlineold{9}{0}{10}{0};\dynkinlineold{10}{0}{11}{0};\dynkinlineold{11}{0}{12}{0};\dynkinlineold{12}{0}{13}{0};\dynkinlineold{13}{0}{14}{0};\dynkinlineold{14}{0}{15}{0};\dynkinlineold{15}{0}{16}{0};\dynkindotold{2}{1} \foreach \x in {1,...,16} {\dynkindotold{\x}{0}} \nn \end{dynkin}\nn
\eea
corresponding respectively to $SO(32)\times SU(2)$ and $SO(34)$. Since a line in the Dynkin diagram means that the new simple root is not orthogonal to the former one, then the Cartan matrix for this situation should have an off-diagonal term in the row corresponding to the new node and the column of the previous node, which according to \eqref{Eenhanced} means that there is a non-zero Wilson line. Thus, no Wilson line (or a line in $\Gamma_{16}$, which is equivalent to no Wilson line) gives the enhancement group $SO(32)\times SU(2)$ and, as explained above, this enhancement works as in the bosonic theory, at $R=1$. The enhancement symmetry group $SO(34)$ is obtained with a Wilson line in the vector or negative-chirality spinor conjugacy classes, and will be presented in detail in section \ref{sec:SO34}. For compactifications on $T^k$, the $k$ extra nodes give as largest enhancement symmetry group $SO(32+2k)$, and this happens when Wilson lines in all directions are turned on. For less symmetric Wilson lines one gets smaller groups, and it is easy to see from the Dynkin diagrams what are all the possible groups. Here we draw all the possibilities for $k=2$ only 
\bea
\begin{dynkin} \dynkinlineold{1}{0}{2}{0}; \dynkinlineold{2}{0}{3}{0}; \dynkinlineold{2}{0}{2}{1}; \dynkinlineold{3}{0}{4}{0}; \dynkinlineold{4}{0}{5}{0};\dynkinlineold{5}{0}{6}{0}; 
\dynkinlineold{6}{0}{7}{0}; \dynkinlineold{7}{0}{8}{0};\dynkinlineold{8}{0}{9}{0};\dynkinlineold{9}{0}{10}{0};\dynkinlineold{10}{0}{11}{0};\dynkinlineold{11}{0}{12}{0};\dynkinlineold{12}{0}{13}{0};\dynkinlineold{13}{0}{14}{0};\dynkinlineold{14}{0}{15}{0};\dynkinlineold{15}{0}{16}{0};\dynkinlineold{16}{0}{17}{0};\dynkindotold{2}{1} \foreach \x in {1,...,17} {\dynkindotold{\x}{0}} \nn \end{dynkin}
\qquad
\begin{dynkin} \dynkinlineold{1}{0}{2}{0}; \dynkinlineold{2}{0}{3}{0}; \dynkinlineold{2}{0}{2}{1}; \dynkinlineold{3}{0}{4}{0}; \dynkinlineold{4}{0}{5}{0};\dynkinlineold{5}{0}{6}{0}; 
\dynkinlineold{6}{0}{7}{0}; \dynkinlineold{7}{0}{8}{0};\dynkinlineold{8}{0}{9}{0};\dynkinlineold{9}{0}{10}{0};\dynkinlineold{10}{0}{11}{0};\dynkinlineold{11}{0}{12}{0};\dynkinlineold{12}{0}{13}{0};\dynkinlineold{13}{0}{14}{0};\dynkinlineold{14}{0}{15}{0};\dynkinlineold{15}{0}{16}{0};\dynkindotold{2}{1} \foreach \x in {1,...,17} {\dynkindotold{\x}{0}} \end{dynkin}\\
\begin{dynkin} \dynkinlineold{1}{0}{2}{0}; \dynkinlineold{2}{0}{3}{0}; \dynkinlineold{2}{0}{2}{1}; \dynkinlineold{3}{0}{4}{0}; \dynkinlineold{4}{0}{5}{0};\dynkinlineold{5}{0}{6}{0}; 
\dynkinlineold{6}{0}{7}{0}; \dynkinlineold{7}{0}{8}{0};\dynkinlineold{8}{0}{9}{0};\dynkinlineold{9}{0}{10}{0};\dynkinlineold{10}{0}{11}{0};\dynkinlineold{11}{0}{12}{0};\dynkinlineold{12}{0}{13}{0};\dynkinlineold{13}{0}{14}{0};\dynkinlineold{14}{0}{15}{0};\dynkindotold{2}{1} \foreach \x in {1,...,17} {\dynkindotold{\x}{0}} \end{dynkin}
\qquad
\begin{dynkin} \dynkinlineold{1}{0}{2}{0}; \dynkinlineold{2}{0}{3}{0}; \dynkinlineold{2}{0}{2}{1}; \dynkinlineold{3}{0}{4}{0}; \dynkinlineold{4}{0}{5}{0};\dynkinlineold{5}{0}{6}{0}; 
\dynkinlineold{6}{0}{7}{0}; \dynkinlineold{7}{0}{8}{0};\dynkinlineold{8}{0}{9}{0};\dynkinlineold{9}{0}{10}{0};\dynkinlineold{10}{0}{11}{0};\dynkinlineold{11}{0}{12}{0};\dynkinlineold{12}{0}{13}{0};\dynkinlineold{13}{0}{14}{0};\dynkinlineold{14}{0}{15}{0};\dynkinlineold{16}{0}{17}{0};\dynkindotold{2}{1} \foreach \x in {1,...,17} {\dynkindotold{\x}{0}} \end{dynkin}\nn
\eea
corresponding respectively to $SO(36)$, $SO(34) \times SU(2)$, $SO(32) \times SU(2)^2$ and $SO(32) \times SU(3)$. 

For the $E_8\times E_8$ heterotic theory, the situation is less rich in the cases in which the dimension of the resulting  group is  larger than that of $E_8\times E_8$. As we explained above, since $\Gamma_{g}^*=\Gamma_8 \times \Gamma_8$,  a Wilson line that preserves the $E_8 \times E_8$ symmetry should be in the lattice, and thus equivalent to no Wilson line. This can also be seen from the Dynkin diagram of $E_8 \times E_8$ 
 \bea
\begin{dynkin} \dynkinlineold{1}{0}{2}{0}; \dynkinlineold{2}{0}{3}{0}; \dynkinlineold{3}{0}{4}{0}; \dynkinlineold{3}{0}{3}{1}; \dynkinlineold{4}{0}{5}{0};\dynkinlineold{5}{0}{6}{0}; 
\dynkinlineold{6}{0}{7}{0};\dynkindotold{3}{1} \foreach \x in {1,...,7} {\dynkindotold{\x}{0}} \end{dynkin} \quad \begin{dynkin} \dynkinlineold{1}{0}{2}{0}; \dynkinlineold{2}{0}{3}{0}; \dynkinlineold{3}{0}{4}{0}; \dynkinlineold{3}{0}{3}{1}; \dynkinlineold{4}{0}{5}{0};\dynkinlineold{5}{0}{6}{0}; 
\dynkinlineold{6}{0}{7}{0};\dynkindotold{3}{1} \foreach \x in {1,...,7} {\dynkindotold{\x}{0}} \nn \end{dynkin}
\eea
where we see immediately that the extra nodes cannot be linked to any of the $E_8$'s, as any extra line would get us away from $ADE$. Then the possible enhancements are groups which are products of the form $E_8\times E_8 \times H$, where $H$ is any semi-simple group of rank $k$, and each $H$ arises at the same point in moduli space as in the compactifications of the bosonic theory on $T^k$ \cite{cgin}.  However, maximal enhancement can still be obtained by breaking one of the $E_8$ to $SO(16)$, and then the richness of the $SO(32)$ case is recovered (e.g. enhancement to $SO(18)\times E_8$). 

If $A \notin \Gamma_g^*$, part or all of the $SO(32)$ or $E_8 \times E_8$ symmetry is broken, and one can  still  see groups that arise from the Dynkin diagrams. For compactifications on $T^k$, a priori any group of rank $16+k$ in the ADE class can arise. However, we need to take into account that there are only $k$ linearly independent Wilson lines that  can be turned on, so not any ADE group is actually achievable. 

 Points of enhancement are fixed points of some $O(k,k+16;{\mathbb Z})$ symmetry. Enhancement groups that are not semi-simple, i.e. that contain $U(1)$ factors, arise at lines, planes or hyper-planes in moduli space. On the contrary, maximal enhancement occurs at isolated points in moduli space. These are fixed points (up to discrete transformations) of the $O_D$ duality symmetry, or more general duality symmetries involving $O_D$. This is developed in detail   in sections \ref{sec:T-duality} and \ref{sec:36} for compactifications on a circle, to which we now turn.
 
\section{Compactifications on a circle}

\label{sec:circle}

All the possible enhancement groups  in $S^1$ compactifications can be obtained from the generalized Dynkin diagrams \cite{go,ginsparg,cv} that we review in Appendix \ref{app:Dyndia}.
Here we list all the possible maximal enhancements for the $\Gamma_{16}$ and $\Gamma_8 \times \Gamma_8$ theories, together with the point in the fundamental region that gives that enhancement ($p, q\in{\mathbb Z}$, $1 \leq p,q \leq 8$) 
\begin{center}
 \begin{tabular}{|c|c|c|}
 \hline
Wilson line & $R^{-2}$ & Gauge group \\
\hline
$\left(0_{8-p},\left(\tfrac{q}{2(p+q)}\right)_{p+q}, \left(\tfrac12\right)_{8-q}\right) $&$8\left(\tfrac{1}{p} + \tfrac{1}{q} \right)$ & $E_{9-p} \times E_{9-q} \times SU(p+q)$\\
 $\left(-\tfrac{q}{2(6+q)},\left(\tfrac{q}{2(6+q)} \right)_{7+q}, \left(\tfrac12\right)_{8-q}\right)$ 
 & 
 $2 - \tfrac{2}{q+9} + \tfrac{8}{q}$ 
 &$SU(9+q) \times E_{9-q}$ \\
$\left(-\tfrac14,\left(\tfrac14\right)_{14},-\tfrac14\right)+\left( 0_{15}, 1 \right)$
&$ 4   $ & $SU(18)$\\
$\left(0_{8+q}, (\tfrac12)_{8-q}\right)$ & $ \tfrac{8}{q}$ & $SO(16 + 2q) \times E_{9-q}$ \\
$\left(0_{15},1\right) $& $2$ &  $SO(34)$ 
\\
\hline
\end{tabular}\\
{Table 1: Maximal  enhancements  for  the $SO(32)$ theory.}
\end{center} 

\begin{center} 
\begin{tabular}{|c|c|c|}
 \hline
Wilson line & $R^{2}$ & Gauge group \\
\hline
$  
\left(0_{8-p},\left(\tfrac{1}{p}\right)_{p}, \left(-\tfrac{1}{q}\right)_{q}, 0_{8-q}\right)+\left(0_7, -1, 1, 0_7 \right)$ &$\tfrac12 \left (\tfrac{1}{p} + \tfrac{1}{q} \right)$ & $E_{9-p} \times E_{9-q} \times SU(p+q)$\\
 $
 \left(-\tfrac16,\left(\tfrac16\right)_{7},\left(-\tfrac{1}{q}\right)_{q},0_{8-q}\right)+\left(0_7, -1, 1, 0_7 \right)$ 
 & $\tfrac12\left(\tfrac{1}{9} + \tfrac{1}{q}\right)$ &$SU(9+q) \times E_{9-q}$ \\
 $ \left(-\tfrac16,\left(\tfrac16\right)_{7}, \left(-\tfrac16\right)_{7},\tfrac16 \right)+\left(0_7, -1, 1, 0_7 \right)
$ & $ \tfrac{1}{9}     $ & $SU(18)$\\
$\left(0_{8}, \left(-\tfrac{1}{q}\right)_{q},0_{8-q}\right)+\left(0_7, -1, 1, 0_7 \right)$ & $ \frac{1}{2q}$ & $SO(16 + 2q) \times E_{9-q}$ \\
$\left(0_{8},\left(-\tfrac16 \right)_{7},\tfrac16\right)+\left(0_7, -1, 1, 0_7 \right)$ & $\tfrac{1}{18}$ &  $SO(34)$ \\
\hline
\end{tabular}\\
Table 2: Maximal  enhancements  for  the $E_8 \times E_8$ theory. \\
\end{center} 
When $p$ and/or $q$ equal $7$ one gets $E_2 = SU(2) \times U(1)$ and the enhancement is not maximal.

In this section we show directly how these groups arise by inspecting the momenta at different points in moduli space. We explicitly work out some examples and show the distribution of the enhancement groups in certain two-dimensional slices of the moduli space, where one can see the rich patterns of gauge symmetries.

The momentum components (\ref{momenta}) are\footnote{From now on, suppressed indices in $p$ are orthonormal indices, i.e. $p_R\equiv p_{Ra}, p_L\equiv p_{La}$.} 
\bea 
p_{R}&=&\frac{1}{\sqrt{2} R} \, \left[n- R^2  w -\pi \cdot A 
-\frac{1}2 |A|^2 w \right]  \, , \nn \\
p_{L}&=&\frac{1}{\sqrt{2} R} \, \left[n+R^2  w -\pi \cdot A 
-\frac{1}2 |A|^2 w \right], \label{momentum} \nn \\
p^{A}&=& \pi^A+w A^{A} \, ,
\eea
where $|A|^2=A^A A^A=A \kappa A^t$.\footnote{\label{foot:a}We are abusing notation, as $|A|^2=A\kappa A$  is not a scalar under reparameterizations of the circle coordinate, i.e. our definition is $|A|^2= A_m^A A_m^A$ where $m$ here is just the circle coordinate. The scalar quantity is ${\mathbb A}^2=|A|^2/R^2$ (see \eqref{a} below). } The massless states, which satisfy ${\bf p_R}=0$,  have left-moving momenta
\beq \label{pLS1}
{\bf p_L}=(\sqrt{2} R w,\pi^A + w A^A) = (\sqrt{2} R w,p^A) \, ,
\eeq
and momentum number on the circle 
\beq \label{n}
{n}= \left({R^2}+\frac{1}2 |A|^2 \right)w+\pi \cdot A \, .
\eeq
The condition $|{\bf p_L}|^2=2$ can be written in the following form, that we shall use 
\beq \label{massformula}
|\pi + w A|^2 = 2(1-w^2 R^2)\, .
\eeq 

In the sector ${\bf p_L}=0$ one has $ n=w=\pi^A=0$, and the massless spectrum corresponds to 
 the common gravitational sector and 18 abelian gauge bosons: 16 from the Cartan sector of $E_8\times E_8$ or $SO(32)$ and 2 KK vectors,
forming the $U(1)^{18}$ gauge group.

The condition $ {\bf p_L}^2=2$ can be achieved in two possible ways: 

1) ${\bf p_L}=(0,p^A)$, with $|p^A|^2=2$,
 
2) ${\bf p_L}=(\pm s,p^A)$, with $0<s\leq \sqrt{2}$\ , \ $s^2+|p^A|^2=2$.

From \eqref{pLS1} we see that sector 1 has $w=0$ and then \eqref{momentum} implies $p^A=\pi^A$. The condition on the norm says that these are the roots of $SO(32)$ or $E_8 \times E_8$.
But as explained in the previous section, one has to impose further that $n \in \mathbb Z$ and thus from \eqref{n}, $\pi \cdot A \in {\mathbb Z}$. We divide the discussion into two cases, one in  which this condition does not break the $SO(32)$ or $E_8 \times E_8$ symmetry, and the second one in which it does. This distinction is useful to understand the enhancement process but, as we will see, is somewhat artificial: all enhancement groups, including those with $SO(32)$ or $E_8 \times E_8$ as subgroups, can be achieved with Wilson lines that are not in the dual lattice by appropiately choosing the radius.

\subsection{Enhancement of $SO(32)$ or $E_8 \times E_8$ symmetry}
\label{sec:enhancement}

If we want the condition $\pi \cdot A \in {\mathbb Z}$ not to select a subset of the possible $\pi^A$ in the root lattice, or in other words not to break the $SO(32)$ or $E_8 \times E_8$ gauge symmetry, we have to impose 
\beq \label{Aenh}
A \in \Gamma^*_{g} \, ,
\eeq
with 
\bea
\Gamma^*_g=\Gamma_8\times\Gamma_8 \ \ {\rm for} \ E_8\times E_8 \ \ \ \ \ \ {\rm or} \ \ \ \ \ \ \ \  \Gamma^*_g=\Gamma_w=\Gamma_{16}+\Gamma_v+\Gamma_c \ \ {\rm for} \ SO(32) \, .\nn
\eea
We restrict to this case now, and leave the discussion of the possible symmetry breakings to the next section. 

Sector 2 contributes states only at radii $R^2=s^2/(2w^2)$.
The momentum number of these states given in \eqref{n} becomes
\beq \label{condsector2}
n=\frac12 \left(\frac{s^2}{w}+ |A|^2 w \right) +\pi \cdot A = \frac{1}{w} \left(1-\frac{|\pi|^2}{2} \right) \in {\mathbb Z}\, ,
\eeq
where in the last equality we have used \eqref{pLS1} and $|{\bf p_L}|^2=2$.

If $A \in \Gamma$,\footnote{By $\Gamma$ we mean $\Gamma_{16}$ or $\Gamma_8 \times \Gamma_8$, according to which heterotic theory one is looking at.} the  condition $|p^A|^2 < 2$ can only be satisfied for $p^A = \pi^A + wA^A = 0$. Then we have $s^2=2$ and the quantization condition is: $\frac{1}{w} + \tfrac12|A|^2 w \in \mathbb{Z}$. One has $\frac12 |A|^2 \in {\mathbb Z}$, and thus the only way to satisfy it is with $w=\pm1$ and $\pi = \mp A$ which gives two extra states  at  $R=1$, with momentum number $n=\pm (1-\tfrac12 |A|^2)$. 

The condition $0\neq|p^A|^2<2$ is only possible if $A$ is not in the root lattice. And as it is required to be in the weight lattice, this possibility arises in the $SO(32)$ heterotic theory only, for $A\in \Gamma_v$ or $A\in \Gamma_c$.
For $A \in \Gamma_v$, $\pi \cdot A \in {\mathbb Z}$ for $\pi \in \Gamma_{g}$ and $\frac12 |A|^2=\frac12$ (mod 1),
so the only option is $s=1$, giving extra states with $w=\pm1$ at $R=1/\sqrt{2}$. These states enhance $SO(32) \times U(1)$ to $SO(34)$.  We present an explicit example of this  case in section \ref{sec:SO34}. 
For $A \in \Gamma_c$, $\pi \cdot A \in {\mathbb Z}$ for $\pi \in \Gamma_{g}$ but now $\frac12 |A|^2 \in {\mathbb Z}$ and thus we cannot satisfy the quantization condition \eqref{condsector2} this way. However $\pi \cdot A=\frac12$ (mod 1) for $\pi \in \Gamma_{s}$ and thus we recover that for these Wilson lines there is  an enhancement to $SO(34)$ at $R=1/\sqrt{2}$ as well, by states with $w=\pm1$. Note that $A\in \Gamma_c$ is equivalent by a $\Lambda$ shift with $\Lambda \in \Gamma_s$ to $A\in \Gamma_v$. As we can see from \eqref{transformZ}, by this shift the winding number remains invariant, while $\pi \in \Gamma_s$ gets shifted to $\pi' \in \Gamma_g$.  

We conclude that in circle compactifications with Wilson lines that do not break the original $SO(32)$ or $E_8 \times E_8$ groups the pattern of gauge symmetry enhancement is (we give here only the groups on the left-moving side):
\begin{itemize}

\item $E_8 \times E_8 \times U(1)\rightarrow E_8 \times E_8 \times SU(2)$ at $R=1$ if $A \in \Gamma_8 \times \Gamma_8$ 

\item $SO(32) \times U(1)\rightarrow SO(32) \times SU(2)$ at $R=1$ if $A \in \Gamma_{16}$,  or 

\item $SO(32) \times U(1)\rightarrow SO(34)$ at $R=\frac1{\sqrt{2}}$ if $A \in \Gamma_{v}$ or $A \in \Gamma_{c}$

\end{itemize}

\begin{figure}[H] 
\begin{subfigure}{.5\textwidth}
\centering
\imagen{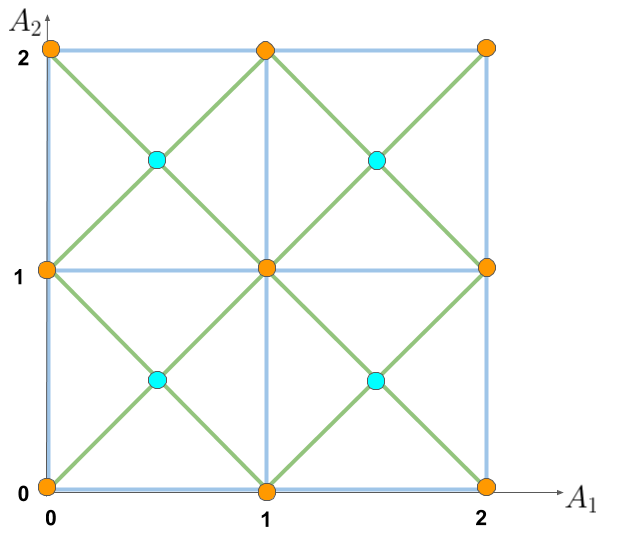}
\caption{$SO(32)$ heterotic }
\label{fig:SO32anyR}
\end{subfigure}
\begin{subfigure} {.5\textwidth}
\centering
\imagen{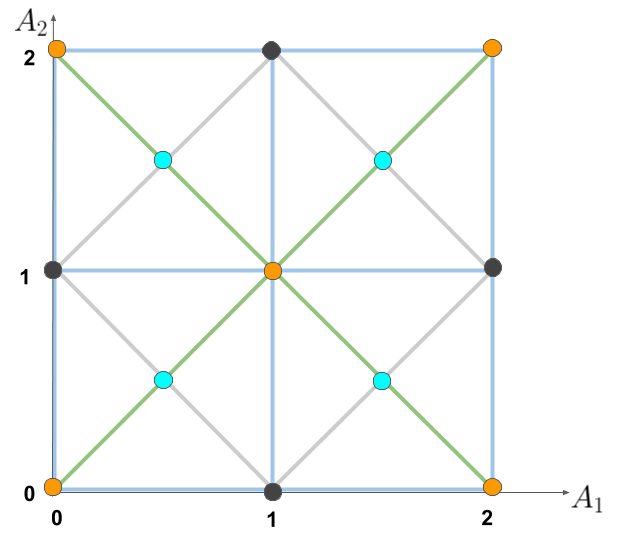}
\caption{$E_8 \times E_8$ heterotic } 
\label{fig:E8anyR}
\end{subfigure}
\caption{Enhancement groups on the left sector of the heterotic theory on the slice of moduli space defined by $A^{3,...,16}=0$, $R=R_0$ with a generic $R_0>1$}
\label{fig:SO32E8anyR}
\end{figure}

\begin{figure}[H] 
\begin{subfigure}{.5\textwidth}
\centering
\imagen{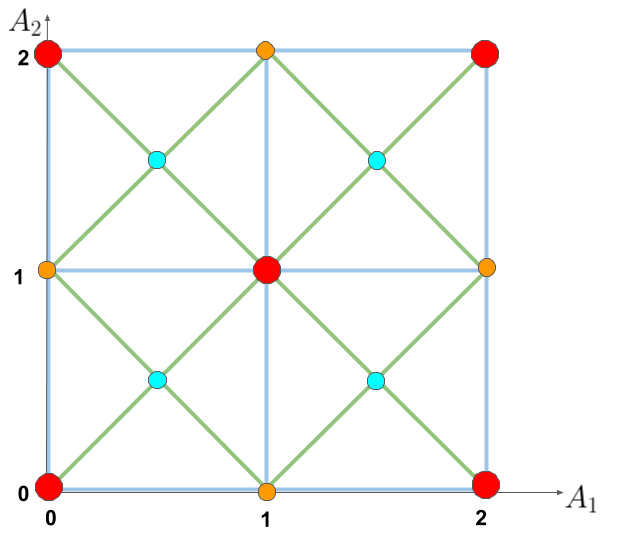}
\caption{ $SO(32)$ heterotic }
\label{fig:SO32R1}
\end{subfigure}
\begin{subfigure} {.5\textwidth}
\centering
\imagen{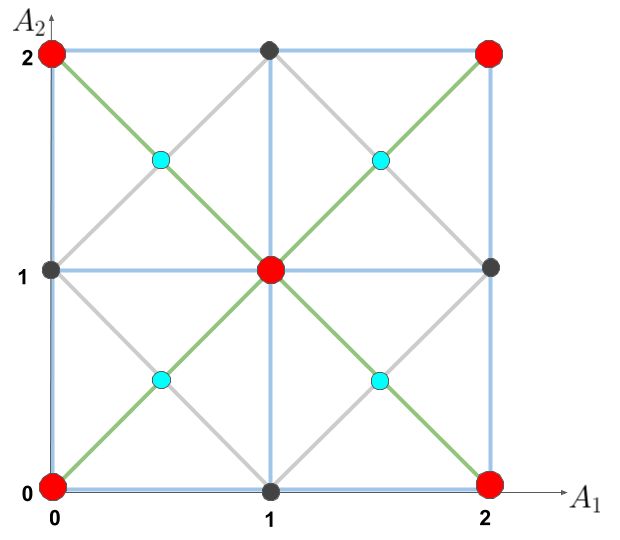}
\caption{$E_8 \times E_8$ heterotic} 
\label{fig:E8R1}
\end{subfigure}
\caption{Enhancement groups on the left sector of the heterotic theory on the slice of moduli space defined by $A^{3,...,16}=0$, $R=1$.}
\label{fig:SO32E8R1}
\end{figure}

\begin{figure}[H] 
\begin{subfigure}{.5\textwidth}
\centering
\imagen{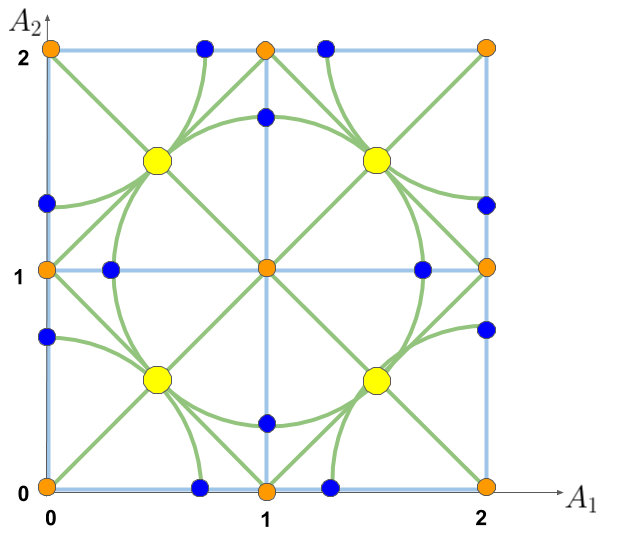}
\caption{$SO(32)$ heterotic }
\label{fig:SO32R3}
\end{subfigure}
\begin{subfigure} {.5\textwidth}
\centering
\imagen{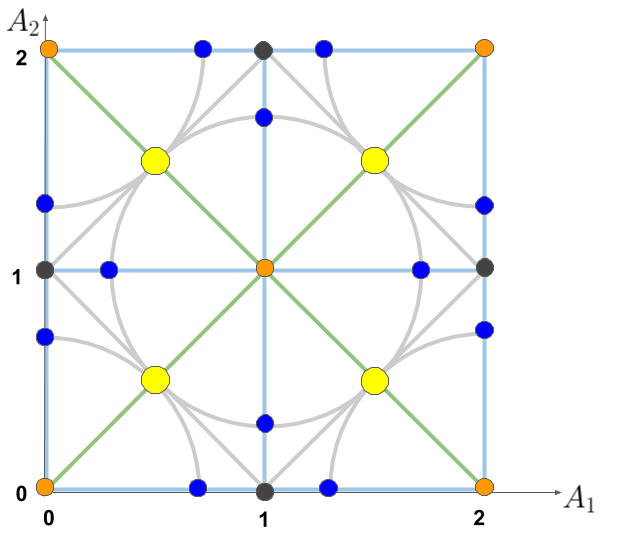}
\caption{$E_8 \times E_8$ heterotic } 
\label{fig:E8R3}
\end{subfigure}
\caption{Enhancement groups on the left sector of the heterotic theory on the slice of moduli space defined by $A^{3,...,16}=0$, $R^2=3/4$.}
\label{fig:SO32E8R3}
\end{figure}

\begin{figure}[H] 
\begin{subfigure}{.5\textwidth}
\centering
\imagen{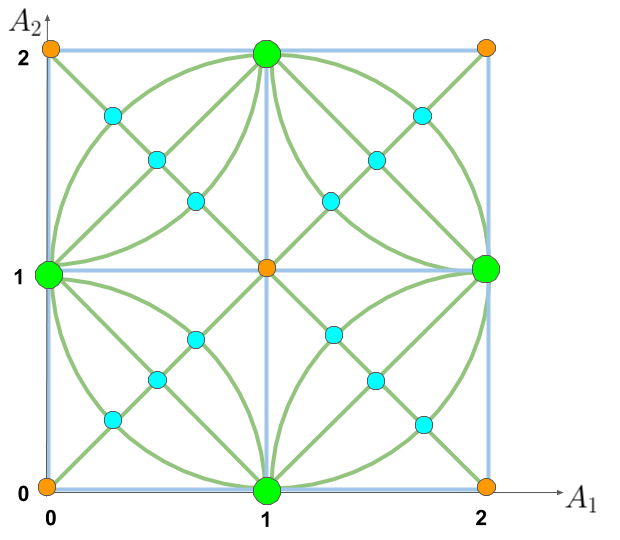}
\caption{$SO(32)$ heterotic }
\label{fig:SO32R2}
\end{subfigure}
\begin{subfigure} {.5\textwidth}
\centering
\imagen{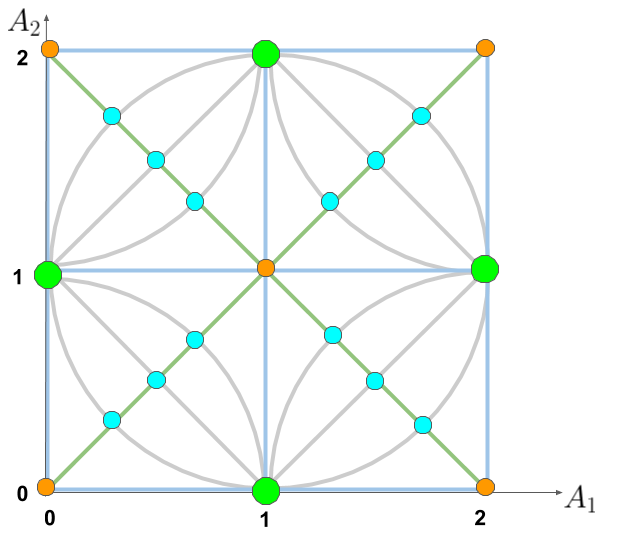}
\caption{$E_8 \times E_8$ heterotic } 
\label{fig:E8R2}
\end{subfigure}
\caption{Enhancement groups on the left sector of the heterotic theory on the slice of moduli space defined by $A^{3,...,16}=0$, $R^2=1/2$.}
\label{fig:SO32E8R2}
\end{figure}

\begin{figure}[H] 
\begin{subfigure}{.5\textwidth}
\centering
\imagen{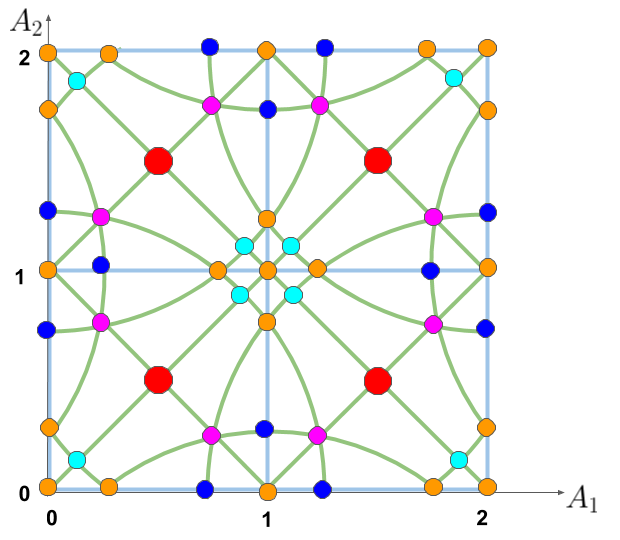}
\caption{$SO(32)$ heterotic }
\label{fig:SO32R4}
\end{subfigure}
\begin{subfigure} {.5\textwidth}
\centering
\imagen{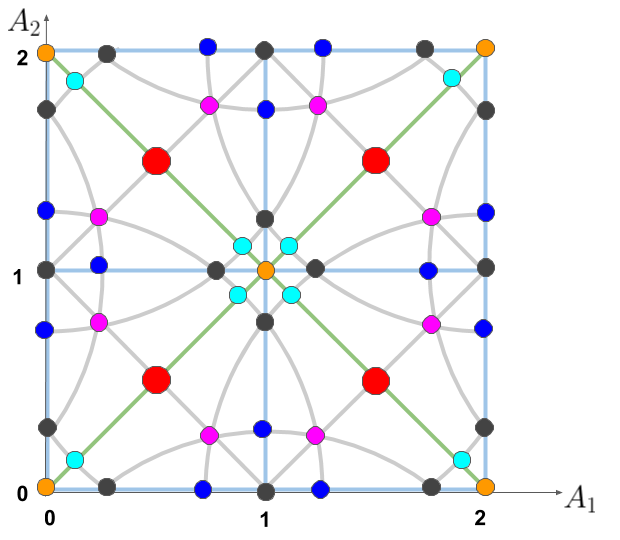}
\caption{$E_8 \times E_8$ heterotic } 
\label{fig:E8R4}
\end{subfigure}
\caption{Enhancement groups on the left sector of the heterotic theory on the slice of moduli space defined by $A^{3,...,16}=0$, $R^2=1/4$.}
\label{fig:SO32E8R4}
\end{figure}

In the following figures we show slices of the moduli space. To exhibit the increase in the number of possible enhancement groups  as the radius decreases and more  winding numbers contribute, as well as the symmetries in the Wilson lines, we present
 figures \ref{fig:SO32E8anyR}, \ref{fig:SO32E8R1}, \ref{fig:SO32E8R3}, \ref{fig:SO32E8R2} and \ref{fig:SO32E8R4} corresponding to compactification on a circle of generic radius $R^2> 1$ and at $R^2=1$, $R^2=\tfrac34$, $R^2=\tfrac12$ and $R^2=\tfrac14$, respectively.\footnote{For the $E_8 \times E_8$ heterotic theory, the Wilson lines chosen do not break the second $E_8$ factor and therefore we display the unbroken gauge group corresponding to the circle and first $E_8$ directions. Figure \ref{fig:E8anyR} can be found in \cite{BGS}.}  The circles in  figures \ref{fig:SO32E8R3}, \ref{fig:SO32E8R2} and \ref{fig:SO32E8R4} reflect the dependence on $|A|^2$ and invariance under rotations.
Two dimensional slices given by one parameter in the Wilson line and the radial direction are shown in figures \ref{fig:SO32E8AR_1} and \ref{fig:SO32E8AR_2}. More figures of slices of moduli space are given in Appendix \ref{app:extrafigures}. 

The first item above corresponds to the red points in figures \ref{fig:E8R1} and \ref{fig:E8AR_1}, while the second and third ones correspond, respectively to the red and green points in figures \ref{fig:SO32R1}, \ref{fig:SO32R2} and \ref{fig:SO32AR_1}. Note that there are also red points in figure \ref{fig:SO32E8R4}, but as we will see, these arise in a different way as above, by a combination of breaking and enhancement. 
In the next section we will show how the enhancement at some of the other special points in the figures arise. 

\begin{figure}[H] 
\begin{subfigure}{.5\textwidth}
\centering
\includegraphics[width=.73\linewidth]{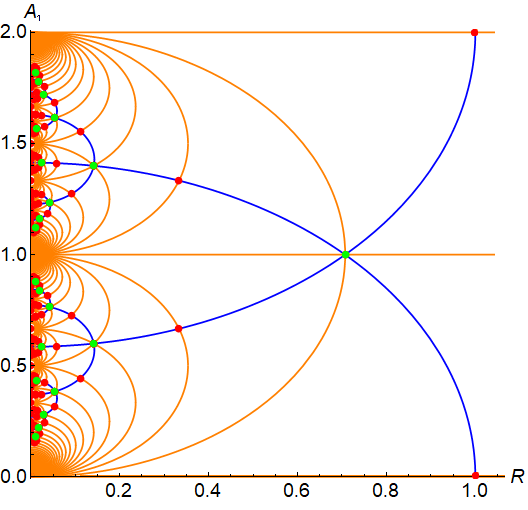}
\caption{$SO(32)$ heterotic }
\label{fig:SO32AR_1}
\end{subfigure}
\begin{subfigure} {.5\textwidth}
\centering
\includegraphics[width=.73\linewidth]{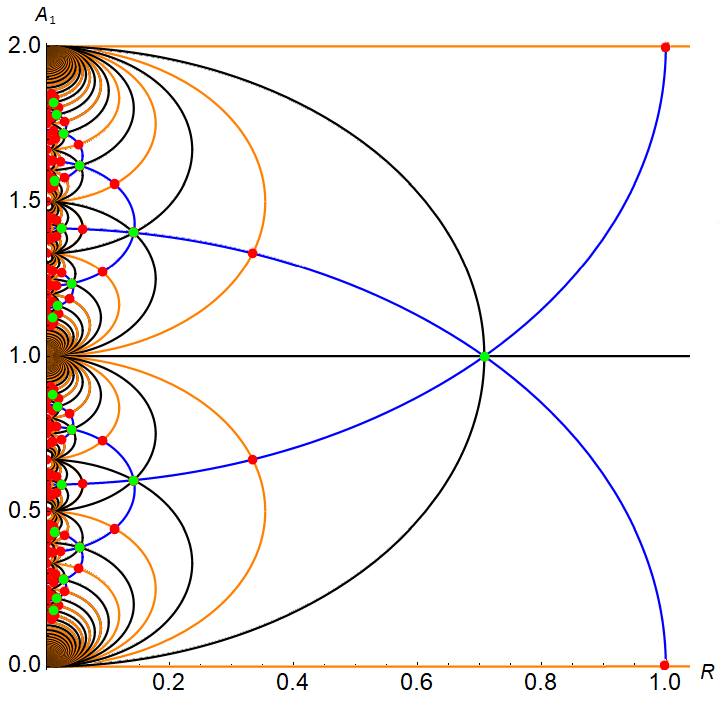}
\caption{$E_8 \times E_8$ heterotic } 
\label{fig:E8AR_1}
 \end{subfigure}
\caption{Enhancement groups on the left sector of the heterotic theory on the slice of moduli space defined by $A^{2,...,16}=0$.}
\label{fig:SO32E8AR_1}
\end{figure}

\begin{figure}[H] 
\begin{subfigure}{.5\textwidth}
\centering
\includegraphics[width=.73\linewidth]{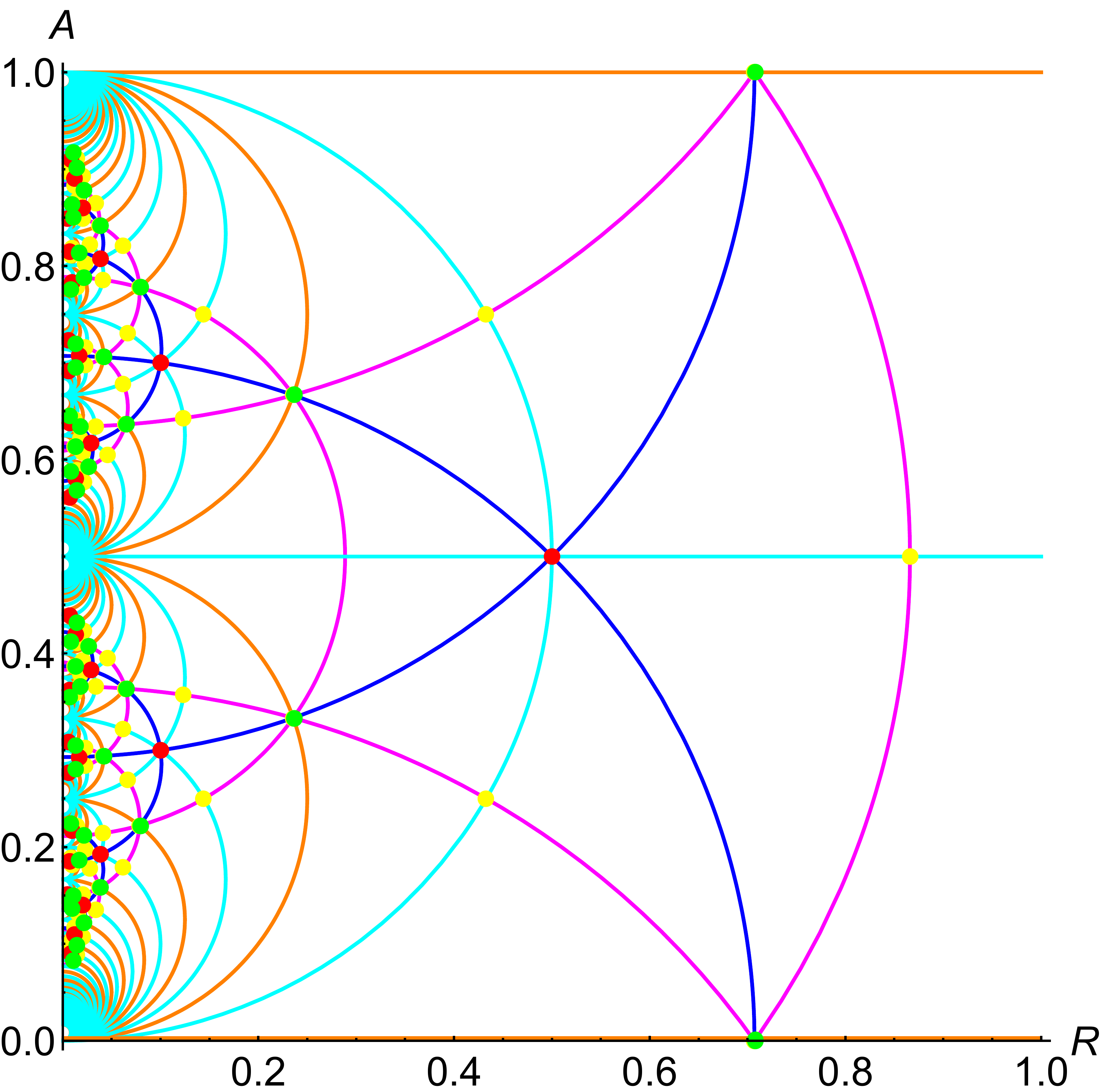}
\caption{$SO(32)$ heterotic }
\label{fig:SO32AR_2}
\end{subfigure}
\begin{subfigure}{.5\textwidth}
\centering
\includegraphics[width=.73\linewidth]{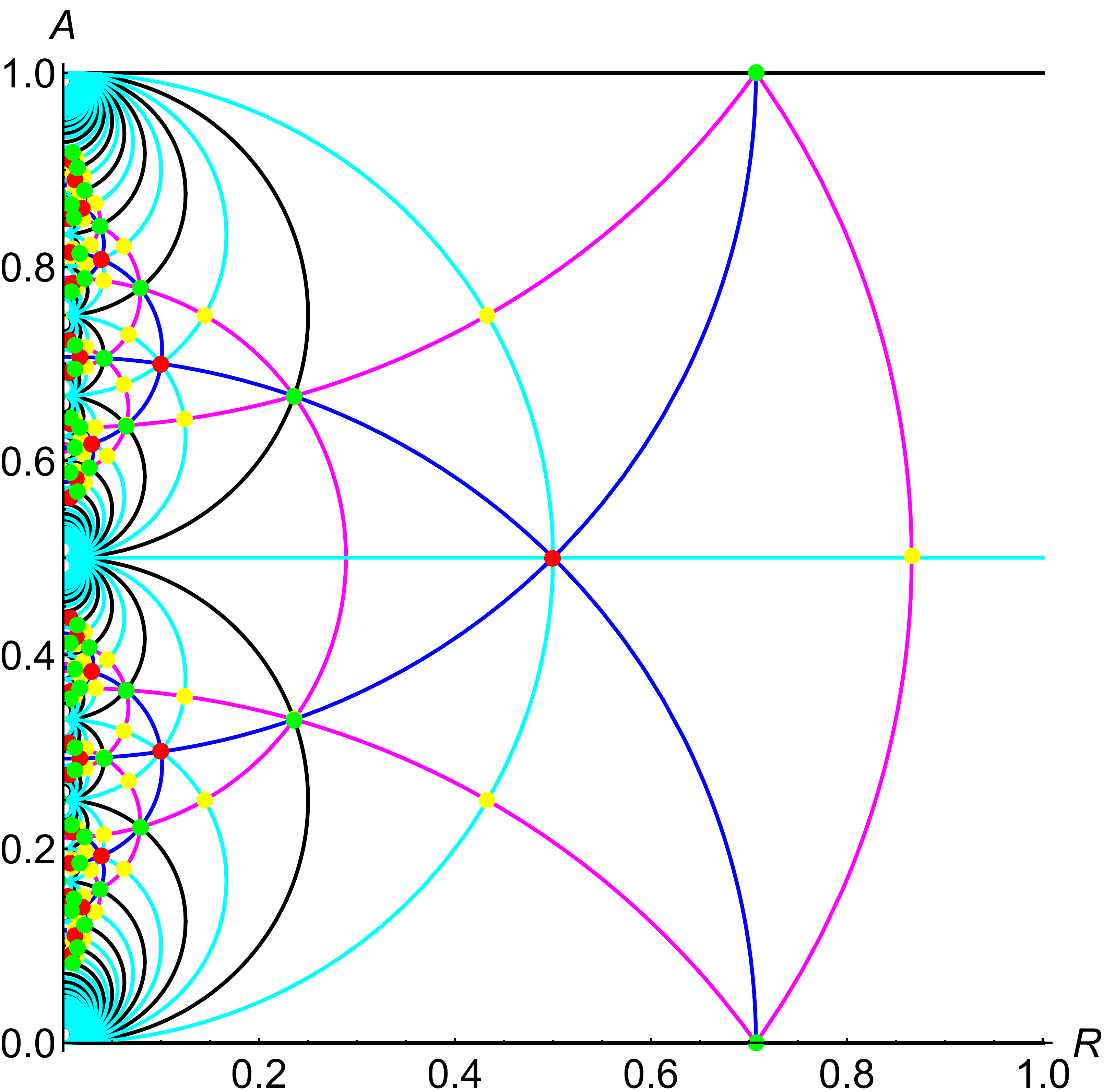}
\caption{$E_8 \times E_8$ heterotic} 
\label{fig:E8AR_2a}
\end{subfigure}
\caption{Enhancement groups on the left sector of the heterotic theory on the slice of moduli space defined by $A, R$, with Wilson line $A^I=\left(A,0_7,A+1,0_7\right)$.} 
\label{fig:SO32E8AR_2}
\end{figure}

\begin{figure}[H] 
\begin{subfigure}{.5\textwidth}
\centering
$SO(32)$ heterotic:
\bleyenda
 \text{(\ref{fig:SO32anyR} to   \ref{fig:SO32R4})}\,\rightarrow \,&\begin{cases}
 \cua{white}\quad U(1)^3 \times SO(28)\\
 \cua{verde2}\quad U(1)^2 \times SU(2) \times SO(28)  \\
\cua{celeste}\quad U(1)^2 \times  SO(30)   
 \end{cases} \\
  \text{(\ref{fig:SO32AR_1})}\,\rightarrow \,\cua{white}&\quad U(1)^2 \times  SO(30) \\
   \text{(\ref{fig:SO32AR_2})}\,\rightarrow \,\cua{white}&\quad U(1)^2 \times SU(2) \times SO(28)\\
\cua{cyan} &\quad U(1) \times SU(2)^2 \times  SO(28)  \\
\cua{magenta} &\quad  U(1) \times  SU(3)\times SO(28)\\
\cua{azul}&\quad U(1) \times SU(2) \times SO(30)  \\
\cua{naranja} &\quad U(1) \times  SO(32)  \\
\cua{yellow} &\quad SU(2) \times SU(3) \times SO(28)\\
\cua{red} &\quad SU(2)\times SO(32)\\
\cua{green} &\quad   SO(34)\\
\text{ }\\
\text{ }
\eleyenda
\end{subfigure}
\begin{subfigure}{.5\textwidth}
\centering
$E_8 \times E_8$ heterotic:
 \bleyenda
   \text{(\ref{fig:E8anyR} to   \ref{fig:E8R4})}\,\rightarrow \,&\begin{cases}
 \cua{white}\quad U(1)^3 \times SO(12) \times E_8\\
  \cua{gris}\quad U(1)^2 \times  SU(2) \times SU(12) \times E_8 \\   
 \cua{celeste}\quad U(1)^2 \times  SO(14) \times E_8 \\   
\cua{verde2}\quad U(1)^2 \times E_7 \times E_8  
 \end{cases} \\
   \text{(\ref{fig:E8AR_1})}\,\rightarrow \,\cua{white}&\quad U(1)^2 \times  SO(14) \times E_8 \\
   \text{(\ref{fig:E8AR_2a})}\,\rightarrow \,\cua{white}&\quad U(1)^2 \times SU(2) \times SO(12) \times E_8  \\
\cua{cyan}  &\quad  U(1) \times SU(2) \times E_7 \times E_8 \\
\cua{magenta}  &\quad  U(1) \times  SU(3)\times SO(12) \times E_8 \\
\cua{blue}  &\quad  U(1) \times  SU(2)\times SO(14)  \times E_8 \\
\cua{black}  &\quad  U(1) \times  SO(16)  \times E_8 \\
\cua{naranja}&\quad U(1) \times  E_8  \times E_8\\
\cua{yellow}  &\quad  SU(3) \times E_7 \times E_8 \\
\cua{red}  &\quad  SU(2) \times E_8 \times E_8 \\
\cua{green}  &\quad SO(18) \times E_8
  \eleyenda
\end{subfigure}
\end{figure}

\subsection{Enhancement-breaking of gauge symmetry}
\label{sec:enh-break}

Whenever the Wilson line is not in the dual root lattice, part or all of the $SO(32)$ or $E_8 \times E_8$ symmetry is broken. However, this does not imply that no symmetry enhancement from the circle direction is possible. The pattern of gauge symmetries can still be rich. We denote these cases enhancement-breaking of gauge symmetry. This nomenclature can be confusing however: for specific values of $R$ and $A$, there is the possibility that the symmetry enhancement is so large that it restores the original $SO(32)$ or $E_8 \times E_8$ symmetry, or even leads to a larger group of rank 17. This means that we can have a maximal enhancement even if the Wilson line is not in the dual root lattice, either to the groups listed at the end of the previous section, or to any other simply-laced, semi-simple group of rank 17, such as for example $SO(18) \times E_8$. 

The massless states for an arbitrary Wilson line are the following:

Sector 1 has $w=0$ (and thus $p^A=\pi^A$) and consists of the roots of $SO(32)$ or $E_8 \times E_8$ satisfying  $\pi \cdot A \in \mathbb Z$, which form a subgroup $H \subset SO(32)$ or $H \subset E_8 \times E_8$.  We give examples of  Wilson lines preserving 
$U(1) \times SU(16) \subset SO(32)$, 
$SO(2p) \times SO(32-2p) \subset SO(32)$, 
$U(1) \times SU(p) \times SO(32-2p) \subset SO(32)$,
$U(1) \times SU(9) \times SO(16)\subset E_8 \times E_8$,
$U(1)^2 \times SU(8) \times SU(8)\subset E_8 \times E_8$,
$SO(16) \times E_8  \subset E_8 \times E_8$,  
$ SU(2) \times E_7  \times E_8 \subset E_8 \times E_8$ 
in the following sections. 


 Sector 2 contains states only at radii $R^2=s^2/(2w^2)$. Quantization of momentum gives the condition \eqref{condsector2}.
 If there are states in this sector, there is an enhacement of $H \times U(1)$ to $H\times SU(2)$ (where the $SU(2)$ can be on the circle direction or along some direction mixing the circle with the heterotic directions)  or to a group that is not a product, like for example enhancement of $SO(16) \times U(1)$ to $SO(18)$, as we will show in detail.

On figures \ref{fig:SO32E8AR_1} to \ref{fig:e8e8_n5_n0_11} sector 1 is represented by the horizontal lines and sector 2 by the curves.

Now 
we show explicitly how the groups mentioned in sector 1 get enhanced respectively to 
$SO(34)$ at $R^2=\tfrac12$;
$SU(18)$ at $R^2=\tfrac14$;
$E_{p+1}\times SO(32-2p)$ at $R^2= 1 - \tfrac{p}{8}$;
$E_{p+1}\times SU(16-p)$ at $R^2= 1 - \tfrac{8}{16-p}$  in the $SO(32)$ theory, and
$SO(34)$ at $R^2 = \tfrac{1}{18}$;
$SU(18)$ at $R^2=\tfrac19$;
$SO(18) \times E_8$ at $R^2=\frac12$;
$SU(2) \times E_8 \times E_8$ at $R^2 = \tfrac14$ in the $E_8 \times E_8$.\\

\noindent{\bf{- Explicit examples for the $\Gamma_{16}$ theory}}

Here we present some examples of symmetry enhancement-breaking. 
The roots of $SO(32)$ are given by 
\beq \label{rootsSO32}
SO(32) \ : \quad (\underline{\pm1,\pm1,0^{14}})\, , 
\eeq
where underline means all possible permutations of the entries.

\subsubsection{$U(1) \times SO(32) \to   SO(34)$}
\label{sec:SO34}

Consider the $SO(32)$ heterotic theory compactified on a circle of radius $R=1/\sqrt{2}$ with a Wilson line $ A=(1,0,\dots , 0) \in \Gamma_v$. 
The states with $p_R=0$ have left-moving momenta 
\bea
{\bf p_L}=(w, \, \pi^A+\delta_1^A w\,  )\, ,
\eea
where the first entry corresponds to the circle direction. In sector 1, with $w=0$,  all the momenta satisfy $|\pi^A|^2=2$ and $\pi \cdot A \in {\mathbb Z}$. The last condition holds for any $\pi^A \in \Gamma_g$, and thus in this sector one has all the root vectors of $SO(32)$ given in \eqref{rootsSO32}. In sector 2 we have $s=1$ and $w=\pm1 $. Here we get massless states coming from three different sectors of the $SO(32)$ weight lattice, namely 
 
2.a) $|\pi|^2=2$, with $\pi^1=\pm 1$ 
\bea
{\bf p_L} =(\pm1, 0,\underline{\pm1,0,0,\dots,0})\qquad 
\eea
(where the signs are not correlated). These are $60$ states with $n=0$.

2.b) $|\pi|^2=0$,
\bea
{\bf p_L}=(\pm1,\pm1,0,\dots,0)\, .\qquad 
\eea
These are 2 states, which have $n=w$.

2.c) $|\pi|^2=4$, with $\pi^1=\pm2$
\bea
{\bf p_L}=(\mp1,\pm1,0,\dots,0)\,  .
\eea
Another $2$ states with $n=-w$. 

We thus get 64 extra states, which together with the Cartan direction of the circle, enhance the $SO(32)$ to $SO(34)$. This point in moduli space  is illustrated in green in figures \ref{fig:SO32R2}, \ref{fig:SO32AR_1} and \ref{fig:SO32AR_2}. In figure \ref{fig:SO32R2} the other green points differ from this by a $\Lambda$-shift, while the other green points in figures \ref{fig:SO32AR_1} and \ref{fig:SO32AR_2}, that appear at a different radii, will be explained in section \ref{sec:A1R}.

\subsubsection{$U(1)^2 \times SU(16) \to   SU(18)$}
\label{sec:SU(18)_gamma16}

We now take the Wilson line  $A = \left( \left(\tfrac14\right)_{15}, -\tfrac34\right)$.
In  sector 1 ($w=0$) we have the roots of $SO(32)$ that obey:
\beq 
\tfrac14\sum_{A=1}^{16} \pi^A - \pi^{16}  \in \mathbb{Z}\, .
\eeq 
Since the sum cannot be a multiple of $4$, it has to vanish. Then we have the roots with two non-zero entries of  opposite signs, that is $SU(16)$. For a generic $R$ this is the gauge group, but if $R^2=\tfrac14$  we get enhancement to the maximal group $SU(18)$. In this case, the mass formula \eqref{massformula} gives
\beq 
\sum_{i=1}^{15} (\pi_i + \tfrac{w}{4})^2 + (\pi_{16} - w + \tfrac{w}{4})^2= \sum_{i=1}^{16} (\hat\pi_i + \tfrac{w}{4})^2 =2 - \tfrac{w^2}{2}  \nn
\eeq 
where we defined $\hat \pi = (\pi_1,\pi_2,\dots,\pi_{15},\pi_{16}-w)$.
If $w$ is even then $\hat \pi$ is in $(0)$ or $(s)$, but if it is odd then $\hat \pi$ is in $(v)$ or $(c)$. We also have the quantization condition:
\beq 
\frac{\tfrac12|\pi|^2 - 1}{w} = \frac{\tfrac12|\hat{\pi}|^2 + \tfrac12 w^2 + w \hat\pi_{16} - 1}{w} = \frac{\tfrac12|\hat\pi|^2  - 1}{w}+ \tfrac{w}{2} + \hat\pi_{16} \in \mathbb{Z}\, .
\eeq 

For $w=1$,  $
-\sum_{i=1}^{16}\hat\pi_i = 2|\hat\pi|^2  -1 $, and
the solutions are $\hat\pi = -\left(\underline{1,0_{15}} \right)$ on $(v)$ and $\hat\pi = -\left(\underline{(\tfrac12)_{15},-\tfrac12} \right)$  on $(c)$. 

For $w=2$, $ 
\sum_{i=1}^{16} (\hat\pi_i + \tfrac{1}{2})^2 = 0
$, with unique solution $\hat\pi = -\left(\left(\tfrac12\right)_{16} \right)$.

They all obey the quantization condition, and add up to $66$ additional states. Together with the $240$ roots of $SU(16)$, they complete the $306$ roots of $SU(18)$.

\subsubsection{$U(1) \times SO(2p) \times SO(32-2p) \to  E_{p+1} \times SO(32-2p) $}
\label{sec:SO(32-2p)SO(2p)}

Now we take a Wilson line $A=\left(\left(\tfrac12\right)_p,0_{16-p}\right)$,  $2 \le p \le 8$, in the  SO(32) theory\footnote{Note that $p>8$ is equivalent,   by a shift  $\Lambda=-\left(\left(\tfrac12\right)_{16}\right)$, to $p'=16-p <8$.}. 

The massless states that survive  in sector 1 ($w=0$)  are those with momentum $\pi^A$ satisfying
\beq 
	\frac{1}{2}\sum\limits_{A=1}^{p}\pi^{A} \in \mathbb{Z}\, .
	\eeq
Then the surviving states have momenta
	\beq \label{wharever1}
	{\bf p_L}&=(0, \underline{\pm 1, \pm 1, 0_{p-2}},0_{16-p}) \longrightarrow  SO(2p) \, ,\\
	{\bf p_L}&=(0,0_p,\underline{\pm 1, \pm 1,0_{14-p}}) \ \  \longrightarrow  SO(32-2p)\, .
	\eeq

For  generic radius there are no states with non-zero winding, and then we get $SO(2p) \times SO(32-2p)$. These points are illustrated for $p=2$  by the cyan dots in figures \ref{fig:SO32anyR}, \ref{fig:SO32R1}, \ref{fig:SO32R2} and \ref{fig:SO32R4}; for $p=7$, on the horizontal cyan line in figure \ref{fig:SO32AR_2} and for other values of $p$, at half-integer values of the horizontal lines of the figures in appendix \ref{app:extrafigures}. 

At  special values of $R$  some states with non-vanishing winding are massless. For example, when $R^2=1-\frac{p}{8}$ for $p<8$, the $U(1) \times SO(2p)$ is enhanced to $E_{p+1}$. In this case, the mass formula \eqref{massformula} is
\beq 
\sum_{i=1}^{p}(\pi_i + \tfrac{w}{2})^2 + \sum_{i=p+1}^{16} \pi_i^2 = 2-2w^2(1-p/8) \le \frac p4\, ,\nn
\eeq 
and then if $p<8$ the lhs must be smaller than $2$. If the $\pi_i$ are semi-integer, then the lhs is always bigger than $2$. Consequently,  $\pi_i$ can only take integer values and  we need
$\sum_{i=p+1}^{16} \pi_i^2 = \bx = 0,\, 1$.

For $w=1$ the solution must be of the form $ 
\left (\underline{(-1)_k,0_{p-k}},\underline{\pm\bx,0_{15-p}} \right)$
and the equation is
 solved for every $p$ if $\bx=0$.
Then we get $ 
\left (\underline{(-1)_k,0_{p-k}},0_{16-p} \right)$.

There is an additional constraint because  $|\pi|^2$ must be even, and then $k$ must be even. The number of states is equal to the way of choosing the value of the first $p$ components. Choosing the first $p-1$ components, the last one is fixed by the constraint. There are $2 \times  2^{p-1} = 2^p$ states with $|w|=1$.

For $w=2$ we get $ 
\sum_{i=1}^{p}(\pi_i + 1)^2 =   p - 6 -\bx $, which is only possible for $p=6,\, 7$. The rhs can only take the values $0$ or $1$. In the first case,  all the $\pi_i$ must  be equal to $-1$. 
Then we get the solutions $
\left ((-1)_7,\underline{\pm 1,0_{8}} \right)$ for $p=7$  and $
\left ((-1)_6,0_{10}\right) $ for $p=6$ .
The second case   is only possible for $p=7$ and $\bx=0$. 
One of the $\pi_i$ can take the value $0$ (or $-2$) and the rest must take the value $-1$: $
\left (\underline{-1\pm 1,(-1)_6},0_9 \right)$ for $p=7$.
In total we have $2$ states with $|w|=2$ for $p=6$ and $2\times(18 + 14)=64$ for $p=7$.

For $w\ge3$  the equation cannot be satisfied.
Then for $p<6$ we get $2^{p}$ states (all with $|w|=1$), while for $p=6$ and $7$ we get 
$2$ and $64$ extra states respectively with $|w|=2$.
\beq 
U(1) \times SO(4) \rightarrow SU(2) \times SU(3) \equiv E_3 \text{ (4 extra states)}\\
U(1) \times SO(6) \rightarrow SU(5) \equiv E_4 \text{ (8 extra states)}\\
U(1) \times SO(8) \rightarrow SO(10) \equiv E_5 \text{ (16 extra states)}\\
U(1) \times SO(10) \rightarrow E_6 \text{ (32 extra states)}\\
U(1) \times SO(12) \rightarrow E_7 \text{ (66 extra states)}\\
U(1) \times SO(14) \rightarrow E_8 \text{ (192 extra states)}
\eeq 

Recalling that $E_2=U(1)\times SU(2)$,  this is also valid for $p=1$, where we get the enhancement at $R^2 = \tfrac78$:
\beq 
U(1)^2 \equiv U(1) \times SO(2) \rightarrow  U(1) \times SU(2) \equiv E_2 \text{ (2 extra states)}\, .
\eeq 
The enhancement group $U(1) \times SU(2) \times SO(30)$, as any non-maximal enhancement, does not arise at an isolated point, but at a line, displayed in blue  in figure \ref{fig:SO32AR_1}.

Applying the statement to $p=8$, appears an enhancement from $U(1) \times SO(16)$ to $E_9$ at $R=0$. Since $E_9$ has infinite dimension, we would need infinite massless states with infinitely many different winding numbers. It is obvious that at $R=0$ winding states do not cost any energy, and thus one can have all the windings. The mass equation is:
\beq 
\sum_{i=1}^{8}(\pi_i + \tfrac{w}{2})^2 = 
2-\bx\, .
\eeq 
We see that for this value of $p$ the rhs is independent of the winding number. If $w=1$ then $\pi = \left (\underline{(-1)_k,0_{8-k}},0_{8}\right)$ is a solution (if $k$ is even). For any other odd value of $w$ we have the solution: $\pi = -\left (\underline{\left(\frac{w+1}{2}\right)_k,\left(\frac{w-1}{2}\right)_{8-k}},0_{8}\right)$. These, together with the states with even $w$, give infinite massless states. 

We can see all these enhancements at the intersections of the  lines at $A=1/2$ in figures \ref{fig:so32_n1} to \ref{fig:so32_n16} that occur at $R^2 = 1-\tfrac{p}{8}$.

 \subsubsection{$U(1)^2  \times SO(2p) \times SU(16-p)  \to  SU(2) \times E_{p+1} \times SU(16-p) $}
\label{sec:SU2_E_SU}

Consider the Wilson line $A=\left(\left(\tfrac{4}{16-p}\right)_p,0_{p}\right)$, with $0 \le p \le 7$.

The massless states that survive  in sector 1 ($w=0$)  are those with momentum $\pi^A$ satisfying $
	\frac{4}{16-p}\sum\limits_{A=1}^{16-p}\pi^{A} \in \mathbb{Z}.$
Then the surviving states have momenta
	\beq \label{wharever1}
	{\bf p_L}&=(0,0_{16-p},\underline{\pm 1, \pm 1,0_{p+2}})   \longrightarrow  SO(2p) \\
	{\bf p_L}&=(0, \underline{1, -1, 0_{14-p}},0_{p}) \ \ \longrightarrow  SU(16-p)
		\eeq

For  generic radii there cannot be states with non-zero winding, and then  the symmetry group is $SO(2p) \times SU(16-p)$. This is illustrated in the white spaces of the figures in  Appendix \ref{app:extrafigures}. 

There are special values of $R$ where some states with non-vanishing winding are massless. For example, when $R^2=1-\frac{8}{16-p}$, the $U(1)^2 \times SO(2p)$ is enhanced to $SU(2) \times E_{p+1}$. To see this, consider the mass formula \eqref{massformula} 
\beq 
\sum_{i=1}^{q}(\pi_i + \tfrac{4w}{q})^2 + \sum_{i=q+1}^{16} \pi_i^2 = 2-2w^2(1-8/q) \, \ {\rm where} \ q\equiv 16-p \ .  \nn
\eeq 
For $w\neq 0$, the rhs is smaller than or equal to $16/q$ and then the lhs must be smaller than $2$. 
If the $\pi_i$ are integer, then we need $
\sum_{i=q+1}^{16} \pi_i^2 = \bx = 0,\, 1$ and
 it follows that
\beq 
\sum_{i=1}^{q}(\pi_i + \tfrac{4w}{q})^2 = 
2-2w^2(1-8/q)-\bx\, . \nn
\eeq
For $w=1$, $
\sum_{i=1}^{q}(\pi_i + \tfrac{4}{q})^2 = 16/q -\bx \le\tfrac{16}{9}$. If one of the $\pi_i$ is different from $0$ or $-1$ then the lhs is larger than $16/q$. So the solution must be of the form $ 
\left (\underline{(-1)_k,0_{16-p-k}},\underline{\pm\bx,0_{p-1}} \right)$ and
then  $k = \bx = 0$. There are only  two states (considering also $w=-1$) with momentum
$
\left (0_{16} \right)
$.

For $w=2$ we get $ 
\sum_{i=1}^{q}(\pi_i + 8/q)^2 =  - 6 + 64/q -\bx$ which is only possible for $q=9,\, 10$ ($p=7,6$). 
If $p=6$ then we need $\bx=0$, the rhs is $\tfrac25$ and we only have the solution $ 
\left ((-1)_{10},0_6\right).$
If $p=7$ then, for $\bx=0$ and $\bx=1$ the rhs takes the values $\tfrac{10}{9}$ and $\tfrac{1}{9}$. The equation for $\bx=0$ is impossible to satisfy, and then we get $ 
\left ((-1)_9,\underline{\pm 1,0_{6}} \right)$. 
In total we have $2$ states with $|w|=2$ for $p=6$ and $2\times(
14)=28$ for $p=7$.

For $w\ge3$ we get $ 
\sum_{i=1}^{q}(\pi_i + 12/q)^2 =   144/q - 16 -\bx \leq 0$.
Then for $q>10$ ($p<6$) there are $2$ states (both with $|w|=1$), while for $p=6$ and $7$ there are
$2$ and $28$ extra $|w|=2$ states respectively.

If the $\pi_i$ are semi-integer, then the last $p$ values have to be $\pm \tfrac12$:
\beq 
\sum_{i=1}^{q}(\pi_i + \tfrac{4w}{q})^2  =\tfrac{q}{4} - 2 - 2w^2(1-8/q) 
\eeq 
For $w=1$, $
\sum_{i=1}^{q}(\pi_i + \tfrac{4}{q})^2  =
\tfrac{(q-8)^2}{4q} \leq 1$ and the $\pi_i$ can only take the values $\pm\tfrac12$. The solutions are of the form
$
\left( \underline{\left(\tfrac12\right)_k, \left(-\tfrac12 \right)_{16-p-k}},\underline{\left(\pm\tfrac12 \right)_{p}} \right)
$, and the equation implies
 $k=0$. Then, for $|w|=1$, we get the $2\times 2^{p+1 + \delta_{p,0}}$ solutions $ 
\left(\left(-\tfrac12 \right)_{16-p},\underline{\left(\pm\tfrac12 \right)_{p}} \right)$.

For $w=2$ we obtain $ 
\sum_{i=1}^{q}(\pi_i + \tfrac{8}{q})^2  =
\tfrac{(q-8)(q-32)}{4q} \leq 0$, and
then there are no states with $|w|>1$.

In total, for $p<6$ we get $2 + 2^{p + \delta_{p,0}}$ states (all of them with $|w|=1$), while for $p=6$ and $7$ we get 
$2$ and $28$ extra states respectively with $|w|=2$.
\beq 
U(1)^2 \rightarrow SO(4) \equiv SU(2) \times E_1 \text{ (4 extra states)}\\
U(1)^3 \equiv U(1)^2 \times SO(2) \rightarrow SU(2) \times SU(2) \times U(1) \equiv SU(2) \times E_2 \text{ (4 extra states)}\\
U(1)^2 \times SO(4) \rightarrow SU(2) \times SU(2) \times SU(3) \equiv SU(2) \times E_3 \text{ (6 extra states)}\\
U(1)^2 \times SO(6) \rightarrow SU(2) \times SU(5) \equiv SU(2) \times E_4 \text{ (10 extra states)}\\
U(1)^2 \times SO(8) \rightarrow SU(2) \times SO(10) \equiv SU(2) \times E_5 \text{ (18 extra states)}\\
U(1)^2 \times SO(10) \rightarrow SU(2) \times E_6 \text{ (34 extra states)}\\
U(1)^2 \times SO(12) \rightarrow SU(2) \times E_7 \text{ (68 extra states)}\\
U(1)^2 \times SO(14) \rightarrow SU(2) \times E_8 \text{ (158 extra states)}\nn
\eeq 

At $p=8$ we seem to get an enhancement from $U(1)^2 \times SO(16)$ to $SU(2) \times E_9$ at $R=0$. 

All of these enhancements can be seen on the intersections of the red and purple curves of figures \ref{fig:so32_n9} to \ref{fig:so32_n16} that occur at $R^2 = 1-\tfrac{8}{q}$.\\

\noindent{\bf - {Explicit examples for the $\Gamma_{8}\times \Gamma_8$ theory}}

The roots of $E_8 \times E_8$ are
\bea \label{rootsE8E8}
E_8 \times E_8 \ : && (\underline{\pm1,\pm1,0^{6}},0^8)\, ,  \  (0^8,\underline{\pm1,\pm1,0^{6}}) \ , \ \\
&& ((\pm\frac12)^8,0^8) \, , \ (0^8,(\pm\frac12)^8) \ , \rm{with \ even \ number \ of \ + \ signs} \nn
\eea

\subsubsection{$U(1)^2 \times SU(9) \times SO(16) \to   SO(34)$}
\label{sec:SO34_gamma8}
Consider the $\Gamma_8 \times \Gamma_8$ theory compactified with Wilson line $A = \left( \left(\tfrac16\right)_7, \tfrac56, 1, 0_7\right)$.
In  sector 1 ($w=0$) we have the roots of $E_8 \times E_8$ that obey:
\beq 
\tfrac16\sum_{A=1}^7 \pi^A + \tfrac56 \pi^{8} + \pi^9 \in \mathbb{Z}
\eeq 
This breaks into two conditions, one for each $E_8$:
\beq 
\tfrac16\sum_{A=1}^7 \pi^A + \tfrac56 \pi^{8} \in \mathbb{Z}\,  \qquad {\rm and}\qquad
\pi^9 \in \mathbb{Z}\, . \label{twoconditions}
\eeq 
For the first condition we have $(0)$ and $(s)$ roots. The $(0)$ roots are vectors of the form $\left (\underline{\pm 1, \pm 1, 0_{6}} \right )$. The condition implies that if $\pi_8=0$ then we need opposite signs for the two non-zero entries. If $\pi^8 = \pm1$ then the other non-zero entry must have the same sign. We get $ 
\left (\underline{1, -1, 0_{5}},0 \right )$ and $
\pm\left (\underline{1, 0_{6}},1 \right )$. These are $42 + 14  = 56$ roots.

The $(s)$ roots are vectors of the form $\left (\underline{\left(\pm \tfrac12\right)_{8}} \right )$ with an even number of minus signs. The condition is $\sum_{A=1}^7 \pi^A + 5 \pi^8 = 0$ mod $6$. The absolute value of the lhs can only be $0$ or $6$. In the first case one of the first $5$ components must have a different sign than the rest, and in the second case all the $8$ components must have the same sign and we get $ 
  \pm \left ( \underline{\left(\tfrac12 \right)_{6},- \tfrac12}, -\tfrac12 \right )$ and $
 \pm \left ( \left(\tfrac12 \right)_{8} \right )$. These are $14 + 2 = 16$ roots.

In total we have the $56+16=72$ roots of $SU(9)$.

The second condition leaves only the integer roots, and then we have $SO(16)$.

For an arbitrary value of $R$ there cannot be states with non-zero winding, and then the gauge group is $SU(9)\times SO(16)$.

Now we  show that when $R^2= \tfrac{1}{18}$ there is enhancement of the gauge symmetry to $SO(34)$. The mass formula \eqref{massformula} is:
\beq 
\sum_{i=1}^7 (\pi_i + \tfrac{w}{6})^2 + (\pi_8 + \tfrac{5w}{6})^2 + (\pi_9 + w)^2 + \sum_{i=10}^{16}\pi_i^2 = 2 - \tfrac{w^2}{9}<2\, .
\eeq Then $\sum_{i=10}^{16}\pi_i^2$ can only take the values $0$, $1$ or $\tfrac74$. In the last case, we also have that $(\pi_9+w)^2\ge\tfrac14$, which  means that there are no spinorial roots in the last $8$ components. The only possibilities are: $\left(-w,0_{7} \right)$ and $\left(-w,0_{7} \right)\pm\left(\underline{1,0_{7}} \right)$. The first (second) case requires $w$ to be even (odd).
Defining $\hat\pi = \left(\pi_1, \pi_2,\dots,\pi_7,-\pi_8-w \right)$,
we have:
\beq 
\sum_{i=1}^8 (\hat\pi_i + \tfrac{w}{6})^2  = 2 - \tfrac{w^2}{9}  
- \tfrac{1-(-1)^w}{2}
\eeq 
but now the condition for the integer vectors is $\sum_{i=1}^{8} \hat \pi_i$ odd (even) when $w$ is odd (even); and for the half-integer vectors we have the $(s)$ conditions if $w$ is odd and the $(c)$ conditions if $w$ is even.

The quantization condition is
\beq
\frac{\tfrac12|\pi|^2 - 1}{w} \in \mathbb{Z} \rightarrow \begin{cases} 
|\pi|^2=0 \text{ mod }2 & \text{ for }|w|=1 \\
|\pi|^2=2 \text{ mod }4 & \text{ for }|w|=2 \\
|\pi|^2=2 \text{ mod }6 & \text{ for }|w|=3 
\end{cases}
\eeq 

If $w=1$, $ 
-\sum_{i=1}^8 \hat\pi_i = 3|\hat\pi_i|^2 -2.
$
The minimum value for $|\hat \pi_i|^2$ is $1$, and in that case we have $\hat\pi = -\left(\underline{1,0_7} \right)$.

$|\hat \pi_i|^2=2$ can only be achieved for the $(s)$ conjugacy class, and then $\hat\pi = -\left(\left(\tfrac12\right)_8 \right)$.

 $|\hat\pi|^2=3$  is for the $(v)$ conjugacy class, $-\sum_{i=1}^8 \hat\pi_i = 7$,
but this cannot be achieved. The same happens for greater values of $|\hat\pi|^2$.

If $w=2$, $
-\sum_{i=1}^8 \hat\pi_i = \tfrac32|\hat\pi_i|^2 -1$.
Then $|\hat \pi|^2$ has to be even. The minimum value is $0$, which could be achieved only on $(0)$, and the equation cannot be solved.
$|\hat \pi|^2=2$ can only be achieved for  $(0)$ and  we get $\left(\underline{(-1)_2,0_6} \right)$. $|\hat \pi|^2=4$ has the solution
 $\hat\pi = -\left(\underline{\tfrac32,(\tfrac12)_{7}}\right)$. And for $|\hat \pi|^2=6$ the equation cannot
 be satisfied.

If $w=3$, $ 
\sum_{i=1}^8 (\hat\pi_i + \tfrac{1}{2})^2  =  0$,
and the only solution is $\hat\pi = \left( \left(-\tfrac12\right)_8\right)$. That is $\pi = \left((-\tfrac12)_7, -\tfrac52, -3, 0_7 \right) + \left(0_8, \underline{\pm1,0_7} \right)$. This has $|\pi|^2 =  12$, $18$ or $24$,  which  do not obey the quantization condition.

If $w=4$, $
 -\sum_{i=1}^8 \hat\pi_i   =  \tfrac34|\hat\pi|^2 + \tfrac16$.
But this equation cannot be solved for integer  $|\hat\pi|^2$.

Defining $8$ more components for a $16$ dimensional $\hat\pi$ such that $\hat\pi_9 = \pi_9 - w$ and the rest equal to the last $7$ components of $\pi$, one can write the additional states $\hat \pi$ for $R^2 = \tfrac{1}{18}$ as $\left(\underline{\pm1,0_7},\underline{\pm1,0_7} \right)$ and $
\left(\pm\left(\tfrac12\right)_8,\underline{\pm1,0_7}\right)$
for $|w|=1$ and $\pm\left(\underline{(1)_2,0_6},0_8 \right)$ and as $
\pm\left(\left(\underline{\tfrac32,(\tfrac12)_{7}}\right),0_{8} \right)$
for $|w|=2$.  The former  are $256 + 32 = 288$ states and the latter $56 + 16= 72
$ states.
In total these  $360$ additional states added to the $184$ roots of $SU(9)\times SO(16)$ give the $544$ roots of $SO(34)$.

In figure \ref{fig:e8e8_n0_n8b_11} we show this maximal enhancement on the intersection between one red, two yellow and one green curves. The integer states with $|w|=1$ and $|w|=2$ give the red curve, the half-integer states with $|w|=1$ give the green curve and the ones with $|w|=2$ are represented by the yellow curve. The additional states without winding are those in the yellow line.

\subsubsection{$U(1)^3 \times SU(8) \times SU(8) \to   SU(18)$}\label{sec:SU18_gamma8}

Consider the  Wilson line $A = \left( \left(\tfrac16\right)_7, \tfrac56, \left(\tfrac16\right)_7, \tfrac56\right)$ in the $\Gamma_8 \times \Gamma_8$ theory.

In  sector 1 ($w=0$) we have the first condition of \eqref{twoconditions} for each of the $E_8$, then we get the $144$ roots of $SU(9) \times SU(9)$. For an arbitrary value of $R$ this is the gauge group. 

For $R^2= \tfrac{1}{9}$ there is enhancement of the gauge symmetry to $SU(18)$. To see this, take the mass formula \eqref{massformula} 
\beq 
\sum_{i=1}^7 (\pi_i + \tfrac{w}{6})^2 + (\pi_8 + \tfrac{5w}{6})^2 + \sum_{i=9}^{15} (\pi_i + \tfrac{w}{6})^2 + (\pi_{16} + \tfrac{5w}{6})^2 = 2 - \tfrac{2 w^2}{9}<2\, .
\eeq 

Defining $\hat\pi = \left(\pi_1, \pi_2,\dots,\pi_7,-\pi_8-w,\pi_9, \pi_{10},\dots,\pi_{15},-\pi_{16}-w \right)$
we have:
\beq 
\sum_{i=1}^{16} (\hat\pi_i + \tfrac{w}{6})^2 = 2 - \tfrac{2w^2}{9}  \, ,\\
\eeq 
but now $\hat \pi$ has to be on the conjugacy classes $(ss)$, $(vv)$, $(sv)$ or $(vs)$ if $w$ is odd and on $(cc)$, $(00)$, $(0c)$, $(c0)$ if $w$ is even.

We also have to obey the quantization condition $\frac{\tfrac12|\pi|^2 - 1}{w} \in \mathbb{Z}$.

If $w=1$, $ -\sum_{i=1}^{16} \hat\pi_i  = 3|\hat\pi|^2  - 4$
and $\hat\pi$ is on $(vv)$, $(ss)$, $(vs)$ or $(sv)$. The minimum value for $|\hat \pi_i|^2$ is $2$, and in that case  $(\hat\pi,\hat\pi') = -\left(\underline{1,0_7},\underline{1,0_7} \right)$.

$|\hat\pi|^2=3$ can only be achieved for the $(vs)$ and $(sv)$ conjugacy classes, and
$\hat\pi = -\left(\underline{1,0_7},(\tfrac12)_8 \right)$, $-\left((\tfrac12)_8,\underline{1,0_7} \right)$.
$|\hat\pi|^2=4$  is for the $(ss)$ and $(vv)$ conjugacy classes, and 
 $\hat\pi = -\left((\tfrac12)_{16} \right)$.
 $|\hat\pi|^2 =5$  is for the $(sv)$ and $(vs)$ conjugacy classes, $-\sum_{i=1}^{16} \hat\pi_i = 11$ which cannot be achieved. 
The same happens for greater values of $|\hat\pi|^2$.
 
 If $w=2$, $-\sum_{i=1}^{16}\hat\pi_i  = \tfrac32|\hat\pi|^2  + 1$ implies $(00)$, $(cc)$, $(c0)$ or $(0c)$. The minimum value for $|\hat \pi_i|^2$ is $0$, but then the equation cannot be solved. 

$|\hat\pi|^2=2$ can only be achieved for $(00)$, $(0c)$ or $(c0)$, but there is no solution. 

$|\hat\pi|^2=4$ implies $-\sum_{i=1}^{16}\hat\pi_i = 7$ and this cannot be achieved.
The same happens for greater values of $|\hat\pi|^2$.

If $w=3$, $\sum_{i=1}^{16} (\hat\pi_i + \tfrac{1}{2})^2   =  0$ has only a solution belonging to $(ss)$, namely $\hat\pi = -\left( \left(\tfrac12\right)_{16}\right)$. 

It can be shown that all of these states obey the quantization condition.
Then, the additional states for $R^2 = \tfrac{1}{18}$ are $\pm\left(\underline{1,0_7},\underline{1,0_7} \right)$, $\pm\left(\underline{1,0_7},(\tfrac12)_8 \right)$ and $\pm\left((\tfrac12)_8,\underline{1,0_7} \right)$ for $|w|=1$ and $\hat\pi = \left( \left(-\tfrac12\right)_{16}\right)$ for $|w|=3$
$\left(\underline{\pm1,0_7},\underline{\pm1,0_7} \right)$ and $
\left(\pm\left(\tfrac12\right)_8,\underline{\pm1,0_7}\right)$
for $|w|=1$ and $\pm\left(\underline{(1)_2,0_6},0_8 \right)$ and $
\pm \left( \left(\tfrac12\right)_{16}\right)$
for $|w|=2$.  The former  are $128 + 32 = 160$ states and the latter $2$ states.
In total these are $162$ additional states, which added to the $144$ roots of $SU(9)\times SU(9)$ give the $306$ roots of $SU(18)$.

In figure \ref{fig:e8e8_n8b_n8b_11} we show this maximal enhancement on the intersection between one red, two yellow and one green curves. The integer states with $|w|=1$ are represented by the red curve, the half-integer states with $|w|=1$ give the yellow curve, the states with $|w|=3$ are represented by the green curve and the additional states with $w=0$ give the yellow horizontal line.

\subsubsection{$U(1) \times SO(16) \times E_8 \to SO(18) \times E_8 $}
\label{sec:E8SO(18)}

Consider the $E_8 \times E_8$ heterotic string compactified on a circle of radius $R=\frac{1}{\sqrt{2}}$, with Wilson line $A = \left(1,0^{7},0^8 \right)$, which is of the form $(v0)$ according to the notation of Appendix \ref{app:lattices} (see \eqref{Gamma16Gamma8} in particular). This Wilson line leaves the second $E_8$ unbroken, while from the first $E_8$, the surviving states in sector 1 are the ones with integer entries, i.e. those in the first line of \eqref{rootsE8E8}. The group $H$ from sector 1 is then $SO(16) \times E_8$ and the corresponding points in moduli space are illustrated by the grey dots in figure \ref{fig:E8anyR}.

In sector 2 we have states with $w=\pm1$ such that $s=1$, $|{\bf p}^A|^2=1$. The surviving states have the following momenta
\bea
&&{\bf p_L}=(0,\underline{\pm 1,\pm 1, 0_6}),\, w=0 , \, |\pi|^2=2 \quad  \text{  $112$ roots} \nn \\
&&{\bf p_L}=(\pm 1,0 ,\underline{\pm 1,0_6}),\, w=\pm 1, |\pi|^2=2 , \ \text{ $28$ roots} \nn \\
&&{\bf p_L}=(\pm 1 , \pm 1, 0_7),\, \ \  \ w=\pm 1, \pi=0,  \qquad \text{$2$ roots} \nn\\ 
&&{\bf p_L}=(\pm 1 , \mp 1, 0_7),\, \ \  \ w=\pm 1, |\pi|^2=4,  \quad \text{$2$ roots}\, , \nn 
\eea
where the first entry corresponds to the circle and the subsequent ones  to the 8 directions along the Cartan of the first $E_8$ factor. The first line contains the states of sector 1. These are the 144 roots of SO(18). This point in moduli space, together with its equivalent ones, are illustrated by the green dots in figure \ref{fig:E8R2}, \ref{fig:E8AR_1} and \ref{fig:E8AR_2a}.

\subsubsection{$U(1) \times SU(2) \times E_7 \times E_8 \to SU(2) \times E_8 \times E_8$}
\label{sec:E7}

This is an interesting example of enhancement-breaking in the $E_8 \times E_8$ heterotic theory, where first the $E_8$ is broken to $SU(2) \times E_7$ by the Wilson line $A=\left(\left (\frac{1}{4}\right )_8,0_{8}\right )$ and then enhanced by the circle direction to $SU(2) \times E_8 $.

The Wilson line leaves the second $E_8$ unbroken, while  the surviving roots from the first $E_8$ have 9-momenta 
\beq
{\bf p_L}&=\pm (0,\underline{1,-1,0_{6}}) \\
{\bf p_L}&= \pm \left(0, \left(\tfrac12\right)_8 \right) \\
{\bf p_L} &= \left(0,\underline{\left(\tfrac12\right)_4,-\left(\tfrac12\right)_4} \right) 
\eeq
This, gives 128 roots, which together with the 8 Cartan directions, gives an unbroken gauge group $H=SU(2) \times E_7  \subset E_8$.

Additionally at $R=\frac12$ there are $114$ states in sector 2: two with $w=\pm 2$ and 112 with $w=\pm 1$ and momentum
\beq \label{112}
{\bf p_L}&=\left(\pm \tfrac{\sqrt{2}}{2},\underline{\mp\left(\tfrac34\right)_2,\pm\left(\tfrac14\right)_6} \right) \\
{\bf p_L}&=\left(\pm \tfrac{\sqrt{2}}{2},\underline{\pm\left(\tfrac34\right)_2,\mp\left(\tfrac14\right)_6} \right) 
\eeq
These states give a total of  114 extra states that add up to the previous 136 states, plus the circle direction, adding up to the 251 states of $SU(2) \times E_8$. So at $R=\frac12$ we get enhancement to $SU(2) \times E_8 \times E_8$, which works very differently than the enhancement occurring at $R=1$, mentioned in section \ref{sec:enhancement}.

In figure \ref{fig:e8e8_n8_n0} we present these maximal enhancements for the $\Gamma_8 \times \Gamma_8$ theory, and we also show a maximal enhancement to $SU(3) \times E_7 \times E_8$. The additional states with $w=0$ are represented by the cyan line and the states with $|w|=1$ together with the ones with $|w|=2$ are represented by the orange curve.

\subsection{Exploring a slice of moduli space}
\label{sec:A1R}

In this section we present a detailed analysis of the slice of moduli space for compactifications of the heterotic theory on a circle at any radius and Wilson line given by
\beq \label{A1only}
 A=(A_1,0_{15}) \ .
 \eeq
 The results of this section are displayed in figure \ref{fig:SO32E8AR_1}. Here we present the main ingredients of the calculations, and leave further details  to Appendix 
\ref{app:enhancement}. 

For this type of Wilson line,  the states with $w=0$ (sector 1) that survive, are those satisfying
\beq
\pi_1 A_1 \in {\mathbb Z}\, .
\eeq 
This preserves all the roots only if $A_1 \in \mathbb{Z}$ for the $\Gamma_{16}$ case, or $A_1 \in 2\mathbb{Z}$  for the $\Gamma_8 \times \Gamma_8$ case. These correspond to the horizontal orange lines in figure \ref{fig:SO32E8AR_1}, where at any generic radius, the gauge symmetry is $U(1) \times SO(32)$, or $U(1) \times E_8 \times E_8$.
If  $A_1$ is an odd number, then the  $SO(32)$ symmetry is unbroken, but the $E_8 \times E_8$ is broken to  $SO(16)\times E_8$, which is depicted with a black line at $A_1=1$ in figure \ref{fig:E8AR_1}.
 
 If $A_1 \notin \mathbb{Z}$, then we have just the roots with $\pi_1 = 0$. That is, the $420$ roots of $SO(30)$ or the $324$ roots of $SO(14)\times E_8$. This corresponds to the white regions in figure \ref{fig:SO32E8AR_1}.  

Now, depending on the value of $R$, we can have additional states in sector 2, i.e. states with non-zero winding\footnote{From now on we take $w>0$, keeping in mind that for every massless state with $w$ there is also a massless state with $-w$.} which momenta satisfy $|{\bf p_L}|^2=2$ and have a quantized momentum number on the circle. Then,  according to \eqref{pLS1} and (\ref{condsector2}), they should obey
\beq \label{condi2}
|\pi + w A|^2 = 2(1-w^2 R^2)\, ,\\
\frac{1}{w}\left(1- \tfrac12|\pi|^2\right )  \in \mathbb{Z} \  .
\eeq 
The first equation implies  $R^{-1}\geq w$, and the simplest solution is
\beq 
\pi = \left (\pm\sqrt{2(1-w^2R^2)} - wA_1,0_{15} \right) \, .\nn 
\eeq  
But $\pi$ is in an even lattice, which implies $\pi_1 = -2q$, $q \in \mathbb{Z}$. The quantization condition for $n$ yields
\beq 
\tfrac{2q^2 - 1}{w} \in \mathbb{Z} \, ,\nn 
\eeq 
so  we have only the  winding numbers that are divisors of the numbers that can be written as $2q^2 - 1$, for some integer $q$.
In terms of $q$, the Wilson lines are of the form
\beq \label{awq}
A_1  = \frac{2q \pm \sqrt{2-2w^2R^2} }{w} \equiv a_{w,q}(R) \ , \quad  \{ w, q, \tfrac{2q^2 - 1}{w} \}  \in {\mathbb Z}\, .
\eeq 

If the radius also satisfies $R<\frac{1}{\sqrt{2}w}<\frac1w$,  we have additional solutions where some of the other components of $\pi$ are non-zero, such that
\beq 
\pi + wA =& \left(\pm\sqrt{1-2w^2R^2},\underline{\pm1,0_{14}} \right) \text{ for $\Gamma_{16}$}\, ,  \nn \\
\pi + wA =& \left(\pm\sqrt{1-2w^2R^2},\underline{\pm1,0_{6}},0_8 \right) \text{ for $\Gamma_{8}\times \Gamma_8$  } \, .\nn 
\eeq 
The quantization conditions are the same as before, but now the Wilson lines have the following behavior as a function of the radius
\beq \label{bwq}
A_1 = \frac{2q+1 \pm \sqrt{1-2w^2R^2} }{w} \equiv b_{w,q}(R) \ , \quad  \{ w, q, \tfrac{2q^2 - 1}{w} \}  \in {\mathbb Z}\, .
\eeq 
 
If additionally $R< (2\sqrt{2}w)^{-1}$ we have yet other possible solutions, but only for the $E_8 \times E_8$ theory, where
\beq 
\pi + wA =& \left (\pm  \tfrac12\sqrt{1 - 8w^2R^2},\underline{(\pm \tfrac12)_{7}},0_8 \right) \text{ for $\Gamma_{8}\times \Gamma_8$. } \nn 
\eeq 
The lines and quantization conditions are:
\beq \label{cwq}
A_1 = \frac{ q+\tfrac12 \pm  \sqrt{\tfrac14 - 2w^2R^2}}{w} \equiv  c_{w,q}(R) \ , \quad  \{ w, q, \tfrac{q(q + 1)}{2w} \}  \in {\mathbb Z}\, ,
\eeq 
where we used $(\pi_1)^2  = |\pi|^2 - \tfrac{7}{4}$ 
and $\pi_1 = -\left ( q + \tfrac12\right )$.

For a given $q$ and $w$, whenever the Wilson line is of the form $a_{w,q}$ in \eqref{awq}, we get $2$ massless states (one for $w>0$ and another one for $w<0$). If there are no more states, then we have enhancement to $U(1) \times SU(2)\times SO(30)$ and $U(1) \times SU(2) \times SO(14) \times E_8$. These correspond to the blue lines in figure \ref{fig:SO32E8AR_1}, where for example in figure \ref{fig:SO32AR_1}, the long blue line going from $(R,A_1)=(0,\sqrt 2)$ to $(1,0)$ corresponds to $a_{1,0}=\sqrt{2(1-R^2)}$, while its mirror one along the axis $A_1=1$ is $a_{1,1}=2-a_{1,0}$. 

For Wilson lines of the form $b_{w,q}$ in \eqref{bwq}, we get $60$ extra states for the $\Gamma_{16}$, and $28$ for $\Gamma_8 \times \Gamma_8$. The former promote the enhancement to $U(1) \times SO(32)$, while the latter to $U(1) \times SO(16) \times E_8$, and they correspond respectively to the orange lines in figure \ref{fig:SO32AR_1} and the black lines in figure \ref{fig:E8AR_1}. The largest curved orange line in the former and black line in the latter going from $(0,0)$ to $(0,2)$ corresponds to $b_{0,1}=1\pm\sqrt{1-2R^2}$, where the plus sign is for the upper half of the curve, and the minus sign for the lower half. 

Finally, Wilson lines of the form $c_{w,q}$ in 
\eqref{cwq} give in the $E_8 \times E_8$ heterotic theory, $2 \times 2^{6} = 128$ states (the sign of one of the seven $(\pm \tfrac12)$ is determined by the sign of the other $6$ and the sign chosen for the Wilson line).
Note that $c_{w,q}(R) = b_{2w,q}(R)$. It is not hard to show that 
a Wilson line that can be written as $c_{w,q}(R)$ can always be written as $b_{2w,q}(R)$, but the function $b$ can also have an odd $w$. 
Wilson lines $b$ that can also be written as $c $ bring then a total of $28+128=156$ states, which corresponds to the enhancement to $U(1) \times E_8 \times E_8$ in the orange lines of figure \ref{fig:E8AR_1}.

There are only two kinds of intersections between lines, and the points of intersection correspond to points of maximal enhancement (see Appendix \ref{app:enhancement} for details): 
\bi 
\item between a blue curve $a(R)$ with $w_1$ and an orange curve $b(R)$ with $w_2$, where the enhancement group is $SU(2) \times SO(32)$ ($SU(2) \times E_8 \times E_8)$ in the $SO(32)$ ($E_8 \times E_8$) theory. These are the red dots of figure \ref{fig:SO32E8AR_1}, and arise at 
\beq
(R,A_1)=\left(\frac{1}{\sqrt{w_1^2+2w_2^2}},\frac{2}{w_1} \left(q\pm w_2 R \right)\right)=\left(\frac1C,\frac{2k}{C}\right) \ ,  \nn 
\eeq
for some integer $k$, with $C=1,3,9,11,...$ are all the integers whose prime divisors are 1 or 3 (mod 8) (see Table 3).

\item between two blue $a(R)$ with $w_1$ and $w_2$ and two orange (black) curves $b(R)$ with $w_3$ and $w_4$, where the enhancement group is $SO(34)$ ($SO(18)\times E_8$) for the $SO(32)$  ($E_8\times E_8$) theory. These are the green dots of figure \ref{fig:SO32E8AR_1}, and arise at\footnote{We get additionally $R=\frac{1}{\sqrt{w_1^2+w_2^2}}=\frac{1}{\sqrt{2} \sqrt{w_3^2+w_4^2}}$.}  
\beq
(R,A_1)=\left(\frac{1}{\sqrt{w_1^2+w_2^2}},\frac{2}{w_1} \left(q\pm \tfrac{1}{\sqrt{2}}w_2 R \right)\right)=\left( 
\frac{1}{\sqrt2 C},\frac{k}{C} \right) \, , \quad  \  \nn 
\eeq
for some integer $k$, with $C=1,5,13,17,...$ are all the integers whose prime divisors are Pythagorean primes (see Table 3)
\ei

In Appendix \ref{app:enhancement} we give the details of the calculations and also prove that these are the only possible intersections for this type of Wilson lines. In Appendix \ref{app:extrafigures} we present other slices of moduli space given by the radius and Wilson lines determined by a single parameter $A$.  In section \ref{sec:36} we show how these points arise as fixed points of a duality symmetry. 

\subsection{T-duality in circle compactifications}
\label{sec:T-duality}

In this section we discuss the action of T-duality in the heterotic string compactified on a circle. By T-duality we mean the action of certain type of transformations 
in $O(1,17, {\mathbb Z})$ that relate a given heterotic theory with 16-dimensional lattice $\Gamma$, compactified on a circle of radius $R$ and Wilson line $A$, to another heterotic theory with lattice $\Gamma'$, compactified on a circle of radius $R'$ and Wilson line $A'$. 
In this section we discuss the usual T-duality exchanging momentum and winding numbers, while in the next section we discuss more general dualities, and their fixed points.  

The duality generated by the matrix $O_D$ is the usual T-duality transformation exchanging momentum and winding numbers
\beq
(w',n',\pi')=(n,w,\pi) \ .
\eeq
Since $\pi$ stays untouched, this duality is possible if $\Gamma'=\Gamma$. Its action on the background fields 
 can be worked out from the generalized metric
\eqref{G}, which for the circle is\footnote{Here we choose the Cartan-Weyl basis where the Killing metric for the Cartan subgroup $\kappa^{IJ}$ is diagonal.} 
\beq
\label{Gcircle}
{\cal M} =\begin{pmatrix} R^2 (1+\tfrac12 {\mathbb A}^2)^2 &-\tfrac12 {\mathbb A}^2 & (1+ \tfrac12 {\mathbb A}^2) A \\  - \tfrac12 {\mathbb A}^2 & \frac{1}{R^2} &-\frac{1}{R^2} \, A \\
(1+ \tfrac12 {\mathbb A}^2)\,  A^t & -\frac{1}{R^2} \, A^t & {\rm I} +\frac{1}{R^2} A^t A \end{pmatrix} \  , 
\eeq
where we have defined the scalar 
\beq \label{a}
{\mathbb A}^2\equiv \frac{|A|^2}{R^2}  \ .
\eeq
The action of $O_{D}$  transforms this into
\beq
{\cal M}'=O_D {\cal M} O^t_D={\cal M}^{-1}= \begin{pmatrix}  \frac{1}{R^2}  &-\tfrac12 {\mathbb A}^2 & -\frac{1}{R^2} \, A  \\  - \tfrac12 {\mathbb A}^2 &R^2 (1+\tfrac12 {\mathbb A}^2)^2 & (1+ \tfrac12 {\mathbb A}^2) A \\
-\frac{1}{R^2} \, A^t & (1+ \tfrac12 {\mathbb A}^2)\,  A^t &{\rm I}  +\frac{1}{R^2} A^t A \end{pmatrix}\, ,
\eeq
and thus we get
\bea \label{A'}
A'=-\frac{A}{R^2 (1+\tfrac12 {\mathbb A}^2)} \ , \qquad
R'=\frac{1}{R \, (1+\tfrac12 {\mathbb A}^2)} \quad (\, \Rightarrow \, \frac{A'}{R'}=-\frac{A}{R} \, )\nn \ 
\eea 
in agreement with the heterotic Buscher rules for scalars \cite{heteroticBuscher}. 
We get that a background has $R'=R$ for
\beq \label{Rsd}
R_{\rm sd}^2=1-\frac12 |A|^2 \  \quad (\, \Rightarrow \, R'=R\, , \, A'=-A \, )
\eeq 
Additionally, if $2A\in \Gamma'$, then $A'=-A\sim A$, and therefore the background is fully self-dual, satisfying  ${\cal M}={\cal M}^{-1}$ up to discrete transformations (these are of the form \eqref{theta}, \eqref{GLk} or \eqref{OLambda}, but for the circle the only non-trivial one is a $\Lambda$-shift \eqref{OLambda}). 

All the examples of 
enhancement discussed in section \ref{sec:enh-break} except for \ref{sec:E7} satisfy the self-duality condition \eqref{Rsd}. By perfoming a $\Lambda$-shift to the Wilson line of  \ref{sec:E7} we can bring it to the equivalent one $A=\left((-3/4)_2,(1/4)_6,0_8\right)$, which satisfies \eqref{Rsd}.

%
 
 For Wilson lines with only one non-zero component, we have that the fixed ``points" of this symmetry correspond actually to lines of non-maximal enhancement symmetry where the Wilson lines are functions of the radius ($A=A(R_{\rm{sd}})$), and are such that  $A\sim A_{\rm{sd}}$, with $|A_{\rm{sd}}|^2=2(1-R^2_{\rm{sd}})$.  
 
 We now discuss  the differences between fixed points of duality symmetries further, exploring more general dualities and their fixed points. 

\subsection{More general dualities and fixed points}
\label{sec:36}

The transformation $O_D$ discussed before is a particular type of transformation that changes the sign of $p_R$ while it rotates ${\bf p_L}$, preserving its norm  (in compactifications of the bosonic theory on a circle, $p_L$ has a single component and $O_D$ just leaves it invariant, but in the heterotic theory $O_D$ rotates the 17-dimensional vector ${\bf p_L}$). It would be very interesting to understand what are all the possible transformations that do this, and obtain their fixed points. Here we do something more modest, namely 
we work out the set of transformations that change the sign of $p_R$ and rotate ${\bf p_L}$, leaving its circle direction component invariant. We thus require
\beq \label{pp}
(p_L', p'^A, p'_R) =(p_L, U^{AB}p^B, -p_R) 
\eeq
with $U \in O(16,{\mathbb Z})$. These transformations generically link a given heterotic theory with lattice $\Gamma$, in a background defined by $(A,R)$ to another heterotic theory with lattice $\Gamma'$ in a dual background with $(A',R')$. The duality transformation depends on the matrix $U$ and we use a convenient parameterization to relate the radii $R$ and $R'$, namely we define a positive number $r$ such that 
\beq \label{r}
R'=\frac{1}{rR} \ .
\eeq
The duality transformation that achieves \eqref{pp} should have the form
\beq \label{O}
 O_{U}=\begin{pmatrix}
-\frac{r|A'|^2}{2} & \frac{1}{r}+A' U A^t +\frac{r|A|^2}{2}\frac{|A'|^2}{2} & \frac{r|A'|^2}{2} A + A' U \\
        r          & -\frac{r|A|^2}{2}                                      &  -rA \\
  -r A'^t       &   U A^t  +\frac{r|A|^2}{2}A'^t        & U+r A'^t A\end{pmatrix} \ .
\eeq
Requiring this to be in $O(1,17;{\mathbb Z})$, we get a set of quantization conditions like for example\footnote{The fact that we get a quantization condition for $|A|$ may sound strange, but it means that if $A$ is not quantized properly there is no duality that leaves the circle direction of ${\bf p_L}$ invariant. If one allows the full ${\bf p_L}$ vector to rotate under the transformation, then we have, as shown, at least the duality $O_D$ discussed in previous section.} (the full set of quantization conditions is given in \eqref{quantOfull})
\beq \label{quantcond}
r  \, ,\, \frac{r|A|^2}{2} \, ,\, \frac{r|A'|^2}{2} \, ,\, \frac{1}{r} + A' U A + r\frac{|A|^2}{2}\frac{|A'|^2}{2} \in \mathbb{Z} \, .
\eeq
It is instructive to decompose the matrices $O_U$  as the product $O_{\Lambda'}O_D O_N O_MO_{\Lambda}$ with $\Lambda=-A, M= r$, $N = U$ and $\Lambda'=A'$,
which allows to interpret the transformations as the following series of operations
\begin{enumerate}
\item $O_{-A}$: eliminates the Wilson line $A$ through a  $\Lambda$-shift,
\item $O_{r}$: rescales $R \rightarrow rR$,
\item $O_{U}$ performs a change of basis in the heterotic directions
\item $O_{D}$: performs a T-duality along the circle ($\eta$),
\item $O_{A'}$: adds the  Wilson line $A'$ through a $\Lambda$-shift.
\end{enumerate}



We divide the discussion into the dualities where $\Gamma=\Gamma'$, and those where the dual lattice is not the original one. To denote the different sublattices that will play a role, it is useful to use the $(0)$, $(v)$, $(s)$ and $(c)$ conjugacy classes of $SO(16)$, corresponding respectively to the root, vector, positive and negative-chirality spinor classes. These are defined in \eqref{a1}-\eqref{a4}. The lattices $\Gamma_{16}$ and $\Gamma_8 \times \Gamma_8$ contain the following vectors (see \eqref{Gamma16Gamma8}-\eqref{decompositionGammas})
\beq \label{decompositionGammast}
 \Gamma_{16} &=(00),(vv),(ss),(cc) \\
\Gamma_8 \times \Gamma_8&=(00),(ss),(0s),(s0) 
\eeq   
One could have chosen different conventions in which some of the $s$ classes are turned into $c$ classes, and doing that build four other lattices, that we denote $\Gamma_{16}^-$,  $\Gamma_8^- \times \Gamma_8^-$, $\Gamma_8^- \times \Gamma_8^+$ and $\Gamma_8^+ \times \Gamma_8^-$. We give these in \eqref{decompositionGammasminus}. Note that a lattice $\Gamma^+$ is equivalent  to a lattice $\Gamma^-$, the choice $(s)$ versus $(c)$ conjugacy class is a convention with no physical relevance. Here it is important however to make the distinction whether a given duality maps, say, $\Gamma^+$ to $\Gamma^+$, or $\Gamma^+$ to $\Gamma^-$.

In the following we write the main results, leaving the details to Appendix \ref{app:B6}. The results for generic Wilson lines, assuming that $r$ is a prime number, are summarized in Table 4. We later concentrate on the situation where the Wilson lines are of the form \eqref{A1only}, i.e. with  only one non-zero component, as we did in section \ref{sec:A1R}, to see what happens when the assumption that $r$ is prime is relaxed. For Wilson lines of this form, the $O(16)$ symmetry is broken to $O(15)$, and there are four inequivalent choices of $U$ that we will analyze in detail
\beq \label{Upm}
U = \pm I \  \quad { \rm or} \quad  
U_{\pm} \equiv \pm \text{diag}(1,-1_{15}) \ .
\eeq

\subsubsection{$\Gamma \leftrightarrow \Gamma$}
\label{sec:gamma=gamma'}

The dualities for which the lattice does not change involve those where $\pi$ is invariant, such as the one discussed in the previous section. But as explained above, one can have more general dualities even when $\Gamma'=\Gamma$, and thus more general fixed points. Fixed points of a duality are those for which $R'=R$ and $A'=A$.\footnote{One could also consider a more general situation where $A'+\Lambda'=A+\Lambda$ with $\Lambda (\Lambda') \in \Gamma (\Gamma')$. Since here $\Gamma=\Gamma'$, then $A \sim A'$. Since we are considering $\Lambda$-shifts as part of the duality transformations, we can restrict without loss of generality to dualities where $A'=A$.} 

To make the analysis tractable for generic Wilson lines, we restrict to the situation where $r$ is a prime number and $U=I$, and relax this assumption only in the setup where the Wilson lines have just one non-zero component. Under the assumption that $r$ is a prime number,
the full set of quantization conditions \eqref{quantOfull} are satisfied if and only if 
(see details in Appendix \ref{app:B6})
\beq
A \in \Gamma \, , \, A' \in \Gamma \, , \quad r =1  \ ,
\eeq
and thus the fixed points of these transformations are at $R=1$ and $A$ any point in the lattice $\Gamma$. They correspond to enhancements to $SU(2) \times SO(32)$ and $SU(2) \times E_8 \times E_8$ discussed in section \ref{sec:enhancement}.
These points appear in the diagonal entries in Table 4. 

Let us now analyze in more detail the fixed points of the dualities for the subset of Wilson lines  of the form \eqref{A1only}, i.e.  with only one non-zero component.
The quantization conditions evaluated at the fixed points turn into (see Appendix \ref{app:B6} for details of the calculation)
\beq 
n \, ,  \,   m \, ,  \frac{2n^2\pm 1}{m} \in \mathbb{Z}  \quad {\rm for} \quad U=\pm I  \quad {\rm  where} \ n =\tfrac12 A_1R^{-1}, m=R^{-1}, \label{condiciondual1t} 
\eeq 
 and  
\beq 
n \, ,  \,   m \, ,  \frac{n^2 \pm 1}{2m} \in \mathbb{Z} \quad {\rm for} \quad U=U_{\pm} \quad{\rm where \ now} \ n = \tfrac1{\sqrt 2} A_1R^{-1}, m=\tfrac{1}{\sqrt{2}}R^{-1}.
 \, \label{condiciondual2t}
\eeq 







We write in Table 3 all the fixed points for $U=I$ and $U=U_+$ where $0 \leq A_1 \leq 1$. The lines $\pm A_1 \text{ mod }2$ are also fixed points\footnote{The other two options  $U=-I$ and $U=U_-$ do not leave the Wilson line invariant. The fixed points of these dualities  are the points where the positive and negative branches of the curves $a(R)$ and $b(R)$, defined in \eqref{awq} and \eqref{bwq}, intersect. These are the points where the arguments in the square roots  are zero. Most of these points do not correspond to points of maximal enhancement. Those that do correspond to $-I A=U_{-} A=-A \sim A$, which are also fixed points for $U=I$ or $U=U_-$.} .

\begin{center}
\boxed{
\begin{array}{c|c}
 U=\text{diag}(1_{16})  & U=\text{diag}(1,-1_{15}) 
 \\
\hline 

\begin{array}{c|c}
R^{-1} & A_1 \\
\hline
 1 & 0 \\
  {3} &  \frac{2}{3} \\
  {9} &  \frac{4}{9} \\
  {11} &  \frac{8}{11}  \\
  {17} &  \frac{10}{17}  \\
  {19} &  \frac{6}{19} \\
  {27} &  \frac{22}{27}  \\
  {33} &  \frac{8}{33},\frac{14}{33}  \\
  {41} &  \frac{30}{41} \\
  {43} &  \frac{16}{43}  \\
  {51} &  \frac{10}{51},\frac{44}{51}  \\
  {57} &  \frac{32}{57},\frac{44}{57}   \\
  {59} &  \frac{36}{59}   \\
  {67} &  \frac{20}{67}  \\
  73 & \frac{12}{73} \\
  81 & \frac{22}{81} \\
  83 & \frac{74}{83} \\
  89 & \frac{40}{89} \\
  97 & \frac{80}{97} \\
  99 & \frac{14}{99},\frac{58}{99} 
 \end{array}

&

\begin{array}{c|c}
R^{-1} & A_1 \\
\hline
  {\sqrt{2}} & 1 \\
  {5 \sqrt{2}} &  \frac{3}{5} \\
  {13 \sqrt{2}} &  \frac{5}{13} \\
  {17 \sqrt{2}} &  \frac{13}{17}\\
  {25 \sqrt{2}} &  \frac{7}{25} \\
  {29 \sqrt{2}} &  \frac{17}{29}\\
  {37 \sqrt{2}} &  \frac{31}{37}\\
  {41 \sqrt{2}} &  \frac{9}{41} \\
  {53 \sqrt{2}} &  \frac{23}{53} \\
  {61 \sqrt{2}} &  \frac{11}{61} \\
  {65 \sqrt{2}} &  \frac{47}{65},\frac{57}{65} \\
  73 \sqrt{2} & \frac{27}{73}   \\
 85 \sqrt{2} &  \frac{13}{85},\frac{47}{85}   \\
 89 \sqrt{2} &  \frac{55}{89}   \\
 97 \sqrt{2} &  \frac{75}{97}   \\
 101 \sqrt{2} &  \frac{91}{101}   \\
 109 \sqrt{2} &  \frac{33}{109}   \\
 113 \sqrt{2} &  \frac{15}{113}   \\
 125 \sqrt{2} &  \frac{57}{125}   \\
 137 \sqrt{2} &  \frac{37}{137}   \\
 145 \sqrt{2} &  \frac{17}{145},\frac{133}{145}   
\end{array} \\
\hline
\textcolor{red}{SU(2) \times SO(32)}  & \textcolor{green}{SO(34)}   \\
{\rm or} & {\rm or}    \\
\textcolor{red}{SU(2) \times E_8 \times E_8}  & \textcolor{green}{SO(18) \times E_8}  
\end{array}
}\\
\vspace{0.3cm}
Table 3: Fixed points of the dualities $O_U$.
\end{center}

These points in moduli space are points of maximal enhancement symmetry. Those in the first column give rise to $SU(2)\times SO(32)$ for $\Gamma_{16}$ or $SU(2)\times E_8 \times E_8$ for $\Gamma_8 \times \Gamma_8$, and are depicted by red dots in figure \ref{fig:SO32E8AR_1}. The second column contains all the points of maximal enhancement groups $SO(34)$ or $SO(18)\times E_8$, and correspond to the green dots in figure \ref{fig:SO32E8AR_1}.


\subsubsection{$\Gamma \leftrightarrow \Gamma'$ }
\label{sec:gamaneqgamma'}

Note that unless $\Gamma=\Gamma_{16}^\pm$ and $\Gamma'=\Gamma_8^\pm \times \Gamma_8^\pm$ (or the other way around, and using any combination of signs) $-$ situations that we analyze separately in the next section $-$ there exists some $U_1 \in O(16, {\mathbb Z})$ such that $\Gamma'=U_1 \Gamma$. In that case, the duality with $\Gamma' \neq \Gamma$, $U_2$ and $A'$ is equivalent to one between $\Gamma$ and $\Gamma'' = \Gamma$, $U'' = U_1 U_2$ and has $A''=A' U_1$. Restricting to diagonal matrices $U$
, we see that the dualities with $U$ and $\Gamma'=\Gamma$ are equivalent to the dualities with $U=I$ but where $\Gamma'$ is
\bea
&&\Gamma'=\Gamma^{\pm}_{16} \text{ for } \Gamma=\Gamma_{16} \text{ and det}(U)=\pm 1\\
&&\Gamma'=\Gamma^{\pm_1}_8\times \Gamma^{\pm_2}_8 \text{ for } \Gamma=\Gamma_8\times \Gamma_8 , \text{ det}_1(U)=\pm_1 1 \text{ and det}_2(U) = \pm_2 1  
\eea
where $\text{det}_1$ ($\text{det}_2$) is the product of the $8$ first (last) diagonal elements and the lattices $\Gamma^{\pm}$ are defined in Appendix \ref{app:lattices}. 
If additionally the Wilson line $A$ is invariant under the action of $U$ (up to a $\Lambda$-shift)
we get exactly the same fixed points that one gets for a duality with $\Gamma=\Gamma'$.  
Since Wilson lines of the type \eqref{A1only} are invariant under the action of a diagonal $U$ such that the first component is +1, we get the same fixed points of section \ref{sec:gamma=gamma'} that correspond to enhancement
to $SO(34)$ or $SO(18)\times E_8$.

Under the assumption that $r$ is a prime number, the quantization conditions are satisfied if and only if 
\beq \label{conlgammagamma'}
A \in (\Gamma\cap&\Gamma')^{*}\backslash \Gamma \, , \, A' \in (\Gamma\cap\Gamma')^{*}\backslash \Gamma' \, , \, r=2 \eeq
and thus the fixed points of these transformations are at $R=\tfrac{1}{\sqrt{2}}$, and correspond to the enhancements 
$SO(34)$ and $SO(18) \times E_8$. The possible Wilson lines for the different choices of $\Gamma$ and $\Gamma'$ are 
given in Table 4.

\subsubsection{$SO(32) \leftrightarrow E_8\times E_8$}
\label{sec:SO32toE8E8}

There is no $U\in O(16,{\mathbb Z})$ that transforms the lattices $\Gamma_{16}$ and $\Gamma_8 \times \Gamma_8$ into each other, and thus the case $\Gamma = \Gamma_{16}$ and $\Gamma' = \Gamma_8 \times \Gamma_8$ is different from the ones considered previously. 

Here, for simplicity, we restrict to $U=1$, namely we analyze dualities such that $({\bf p_L}',p'_R)=({\bf p_L},-p_R)$. The quantization conditions under the assumption that $r$ is a prime number, are given in \eqref{conlgammagamma'}. For $\Gamma=\Gamma_{16}$ and $\Gamma'=\Gamma_8 \times \Gamma_8$, the possible Wilson lines are the following
\beq \label{WL}
A \in (\Gamma\cap\Gamma')^{*}\backslash \Gamma = (0s),(s0),(vc),(cv)\\
A' \in (\Gamma\cap\Gamma')^{*}\backslash \Gamma' = (vv),(cc),(vc),(cv) \\
\eeq
However, there is something very curious here: the fixed points of these dualities, corresponding to $R=\frac{1}{\sqrt{2}}$, are not points of maximal enhancement but points of enhancement $U(1) \times SO(16) \times SO(16)$. Furthermore, this enhancement group arises at any radius, so Wilson lines of the form \eqref{WL} give rise to lines in moduli space, and as such are also ``fixed points" of dualities that do not involve $O_D$. 

Let us illustrate this better with an example:
Take $A=((\frac12)_8,0_8)\in (s0)$ and $A'=(1,0_7,1,0_7)\in (vv)$. For the time being, we take $r=2$, i.e. $R'=1/(2R)$, but we do not necessarily stand at the self-dual radius. 

The Wilson line $A$ breaks the $SO(32)$ gauge symmetry to 
$SO(16)\times SO(16)$, as shown in section \ref{sec:SO(32-2p)SO(2p)}.
For this Wilson line, one has additionally states which are neutral under $SO(16)\times SO(16)$, i.e. with $p^A=0$. Since these should have $\pi=-w A$, then only states with
$w=2m$, $m\in\mathbb Z$ are allowed. These states have left and right-moving momenta on the circle
\bea
p_L=\frac1{\sqrt{2}R}\left(\tilde n+2R^2m\right)\, , \qquad
p_R=\frac1{\sqrt{2}R}\left(\tilde n-2R^2m\right)\, , \label{so32}
\eea
where $\tilde n=n+w$. 
Let us pause for a second to show that there is no enhacement to $SU(2) \times SO(16) \times SO(16)$ with this Wilson line. We have shown in section \ref{sec:SO(32-2p)SO(2p)} that there are no additional massless states charged under $SO(16) \times SO(16)$, i.e. with non-zero winding number and $p^A\neq 0$. Regarding extra neutral massless states, it is very easy to see from \eqref{so32} that there are none of this form: states with momenta $(p_L,p^A,p_R)=(\sqrt{2},0,0)$, satisfy $2R^2m=\tilde n$, while requiring at the same time $p_L=\sqrt{2}$ would lead to $\tilde n m=\tfrac12$, which has no solution. Thus, the compactification of the $SO(32)$ heterotic string with Wilson line $A=((\frac12)_8,0_8)$ leads to $U(1) \times SO(16) \times SO(16)$ at any radius.  

The Wilson line $A'=(1,0_7,1,0_7)$ breaks the $E_8 \times E_8$ symmetry also to
$SO(16) \times SO(16)$. There are also states which are neutral under $SO(16)\times SO(16)$, of the same form as before, i.e. with momenta 
\bea 
p'_L=\frac1{\sqrt{2}R'}\left(\tilde n'+2R'^2m'\right)\, , \qquad
p'_R=\frac1{\sqrt{2}R'}\left(\tilde n'-2R'^2m'\right)\, , \label{e8}
\eea
where $w'=2m'$ and $\tilde n'=n'+w'$.

Comparing \eqref{e8} and (\ref{so32}), we see that $(p'_L,p_R)=(p_L,-p_R)$ if
$(\tilde n', m')=(m, \tilde n)$ and $RR'=\frac12$. This is true for any value of $R$.

\vspace{1cm}

In the following table we write the fixed points of the dualities between a theory with lattice $\Gamma$ (row) and another one with $\Gamma'$ (column) for the smallest value of the parameter $r$ defined in \eqref{r}, which are $r=1$ or $r=2$. We indicate the conjugation classes of the possible Wilson lines (for a given row and column, any $A$ given can be dualized to any $A'$), and the enhancement group arising at the fixed point of the duality.

\hspace{-6em}
\begin{tiny}
\begin{tabular}{|l|c|c|c|c|c|c|}
\hline
& $\Gamma_{16}$ & $\Gamma^{-}_{16}$ & $\Gamma_{8} \times \Gamma_{8}$& $\Gamma_{8} \times \Gamma^-_{8}$ & $\Gamma^-_{8} \times \Gamma_{8}$ & $\Gamma^-_{8} \times \Gamma^-_{8}$\\
\hline
$\Gamma_{16} $ &$SO(32)\times SU(2)$& $SO(34)$ & $SO(16)\times SO(16)\times U(1)$&-&-&$SO(16)\times SO(16)\times U(1)$\\
  &$R=1$& $R=1/\sqrt{2}$ & $R=1/\sqrt{2}$ & & & $R=1/\sqrt{2}$\\
    &$A=00,ss,vv,cc$& $A=0v, v0, cs, sc$ &$A=vc, cv,0s, s0$ & & &$A=vs,sv,0c,c0$\\
      &$A'=00,ss,vv,cc$&$A'=0v, v0, ss, cc$ &$A'= vc, cv,vv, cc$ & & &$A'=vs,sv,vv,ss$\\
\hline
$\Gamma^{-}_{16}$ &  &$SO(32)\times SU(2)$&-&$SO(16)\times SO(16)\times U(1)$&$SO(16)\times SO(16)\times U(1)$& -\\
 & &$R=1$& & $R=1/\sqrt{2}$ & $R=1/\sqrt{2}$&\\
  &  &$A=00,ss,sc,cs$& &$A=vs, cv,s0, 0c $  &$A=sv, vc,0s, c0 $ &\\
  &    &$A'=00,ss,sc,cs$&  &$A'=vs, cv,cs, vv $ &$A'=sv, vc,sc, vv $ &\\
\hline
 $\Gamma_{8} \times \Gamma_{8}$&  & &$E_8 \times E_8 \times SU(2)$&$E_8 \times SO(18)$ & $SO(18) \times E_8$ & \text{-}\\
& & &$R=1$& $R=1/\sqrt{2}$& $R=1/\sqrt{2}$ & \\
&  &  &$A=00,ss,0s,s0$&$A=0v,sv,0c,sc$ & $A=v0,vs,c0,cs $ & \\
&  &    &$A'=00,ss,0s,s0$&$A'=0v,sv,0s,ss$  & $A'=v0,vs,s0,ss $& \\
\hline
$\Gamma_{8} \times \Gamma^-_{8}$&  & & &$E_8 \times {E_8}^{-} \times SU(2)$& - &$SO(18)\times {E_8}^{-}$\\
&& & &$R=1$ &  & $R=1/\sqrt{2}$\\
& & &  &$A=00,sc,0c,s0$&   & $A=v0, vc, c0, cc$ \\
&  &&    &$A'=00,sc,0c,s0$  & & $A'=v0, vc, s0, sc$\\
\hline
 $\Gamma^-_{8} \times \Gamma_{8}$&  & & & &${E_8}^{-} \times E_8  \times SU(2)$&${E_8}^{-}\times SO(18)$\\
&&& & &$R=1$ & $R=1/\sqrt{2}$ \\
&& & &  &$A=00,cs,0s,c0$&  $A=0v, cv, 0c, cc$  \\
&&  &&    &$A'=00,cs,0s,c0$   &$A'=0v, cv, 0s, cs$ \\
\hline
 $\Gamma^-_{8} \times \Gamma^-_{8}$&  & & & & &${E_8}^{-} \times {E_8}^{-} \times SU(2)$\\
&&&& & &$R=1$   \\
&&& & &  &$A=00,cc,0c,c0$  \\
&&&  &&    &$A'=00,cc,0c,c0$    \\
\hline
\end{tabular}
\end{tiny}
\begin{center} Table 4: Points of symmetry enhancement as fixed points of duality symmetries  
\end{center}

\section{Effective action and Higgs mechanism}

Now that we saw the rich structure of duality symmetry, we turn to its explicit target
space realization. 
The global duality symmetry of the dimensionally reduced heterotic supergravity action   has been deeply investigated in the seminal papers by J. Maharana and J. Schwarz \cite{ms} and N. Kaloper and R. Myers \cite{km},
   and more recently in \cite{hsz}. If the gauge fields are truncated to the Cartan subsector
of the $E_8\times E_8$ or $SO(32)$  gauge group, the dimensional reduction of heterotic supergravity from 10 to $10 - k$ dimensions produces a theory  with $U(1)^{2k+16}$ abelian gauge symmetry  and  a continuous global $O(k,k+16;\mathbb R)$ symmetry.  If the reduction includes the full set of $E_8\times E_8$ or $SO(32)$ gauge fields and no Wilson lines, the global symmetry  reduces to $O(k,k;\mathbb R)$, while a compactification with Wilson lines for the Cartan gauge fields of a rank $16-r$ subgroup of the rank 16 gauge group $G_L$, gives an effective field theory with
global $O(k,k + 16-r;\mathbb R)$ duality symmetry \cite{hsz}. The analysis of  \cite{hsz}
is based
on string-theoretic arguments and holds to any order in the $\alpha'$ expansion of the heterotic string effective field theory action involving all the massless string states, except those that become massless at self-dual points of the moduli space.

Including the massless states
with nonzero winding or momentum number on $T^k$  in  the effective field theory of the toroidally compactified heterotic string  is not difficult, as it  is a gauged supergravity. The action
 with at most two derivatives
of the massless fields is then completely determined by the gauge group.  Therefore, although the field theoretical Kaluza-Klein  reduction of heterotic supergravity cannot describe
 the  string modes that give rise to maximally enhanced gauge symmetry,  the action is entirely fixed.

Nevertheless, we will see in the forthcoming sections that the explicit construction of  the (toroidally compactified) heterotic string  effective action  from the scattering amplitudes of massless string modes at self dual points of the moduli space, and its  manifestly duality-covariant reformulation, give  important information. In particular, 
we will obtain novel relations between the $SO(32)$ and $E_8\times E_8$ theories. 
We will also consider  the light states that acquire mass when slightly perturbing the background fields and  revisit the gauge symmetry breaking and Higgs mechanism, both from the field theory and the string theory  viewpoints.

\subsection{Effective action of massless states}

The three-point functions of all the (toroidally compactified) heterotic string massless vertex operators  are reviewed in Appendix \ref{app:3point}, where we also compute the four point function of the massless scalars. 
 These amplitudes are reproduced from the  S-matrix of the following effective action 
\bea
S&=&\frac{1}{{2\kappa_d^2}} \int d^d x {\sqrt {-G}}e^{-2{\varphi}}\left ({ R}+ 4\partial_{\mu}\varphi \partial^{\mu}\varphi  - \frac1{12 }H_{\mu\nu\rho}H^{\mu\nu\rho}  -\frac{1}{4 } F_{\mu\nu}^{ \Gamma}F^{ \mu\nu}_{\Gamma} -\frac{1}{4 }\bar F_{\mu\nu}^{ m} \bar F^{ \mu\nu}_{m}
 \right.\nn\\
 &&\ \  \ \ \ \ \ 
-\frac14D_{\mu} {\rm S}^{ mn} D^{\mu} {\rm S}_{ mn}
  -\frac1{{2}}  {S}_{\Gamma m}F^{ \Gamma}_{\mu\nu} \bar F^{ m\mu\nu}    - \ \frac14 \
 {\rm S}_{\Gamma m} {\rm S}_{\Gamma'}{}^m {\rm S}_{\Lambda n} {\rm S}_{\Lambda'}{}^n f^{\Gamma \Lambda \Pi}f^{\Gamma' \Lambda'}{}_{\Pi} \Bigg )  , \ \ \ \ \  \label{accionefectiva}
\eea
which also contains terms from higher point functions  that we have not computed but need to be included on the basis of gauge symmetry.
Here $\kappa_d$ is the effective Planck   coupling constant (related to the  gauge coupling $ g_d$  as $ g_d=\sqrt2 \kappa_d$) and \footnote{
We have rescaled the polarizations introduced in Section 2 as  $
G_{\mu\nu}=\eta_{\mu\nu}+2\kappa_d\epsilon_{(\mu\nu)} \longrightarrow e^{-\frac{2 \kappa_d}{\sqrt{d-2}}D} G_{\mu\nu}$, $\epsilon_{[\mu\nu]}\longrightarrow \frac{B_{\mu\nu}}{2\kappa_d}$, $A_\mu^\Gamma\longrightarrow \frac{A_\mu^\Gamma}{g_d}$, $\bar A_\mu^m\longrightarrow \frac{\bar A_\mu^m}{g_d}$, ${ S}_{mn}\longrightarrow \frac{{S}_{mn}}{2\kappa_d}$, ${ S}_{Im}\longrightarrow \frac{{ S}_{Im}}{g_d}$, $S_{\alpha m}\longrightarrow \frac{S_{\alpha m}}{g_d}$. We also redefined the dilaton $
D =\frac{2 }{\kappa_d\sqrt{d-2}}(\varphi-\varphi_0)$, so that
 $ \kappa_d \longrightarrow e^{-\varphi_0}\kappa_d$ and $  g_d\longrightarrow e^{-\varphi_0}g_d$.
} 
\bea 
H_{\mu\nu\rho}&=&3\left(\partial_{[\mu}B_{\nu\rho]}+A^\Gamma_{[\mu}\partial_\nu A_{\rho]\Gamma}+\frac13 f_{\Gamma\Lambda\Omega}A_\mu^\Gamma A_\nu^\Lambda A_\rho^\Omega -\bar A^m_{[\mu} \partial_\nu \bar A_{\rho ]m}\right)\, ,\label{h}\\
F_{\mu\nu}^\Gamma &=&\partial_{\mu}A_{\nu}^{\Gamma} -\partial_{\nu}A_{\mu}^{\Gamma} +  f^{\Gamma}{}_{\Lambda\Omega}A_{\mu}^{\Lambda} A_{\nu}^{\Omega}\, ,
\quad \bar F_{\mu\nu}^m =\partial_{\mu} \bar A_{\nu}^m-\partial_{\nu} \bar A_{\mu}^m \, ,\nn\\
D_\mu {\rm S}^{\Gamma m}&=&\partial _\mu {\rm S}^{\Gamma m}+f^{\Gamma}{}_{\Lambda\Omega}A_\mu^{\Lambda} {\rm S}^{\Omega m}\, , \label{accionefectivadef}
\eea
with
${\rm S}_{\Gamma m}=({\rm G}_{mn}, {\rm B}_{mn},{\rm A}_{I m}, {\rm A}_{\alpha m})$ denoting the scalar fields. The indices $m,n=1, \dots, k$ correspond to the dimensions on $T^k$ and $\Gamma, \Lambda=1,...,N$ are the adjoint indices of the Lie algebra associated to the gauge group $G_L$ of dimension $N$ and structure constants  $ f^{\Gamma}{}_{\Lambda\Omega}$.

For ten external dimensions (i.e. when there are no compact internal dimensions other than the 16 chiral ``heterotic''  ones), $d=10$, the gauge group is  $E_8\times E_8$ or $SO(32)$ and  $N=496$. There are neither scalar ${\rm S}_{\Gamma m}$ nor   vector $A_\mu^m, \bar A_\mu^m$  fields. Then the action reduces to the first four  terms in \eqref{accionefectiva}, with $\Gamma=(I,\alpha)=1, \dots, 496$, and the last term in $H_{\mu\nu\rho}$ vanishes. 

For compactifications on $T^k$, $d=10-k$, at generic values of the background fields, the gauge group is $U(1)^{16+k}_L\times U(1)_R^k$,  $N=16+k$, and the index $\Gamma \equiv \hat I=1, \dots , 16+k$. The vectors and scalars are only those in sector 1 of section \ref{sec:masslessspectrum}. We denote the gauge fields as the polarization vectors in the vertex operators $( A_\mu^m, A_\mu^I, \bar A_\mu^m)$ and the scalar fields are ${\rm G}_{mn}={G}_{mn} +{S_{(mn)}(x)}$, ${\rm B}_{mn}=B_{mn} +{S_{[mn]}(x)}$, ${\rm A}_{I m}=A_{Im} +{S_{I m}(x)}$, where  the fluctuations are  denoted like the polarizations of the vertex operators creating the string scalar states. In this case,  (\ref{accionefectiva}) agrees with the effective action obtained  in \cite{ms} from dimensional reduction of heterotic supergravity with gauge group  truncated to the Cartan subgroup. The theory  has a global $O(k, k+16; \mathbb R)$ symmetry. 

At the specific points in moduli space where the gauge symmetry is enhanced,  it is convenient to split the index $\Gamma=({\hat I, \alpha=\overline \alpha, \underline \alpha})$, where $\hat I=1,...,16+k$ denotes the Cartan generators and $\underline \alpha$ $ (\overline \alpha)$ are the positive (negative) roots  of $G_L$. The vectors $A_\mu^{\hat I}$ and $\bar A_\mu^m$ correspond to the left and right Cartan generators in sector 1, respectively, while $A_\mu^{\alpha}$ correspond to the vectors of sector 2, as defined  in section 2.3. The scalars ${\rm S}^{\hat I m}$
correspond to the $(16+k)\times k$ scalars in sector 1, while the ${\rm S}^{\alpha m}$ correspond to the scalars in sector 2.  In this case,  ${\rm G}_{mn}={G}^{sd}_{mn} +{S_{(mn)}(x)}$, ${\rm B}_{mn}=B^{sd}_{mn} +{S_{[mn]}(x)}$, ${\rm A}_{I m}=A^{sd}_{Im} +{S_{I m}(x)}, {\rm A}_{\alpha m}={S_{\alpha m}(x)}$, and the superindex $sd$ refers to the self-dual values of the background fields. The algebra in the Cartan-Weyl basis is
\bea \label{strconstants}
\left[J^{\hat I}, J^\alpha\right]=\alpha^{\hat I} J^\alpha\, , \ \ \  \  \left[J^\alpha, J^\beta\right]=\left\{\begin{matrix}
\varepsilon({\bf{\alpha,\beta}}) J^{\alpha +\beta} & {\rm if} \ {\bf \alpha} + {\bf \beta} \  {\rm is \ a \ root}\\
{\alpha}_{\hat I}\ J^{\hat I}& {\rm if} \ {\bf \alpha} =-{\bf \beta}\\
0&{\rm otherwise} 
\end{matrix}\right. \, , \nn
\eea
where $\varepsilon(\alpha,\beta)=\pm1$ for simply-laced algebras. Note that it is completely determined by the vertex operators of the vector states: the roots $\alpha^{\hat I}$ are the momenta of the string states and $\varepsilon(\alpha,\beta)$ is given by the cocycle factors in the currents \eqref{cocycles} $c_\alpha c_\beta=\varepsilon(\alpha,\beta)c_{\alpha+\beta}$.
When the gauge group $G_L$ is a product,  the structure constants (and the indices $\Gamma, \hat I, \alpha)$ split into those of each factor, e.g. for  $SO(32)\times H$, $\Gamma =(\Gamma_{SO(32)},\Gamma_{H})$ with $\Gamma_{SO(32)}=( I=1,\dots, 16 ; \alpha=1,\dots,480)$ and $\Gamma_{H}=( m=1,\dots, k; \alpha=1,\dots, N-496-k)$, while for $SO(32)\times U(1)_L^k$ or $SO(34)$ they are only those of the non-Abelian piece. The Cartan-Killing metric is defined to be a block diagonal matrix containing the Cartan-Killing metrics of the groups $\kappa= {\rm diag}(\kappa_{SO(32)}, \kappa_{H})$ or $\kappa= {\rm diag}(\kappa_{SO(32)}, 1_{k\times k})$. 


For gauge groups of the form $G_L\times U(1)_L^{k}\times U(1)_R^{k}$, the  action (\ref{accionefectiva}) agrees with the dimensionally reduced heterotic supergravity action obtained in \cite{hsz},  including the scalar potential (although the reduction of \cite{hsz} contains an additional term with six scalars that we have not computed)\footnote{The redefinitions $A_{\mu}^{(1)m}=\frac1{\sqrt2}(A_{\mu}^m+\bar A_{\mu}^m),
A_{m\mu}^{(2)}=\frac1{\sqrt2}G_{mn}(A_{\mu}^m-\bar A_{\mu}^m)$ and $B_{\mu\nu}=-b_{\mu\nu}, B_{mn}=-b_{mn}$
 are necessary to compare with \cite{hsz}. Note that the KK reductions of the metric and ${\rm B}$ field, $A_\mu^{(1)m}$ and $A_{m\mu}^{(2)}$, having the internal indices up and down repectively, cannot couple through one scalar field, unlike the left and right vector fields $A_\mu^\Gamma$ and $\bar A_\mu^m$ in \eqref{accionefectiva}.  See the next section and the equivalent discussion in \cite{agimnr}.
}. It possesses  $O(k,k;\mathbb R)$ global symmetry. 

In the case of enhanced gauge groups of the form $G_L\times U(1)_R^k$, in which the $k$ left-moving Cartan generators are absorbed by the Cartan subgroups of the non-abelian group $G_L$,  the structure constants  completely break the  global  symmetry. However, 
 (\ref{accionefectiva})  can be  rewritten in $O(k,n)$ covariant form, where $n$ equals the dimension of the full gauge group.  We review this rewriting  in the next section, where we also  present an alternative reformulation of \eqref{accionefectiva} from a generalized Scherk-Schwarz compactification of double field theory. This will allow us to obtain novel relations between the $E_8\times E_8$ and $SO(32)$ heterotic theories.

From \eqref{accionefectiva} one can see some of the features of the spontaneous breaking of gauge symmetry that occurs away from the enhancement points. An effective stringy Higgs mechanism is already encoded in the string theory computation, which can
be interpreted as triggered by the vacuum expectation values of the scalar fields in the Cartan sector ${\rm S}_{\hat I m}$, which give mass to the vectors in the non-Cartan sector from the covariant derivatives in the kinetic terms,  while the scalars without legs in the Cartan sector acquire mass from the scalar potential.
We present the relevant details in the forthcoming sections. 

\subsection{Higgs mechanism in string theory}

When moving away from the points in moduli space where the gauge symmetry is enhanced,  ${\bf p_R}\ne 0$ and the extra massless vectors and scalars in sector 2 acquire mass. The dependence of the vertex operators on the background fields  is contained in the exponential factors of the internal coordinates, which  become 
\bea
J^\alpha =c_\alpha e^{i \pi_{(\alpha)\hat I}  Y_L^{\hat I}(z)}\rightarrow J^{p_L,p_R}(z,\bar z)=c'_\alpha e^{i p_{L\hat I}  Y_L^{\hat I}(z)+ip_{Rm}Y_R^m(\bar z)}\, , \label{jota}
\eea
where $c'_\alpha =c_\alpha$, as  we will see later.
In particular, the $[U(1)_L]^{k+16}\times [ U(1)_R]^k$ charges of these states,   $(q^{\hat I},\bar q^m)=(p_L^{\hat I}, p_R^m)$, are generated by $J^{\hat I}\otimes J^m$.

The OPE of the energy-momentum tensor with the massive vector boson vertex operators
develop  a cubic pole, and it is necessary to combine  these operators  with those of the massive scalars in order to cancel the anomaly. As discussed in \cite{agimnr}, the vertex operators of the massless vectors ``eat'' the scalars $S^{\alpha m}$  and the conformal anomalies can be canceled when redefining
\bea
A_{(0)}'\sim J^{p_L,p_R}(z,\bar z) \left(A^{\alpha}_{\mu}(k)\bar\Upsilon'^{\mu}(\bar z)-\xi S^{\alpha m}(k) \bar\Upsilon'_m(\bar z)\right) e^{i k \cdot X(z,\bar z)}\, ,
\eea
with
\bea
{\bar\Upsilon}'^\mu=i\sqrt2\bar\partial X^\mu + \frac{1}{\sqrt{2}} k\cdot\bar\psi\bar\psi^\mu - p_{Rn}\bar\chi^n\bar\psi^\mu  \, , \qquad
{\bar\Upsilon}'^m =i\bar\partial Y^m + \frac{1}{\sqrt{2}} k\cdot\bar\psi\bar\chi^m - p_{Rn}\bar\chi^n\bar\chi^m  \, ,\nn
\eea
 if
\bea
k\cdot A_\alpha-\xi p^m_{R}S_{\alpha m}=0\, ,
\eea
where $\xi$ is some coefficient.
In terms of fields, this is
\bea
\partial_\mu A_\alpha^\mu+i\xi p_{R}^mS_{\alpha m}=0\, ,
\eea
corresponding to the $R_\xi$ t'Hooft gauge condition where $p_R$ can be identified with a non vanishing vev.
Then the physical massive vector boson vertices are actually $A'$, and the scalars $S_{\alpha m}$ disappear from the spectrum.

Note that the fields associated to $A'$ have well defined charges $(\bf{p_L}, \bf{p_R})$, and since $m^2=-k^2$, the gauge condition can be written as 
\bea
k\cdot \left(A_\alpha+k\xi \frac1{2p_R^2}p_R^mS_{\alpha m}\right)=0\, ,
\eea
implying an effective polarization
\bea
A'_{\alpha\mu}( p_L, p_R, k)=A_{\alpha\mu}-\xi \frac{k_\mu}{2p_R^2} p_R^mS_{\alpha m}\, .
\eea
This leads to a massive vector of the form
\bea
A'_{\alpha\mu}=A_{\alpha\mu}-\xi \frac{1}{2p_R^2} p_R^m\partial_\mu S_{\alpha m}\, ,
\eea
where $p_R^2\ne 0$ is related to the vevs. This is the usual massive vector field incorporating the would-be Goldstone bosons $p_R^m S_{\alpha m}$ that provide the longitudinal polarization.

Unlike the case of the toroidally compactified bosonic string, in the heterotic string  all the massive scalars are Goldstone bosons. Since the   gauge group in the supersymmetric  right sector  is abelian, there are no other massive scalars from the compactification of the massless states.

The non-vanishing three point functions involving massless and light  states, i.e. states that are massless at the self-dual points and become massive when perturbing the background fields,  are listed in Appendix C, and they lead to the following effective action 
 \bea 
  S'&=& \frac{1}{2\kappa_d^2}\int d^d x {\sqrt G} {e^{-2\varphi}} 
\Big({R}+4(\partial_\mu\varphi)^2-\frac 1{12}
H'_{\mu\nu\rho}H'^{\mu\nu\rho}-\frac{1}{4 }F_ { \mu\nu}^{\hat I}F^{\mu\nu }_{\hat I}-\frac14\bar F_
{\mu\nu}^{m}\bar F^{\mu\nu }_m
\nn\\
&&  \ \ \ \ \ \ \ \      
 -\frac14 {F'}_{\mu\nu}^{p}F^{'-p\mu\nu} -\frac12p_R^2 A'^p_\mu A'^{-p}_\nu G^{\mu\nu} -\frac14 \partial_{\mu} {\rm S}_{\hat I n} \partial^{\mu} {\rm S}^{\hat I n}
 +  \sqrt{2} p_L^{\hat I}p_R^{m}
A'^{p \mu}_{}A'_{p\mu} S_{\hat Im}\nn\\
&&\ \ \ \  \ \ \ \  \left.  - \frac 1{{2}}
F^{\hat I}_{\mu\nu} \bar F^{m \mu\nu}   S_{\hat I m}+\frac i2  p_L^{\hat I} A'^{p}_{\mu}A'^{-p}_{\nu}F^{\mu\nu}_{\hat I}
-\frac i2 p_R^{m} A'^{p}_{\mu}A'^{-p}_{\nu} \bar F_{\mu\nu}^{m}
\right) \label{actionnsdr}
 \eea
  with 
\bea
 F'^{p_1}_{\mu\nu}
&=&2\partial_{[\mu}A'^{p_1}_{\nu]}
 +\varepsilon(p_1,p_2)A'^{-p_2}_{[\mu}A'^{p_1+p_2}_{\nu]} -2i {p_{1L}}_{\hat I}A_{[\mu}^{\hat I}A'^{p_1}_{\nu]}
  -2i  {p_{1R}}_{m}\bar A_{[\mu}^{m}A'^{p_1}_{\nu]}\nn\\
  F_{\mu\nu}^{\hat I}
&=&2\partial_{[\mu}A_{\nu]}^{\hat I}\, ,
  \qquad 
 \bar F_{\mu\nu}^{m}=2\partial_{[\mu}\bar A_{\nu]}^{m}\nn\\
H'_{\mu\nu\rho}&=&3\left(\partial_{[\mu}B_{\nu\rho]}+A^{\hat I}_{[\mu}\partial_\nu A_{\rho]\hat I}+A'^p_{[\mu}\partial_\nu A'_{\rho]p}+\frac 13 \varepsilon(p_1,p_2)A'^{p_1}_\mu A'^{p_2}_\nu A'^{-p_1-p_2}_\rho\right.\nn\\
&&\left.\ \ \ \ \ \ \ \ \ \ \ \ \  -i \  p_{L\hat I}A'^{p}_{[\mu} A'^{-p}_{\nu} A_{\rho]}^{\hat I}-i p_{Rm} A'^{p}_{[\mu} 
A'^{-p}_{\nu} A_{\rho]}^{m}-\bar A^m_{[\mu} \partial_\nu \bar A_{\rho ]m}
\right)\, , \label{masac}
\eea

The S-matrix of this massive gauge field theory coupled to gravity  reproduces the string theory three-point amplitudes. The non-Abelian pieces in the field strength of the massive gauge fields and in the Chern-Simons terms in $H'_{\mu\nu\rho}$ correctly appear in terms of the charges of the corresponding fields $(q^{\hat I}, q^m)=(p^{\hat I}_L, p^m_R)$. These charges determine  the coefficients of the vector boson three-point functions, which can be identified with structure constants 
\bea
f^m{}_{\underline p \ \overline p}=ip_R^m\, ,\qquad  f^{\hat I}{}_{\underline p \ \overline p}=ip_L^{\hat I}\, , \qquad f^{p_1+p_2}{}_{p_1p_2}=\varepsilon(p_1,p_2)\, , \label{sc}
\eea
  reflecting the fact that the gauge interactions in string theory are a manifestation of an underlying affine Lie algebra.  This 
algebra is  isomorphic   to that of  the enhanced $G_L$ group  \cite{aamp}, which  justifies the identification  $c'_\alpha =c_\alpha$ used  in (\ref{jota}) (we will comment further on this result in the next section).

 Not all the terms in the action can be obtained from the three-point functions, but we have completed the expressions so that they correctly reproduce the massless case when $p_R=0$ and $p_L\in\Gamma$.

All the terms of the scalar potential of the massless theory (\ref{accionefectiva}) are absorbed by the field strengths of the massive vectors or by interaction terms containing massive vectors.

\section{Heterotic double field theory }
\label{sec:DFT&SS}

Although   the action \eqref{accionefectiva} can be generically obtained  by dimensional reduction of  heterotic supergravity from 10 to  $10-k$ dimensions, not all the effective actions of massless fields obtained from toroidally compactified heterotic string theory can be uplifted to  higher dimensional supergravities. In particular, the states with nonzero winding or momentum number on $T^k$ cannot be captured by field theoretical Kaluza Klein compactifications.
To find the higher dimensional description of these string modes, one has to refer to gauged double field theory (DFT) \cite{hk,GM,gmnp},  an $O(D,D+N;\mathbb R)$ covariant rewriting of heterotic supergravity, with $D$ the dimension of space-time and $N$ the dimension of the  gauge group. 

In this section we review this construction and show that the effective action \eqref{accionefectiva} can be rewritten in terms of $O(k,N)$ multiplets.  The reformulation is achieved essentially assembling the $N+k$ gauge fields  as a vector, the $Nk$ moduli scalars as part of a  symmetric tensor and the  structure constants of the non-abelian gauge groups  as an antisymmetric three-index tensor under $O(k,N)$ transformations. The procedure generalizes the  analysis of \cite{hsz} by including all the massless string modes at self-dual points of the moduli space, in which the $k$ left Kaluza-Klein vector fields become part of the Cartan subgroup of the maximally enhanced gauge group.

Furthermore, using the equivalence between gauged DFT and generalized Scherk-Schwarz (gSS) compactifications \cite{GM}, we present an explicit realization of the internal generalized vielbein which reproduces the structure constants of all the enhanced gauge groups under generalized diffeomorphisms. In particular, we show that the structure constants of the $E_8\times E_8$ and $SO(32)$ groups can be obtained from the same deformation of the generalized diffeomorphisms and then the $E_8\times E_8$ and $SO(32)$ theories  can be described as different solutions of the same heterotic  DFT.

\subsection{Gauged double field theory }

The frame-like DFT action reproducing  heterotic supergravity   was  originally introduced in \cite{siegel} and further developed in \cite{hk}. 
 The theory has a global $G = O(D,D+N;\mathbb{R})$ symmetry, a local double-Lorentz $H=O(D-1,1;\mathbb{R}) \times O(1, D-1 + N;\mathbb{R})$ symmetry, and a gauge symmetry generated by a generalized Lie derivative 
\bea
{\cal L}_\xi V_{\cal M} &=&\xi^{\cal P} \partial_{\cal P} V_{\cal M} + \left(\partial_{\cal M} \xi^{\cal P} - \partial^{\cal P} \xi_{\cal M}\right) V_{\cal P}\, .
\eea
The infinitesimal generalized parameter $\xi^{\cal M}$, with ${\cal M} = 1,\dots,2D +N$,
  transforms in the fundamental representation of $G$, and $H$-transformations are generated by an infinitesimal parameter $\Lambda_{\cal A}{}^{\cal B}$, with ${\cal A,B} = 1,\dots,2D +N$.

The constant symmetric and invertible metrics $\eta_{\cal M N}$  and $\eta_{\cal AB}$ raise and lower the indices that are rotated by $G$ and $H$, respectively. In addition there is a constant symmetric and invertible $H$-invariant
metric ${\cal H}_{\cal A B}$ 
constrained to satisfy
\bea
{\cal H}_{\cal A}{}^{\cal C} {\cal H}_{\cal C}{}^{\cal B} = \delta_{\cal A}^{\cal B} \ . \label{flatconstraint}
\eea
The three metrics $\eta_{\cal M N}$, $\eta_{\cal A B}$ and ${\cal H}_{\cal A B}$ are invariant under the action of ${\cal L}$, $G$ and $H$.

The fields of the theory are a generalized vielbein $E^{\cal A}{}_{\cal M}$ and a generalized dilaton $d$. The former is constrained to relate the metrics $\eta_{\cal A B}$ and $\eta_{\cal M N}$, and allows to define a generalized metric  ${\cal H}_{\cal M N}$ from ${\cal H}_{\cal A B}$
 \bea
\eta_{\cal M N} = E^{\cal A}{}_{\cal M}{}\eta_{\cal A B} E^{\cal B}{}_{\cal N}     \ , \ \ \ \ \ \      {\cal H}_{\cal M N} = E^{\cal A}{}_{\cal M} {\cal H}_{\cal A B} E^{\cal B}{}_{\cal N} \ . \label{flattocurve}
\eea
The theory is defined on a $2D+N$ dimensional space  but the coordinate dependence of fields and gauge parameters is restricted by a strong constraint
\bea
\partial_{\cal M} \partial^{\cal M} \dots = 0 \ , \ \ \ \ \ \partial_{\cal M} \dots \ \partial^{\cal M} \dots = 0 \ , \label{StrongConstraint}
\eea
the derivatives $
\partial_{\cal M}$ transforming in the fundamental representation of $G$ and the dots representing arbitrary products of fields.

DFT can be deformed in terms of so-called fluxes or gaugings $f_{\cal M N P}$ \cite{hk}, a set of constants that satisfy linear and quadratic constraints
\bea
f_{\cal M N P} = f_{[\cal M N P]} \ , \ \ \ \ f_{[\cal M N}{}^{\cal R} f_{\cal P] R}{}^{\cal Q} = 0 \ , \label{Constraintsfs1}
\eea
and the following additional constraint is required to further restrict the coordinate dependence of fields and gauge parameters
\bea
f_{\cal M N}{}^{\cal P} \, \partial_{\cal P} \dots = 0 \ . \label{fderivative}
\eea

The generalized dilaton and frame transform under generalized diffeomorphisms and $H$-transformations as follows
\bea
\delta d &=& \xi^{\cal P} \partial_{\cal P} d - \frac 1 2 \partial_{\cal P} \xi^{\cal P}  \ \ \ \ \Leftrightarrow \ \ \ \ \delta e^{-2d} = \partial_{\cal P} \left(\xi^{\cal P} e^{-2 d}\right) \, , \label{vard}\\
\delta E^{\cal A}{}_{\cal M} &=& \widehat {\cal L}_\xi E^{\cal A}{}_{\cal M} + \delta_\Lambda E^{\cal A}{}_{\cal M}   \ , \label{varE}
\eea
where
\bea
\widehat {\cal L}_\xi E^{\cal A}{}_{\cal M} &=&{\cal L}_\xi E^{\cal A}{}_{\cal M} + f_{\cal M P}{}^{\cal Q} \xi^{\cal P} E^{\cal A}{}_{\cal Q} \ , \label{liederiv}\\
\delta_\Lambda E^{\cal A}{}_{\cal M} &=& E^{\cal B}{}_{\cal M} \Lambda_{\cal B}{}^{\cal A} \ . \label{DeltaLambdaE}
\eea

The DFT action can be expressed in terms of the generalized fluxes
\bea
{\cal F}_{\cal A B C} &=& 3 \partial_{[{\cal A}} E_{\cal B}{}^{\cal N} E^{\cal P}{}_{\cal C]} \eta_{\cal N P} + f_{\cal M N P} E^{\cal M}{}_{\cal A} E_{\cal B}{}^{\cal N} E_{\cal C}{}^{\cal P}
\ ,
\label{genflux}\\
{\cal F}_{\cal A} &=& 2 \partial_{\cal A} d - \partial_{\cal B}E_{\cal AM} E^{\cal BM}
\ ,\label{genfluxd}
\eea
as \cite{gmnp}
\bea
S &=&\int dX e^{-2d}\left[{\cal F}_{ \cal A  B  C}{\cal F}_{\cal  D E F}\left (\frac14
{\cal H}^{\cal  A D}\eta^{\cal B E}\eta^{\cal C F}-\frac1{12}
{\cal H}^{\cal AD}{\cal H}^{ \cal B E}{\cal H}^{\cal CF} -\frac16
\eta^{\cal AD}\eta^{\cal BE}\eta^{\cal CF}\right)\right.\nn\\
&&\qquad \qquad \qquad \left.+ \left(2\partial_{\cal A}{\cal F}_{\cal B}-{\cal F}_ {\cal A}{\cal F}_{\cal B} \right)\left({\cal H}^{\cal AB}-\eta^{\cal AB}\right)\right]\, ,\label{gfs}
\eea
and it
 is fixed by demanding $H$-invariance, since the generalized fluxes are not $H$-covariant.

\subsection{Parameterization and choice of section}

Choosing specific global and local groups  and parameterizing the fields in terms of  metric, two-form, vector and scalar fields one can make contact with the (toroidally compactified) heterotic string modes and effective actions of the previous sections. To this aim, we first consider the theory at  points of the moduli space in which the gauge group is $G_L\times U(1)_L^k\times U(1)_R^k$, and in the next subsection extend the construction to account for the maximally enhanced gauge groups $G_L\times U(1)_R^k$.

 Taking the space-time dimension $D=d+k$ and the gauge group $G_L\times U(1)_L^k\times U(1)_R^k$,  the $G$ indices split as $V^{\cal M}=(V^\mu, V_\mu,  V^M)$ and the $H$ indices split as $V_{\cal A}=
(V_{\overline{\mathbb A}}, V_{\underline{\mathbb A}}, V_{\bf a})$, where $\mu, \overline{\mathbb A}, {\underline{\mathbb A}} = 1,\dots , d$ (external)\footnote{ This notation for the right and left indices should be distinguished from the notation $\bar \alpha$ and $ \underline \alpha$  used for the positive and negative roots of the gauge algebra   in the previous section.} and $M,{\bf a}=1,\dots,2k+ N$ (internal), $N$ being the dimension of $G_L$. The splitting breaks $G$ and $H$ into external and internal pieces 
\bea
G\rightarrow G_e\times G_i\, , \qquad \qquad \qquad  H\rightarrow H_e\times H_i
\eea
where
\bea
G_e&=&O(d,d;\mathbb R)\, , \qquad \qquad \qquad \qquad \qquad \ \ \ G_i=O(k,k+N;\mathbb R)\, ,\nn\\
H_e&=&O(d-1,1;\mathbb R)\times O(1, d-1;\mathbb R)\, , \qquad H_i=O(k;\mathbb R)\times O(k+N;\mathbb R)\, .\nn
\eea
Then the $G$-vector $V^{\cal M}$ contains a $G_e$-vector $(V^\mu, V_\mu)$ and a $G_i$-vector $V^M=(V^m, V_m, V^\Gamma)$ and the $H$-vector $V_{\cal A}$ contains a $H_e$-vector $(V_{\overline{\mathbb A}}, V_{\underline{\mathbb A}})$ and a $H_i$-vector $V_{\bf a}=(V_{\overline a}, V_{\underline a}, V_G)$. Under this decomposition, the degrees of freedom can be decomposed as
\bea 
{\rm dim }\ (G/H)= D(D+N)&=& \frac {d (d+1)} 2 + \frac {d (d-1)} 2 + ~ d (2k+ N) ~ + ~ k\times (k+ N) \nn\\
&& \ \ \ \  G_{\mu \nu} \ \ \ \ \ \ \ \ \ \ \   B_{\mu \nu} \ \ \ \ \ \ \ \ \  \  {\cal A}_\mu{}^M \ \ \ \ \  \  \ \  \ \  \ \ {\cal E}_M{}^{\bf a}\nn
\eea
where ${\cal E}_M{}^{\bf a}$ parameterizes the coset $G_i/H_i$. The $G$ and $H$ invariant metrics are
\bea
\eta_{\cal MN}=\begin{pmatrix}0&\delta_\mu{}^\nu&0&0&0\\
\delta^\mu{}_\nu&0&0&0&0\\
0&0&0&\delta_m{}^n&0\\
0&0&\delta^m{}_n&0&0\\
0&0&0&0&\kappa_{\Gamma\Lambda}\end{pmatrix}\, , \quad \begin{matrix}
\eta_{\cal AB}=\ diag(-g_{\overline{\mathbb {AB}}},
g_{\underline{\mathbb {AB}}},-\delta_{\overline a\overline b},\delta_{\underline a\underline b},\delta_{FG})\, ,\\
\\
\\
{\cal H}_{\cal AB}=\ diag(g_{\overline{\mathbb {AB}}},
g_{\underline{\mathbb {AB}}},\delta_{\overline a\overline b},\delta_{\underline a\underline b},\delta_{FG})\, .
\end{matrix}
\eea

We can parameterize  the generalized frame  in terms of the $d$-dimensional fields as
\bea
E^{\cal A}{}_{\cal M} = \frac1{\sqrt2}\begin{pmatrix} 
-e_{\overline {\mathbb A}\mu} -\hat e_{\overline{\mathbb A}}{}^\nu {\cal C}_{\nu\mu}& \hat e_{\overline{\mathbb A}}{}^\mu & -\hat e_{\overline{\mathbb A}}{}^\rho{\cal A}_{\rho M}\\ 
e_{\underline{\mathbb A}\mu} -\hat e'_{\underline{\mathbb A}}{}^\nu {\cal C}_{\nu\mu}&\hat e_{\underline{\mathbb A}}{}^\mu & -\hat e_{\underline{\mathbb A}}{}^\nu{\cal A}_{\nu M} \\
\sqrt2{\cal E}^{\bf a}{}_N{\cal A}^N_{\mu}&0&\sqrt2{\cal E}^{\bf a} {}_M\end{pmatrix}\, , \label{e0}
\eea
where the vielbeins $e_\mu{}^{\overline{\mathbb A}}$ and $e_\mu{}^{\underline{\mathbb A}}$ for the right and left sectors define the same space-time metric ${\cal G}_{\mu\nu}=e_\mu{}^{\overline{\mathbb A}}g_{\overline {\mathbb A}\overline{\mathbb B}}e_\nu{}^{\overline{\mathbb B}}=e_\mu{}^{\underline{\mathbb A}}g_{\underline {\mathbb A}\underline{\mathbb B}}e_\nu{}^{\underline{\mathbb B}}$ and ${\cal C}_{\mu\nu}={\cal B}_{\mu\nu}+\frac12{\cal A}_\mu^M {\cal A}_{\nu M}$. 

The internal part of the generalized vielbein ${\cal E}^{\bf a}{}_M$ can be written in terms of the background fields 
and perturbations  as ${\cal E}^{\bf a}{}_M={\cal E}_0^{\bf a}{}_M+\delta {\cal E}^{\bf a}{}_M$, with
\beq \label{genframeLR01}
 \begin{pmatrix}{\cal E}_{0\overline a} \\  {\cal E}_{0\underline a}\\  {\cal E}_{0A} \end{pmatrix}=\frac{1}{\sqrt2}\begin{pmatrix} 
-e_{\overline a m} -\hat e_{\overline a}{}^n {\cal C}_{nm}&\hat e_{\overline a}{}^m & -\hat e_{\overline a}{}^n {\cal A}_{n}^I\\
e_{\underline a m} -\hat e_{\underline a}{}^n {\cal C}_{nm}&\hat e_{\underline a}{}^m & -\hat e_{\underline a}{}^n {\cal A}_{n}^I \\
\sqrt2\tilde e_A{}^J {\cal A}_{J m} &0 &\sqrt2\tilde e_A{}^{I}
\end{pmatrix} \ ,
\eeq
where $e^{\overline a}{}_m$ and $e^{\underline a}{}_m$ are two different frames for the same  background metric ${\cal G}_{mn}$,  $\hat e_{\overline a}{}^m, \hat e_{\underline a}{}^m$ are the inverse frames and ${\cal C}_{mn}={\cal B}_{mn}+\frac12{ \cal A}_m^I {\cal A}_{n I}$. 

Then
the generalized metric  is
\bea
{\cal H}_{\cal M N} = \left(\begin{matrix}  {\cal G}_{\mu \nu} +  {\cal C}_{\rho \mu}  {\cal C}_{\sigma \nu}{\cal  G}^{\rho \sigma} +  {\cal A}_\mu{}^P  {\cal M}_{PQ}  {\cal A}_{\nu}{}^Q   &   -  {\cal G}^{\nu \rho}  {\cal C}_{\rho \mu} &  {\cal C}_{\rho \mu}  {\cal G}^{\rho \sigma} {\cal A}_{\sigma N} +  {\cal A}_{\mu}{}^P  {\cal M}_{PN}  \\ -  {\cal G}^{\mu \rho} {\cal C}_{\rho \nu}
                 &  {\cal G}^{\mu \nu} &  - {\cal G}^{\mu \rho}  {\cal A}_{\rho N}
                  \\
 {\cal C}_{\rho \nu}  {\cal G}^{\rho \sigma} {\cal  A}_{\sigma M} + {\cal A}_\nu{}^P {\cal  M}_{MP}
                 &  -  {\cal G}^{\nu \rho}  {\cal A}_{\rho M}  &  {\cal M}_{MN} + {\cal  A}_{\rho M}  {\cal G}^{\rho \sigma}  {\cal A}_{\sigma N}\end{matrix}\right)\, , \nn
\eea
and the symmetric and $G_i$-valued  matrix   ${\cal M}_{MN}={\cal E}^{\bf a}{}_M\delta_{\bf ab}{\cal E}^{\bf b}{}_N\in O(k,k+N;\mathbb R)$ is
\beq
\label{G2}
{\cal M}_{MN}=\begin{pmatrix} {\cal G}_{mn}+{\cal C}_{lm}{\cal G}^{lk}{\cal C}_{kn} +{\cal A}_{m}^\Gamma {\cal A}_{\Gamma n} & -{\cal G}^{nk}{\cal C}_{km} & {\cal C}_{km}{\cal G}^{kl}{\cal A}_{l\Lambda}+{\cal A}_{m\Lambda} \\  -{\cal G}^{mk}{\cal C}_{kn}  & {\cal G}^{mn} &-{\cal G}^{mk}{\cal A}_{k\Lambda}\\
{\cal C}_{kn}{\cal G}^{kl}{\cal A}_{l\Gamma}+{\cal A}_{nJ}&-{\cal G}^{nk}{\cal A}_{k\Gamma}&\kappa_{IJ}+{\cal A}_{k\Gamma} {\cal G}^{kl}{\cal A}_{l\Lambda} \end{pmatrix} \, ,
\eeq
where the fields depend on the external coordinates.

With this parameterization in (\ref{genflux}), taking $e^{-2d}=\sqrt{-{\cal G}}e^{-2\varphi}$ in (\ref{genfluxd}) and resolving the strong constraint \eqref{StrongConstraint} in the supergravity frame, after  integrating (\ref{gfs}) along the internal coordinates one gets an action of the form of (the electric bosonic sector of)  half-maximal gauged supergravity \cite{Arg}
\bea
S &=& \int d^d X \sqrt{-{\cal G}} e^{-2 \varphi} \left[ R +  4 D_\mu \varphi D^\mu \varphi - \frac 1 {12} H_{\mu \nu \rho}  H^{\mu \nu \rho}- \frac 1 4 {\cal F}_{\mu \nu}{}^M {\cal F}^{\mu \nu N} {\cal M}_{MN}\right. \nn\\
&&\left. \qquad\qquad \qquad\qquad +\frac 1 8 D_\mu {\cal M}_{MN} D^\mu {\cal M}^{MN}-V
   \right] \ , \label{ActionGaugedSugra}
\eea
where 
\bea \label{covariant}
H_{\mu\nu\rho}&=&3\left(\partial_{[\mu}{\cal B}_{\nu\rho]}-{\cal A}^M_{[\mu}\partial_\nu {\cal A}_{\rho]M} -\frac13f_{MNP}{\cal A}^M_{[\mu}{\cal A}^N_{\nu}{\cal A}^P_{\rho]}\right)\, ,\nn\\
   {\cal F}^{M}_{\mu\nu} &=& 2\partial_{[\mu} {\cal A}_{\nu]}^M+f^{M}{}_{NP}{\cal A}^N_{\mu}{\cal A}^P_{\nu}\ .\nn\\
 D_\mu {\cal M}_{MN}&=&\partial_\mu{\cal M}_{MN}+f_{MP}{}^{Q}{\cal A}^P_\mu{\cal M}_{QN}+f_{NP}{}^{Q}{\cal A}^P_\mu{\cal M}_{MQ}
\eea
and the scalar potential is
\beq \label{Vf}
V= \frac 1 {12} f_{MP}{}^R f_{NQ}{}^S {\cal M}^{MN} {\cal M}^{PQ} {\cal M}_{RS}+\frac 1 4 f_{MP}{}^Q f_{NQ}{}^P {\cal M}^{MN}+\frac 1 6 f_{MNP} f^{MNP}
\ .
\eeq

This action reproduces heterotic supergravity  in ten external dimensions  for $k=0$ and $G_L=SO(32)$ or $E_8\times E_8$, with the following identifications.
The  scalar frame is only non-vanishing for ${\cal E}_A{}^M =\tilde e_A{}^{M}$, and then ${\cal M}_{MN} = \kappa_{MN}$ is the constant Killing metric of $G_L$   with  $M, N=\Gamma, \Lambda=1,\dots, 496$,  and the second line in (\ref{ActionGaugedSugra}) vanishes.  The   gaugings are non-vanishing only in the internal directions associated to the gauge group
\bea
f_{\cal M N P} = \left\{ \begin{matrix} f_{\Lambda\Gamma\Omega} \ \ \ \ {\rm if}\  ({\cal M,N,P}) = (\Lambda,\Gamma,\Omega) \\
                                    0 \ \ \ \ \ \  {\rm otherwise}\ \ \ \ \ \ \ \ \ \ \ \ \ \ \ \ \end{matrix}\right. \ , \label{gaugings}
\eea
and are taken to be the structure constants of $G_L$, satisfying the linear and quadratic constraints (\ref{Constraintsfs1}) 
 \bea
f_{\Lambda\Gamma\Omega} = f_{[\Lambda\Gamma\Omega]} \ , \ \ \ \ f_{[\Lambda\Gamma}{}^\Pi f_{\Omega]\Pi}{}^\Delta = 0 \ . \label{Constraintsfsmall}
\eea
Identifying ${\cal A}^{\Gamma}_{\mu} =A_{\mu}^\Gamma, {\cal B}_{\mu\nu}=-B_{\mu\nu}, {\cal G}_{\mu\nu}=G_{\mu\nu}$ one gets the ten dimensional heterotic string low energy effective action \eqref{accionefectiva}.

For $k\ne 0$ and generic values of the background fields,  the gauge group is $U(1)^{2k+16}$ and then  there are no gaugings. In this case, \eqref{ActionGaugedSugra} reproduces (\ref{accionefectiva}) when identifying the generalized gauge fields with the string theory fields as
\bea
{\cal A}_\mu{}^M&=&{\cal E}_{0\bf a}{}^M{\cal A}_\mu{}^{\bf a}= {\cal E}_{0\overline a}{}^M{\cal A}_\mu{}^{\overline a}+ {\cal E}_{0\underline a}{}^M{\cal A}_\mu{}^{\underline a}+ {\cal E}_{0A}{}^M{\cal A}_\mu{}^{A}\nn\\
&=&\frac1{\sqrt2}\begin{pmatrix}A_\mu^m+\bar A_\mu^m\\
G_{mn}(A_\mu^n-\bar A_\mu^n)+C_{mn}(A_\mu^n+\bar A_\mu^n)+\sqrt2A_m^I A_{I\mu}\\
-A_n^I (A_\mu^n+\bar A_\mu^n)+\sqrt2
A_\mu^I\end{pmatrix}\ , \label{cala}
\eea
with $M,N=1,\dots , 2k+16$. The components of the generalized scalar matrix ${\cal M}_{MN}$ are related with the background fields and massless modes of the string theory as ${\cal G}_{mn}=G_{mn}+{S_{(mn)}}$, ${\cal B}_{mn}= -B_{mn}-{S_{[mn]}}, {\cal A}_{I m}=A_{I m}+{S_{Im}}$. 

To make contact with (\ref{accionefectiva}) at  the points of the moduli space giving $G_L\times U(1)^{2k}$  enhanced gauge symmetry, one simply extends
the frame $\tilde e_A{}^{I}$ in \eqref{genframeLR01}  to $ \tilde e_G{}^{\Gamma}$, the gauge  fields $ {A}_{\mu}^I$  in \eqref{cala} to include the non-abelian sector 2 of section 2.3,  i.e.  $ {A}_{\mu}^I\rightarrow  {A}_{\mu}^\Gamma$, and the scalars in \eqref{G2} to $  {\cal  A}_{m}^\Gamma=(A^I_{m}+{S^I_{m}}, S^\alpha_m)$,  where the indices ${\Gamma, \Lambda, F, G}=1, \dots, N$. Plugging all this in \eqref{ActionGaugedSugra} and taking for $f_{MNP}$ the structure constants of $G_L$, one recovers \eqref{accionefectiva}.

 In the cases of maximal enhancement, we can take $G_i= O(k, N)$, with $N$ being the dimension of a simply-laced group of rank $16+k$. The $k$  left  internal dimensions become part of the dimensions associated to the Cartan subgroup of the enhanced gauge group, the left KK gauge fields $A_\mu^m$ become Cartan components of the non-abelian gauge fields ${A}_\mu^\Gamma$ 
 and the  gaugings are the structure constants of the gauge group. 
 In the next section we deal with these cases in full detail and we also show that  the action \eqref{ActionGaugedSugra}  reproduces the right patterns of symmetry breaking when moving away from a point of enhancement.

\subsection{Generalized Scherk-Schwarz reductions}
\label{sec:SS}
We have seen that appropriately choosing the global and local symmetry groups and the gaugings deforming the generalized Lie derivative (\ref{liederiv}), one can account for both the un-compactified and the toroidally compactified versions of the heterotic string effective low energy theory with gauge group $G_L\times U(1)_L^k\times U(1)_R^k$.   To describe the effective theory with maximally enhanced gauge group $G_L\times U(1)_R^k$, we perform a generalized Scherk-Schwarz (gSS) compactification of DFT. 
 Recall that  the result of gauging the theory and parameterizing the generalized fields in terms of the degrees of freedom of the lower dimensional theory  is effectively equivalent to a gSS reduction of DFT  \cite{GM}, which has the advantage of providing  an
 explicit realization of the generalized vielbein $E_{\cal A}{}^{\cal M}$ giving rise to  the  enhanced gauge algebra under the generalized diffeomorphisms (\ref{genflux}) \cite{agimnr,cgin}. 
 In this section we extend the  construction  to the heterotic case, and in particular, we will show that the formulation of \cite{cgin} allows to describe the 
 $E_8\times E_8$ and $SO(32)$ theories  as two solutions of the same heterotic DFT, even before compactification.

The generalized vielbein in gSS reductions  is a product of two pieces, one depending on the $d$ external coordinates $x^\mu$ and the other one depending on the internal ones, $y_L, y_R$:
\beq \label{gss}
E_{\cal A}(x, y_L, y_R)= {\Phi}_{\cal A}{}^{\cal  A'}(x) \mathbb E_{ \cal A'}(y_L, y_R) \ .
\eeq
 The matrix ${\Phi}$ parameterizes the scalar, vector and tensor fields of the reduced $d$-dimensional action and  the twist $\mathbb E$ characterizes the background. 

Let us concentrate on the internal part of the vielbein
\beq
 {\mathcal E}_{\bf a} {}^{M}(x,y_L)=\Phi_{\bf a} {}^{\bf b} (x)  \mathbb E_{\bf b} {}^M(y_L) \ , \label{gssv}
 \eeq
where $M={\bf a} =1,\dots,N+k$, and $N$ is now the dimension of $G_L$ (i.e. the $k$ left internal dimensions are absorbed by the Cartan directions of $G_L$). The matrix $\Phi_{{\bf a} }{}^{{\bf b} }$ describes the fluctuations over the background and  the twist  $\mathbb E_{\bf b}{}^M$  is an element of the coset $\frac{O(k,N)}{O(k)\times O(N)}$, generating the constant fluxes $f_{{\bf ab}}{}^{\bf c}$
which gauge (a subgroup of) the global $O(k,N)$ symmetry. We take $\mathbb E_{\bf b}{}^M$  to depend on $y_L$ only, as this is the only sector with a non-Abelian gauge group. 

The scalar matrix can be written as
\begin{equation} \label{HMHA}
\mathcal M^{MN}=  \delta^{{\bf ab}} {\mathcal E}_{\bf a}{}^{M} {\mathcal E}_{\bf b}{}^{N}= {\bf M}^{\bf ab}(x) \mathbb E_{\bf a}{}^{M}(y_L)\mathbb  E_{\bf b}{}^{N}(y_L),
\end{equation}
 with
\begin{equation}
\label{HAB}
{\bf M}^{{\bf ab}}(x)=\delta^{{\bf cd}} \Phi_{{\bf c}}{}^{{\bf a}}(x) \Phi_{{\bf d}}{}^{{\bf b}} (x)\ .
\end{equation}

We now  expand on the explicit parameterization of $\Phi(x)$ in terms of fluctuations that can be identified with the string theory fields and on
the twist $\mathbb E_{\bf a}{}^M (y_L)$   realizing the enhanced gauge algebra.

\subsubsection{Fluctuations around generic points in moduli space}
\label{sec:torus}
In order to identify the massless vector and scalar fields of the reduced theory with the corresponding string states  at a generic point in moduli space,
we first consider a reduction 
on an ordinary $2k+16$ torus ($i.e.$ no twist).
 There are no gaugings and therefore we get an ungauged action with $2k+16$ abelian $U(1)_L^{k+16} \times U(1)^{k}_R$ vectors  $ {A}_{\mu}^{m}, {A}_{\mu}^{I}, {\bar A}_{\mu}^{m}$ and $(k+16) \times k$ scalars  encoded in ${\bf M}_{\bf ab}$. The  vectors  and  scalars contain the 16 Cartan generators, the $2k$ KK fields and the fluctuations of  the metric, $B$-field  and  Wilson lines on the torus, corresponding to the string states $a_\mu^m, a_\mu^{  I}, \bar a_\mu^m,   g_{mn}, b_{mn}, a_m^I$  in sector 1. To get the precise relation, consider an expansion around a given point in moduli space corresponding to constant background metric $G$, $B$-field $B$ and 
Wilson line $A$. 

The internal part of the generalized vielbein in the left-right basis reads, at first order,
\beq \label{genframeLR0}
 \begin{pmatrix}E_{R} \\  E_{L}\\  E_A \end{pmatrix}=\frac{1}{\sqrt2}\begin{pmatrix} 
-e_0 -\hat e_0 C_0&\hat e_0 & -\hat e_0 A_0\\
 e_0 -\hat e_0 C_0 &\hat e_0& -\hat e_0A_0 \\
\sqrt2\tilde e A_0^t &0 &\sqrt2\tilde e 
\end{pmatrix}+\frac1{\sqrt2}\delta
\begin{pmatrix} -e -\hat e C &\hat e& -\hat e A\\ 
e -\hat e C & \hat e  & -\hat eA \\
\sqrt2\tilde e A^t & 0 &0
\end{pmatrix} \ ,
\eeq
where now we denote $e_0$  and $\hat e_0$ the frames and inverse frames for $G$  to lighten the notation. Note we are not varying the frame for the Killing metric $\tilde e$.
Performing this expansion and accommodating the terms so that it has the form of a gSS reduction 
$
\mathcal E
=  \Phi(x) \mathbb   E$, where now the twist $\mathbb E$ is constant,
one gets  
\beq
\Phi(x)=\rm{Id}+\frac12 \begin{pmatrix}    \varphi_+ +\phi  &
\varphi_- - \phi & -\sqrt{2}(\hat e_0 + \delta \hat e) \delta A \hat{\tilde e}^t  \\
\varphi_- + \phi & \varphi_+ - \phi & -\sqrt{2}(\hat e_0 + \delta \hat e) \delta A \hat{\tilde e}^t  \\
-\sqrt{2}\tilde e \delta A^t \hat e_0^t &  \sqrt{2}\tilde e \delta A^t \hat e{}_0^t  & 0   \end{pmatrix}\nn
\eeq
with
\bea
\varphi_{\pm} &\equiv& \delta \hat e  e_0^t \pm \delta e \hat e_0^t\, , \qquad
\phi \equiv (\hat e_0 + \delta \hat e) (\delta C - \delta A A_0^t )\hat e_0^t\, ,\nn
\eea

The matrix of  $\Phi$ is an element of 
$SO^+(k,k+16;\mathbb R)$, the component of $O(k,k+16;\mathbb R)$ connected to the identity.
Inserting this into \eqref{HAB} we get, up to second order\footnote{We actually get a second order piece in the off-diagonal terms, namely instead of $M$, one gets $M+Q$, where $Q$ contains terms of the form $\delta e^t \delta \hat e, \delta B' \delta B'$, etc., but  this second order piece is not needed for our purpose of computing the action up to quartic order.},
\beq \label{HM}
{\bf M}^{\bf ab}= \delta^{\bf cd} \Phi_{\bf c}{}^{\bf a} \Phi_{\bf d}{}^{\bf b} 
 =\begin{pmatrix} 
I_{\overline a\overline b}+ \frac12( M^t M)_{\overline a\overline b}  & M^t_ {\overline a\underline b} &  M^t_ {\overline aA}\\
 M_{\underline a\overline b}  &  I_{\underline a\underline b} + \frac12 (M  M^t)_{\underline a\underline b}&0\\
M_{A\overline b}&0&I_{AA}+\frac12(MM^t)_{AA} \end{pmatrix}\, ,
\eeq
The $k\times k$ matrix $M_{\underline a\overline b}$ is
\beq \label{MgBA}
M_{\underline {a}\overline {b}}&=& - \hat e_{0\underline{a}}{}^{m} \hat{e}_{0 \overline{ b}}{}^{n}\left(\delta G_{mn}-\delta B'_{mn}\right)  \ , \quad \delta B' \equiv \delta B  +\frac{\delta A A_0^t -  A_0 \delta A^t}{2} \ \ ,
\eeq
where  $\delta G=\delta e^t e_0 + e^t _0\delta e + \delta e^t \delta e$ and $\delta B'$ is the variation of $\delta B$ under  an $O_{\Theta}$ shift \eqref{theta} with $\Theta=\delta B$ and an $O_{\Lambda}$ shift \eqref{OLambda} with $\Lambda=\delta A$ (see footnote \ref{foot:lambdashiftB}).  The $16\times k$ matrix $M_{A\overline a}$ is 
\bea
M_{A\overline a}=\sqrt{2} \tilde e_A{}^I{}\delta A_{Im}\hat{e}_0^m{}_{\overline a}
\eea


%

The fluxes $f_{\bf ab}{}^{\bf c}$  computed from \eqref{genflux}  
vanish as the twist $\mathbb E$ is  constant and the theory is not deformed. Then,
taking the abelian  field strengths $F^{\bf a}_{\mu\nu}=(F_{\mu\nu}^{\overline a}, F_{\mu\nu}^{\underline a}, F_{\mu\nu}^A)$ for the $U(1)_R^k$  and $U(1)^{k+16}_L$  vector fields, we get (up to first order in fluctuations)
\bea
-\frac14{\cal F}_{\mu\nu}^M{\cal M}_{MN}{\cal F}^{N\mu\nu}&=&-\frac14{ F}_{\mu\nu}^{\overline a}\delta_{\overline {ab}}{ F}_{\mu\nu}^{\overline b}-\frac14{ F}_{\mu\nu}^{\underline a}\delta_{\underline {ab}}{ F}_{\mu\nu}^{\underline b}
-\frac14{ F}_{\mu\nu}^{A}\delta_{AB}{ F}_{\mu\nu}^{B}\nn\\
&&-\frac12{ F}_{\mu\nu}^{\overline a}M_{ \overline a\underline b}{  F}_{\mu\nu}^{ \underline b}-\frac1{2}{ F}_{\mu\nu}^{A}M_{ A\overline b}{ F}_{\mu\nu}^{ \overline b}\nn
\eea
and 
\bea
\frac18D_\mu {\cal M}_{MN}D^\mu {\cal M}^{MN}&=&\frac 14\partial_\mu \delta G_{mn}\partial^\mu { \delta G}^{mn}-\frac 14\partial_\mu \delta B_{mn}\partial^\mu { \delta B}^{mn}-\frac 12\partial_\mu \delta A_{Im}\partial^\mu { \delta A}^{Im}\nn
\eea
 Plugging these terms in \eqref{ActionGaugedSugra},
the effective action \eqref{accionefectiva} derived from toroidally compactified string theory is reproduced  if we identify, as in the previous section, $F^{\overline a}_{\mu\nu}=e_0^{\overline a}{}_m\bar F_{\mu\nu}^m, F^{\underline a}_{\mu\nu}=e^{\underline a}_0{}_{m} F_{\mu\nu}^m, F^{A}_{\mu\nu}=\tilde e^{A}{}_{I} F_{\mu\nu}^I$, $\delta G_{mn}=S_{(mn)}, \delta B_{mn}=S_{[mn]}, \delta A_{Im}=S_{Im}$, where the vector and scalar fields correspond to the string theory states $ \bar a_\mu^m, a_\mu^m,a _\mu^I, S_{\hat Im}$ in sector 1 of section 2.3.

\subsubsection{Symmetry enhancement}
\label{sec:twistedtorus}

In order to incorporate the massless degrees of freedom that enhance the $U(1)^{16+k}$ gauge symmetry to   a $N$-dimensional group $G_L$ of rank $16+k$, we identify the $16+k$ torus  with the maximal torus of the enhanced symmetry group, so  that the $O(k,k+16;{\mathbb R})$ covariance of the abelian theory is promoted to $O(k,N;{\mathbb R})$. The $k+16$ left-moving vectors of the previous section combine with the extra massless vector states in sector 2 of section \ref{sec:masslessspectrum}, giving a total of $N$ left-moving massless vectors $a^\Gamma_\mu$, which together with the $k$ right-moving vectors $\bar a^m_\mu$ transform in the fundamental representation of $O(k,N)$. 

The $(N+k) \times(N+k)$   matrix  ${\bf M}^{\bf ab}$ is expanded as in \eqref{HM},  where the scalar fluctuations $S_{\hat I n}, S_{\alpha n}$ are now combined in the  $N \times k$ block $ M_{G\overline a}= \tilde e_G{}^\Gamma{}S_{\Gamma m}\hat{e}^m{}_{\overline a}$     as
\beq \label{HM2}
{\bf M}^{\bf ab}
 =\begin{pmatrix} 
I_{\overline a\overline b}+ \frac12( M^t M)_{\overline a\overline b}  & M^t_ {\overline aG}  \\
 M_{F\overline b}  &  I_{FG} + \frac12 (M  M^t)_{FG}
 \end{pmatrix}\, .
\eeq

The effective action is formally as \eqref{ActionGaugedSugra}, where now the non-abelian left vector fields $A_\mu^G$ and scalars $S_{Gm}$ absorb the KK left vector and scalar fields, yielding 
\bea
-\frac14{\cal F}_{\mu\nu}^M{\cal M}_{MN}{\cal F}^{N\mu\nu}&=&-\frac14{ F}_{\mu\nu}^{\overline a}\delta_{\overline {ab}}{ F}_{\mu\nu}^{\overline b}-\frac14{ F}_{\mu\nu}^{F}\delta_{FG}{ F}_{\mu\nu}^{G}
-\frac12{  F}_{\mu\nu}^{ G} M_{G \overline a}{ F}_{\mu\nu}^{\overline a}\, ,\nn
\eea
with $F_{\mu\nu}^{G}=\tilde e_\Gamma^{G}F_{\mu\nu}^\Gamma=\tilde e_\Gamma^{G}(2\partial_{[\mu}A_{\nu]}^\Gamma+f^\Gamma{}_{\Lambda\Omega}A_\mu^\Lambda A_\nu^\Omega)$ and 
\bea
\frac18D_\mu {\cal M}_{MN}D^\mu {\cal M}^{MN}&=&\frac 14D_\mu M_{G\overline a}D^\mu { M}^{G\overline a}\, ,\nn
\eea
with $
D_\mu M_{G\overline a}=\partial_\mu M_{G\overline a}+f^H{}_{FG}A_\mu^FM_{H\overline a}$.

The structure constants in the field strengths, covariant derivatives and scalar potential can be explicitly computed from the twist $\mathbb E_{\bf a}{}^{M}$, generalizing the procedure introduced for the bosonic string in  \cite{agimnr,cgin} (see also \cite{hm}).  Namely,  the extra massless vectors with non-trivial momentum and winding can  be thought of as coming from a {\it{metric}}, a {\it B}-field and a Wilson line defined in an extended tangent space, with extra  dimensions. The fields in this fictitious manifold depend on a  set of coordinates dual to the components of  momentum and  winding along the compact directions. 
Promoting
 the internal piece of the vielbein $\mathbb E_{\bf a}{}^{M}$ to an element in  $O(k,N;\mathbb{R})$,
 the fluxes computed  from the deformed generalized Lie derivative by 
\begin{equation}
{f}_{\bf abc} = 3\mathbb E_{[\bf a}{}^M \partial_M \mathbb E_{\bf b}{}^N
\mathbb E_{\bf c]}{}^
P\eta_{NP} + \Omega_{\bf abc}  \, ,\label{gfng}
\end{equation}
reproduce the structure constants of the  enhanced gauge algebra, with the deformation $\Omega_{\bf abc}$  defined below. A dependence on the left internal coordinates is therefore mandatory, but we restrict it to dependence only on the Cartan subsector, namely on the  $k+16$ coordinates $y_L^{\hat I}$, $\mathbb E_{\bf a}{}^{M}=\mathbb E_{\bf a}{}^{M}(y_L^{\hat I})$.\footnote{Note that the space itself is not extended further than the 16-dimensional torus and the double torus of dimension $2k$. The derivative in \eqref{gfng}  along ``internal directions" has only non-zero components along the $k+16$ Cartan directions of the $p+k$-dimensional tangent space.}

To be specific, start with the  generalized vielbein
\beq \label{genframe2}
\mathbb E_{\bf a}=\begin{pmatrix}-\hat e {C} & \hat e & -\hat e{A}&\\ 
 e & 0 &0\\
\tilde e{A}^t&0&{\tilde e}
\end{pmatrix}\begin{pmatrix}dy^m\\
\partial_{y^m}\\
dy_I
\end{pmatrix}\, ,
\eeq
where $e, \hat e, \tilde e, {A}$ and ${C}$   are the fields on the torus at the point of enhancement. Then, identify $\partial_{y^m} \leftrightarrow d\tilde y_m$, rotate to the left-right basis on the spacetime indices and bring the generalized vielbein to a block-diagonal form rotating the flat indices, which leads to 
\begin{eqnarray}
\mathbb E_{RL}
&=&{\sqrt{2}}
\left(
\begin{array}{cccc}
-e&0&0\\
0&e&0\\
	0& 0 & {\tilde e}
\end{array}
\right)\begin{pmatrix}dy_R\\
dy_L\\
d\tilde y^I
\end{pmatrix}\, ,\label{genviel2}
\end{eqnarray}
where
\bea
dy_L^m&=&\frac12G^{mn}[(G-C)_{nl}dy^l+d\tilde y_n-A_n^Idy_I]\, , \  dy_R^m=\frac12G^{mn}[(G+C)_{nl}dy^l-d\tilde y_n+A_n^Idy_I]\nn\\
d\tilde y^I&=& \kappa^{IJ} dy_J +\frac1{\sqrt2}A_m^I dy^m\nn \ .
\eea

Finally, we extend this $(2k+16)\times (2k+16)$ matrix so that it becomes an element of $O(k,N)$
of the form
\begin{eqnarray}
\mathbb E^M_{RL}(y_L)
&=&{\sqrt{2}}
\left(
\begin{array}{cccc}
-e & 0 & 0 &0 \\
0&   e &0&0\\
	0& 0 & 1_{16\times 16} & 0 \\
  0&0	& 0 &  \frac1{\sqrt2}{\cal J}
\end{array}
\right)\, ,\label{genviel}
\end{eqnarray}
 where the index $M=1, \cdots , k+N$ and the  $(N-(k+16)) \times (N-(k+16))$ diagonal block ${\cal J}$ contains the left-moving ladder currents associated to the $\alpha_i$ roots of the enhanced gauge group, 
$
{\cal J}_{\alpha}{}^{i}(y^1_L,...,y^{k+16}_L)=\delta_{\alpha}{}^i e^{i \sqrt2\alpha_i\cdot  y_L}$. Note that the $(N+k)\times (N+k)$  matrix 
(\ref{genviel}) depends only on the coordinates associated to the Cartan directions of the algebra. In case the gauge symmetry is enhanced to a product  of groups, ${\cal J}$ contains the currents of all the factors, each set of currents depending on the corresponding Cartan directions. 

Taking for the deformation 
 \bea
\Omega_{\bf abc}=\left\{\begin{matrix} \varepsilon(\alpha,\beta)\;\delta_{\alpha+\beta+\gamma} &\ \ {\rm if \ two \ roots \ are \ positive,}\\
-\varepsilon(\alpha,\beta)\;\delta_{\alpha+\beta+\gamma} &\ \ {\rm if \ two \ roots \ are \ negative,} \\
  \end{matrix}\right.\,\nn
 \eea 
 if ${\bf a,b,c}$ are associated with roots, and zero if one or more indices correspond to Cartan generators,
 all the structure constants can be obtained replacing \eqref{genviel} in (\ref{gfng}).
The deformation accounts for the cocycle factors that were excluded from  the CFT current operators in (\ref{genviel}) but  are necessary  in order to compensate for the minus sign in the OPE $ J^\alpha(z) J^\beta(w)$ when exchanging the two currents and their insertion points $z \leftrightarrow w$ ($c_\alpha c_\beta=\varepsilon(\alpha,\beta)c_{\alpha+\beta}$). It was conjectured in \cite{hhz} that such factors would also appear in the gauge and duality transformations of double field theory, and actually, they can be included without spoiling
 the local  covariance of the theory. Indeed,  the cocycle tensor $\Omega_{\bf abc}$ satisfies the consistency constraints of gauged DFT, \eqref{Constraintsfs1} and \eqref{fderivative},
and it  breaks the $O(k,N)$  covariance of (\ref{ActionGaugedSugra}) to $O(k,k+16)$. 
In this way, all
 the structure constants can be obtained  from (\ref{gfng}) using  the  expression (\ref{genviel}) for the generalized vielbein with the appropriate currents corresponding to the enhanced gauge groups. 
All the gaugings obtained in this way satisfy
 the quadratic constraints \eqref{Constraintsfsmall}, and therefore the construction  is consistent.

It is interesting to note that the deformation $\Omega_{{\bf abc}}$ can be chosen to be the same one for the $E_8\times E_8$ and $SO(32)$ groups. Indeed, we show in Appendix \ref{app:X}  that both groups have 26880 non-vanishing structure constants of the form $f_{\alpha\beta}{}^{\alpha+\beta}$, half of which can be chosen to be $+1$ and the other half $-1$, so that one unique deformation accounts for both heterotic theories.  The generalized vielbeins giving the remaining structure constants which involve one Cartan index can be obtained from (\ref{genviel}) with ${\cal J}$ containing the $E_8\times E_8$ or the $SO(32)$ currents. Choosing the former or the latter
 amounts to choosing a background, and in this sense the two  heterotic theories  can be considered as two solutions of the same gauged DFT, even before compactification. 

Plugging all this in \eqref{ActionGaugedSugra}, we get precisely the effective action \eqref{accionefectiva} derived from the string amplitudes,
where  the  potential is, to lowest order in $M$, 
\beq \label{Vf}
V= \frac{1}{16}M_{F \bar a}M_{F'}{}^{\bar a} M_{G \bar b}M_{G'}{}^{\bar b} f^{F G H}f^{F' G'}{}_{H} 
\ .
\eeq

Note that unlike in the bosonic theory, there is neither a cosmological constant nor  a cubic piece in the potential, which is now bounded from below.  Additionally, the quadratic piece cancels. There is also a sixth-order potential, but in order to get its explicit form we would need to expand $\bf M$ in \eqref{HM} to quartic order in the fields.\footnote{This is not necessary for the quartic order  as  the $n$-th order contributions  to $M_{\bf ab}$ cancel  in the $n$-th order contribution to the potential. }

\subsection{Away from the self-dual points}

In this section we show that moving away from a point of enhancement corresponds to giving a vacuum expectation value to $M^{\hat A \bar a}$,  the 
piece in the matrix of scalar fields that belongs to the Cartan subsector, corresponding to the KK scalars for the metric, $B$-field and Wilson lines. In the next section we show that the mass acquired by the vectors and scalars that are not in the Cartan directions agree with the string theory masses.

In the neighborhood of a given point of enhancement, the scalars in the Cartan subsector acquire a vacuum expectation value $v^{\hat A \bar a}$. 
Then we  redefine
\beq \label{vevs}
M^{\hat A \bar a} \to  v^{\hat A \bar a} + M^{\hat A \bar a}\, ,
\eeq
so that $\left\langle M^{G \bar a} \right\rangle=0$ for all indices $G,\bar a$. 
These vevs spontaneously break the enhanced symmetry:
some or all of the left-moving vectors in non-Cartan directions $A_{\mu}^{\alpha}$ get  a mass  from the covariant derivative of the scalars, given by  
\beq \label{Malpha}
m^2_{A^\alpha}= -f^{\overline \alpha}{}_{\overline \alpha \hat A} f_{\overline\alpha\underline\alpha \hat B} v^{\hat Aa} v^{\hat Ba}=  \alpha_{\hat A} v^{\hat A \bar a} \alpha_{\hat B}  v^{\hat B\bar a}=  |\alpha \cdot v|^2 \ .
\eeq

Note that, as expected,  this is always positive, unlike in the bosonic theory.

We discuss now in more detail the process of spontaneous symmetry breaking. It is simpler for this to use the Chevalley basis for the Cartan generators, where the Killing form is equal to the Cartan matrix $\kappa_{\hat I \hat J}={\cal C}_{\hat I \hat J}$, and   the components of a simple root $\alpha^{\hat I}$ 
(where the subscript $\hat I$ labels the root) are $(\alpha^{\hat I})_{\hat J}=\delta^{\hat I}{}_{\hat J}$. 

We thus have for simple roots $\alpha^{\hat I}$ and non-simple roots $\beta=(n^{\beta})_{\hat I} \alpha^{\hat I}$,
\beq \label{Mvectors}
m^2_{A^{\alpha^{\hat I}}}= |v^{\hat I}|^2  \ , \quad
m^2_{A^{\beta}} =|\beta \cdot v|^2= |n^{\beta} \cdot v|^2 \, .
\eeq

We see that by giving arbitrary vevs to all scalars in the Cartan subsector, all the gauge vectors corresponding to ladder generators acquire mass  and the gauge symmetry is spontaneously broken to $U(1)_L^{k+16} \times U(1)_R^k$. 
Similarly, if $v$ has a row with all zeros, let's say the row $\hat I_0$, then the corresponding (complex) vector $A^{\alpha^{\hat I}_0}$ remains massless, and there is at least an $SU(2)$ subgroup of $G_L$ that remains unbroken.
 The converse is also true, namely
\bea
v^{\hat I_0 \bar a}=0 \ \  \forall \bar a \qquad \Leftrightarrow \qquad m^2_{A^{\alpha^{\hat I_0}}} =0  .
\label{converse}
\eea
For the vectors associated to non-simple roots $\beta$ the situation is more tricky as it depends on which integers $n^{\beta}_{\hat I}$ are non-zero. $A^{\beta}$ remains massless if $v^{\hat I \bar a}=0$ for all $\hat I$ such that $n^{\beta}_{\hat I}\neq 0$ and for all $\bar a$. 

Note that one cannot give masses only to the vectors corresponding to non-simple roots: if all the vectors corresponding to simple roots are massless, then necessarily $v=0$ and  there is no symmetry breaking at all. This implies that the spontaneous breaking of symmetry always involves at least one $U(1)$ factor, corresponding to the Cartan of the $SU(2)$ associated to the simple root whose vector becomes massive. Thus we cannot go from one point  of maximal enhancement  in moduli  space (given by a semi-simple group) to another point of maximal enhancement by a spontaneous breaking of symmetry.       

Regarding the scalars, introducing the vevs for those in the Cartan subsector in the potential \eqref{Vf}, we get at quadratic order in the scalar fields  
\bea
\frac1{16}f^{FGH} f^{F'G'}{}_H({M^tM})_{FF'}({M^tM})_{GG'}&\rightarrow& \frac18\sum_{\alpha, \bar b,\bar c}\left(f^{\hat A\alpha H} f^{\hat A' \alpha'}{}_{ H} \ v_{\hat A\bar b}v^{\bar b}{}_{\hat A'}M_{\alpha \bar c}M_{\alpha'}{}^{\bar c}\right.\nn\\
&&\ \ \ \  \ \ \  \left.+ \ 2f^{\hat A\alpha }{}_{H} f^{{\alpha'}\hat B' H} v_{\hat A\bar b}v_{\bar c\hat B'}M_{\bar b\alpha'}M_{\alpha \bar c} \right)\, .
\ \ \ \ \ \ \ \ \ 
\label{MscalarDFT1}
\eea
The first term gives
\beq \label{MscalarDFT}
-\frac14 \sum_{\rm{all\,  roots} \, \alpha, \bar c} m^2_\alpha   | M^{\alpha \bar c} |^2 \ , \quad {\rm where} \ \ m^2_\alpha= \sum_{\bar b}(\alpha^{\hat A} v_{\hat A\bar b})^2  \, ,
\eeq
and then replacing it in the action \eqref{ActionGaugedSugra}, we see that the mass of the scalar fields agrees with the mass of the vectors \eqref{Malpha}.  The second term can be written as
\beq \label{MscalarDFT2}
\frac14 \sum_{\rm{all\,  roots} \, \alpha}\left(\sum_{\bar b} m_{\bar b}    M^{\alpha \bar b} \right)^2  \ ,\quad {\rm where} \ \ m_{ \bar b}= \sum_{\hat A} \ \alpha^{\hat A} v_{\hat A\bar b}  \ 
\eeq
and $M^\alpha=\sum_{\bar b} m_{ \bar b}   M^{\alpha\bar b} $ is the Goldstone boson contribution which is eaten by the vectors to become massive.  This agrees with the results in \cite{aamp} on which we expand  in the next section.

 We thus get that for arbitrary vevs, all vectors and scalars except  those along Cartan directions acquire masses, and the symmetry is broken to $U(1)_L^{k+16} \times U(1)_R^k$. If $v^{\hat I_0 \bar a}=0$ for a given $\hat I_0$ and for all $\bar a$, while all other vevs are non-zero, then the remaining symmetry is at least $(SU(2) \times U(1)^{k+15})_L \times U(1)_R^k$ where the $SU(2)_L$ factor corresponds to the root $\alpha^{\hat I_0}$,   
and the massless scalars are, besides those purely along Cartan directions, at least all those of the form $M^{\alpha^{\hat I_0}\bar a}$.

\subsubsection{Comparison with string theory}

Let us compare the vector and scalar masses that we got in the previous section from the double field theory effective action, to those of string theory given by \eqref{MZhZ}. 

We decompose the generalized metric ${\cal M}$ as in \eqref{HMHA}, where $\mathbb{E}$ is the twist containing the information on the background  at the point of enhancement and ${\bf M}^{ab}$ represents the fluctuations from the point, parameterized as in \eqref{HM} in terms of the matrix $M$ in \eqref{MgBA} \footnote{Note that $M$ here is a $(k+16)\times k$ matrix spanning along the Cartan directions only, as in section \ref{sec:torus}.  }. Inserting this in the mass formula (\ref{MZhZ}) we get 
\beq \label{MZhZE}
m^2=2 \left(N+\bar N-\left\{\begin{matrix}1&{\rm R \ sector}\\
\frac32&{\rm NS \ sector}\end{matrix}\right.\right) +  Z^t \mathbb{E}^t  \begin{pmatrix} I_{k}+\frac12 M^t  M &
 M^t\\
M  & I_{k+16}+\frac12 M M^t \end{pmatrix}\mathbb{E} Z \ ,
\eeq
 
On the other hand, from Eq. \eqref{pE} 
\beq
 \mathbb{E} \, Z =  \begin{pmatrix} p_{aR} \\ p_{aL} \\ p^A_L \end{pmatrix} \equiv  \begin{pmatrix} p_{R} \\ p_{L}  \end{pmatrix} \ .
\eeq
We thus get
\bea \label{MM}
m^2&=&2 \left(N+\bar N-\left\{\begin{matrix}1&{\rm R \ sector}\\
\frac32&{\rm NS \ sector}\end{matrix}\right.\right) +   p^t_{R} (I_{k}+\tfrac12 M^t  M) p_R +    p^t_{L}  ( I_{k+16}+\frac12 M M^t )  p_{L}   \nn \\
&& \qquad \qquad \qquad \qquad \qquad \qquad \quad \, +  p^t_{R} M^t p_{L} +  p^t_{L}  M p_R\ . 
\eea
The bosonic states that are massless at the point of enhancement (when $M=0$) have $p_R=0$ and $\bar N = \frac12$ in the $NS$ sector. 

The left-moving vectors have either $N=1$ and $p_L=0$, or $N=0$ and ${p}_L=\alpha $ with $\alpha$ a root of the enhanced gauge algebra (and thus $|\alpha|^2=2$). The former vectors (Cartan) are massless for any $M$, while, according to \eqref{MM}, the latter have mass
\beq
m_{A^{\alpha}}^2=\tfrac12 \alpha^t M^t M \alpha \ .
\eeq 
On the right sector the only massless vectors are the Cartan, which are massless for any $M$.
This agrees with the  masses \eqref{Malpha} if we identify 
\beq \label{v}
v= -\hat e_0 \big (  (\delta G - \delta B ') \hat e_0^t ,\sqrt{2} \delta A\hat{\tilde e}^t \big ) \ .
\eeq


The scalars in sector 1 (both legs along Cartan directions) have $p_L=0$, $N=1$ and $\bar N=\frac12$, and are massless for any $M$. The scalars in sector 2 have $N=0$, $\bar N=\frac12$ in the $NS$ sector,  and $p_L=\ax$. Their masses are thus exactly those of the vectors corresponding to the same root, namely 
\beq
m^2_{M^{\alpha \bar a}}=m^2_{A^{\alpha}}  
\eeq
in agreement with what we have found from DFT, Eq. \eqref{MscalarDFT}, confirming that these are the Goldstone bosons of the spontaneous breaking of symmetry. 


It is interesting to recall that the combinations
$ \tilde f_{\bar a}{}^{\underline \alpha\overline \alpha}\equiv f^{\hat A\underline\alpha\overline\alpha}v_{\hat A\bar a}$ appearing in the vector and scalar masses \eqref{Malpha}  and \eqref{MscalarDFT1}
agree with the coefficients of the  string theory three-point functions   involving  one massless right or left vector and two massive left vectors. Then following \cite{aamp}, one could identify the DFT fluxes with the string theory three-point amplitudes and conclude that the fluxes depend on the moduli.  Actually, 
from a gSS DFT point of view,  the vevs can be thought of  as  being encoded either in the twists $\mathbb E_{\bf a}{}^M(y_L)$ or in the fluctuations $\Phi_{\bf a}{}^{\bf b}(x)$. In this section we have developed the latter identification, i.e. the fluxes  $f^{\hat A\underline\alpha\overline\alpha}$ are computed from \eqref{gfng} with the twist \eqref{genviel} containing the currents corresponding to the enhanced gauge group, and the  symmetry is broken by the vevs  shifting the fluctuations in \eqref{vevs}.
In the former case, i.e. to get moduli dependent  fluxes,  one can replace the currents in (\ref{genviel})    by those of the massive vectors  in (\ref{jota}), and then the twists  depend on both the left- and the right-moving internal coordinates, $\mathbb E_{\bf a}{}^M(y_L, y_R)$. In this way, the fluxes computed   from the deformed generalized Lie derivative \eqref{gfng} get mixed indices from the  left and right moving sectors, reproducing   the coefficients of the string theory three-point functions which involve massive vectors \eqref{sc}. One could then interpret that the fluxes  $\tilde f_{\bar a}{}^{\underline \alpha\overline \alpha}$ encode the information about the background through the vertex operators creating the string theory vector and scalar states.

\section{Summary and Outlook}

In this paper we have analysed compactifications of the heterotic string on $T^k$, focusing on the phenomenon of symmetry enhancement arising at special points and curves in moduli space.   The  $O(k,k+16)$ covariant formulation and  the rich structure of  the moduli space of these compactifications were reviewed in section 2.
At special points in moduli space, the abelian $U(1)^k_R \times U(1)^{k+16}_L$   symmetry that arises at  generic points  is enhanced on the left-moving sector to finite groups or product of groups   of rank $k+16$  in ADE. While the symmetry group is maximal (i.e. has no $U(1)$ factors) at isolated points,   non semi-simple groups arise at higher-dimensional subspaces of the moduli space.

The 17-dimensional moduli space of $S^1$ compactifications, involving  the radius of the circle and the 16 components of the Wilson line along the Cartan directions of the $SO(32)$ or $E_8 \times E_8$ gauge group,  was studied in detail in section 3. 
 We found all the possible maximal enhancements  from the generalized Dynkin diagram of the Narain lattice $\Gamma^{1,17}$. These are displayed in Tables 1 and 2. In particular, we showed that the same enhancements can be achieved in both heterotic theories (e.g.  $SO(34)$ enhancement  from the $E_8 \times E_8$ string) and briefly explained how to obtain them in Appendix \ref{app:Dyndia}. 

The discussion of the explicit enhancement process is  split into  compactifications with $\pi \cdot A \in \mathbb{Z}$ and $\pi \cdot A \notin \mathbb{Z}$. Although  all the enhancements can be obtained  with Wilson lines that are not on any lattice by appropriately choosing $R$ (including those with $SO(32)$ or $E_8 \times E_8$ subgroups), the distinction is useful to understand the enhancement process. When the Wilson line has zero vacuum expectation value, or equivalently when the vev is on the root lattice $\Gamma_{g}$, the gauge group of the uncompactified theory is unbroken at generic radius, and the total gauge group on the external space is $U(1)_R \times \left( U(1) \times SO(32)\right)_L$ or $U(1)_R \times \left( U(1) \times E_8 \times E_8 \right)_L$. At $R=1$, there are additional states with momentum and winding that become massless and enhance the $U(1)_L$ to  $SU(2)_L$. For other values of Wilson lines and generic $R$, the  gauge symmetry is determined by the subset of heterotic momenta $\pi$ that have integer inner product with the Wilson line. In the $SO(32)$ theory, one has the interesting possibility of a Wilson line that has integer inner product with all $\pi$, i.e. a Wilson line in the dual root lattice, but which is not in the lattice, namely $A\in \Gamma_v$ or $A\in \Gamma_c$. These two possibilities lead to an unbroken $SO(32)$ gauge symmetry at any radius, while at $R^2=\tfrac{1}{2}$ there are extra massless states with non-zero momentum and/or winding number on the circle, giving a total 17-component left moving momentum with mixed circle and chiral heterotic directions which enhance the gauge symmetry to $SO(34)$.

We developed a method for computing and drawing two dimensional slices of the $17$-dimensional moduli space which neatly exhibit the distribution of the enhanced groups. The family of functions  corresponding to each of the curves and the heterotic momentum of the additional massless states can be obtained from this analysis. While 
 non maximal enhacement occurs at lines, maximal enhacement occurs at isolated points.     More interesting figures arise at smaller radii, and the smaller the radius, the richer the pattern of enhanced gauge symmetries, as there are more winding numbers that lead to massless states. Moreover, we were able to univocally relate the intersections of the curves in the figures with the enhanced groups  obtained from the generalized Dynkin diagram. An interesting output of the construction  is that, in order to obtain groups that contain $SO(32)$ from the $\Gamma_8 \times \Gamma_8$ theory or groups that contain $E_8 \times E_8$ from the $\Gamma_{16}$ theory it is necessary to choose a slice where, for a generic point, the group is $SO(16)\times SO(16)$ or a subgroup of it.

The points of enhacement are fixed points of T-duality symmetries. In section \ref{sec:T-duality}, we presented  the action of the standard T-duality exchanging momentun and winding number, and studied its fixed points, which are at $R^2=1-\frac12 |A|^2$. At these points, the dual background has the same radius and opposite Wilson line, $A'=-A$. If $2A$ is in the root lattice, then $A'=-A\sim A$ and the full background is self-dual. 
 For Wilson lines with only one non-zero component, as those explored in section \ref{sec:A1R}, the fixed ``points" of the T-duality symmetry are not really points, but in this two-dimensional subspace of moduli space they correspond to lines of non-maximal enhancement symmetry, where the Wilson line is a function of the radius ($A=A(R_{\rm{sd}})$), and is such that  $A\sim A_{\rm{sd}}$, with $|A_{\rm{sd}}|^2=2(1-R^2_{\rm{sd}})$.

 More general dualities were studied   in section  \ref{sec:36}, in which the dual spectrum, defined by the 17+1-dimensional vector of left and right-moving momentum of the states, has a minus sign on the right-moving component, while on the left-moving part of the vector, it leaves the circle direction invariant, while it inverts 0, 1, 15 or 16 of its components along the heterotic directions. For generic Wilson lines, we studied the fixed points of these symmetries that have the largest radius. The results are given in Table 4.  The columns and rows in this table correspond respectively to the ``theories" before and after the duality (the dualities that invert an odd number of components of the left-moving vector are dualities  between a theory with a given lattice and a theory with another lattice, for example between the $SO(32)$ theory (denoted by $\Gamma_{16}$) and what we called $SO(32)^-$ ($\Gamma_{16}^-$),  differing by the choice of chirality of the spinor representation).  We indicate the radius of the fixed point and the possible Wilson lines before and after the duality transformation.  
 
 We then concentrated on the situation in which the Wilson line has only one non-zero component, in order to study in full generality the pattern of fixed points. We found that the fixed points of the symmetries that do not invert the momentum along the circle or the direction where the Wilson line has a vev, the enhancement of symmetry is maximal, given by $SU(2) \times SO(32)$ ($SU(2) \times E_8 \times E_8$) for dualities that leave the full vector ${\bf p_L}$ invariant, and $SO(34)$ ($SO(18) \times E_8$) for dualities that leave invariant only the momentum along the circle and the direction where the Wilson line is, while invert the other directions. In the former case, the fixed points are those that satisfy the quantization conditions $\{ R^{-1}, \frac12 R^{-1}A, (\frac12 \frac{A^2}{R}+R) \} \in {\mathbb Z}$, while for the latter the requirements are  $\{ \tfrac{1}{\sqrt2}R^{-1}, \tfrac{1}{\sqrt2} R^{-1}A, \tfrac{1}{\sqrt2} (\frac12 \frac{A^2}{R}+R) \} \in {\mathbb Z}$. We collect all the points satisfying these constraints, together with the corresponding Wilson lines, in Table 3.
 
The effective field theory reproducing the three and (some) four point functions of massless states at the enhancement points was constructed in section 4 and reformulated in terms of $O(k, N)$ multiplets, with $N$ the dimension of the gauge group.  We verified that  the string theory results are encoded in  a generalized Scherk Schwarz (gSS) compactification of heterotic DFT, not only at the special points in moduli space giving maximally enhanced gauge symmetries,  but also when moving slightly away from the selfdual points, where many of the fields acquire  mass. 
In the process of symmetry breaking, there is always a $U(1)$ factor in the unbroken symmetry group at any point in the neighborhood, reflecting that non maximal enhancement appears at lines rather than isolated points in the slices of  moduli space represented in the figures of the preceeding sections.

The equivalence between gSS compactifications and gauged DFT was used to show that  a deformation of the generalized Lie derivative,  defined by the cocycles of the  gauge algebra, provides a gauge principle that determines the low energy effective field theory of the toroidally compactified  string at the enhancement points.  The construction of   \cite{cgin} was  extended to obtain the toroidally compactified heterotic string effective field theory  in arbitrary dimensions.  In particular, we have shown that the $SO(32)$ and  $E_8\times E_8$ algebras have the same cocycles, and hence a unique gauged DFT describes the two heterotic theories with these gauge groups  in any dimension, and even before compactification. This is an interesting result, which extends the known equivalence of both heterotic string theories on $T^k$ to the (gauged) supergravity limits, even of the uncompactified theory.
As a consequence, the low energy effective field theories with $SO(32)$ or  $E_8\times E_8$ non-abelian symmetry in any dimension can be considered as two solutions of the same gauged DFT. In this theory, the generalized vielbeins producing the  structure constants from the generalized Lie derivative are parameterized
in terms of  the currents   in the vertex operators of the vector fields. The problem of finding a generalized vielbein or a bracket that gives rise to  the full algebra without adding cocycles  is left for future investigation.

An obvious  natural extension of our work  is to consider in more detail toroidal  compactifications to lower dimensions. 
Not only should it be possible to find more appealing models from a phenomenological point of view, but a richer structure of gauge symmetries will certainly appear. Already in the next simplest step, that of examining the structure of the moduli space in compactifications on $T^2$, a non-vanishing $B$ field  background  turns the analysis more challenging but also more interesting, and important applications to F-theory are expected.

The possibility to construct a low energy effective action invariant under the discrete $O(k, k+16;\mathbb Z)$ duality group, raised in \cite{Giveon:1990mw},  is another interesting question  to address. Since string backgrounds related by duality yield the same physics, one expects that an $O(k, k+16;\mathbb Z)$-invariant low energy effective theory  exists. 
Generic $O(k, k+16;\mathbb Z)$ transformations map states that are outside the Cartan subalgebra of the enhanced gauge group into massive modes, and the typical orbit of a string state  has an infinite number of points.  This theory  should contain all the fields which correspond to string states that become massless at some point of the moduli space. The number of such fields is infinite, and on a given background all except a finite number of them are  massive. Previous work in this direction includes the duality invariant low energy effective action for the ${\cal N} = 4$ heterotic string  constructed in \cite{Giveon:1990mw,Giveon:1990er}, the  description of the entire moduli space from compactifications on higher dimensions performed in  \cite{cgin}  and the introduction of a non-commutative
product  on the compact target space as well as new  vector and scalar fields depending on double periodic coordinates that was suggested in \cite{Aldazabal:2018uzm}.

 The emphasis in our work has been to study gauge symmetry enhancement  in   toroidal compactifications of perturbative heterotic string theory, both  for the characterization of the string theory moduli space and as a symmetry of the low energy effective theory.  Clearly, it would also be desirable to explore extensions and generalizations to other internal spaces, as well as to include non-perturbative effects, where the physics of symmetry enhancement plays an important
part. In particular, winding heterotic $ E_8 \times E_8$ states are related to the dynamics of $D$-particles in the presence of $D8$-branes and orientifold planes in type I’ superstring theory, and have been crucial in the understanding of  subtle aspects of  the Type I/heterotic duality \cite{BGS,cv,gaber}. We hope that the methods developed here are useful to analyze these questions further.

\subsection*{Acknowledgements}
We thank D. Berman, Y. Cagnacci, A. Font, M. Green, S. Iguri, G. Inverso, D. Marqu\'es, S. Massai and R. Minasian for interesting comments and valuable insights, and A. Capdepon for hospitality during the completion of the project.
This work was partially supported
by PIP-CONICET- 11220150100559CO (2015-2017), UBACyT 2014-2017,  ANPCyT- PICT-2016-1358 (2017-2020)  and ANR grant Black-dS-String.


\appendix


\section{Lie algebras and lattices}\label{app:lattices}

Modular invariance of the one-loop partition function of the heterotic string implies that the 16-dimensional internal momenta must take values in an  even self-dual Euclidean lattice, $\Gamma=\Gamma^*$, of dimension 16. 
There are only two of these: $\Gamma_8\times\Gamma_8$, where $\Gamma_8$ is the root lattice of $E_8$, and $\Gamma_{16}$, which is the root lattice of $SO(32)$
 in addition to the  $(s)$ or $(c)$ conjugacy class 
 \bea
 \Gamma_8 \times \Gamma_8 &=&\Gamma_g \qquad  \quad \, \rm{for} \ E_8 \times E_8 \\
\Gamma_{16}&=& \Gamma_g + \Gamma_s \quad  \rm{for} \ SO(32)\nn 
\eea
 
 In this Appendix we summarize some basic notions on these lattices, which are named Narain lattices.

Given a Lie algebra $g$ of rank $n$, taking  arbitrary integer linear combinations of root vectors, one generates an $n$-dimensional Euclidean lattice $\Gamma_g $, called the root lattice. E.g., for  the  rank $n$ orthogonal groups $SO(2n)$,
the $n$ component simple root vectors  are
\bea 
(\pm 1, \pm 1, 0,\dots) \ \ {\rm all \ other \ entries \ zero,}
\eea
and all permutations of these. For $E_8$,  the eight component vectors 
\bea
\begin{matrix}(\pm 1, \pm 1,0,0,0,0,0,0)\ +\ {\rm permutations}\\
\\
\left(\pm\frac12,\pm\frac12,\pm\frac12,\pm\frac12,\pm\frac12,\pm\frac12,\pm\frac12,\pm\frac12\right) \ {\rm even \ number \ of \ "-" \ signs}\end{matrix}
\eea
contain the 240 roots, i.e. the 112 root vectors of $SO(16)$ and 128 additional vectors.

Any Lie group $G$ has infinitely many irreducible representations which are characterized by their weight vectors. Irreducible representations fall into different conjugacy classes, and $\Gamma_g$ can be thought of as the $(0)$ conjugacy class.
Two different representations are said to be in the same conjugacy class if the difference between their weight vectors is a vector of the root lattice.

While $E_8$ has only one conjugacy class, namely $(0)$, the $SO(2n)$ algebras have four inequivalent conjugacy classes:
\begin{itemize}
\item  The $(0)$ conjugacy class, i.e. the root lattice, contains vectors of the form
\bea
(n_1, \dots,n_{n})\, , \quad n_i\in {\mathbb Z}\, , \quad \sum_{i=1}^n n_i=0 \ {\rm mod}\ 2\, . \label{a1}
\eea
\item  The vector conjugacy class, denoted by (v), contains  vectors of the form
\bea
(n_1, \dots,n_{n})\, , \quad n_i\in {\mathbb Z}\, , \quad \sum_{i=1}^n n_i=1 \ {\rm mod}\ 2\, .\label{a2}
\eea
\item The spinor conjugacy class, denoted by (s), contains  vectors of the form
\bea
(n_1+\frac12, \dots,n_{n}+\frac12)\, , \quad n_i\in {\mathbb Z}\, , \quad \sum_{i=1}^n n_i=0 \ {\rm mod}\ 2\, .\label{a3}
\eea
\item The (c) conjugacy class contains  vectors of the form
\bea
(n_1+\frac12, \dots,n_{n}+\frac12)\, , \quad n_i\in {\mathbb Z}\, , \quad \sum_{i=1}^n n_i=1 \ {\rm mod}\ 2\, .\label{a4}
\eea
\end{itemize}

The weight lattice $\Gamma_w$ is formed by all weights of all conjugacy classes including the root lattice itself. Clearly $\Gamma_g\subset \Gamma_w$, and for a simply-laced Lie algebra, which roots have squared modulus 2,  it can be shown that $\Gamma_g=\Gamma_w^*$.
   Therefore, the weight lattice of $E_8$ contains the weights of the form
\bea \label{weightE8}
\Gamma^8_w: \left\{\begin{matrix}(n_1, \dots,n_8)&& \\
\left(n_1+\frac12,\dots,n_8+\frac12\right) \, , && \sum_{i=1}^8n_i= {\rm even \ integer}\end{matrix}\right.
\eea
with $n_i\in{\mathbb Z}$,  is identical to its root lattice, which implies that it is even self-dual. It is also identical to the $SO(16)$ lattice with the $(0)$ and $(s)$ conjugacy classes

A necessary condition for a self-dual lattice is that it be unimodular. 
The $SO(2n)$ Lie algebra lattices are unimodular if they contain two conjugacy classes. The weight lattice of $Spin(32)/{\mathbb Z}_2$ is identical to the $SO(32)$ lattice with the $(0)$ and $(s)$ conjugacy classes. It is even self-dual and it's vectors are:
\bea
\Gamma^{16}_w: \left\{\begin{matrix}(n_1, \dots,n_{16})&& \\
\left(n_1+\frac12,\dots,n_{16}+\frac12\right) && \sum_{i=1}^{16}n_i= {\rm even \ integer}\end{matrix}\right.
\eea

 Both the root lattice of $E_8\times E_8$  and the weight lattice of $Spin(32)/{\mathbb Z}_2$ contain 480 
vectors of (length)$^2=2$ which are the roots of $E_8\times E_8$ and $SO(32)$, respectively.

It is convenient to write the conjugacy classes of $SO(32)$ in terms of conjugacy classes of  
representations of $SO(16)\times SO(16)$. We denote by $(xy)$ a vector with the first eight components in the conjugacy class $(x)$ of $SO(16)$ and the last eight in the class $(y)$. $x$ and $y$ can be $0$, $s$, $v$ or $c$. We then have $16$ conjugacy classes $(xy)$.
The $SO(32)$ conjugacy classes correspond to the following $SO(16)\times SO(16)$ pairs 
\beq \label{Gamma16Gamma8}
(0) &= (00),(vv)  \\
(s) &= (ss),(cc)  \\
(c) &= (sc),(cs)\\
(v) &= (0v),(v0) 
\eeq
We have then
\beq \label{decompositionGammas}
 \Gamma_{16} &= \Gamma^{16}_{0} + \Gamma^{16}_{s}=(00),(vv),(ss),(cc) \\
\Gamma_8 \times \Gamma_8\equiv \Goo  &= (\Gamma^8_{0}+\Gamma^8_{s})\times(\Gamma^8_{0}+\Gamma^8_{s})=(00),(ss),(0s),(s0) 
\eeq
The dual to the root lattice of $SO(32)$ is 
\beq \label{duallattice16}
(\Gamma_0^{16})^*&=\Gamma_g^*=(00),(vv),(ss),(cc),(0v),(v0),(sc),(cs) . 
\eeq
We also use the following properties of the lattices
 \beq \label{propertiesGammas}
   \Goo \backslash \Gamma_{16} &= (0s),(s0) \\
   \Gamma_{16} \backslash \Goo&= (vv),(cc) \\
    \Gamma_{16} \cap \Goo&=  (00),(ss)  \\
( \Gamma_{16} \cap \Goo)^*&= (00),(ss),(vv),(cc),(vc),(cv),(0s),(s0)  \\
 (\Gamma_{16}\cap\Goo)^{*} \backslash (\Gamma_{16} \cup \Goo) &=(vc),(cv)  \ .
 \eeq

Note that both for $SO(32)$ (or rather $Spin(32)/{\mathbb Z}_2$) and for $E_8$, one could have chosen the opposite chirality, namely the (c) class instead of (s). We will denote this choice  $SO(32)^-$ and $E_8^-$. We can then build the following pairs 
\beq \label{decompositionGammasminus}
 \Gamma_{16}^- &= \Gamma^{16}_{0} + \Gamma^{16}_{c}=(00),(vv),(sc),(cs) \\
\Gamma_8^- \times \Gamma_8^- &= (\Gamma^8_{0}+\Gamma^8_{c})\times(\Gamma^8_{0}+\Gamma^8_{c})=(00),(cc),(0c),(c0) \\
\Gamma_8^+ \times \Gamma_8^- &= (\Gamma^8_{0}+\Gamma^8_{s})\times(\Gamma^8_{0}+\Gamma^8_{c})=(00),(sc),(0c),(s0) \\
\Gamma_8^- \times \Gamma_8^+ &= (\Gamma^8_{0}+\Gamma^8_{c})\times(\Gamma^8_{0}+\Gamma^8_{s})=(00),(cs),(0s),(c0) 
\eeq

\section{Generalized Dynkin diagram of $\Gamma^{1,17}$}\label{app:Dyndia}

The  equivalence of the two heterotic strings  on $S^1$ is determined by the uniqueness of  the  Lorentzian $\Gamma^{1,17}$  root lattice. The  generalized Dynkin diagram of $\Gamma^{1,17}$ is obtained  by adding  roots associated with the crosses in the following extension of the $SO(32)$ and $E_8 \times E_8$ Dynkin diagrams  respectively
 \bea 
 \begin{dynkin} 
\dynkinline{1}{0}{2}{0};
\dynkinline{2}{0}{3}{0};  
\dynkinline{3}{0}{4}{0}; 
\dynkinline{4}{0}{5}{0};
\dynkinline{5}{0}{6}{0}; 
\dynkinline{6}{0}{7}{0};
\dynkinline{7}{0}{8}{0};
\dynkinline{8}{0}{9}{0}; 
\dynkinline{9}{0}{10}{0};
\dynkinline{10}{0}{11}{0};
\dynkinline{11}{0}{12}{0};
\dynkinline{12}{0}{13}{0};
\dynkinline{13}{0}{14}{0};
\dynkinline{14}{0}{15}{0};
\dynkinline{14}{0}{14}{1};
\foreach \x in {1,2,3,4,5,6,7,8,9,10,11,12,13,14,15}{\dynkindot{\x}{0}}
\dynkindot{14}{1}
\dynkincross{2}{1} 
\dynkincross{14}{2}
\dynkincross{2}{2} 
 \end{dynkin}\\  
 \begin{dynkin} 
\dynkinline{1}{0}{2}{0};
\dynkinline{2}{0}{3}{0};  
\dynkinline{3}{0}{4}{0}; 
\dynkinline{4}{0}{5}{0};
\dynkinline{5}{0}{6}{0}; 
\dynkinline{10}{0}{11}{0};
\dynkinline{11}{0}{12}{0};
\dynkinline{12}{0}{13}{0};
\dynkinline{13}{0}{14}{0};
\dynkinline{14}{0}{15}{0};
\dynkinline{14}{0}{14}{1};
\dynkinline{2}{0}{2}{1};
 \dynkinline{14}{1}{14}{2};
\dynkinline{2}{1}{2}{2};
\foreach \x in {1,2,3,4,5,6,10,11,12,13,14,15}{\dynkindot{\x}{0}}
\foreach \x in {7,8,9}{\dynkincross{\x}{0}}
\dynkindot{14}{1}
\dynkindot{2}{1} 
\dynkindot{14}{2}
\dynkindot{2}{2} 
 \end{dynkin}
\eea
The $17$-dimensional moduli space of  inequivalent compactifications can be chosen to be delimited by $19$ boundaries, each of them associated with one of the nodes of the generalized Dynkin diagram
 \bea \label{GDD}
\begin{dynkin} 
\dynkinline{1}{0}{2}{0};
\dynkinline{2}{0}{3}{0};  
\dynkinline{3}{0}{4}{0}; 
\dynkinline{4}{0}{5}{0};
\dynkinline{5}{0}{6}{0}; 
\dynkinline{6}{0}{7}{0};
\dynkinline{7}{0}{8}{0};
\dynkinline{8}{0}{9}{0}; 
\dynkinline{9}{0}{10}{0};
\dynkinline{10}{0}{11}{0};
\dynkinline{11}{0}{12}{0};
\dynkinline{12}{0}{13}{0};
\dynkinline{13}{0}{14}{0};
\dynkinline{14}{0}{15}{0};
\dynkinline{14}{0}{14}{1};
\dynkinline{2}{0}{2}{1};
 \dynkinline{14}{1}{14}{2};
\dynkinline{2}{1}{2}{2};
\foreach \x in {1,2,3,4,5,6,7,8,9,10,11,12,13,14,15}{\dynkindot{\x}{0}}
\dynkindot{14}{1}
\dynkindot{2}{1} 
\dynkindot{14}{2}
\dynkindot{2}{2} 
\foreach \x in {1,2,3,4,5,6,7,8,9,10,11,12,13,14,15}{\dynkinlabel{\x}{-1}{\x}}
\dynkinlabel{15}{1}{16}
\dynkinlabel{1}{1}{17}
\dynkinlabel{15}{2}{18}
\dynkinlabel{1}{2}{19}
 \end{dynkin}
 \eea 

A possible fundamental region for the moduli space is determined by the points satisfying all of the following inequalities
\begin{center}
 \begin{tabular}{|c|c|c|}
\hline 
Node  &  Fund region for $\Gamma_{16}$&  Fund region $\Gamma_{8}\times \Gamma_8$\\
\hline\hline 
$1\le i\le 6$ & $  A_i \leq A_{i+1}  $ & $  A_i \leq A_{i+1} $\\
7  &$  A_7 \leq A_8  $& $ A_7 \leq A_8 + 1 $\\
8  &$  A_8 \leq A_9  $& $  \sum_{i=1}^{16}A_i^2  \geq 2-2R^2 $\\
9  &$  A_9 \leq A_{10}  $& $ A_9 \leq A_{10} + 1$\\
$10\le i\le 15 $  &$  A_{i} \leq A_{i+1}  $& $  A_{i} \leq A_{i+1} $\\
16  &$  A_{16} \leq 1-A_{15}  $& $  A_{16} \leq -A_{15}$\\
17  & $  -A_{2} \leq A_{1}    $& $ -A_2  \leq A_1 $\\
18  &$    \sum_{i=1}^{16}A_i^2 \geq 2-2R^2  $& $  
\sum_{i=9}^{16} A_i \geq 0$\\
19  &$   \sum_{i=1}^{16}(A_i - \tfrac12)^2 \geq 2-2R^2   $& $\sum_{i=1}^{8} A_i \leq 0 $ \\
\hline
\end{tabular}
\end{center}

This defines a 17-dimensional surface resembling a chimney \cite{cv}. In $\Gamma_{16}$, the first $17$ nodes define walls parallel to the $R$ direction and the last two nodes define hyperspheres which delimit the bottom of the chimney. In $\Gamma_8 \times \Gamma_8$, there are $18$ walls and only one hypersphere at the bottom defined by the $8$th node . 
At the borders of the fundamental region, where some equalities are saturated, the gauge symmetry is enhanced. The enhanced gauge group is   obtained by removing all the nodes of the extended Dynkin diagram except  those with saturated inequality. Hence, the  maximally enhanced symmetries saturate all but $2$ of the inequalities\footnote{Actually, if the group has one or two $E_2$,  3 or 4 nodes have to be removed  instead of 2.}. It can be shown that all the possible combinations of  saturated inequalities  produce Dynkin diagrams of the ADE classification.

Some sections of the bottom of the chimney are represented below  in figures \ref{fig:so32_n1} to \ref{fig:so32_n16}  by the red curves that intersect the horizontal axis and the purple curves that intersect the $A=\tfrac12$ line.  These are the sections of the hypersphere associated respectively to the nodes $18$ and $19$ in the $\Gamma_{16}$ case. The absence of  purple curves in the first eight figures is related to the fact that for Wilson lines with more than $7$ zeros there are no spinorial roots which makes the  inequality of the $19$th node impossible to saturate.

All the maximally enhanced groups of the heterotic string on $S^1$ are  listed in the tables of section 3, where we give the point in moduli space lying in the fundamental region where these arise. 

\section{Maximal enhancement points for $A=(A_1,0_{15})$}\label{app:enhancement}

In this Appendix we show how to obtain the maximal enhancement points for the particular case of Wilson lines with only one non-zero component, treated in section \ref{sec:A1R}. We also prove that the only possible maximal enhancements for Wilson lines with only one non-zero entry are to  $SU(2)\times SO(32)$, $SO(34)$, $SU(2)\times E_8 \times E_8$ and $SO(18)\times E_8$. 

The maximal enhancement points are those where two or more curves intersect. There are three types of intersections:
$a_{w_1,q_1}(R)=a_{w_2,q_2}(R),
b_{w_1,q_1}(R)=b_{w_2,q_2}(R)$ and
$a_{w_1,q_1}(R)=b_{w_2,q_2}(R)$, that we treat separately.
In the case of $\Gamma_{8}\times \Gamma_{8}$, the curves $b$ can in principle have a curve $c$ on top of them.

\subsection{${\bf a_{w_1,q_1}(R)=a_{w_2,q_2}(R)}$}
\beq 
a_{w_1,q_1}(R)= \tfrac{2q_1 \pm_1 \sqrt{2-2w_1^2R^2}}{w_1},\quad \tfrac{2q_1^2 -1}{w_1} \in \mathbb{Z}\\
a_{w_2,q_2}(R)= \tfrac{2q_2 \pm_2 \sqrt{2-2w_2^2R^2}}{w_2},\quad \tfrac{2q_1^2 -1}{w_2} \in \mathbb{Z}
\eeq 
imply
\beq 
\mp_1  w_2\sqrt{2-2w_1^2R^2} \pm_2 w_1\sqrt{2-2w_2^2R^2}  =2q_1 w_2 - 2q_2w_1 \equiv C'=2C \in 2\mathbb{Z}\, .
\eeq 
The case $C = 0$ is trivial, so we must assume $C \neq 0$, which leads to
\beq 
R^2 = \frac{2}{C'^2} - \frac{(2w_1^2+2w_2^2 -C'^2)^2  }{8w_1^2 w_2^2 C'^2}\, . \label{formula_R}
\eeq 
Defining $N = \tfrac{(1 - 2q_1^2)}{w_1} w_2 + \tfrac{(1- 2q_2^2)}{w_2} w_1  + 4q_1 q_2 \in \mathbb{Z}$, we can rewrite \eqref{formula_R} as
\beq 
N^2   = 4 - 2C'^2 R^2\, .
\eeq 
 Since $(1-2q_i^2)$ and  $w_i$ are odd,   $N$  is even. Also, since $C'$ and $R$ are non-zero we get $N^2 < 4$, which implies $N=0$, then $R^2 = \frac{2}{C'^2}$.
Then the radius where a curve $a$ with winding $w_1$ intersects another curve $a$ with winding $w_2$ is
\beq 
R^{-2} = w_1^2 + w_2^2\, .
\eeq 
The constraint
\beq 
|q_1 w_2 - q_2 w_1| = \sqrt{\tfrac{w_1^2 + w_2^2}{2}}\implies \tfrac{w_1^2 + w_2^2}{2}\ \ {\rm must \ be \ a \ perfect \ square}\ .\nn
\eeq 
If $w_1 = w_2 = w$, then  $q_1 = q_2 \pm 1$. The winding must be a divisor of both $2q_1^2 - 1$ and $2q_2^2 - 1$, but these numbers are coprime $\forall q_1$.
Then the only possible value of $w$ is $1$.
In conclusion, the only curves $a$ with the same winding number that intersect are $a_{1,q}(R)$ and $a_{1,q\pm 1}(R)$. And the intersection is on  $R = \tfrac{1}{\sqrt{2}}$.

\subsection{${\bf b_{w_1,q_1}(R)=b_{w_2,q_2}(R)}$}

\beq 
b_{w_1,q_1}(R) = \tfrac{2q_1+1 \pm_1 \sqrt{1-2w_1^2R^2}}{w_1},\quad \tfrac{2q_1 (q_1+1)}{w_1} \in \mathbb{Z}\\
b_{w_2,q_2}(R) = \tfrac{2q_2+1 \pm_2 \sqrt{1-2w_2^2R^2}}{w_2},\quad \tfrac{2q_2 (q_2+1)}{w_2} \in \mathbb{Z}
\eeq 
In this case,
\beq
\mp_1 w_2 \sqrt{1-2w_1^2R^2} \pm_2 w_1 \sqrt{1-2w_2^2R^2} = (2q_1+1)w_2 - (2q_2+1)w_1 \equiv C\in \mathbb{Z}\, .
\eeq
 If $C=0$, then $w_1 = w_2$ and $q_1 = q_2$.  If $C \neq 0$
\beq 
R^2 = \tfrac{1}{2  C^2} - \tfrac{(w_1^2+w_2^2-C^2)^2 }{8 w_1^2 w_2^2 C^2} \, .\label{radius_bb}
\eeq 
Defining $N = \tfrac{2q_1(q_1+1)}{w_1} w_2 + \tfrac{2q_2(q_2+1)}{w_2} w_1 - (2q_1+1)(2q_2+1) \in \mathbb{Z}$, we get
\beq 
N^2 = 1 - 2R^2 C^2 < 1\implies N=0\, ,
\eeq 
and then $R^2 = \tfrac{1}{2C^2}$.
Replacing in \eqref{radius_bb},
$
C^2 = w_1^2 + w_2^2$, and  then the radius where  curve $b$ with winding $w_1$ intersects curve $b$ with winding $w_2$ is
$R^{-2} = 2(w_1^2 + w_2^2)$. 

The constraint
\beq 
|(2q_1+1)w_2 - (2q_2+1)w_1 |= \sqrt{w_1^2 + w_2^2} \implies w_1^2 + w_2^2 \ {\rm  is \ a\  perfect \ square}\, .
\eeq 
If  $w_1 = w_2 = w$ then 
$
|2(q_1 + q_2)| = \sqrt{2} w$. The l.h.s. is integer and the r.h.s. is irrational, then there is no winding such that $b_{w,q_1}(R) = b_{w,q_2}(R)$.


\subsection{${\bf a_{w_1,q_1}(R)=b_{w_2,q_2}(R)}$}

\bea
a_{w_1,q_1}(R)&=& \tfrac{2q_1 \pm_1 \sqrt{2-2w_1^2R^2}}{w_1},\quad \tfrac{2q_1^2 -1}{w_1} \in \mathbb{Z}\\
b_{w_2,q_2}(R) &= &\tfrac{2q_2+1 \pm_2 \sqrt{1-2w_2^2R^2}}{w_2},\quad \tfrac{2q_2 (q_2+1)}{w_2} \in \mathbb{Z}
\eea
\beq 
  \mp_1  w_2\sqrt{2-2w_1^2R^2} \pm_2 w_1\sqrt{1-2w_2^2R^2} =2q_1 w_2 - (2q_2 + 1)w_1 = C\in \mathbb{Z}\, .
\eeq 
Since $w_1$ is always odd, then $C$ is also odd (in particular it is non-zero). Then 
\beq 
R^2 = \frac{1}{C^2} -\frac{(w_1^2+2w_2^2 -C^2)^2  }{8w_1^2 w_2^2 C^2} \label{radius_ab}\quad {\rm and} \ \
N^2 = 2-2C^2 R^2\, ,
\eeq 
where $N = \tfrac{(1-2q_1^2)}{w_1}w_2  - \tfrac{2q_2(q_2+1)}{w_2} w_1 + q_1(2q_2+1) \in \mathbb{Z}$, and then
$N=0$ or $1$, which give $R^2 = \tfrac{1}{C^2}$ or $R^{-2} = \tfrac{1}{2C^2}$. From \eqref{radius_ab} we obtain $C^2 = w_1^2 + 2w_2^2$ or $C^2 = (w_1 - w_2)^2 + w_2^2$.
Then the radii where a curve $a$ with $w_1$ intersects another  curve $b$ with  $w_2$ intersect are:
\beq 
R^{-2} = w_1^2 + 2w_2^2\quad {\rm or} \ \
R^{-2} = 2((w_1-w_2)^2 + w_2^2)
\eeq 
For each case we have one of these constraints:
\beq 
|2q_1 w_2 - (2q_2 + 1)w_1| = \sqrt{w_1^2 + 2w_2^2} \quad {\rm or} \
|2q_1 w_2 - (2q_2 + 1)w_1| = \sqrt{(w_1-w_2)^2 + w_2^2} \nn
\eeq 
and then $w_1^2 + 2w_2^2$ or $(w_1-w_2)^2 + w_2^2$ must be a perfect square. 
If  $w_1 = w_2 = w$ we get the constraints:
\beq 
|2q_1  - (2q_2 + 1)| = \sqrt{3}  \quad {\rm or} \ \ 
|2q_1  - (2q_2 + 1)| = 1 
\eeq 
leaving only the second case, with  $q_2= q_1 $ or $q_1 - 1$. The quantization conditions imply that $w$ must be a divisor of both $2q_1^2 - 1$ and $2q_1(q_1\pm 1)$.
But it can be shown that these numbers are coprime, and then $w=1$.
The only curves  with the same windings that intersect are $a_{1,q}(R)$ and $b_{1,q}(R)$ or $b_{1,q-1}(R)$. The intersections are at $R=\tfrac{1}{\sqrt{2}}$.

Summarising, we have:

\beq 
a_{w_1,q_1} =  a_{w_2,q_2} \iff R^{-2} =& w_1^2 + w_2^2 = C^2\\
b_{w_1,q_1} = b_{w_2,q_2}\iff R^{-2} =& 2(w_1^2 + w_2^2)=2 C^2 \\
a_{w_1,q_1} = b_{w_2,q_2} \iff R^{-2} =& \begin{cases} w_1^2 + 2w_2^2 = C^2 \\
 2((w_1 - w_2)^2 + w_2^2) =2 C^2 \end{cases} 
\eeq 

The winding numbers on $b$ can in principle be any positive integer and those on $a$ can only be the divisors of some number of the form $2q^2 -1$, $q \in \mathbb{Z}$.

\subsection{Enhancements to $SO(34)$ or $SO(18)\times E_8$ }

Here we prove that $a_{w_1,q_1}(R) =  a_{w_2,q_2}(R) $ implies that there exist integers $w_3$, $q_3$, $w_4$ and $q_4$ such that $a_{w_1,q_1}(R) =  b_{w_3,q_3}(R) = b_{w_4,q_4}(R) $.

We start with $R^{-2} = w_1^2 + w_2^2$. If $w_1 > w_2$, there are integers $w_3$ and $w_4$ such that $w_1 = w_3+w_4$ and $w_2 = w_3 - w_4$,  because $w_1$ and $w_2$ are odd numbers. Then 
\beq 
R^{-2} = w_1^2 + (2w_3 - w_1)^2 = 2 (w_1^2- 2w_3 w_1 +  w_3^2   + w_3^2) = 2((w_1 - w_3)^2 + w_3^2)
\eeq
Since $ 
R^{-2} =  2((w_1 - w_4)^2 + w_4^2)$ as well,
there exist integers $w_3$, $w_4$, $q_3$ and $q_4$ such that $a_{w_1,q_1}(R) =  b_{w_3,q_3}(R) = b_{w_4,q_4}(R)$. Note that we can always find $q_3$ and $q_4$ because the functions $b$ admit any value of $w$.

Replacing $w_3 = \tfrac{1}{2} (w_1+w_2)$ and $w_4 = \tfrac{1}{2} (w_1 - w_2)$ we get 
\beq 
a_{w_1,q_1}(R) =  a_{w_2,q_2}(R) \implies a_{w_1,q_1}(R) =  a_{w_2,q_2}(R)=b_{(w_1+w_2)/2,q_3}(R) = b_{(w_1-w_2)/2,q_4}(R)\nn
\eeq 
Note that we can also write the radius as $2(w_3^2 + w_4^2)$.
We want to satisfy
\beq 
(\sqrt{2}R)^{-1} = |2q_1 w_3 - (2q_3 + 1) w_1 | = |2q_1 w_4 - (2q_4 + 1) w_1 | = |(2q_3+1)w_4 - (2q_4 + 1)w_3|\, ,\nn
\eeq 
and we have that
\bea
(\sqrt{2}R)^{-1} &=& |q_1 w_2 - q_2 w_1| = |2 q_1 w_3  - (q_1 + q_2) w_1| =|2 q_1 w_4  - (q_1 - q_2) w_1| \nn\\
&=&  |(q_1 + q_2) w_4-(q_1-q_2) w_3|\, . \nn
\eea
Then we need to identify
$q_1 + q_2 = 2q_3 + 1\, , 
q_1 - q_2 = 2q_4 + 1$.

We still have to prove that $2q_3(q_3+1)$ and $2q_4(q_4+1)$ are divisible by $w_3$ and $w_4$, respectively, which amounts to proving that 
\beq 
w_i \text{ is a divisor of } 2q_i^2 - 1 \text{ and } |q_1 w_2 - q_2 w_1| = \sqrt{\tfrac{w_1^2 + w_2^2}{2}}\\ \implies w_1\pm w_2 \text{ is a divisor of } (q_1 \pm q_2)^2 - 1
\eeq
We checked that this is satisfied for the first 300 values of $q_i$.
 
Then we have that
\bea
 a_{w_1,q_1}(R) &=&  a_{w_2,q_2}(R)\implies b_{(w_1+w_2)/2,(q_1 + q_2 - 1)/2}(R) = b_{(w_1-w_2)/2, (q_1 - q_2 - 1)/2}(R)\, .\nn
\eea

To prove  that $b_{w_3,q_3}(R) =  b_{w_4,q_4}(R)$ implies that there exists integers $w_1$, $q_1$, $w_2$ and $q_2$ such that $b_{w_3,q_3}(R) =  a_{w_1,q_1}(R) = a_{w_2,q_2}(R) $, we start with $R^{-2} = 2(w_3^2 + w_4^2)$. Define integers $w_1$ and $w_2$ such that $w_3 = \tfrac{1}{2} (w_1+w_2)$ and $w_4 = \tfrac{1}{2} (w_1 - w_2)$ (we assume $w_3 > w_4$), 
\beq 
R^{-2} = 2( (w_1 - w_3)^2 + w_3^2) \quad  {\rm and} \quad 
R^{-2} = 2( (w_2 - w_3)^2 + w_3^2) \, . \nn
\eeq 
But we still need to satisfy the constraint that $w_1$ and $w_2$ are divisors of $2q_1^2-1$ and $2q_2^2 -1$ for two integers $q_1$ and $q_2$.
With the identifications $q_1 + q_2 = 2q_3 + 1\, , 
q_1 - q_2 = 2q_4 + 1$, we get the correct radius
\beq 
R^{-1} =  \sqrt{2}|(2q_3 + 1) w_4 - (2q_4 + 1)w_3| = \sqrt{2} |2q_1 w_3 - (2q_3+1) w_1|\, , \nn
\eeq 
\beq 
b_{w_3,q_3}(R) =  b_{w_4,q_4}(R) \implies b_{w_3,q_3}(R) =  b_{w_4,q_4}(R)=a_{w_3+w_4, q_3 + q_4 +1  }(R) = a_{w_3-w_4, q_3 - q_4}(R)\, . \nn
\eeq 
We still have to prove that $2q_1^2 -1$ and $2q_2^2-1$ are divisible by $w_1$ and $w_2$, respectively. This is the same as proving that
\beq 
q_i \text{ is a divisor of } 2q_i(q_i+1) \text{ and } |(2q_3 + 1) w_4 - (2q_4 + 1) w_3|= \sqrt{w_3^2 + w_4^2} \\ \implies w_3\pm w_4 \text{ is a divisor of } 2\left [\left ( q_3 + 1/2 \right ) \pm \left ( q_4 + 1/2 \right )\right ]^2 - 1\, ,
\eeq 
which we checked is satisfied. 


In conclusion, we have that, for $R^{-2} = w_1^2 + w_2^2$,  $a_{w_1,q_1}(R) =  a_{w_2,q_2}(R) \iff $
\beq 
a_{w_1,q_1}(R) =  a_{w_2,q_2}(R)=b_{(w_1+w_2)/2,(q_1 + q_2 - 1)/2}(R) = b_{(w_1-w_2)/2, (q_1 - q_2 - 1)/2}(R)\nn\\
\iff b_{(w_1+w_2)/2,(q_1 + q_2 - 1)/2}(R) = b_{(w_1-w_2)/2, (q_1 - q_2 - 1)/2}(R)\, .
\eeq

The Wilson lines that give this enhancement can be written in four different ways
\beq 
\frac{2q_1}{w_1} \pm_1 \sqrt{2}R \frac{w_2}{w_1} = \frac{2q_2}{w_2} \pm_2 \sqrt{2} R \frac{w_1}{w_2} = \frac{2q_3+1}{w_3} \pm_3 \sqrt{2} R \frac{w_4}{w_3} = \frac{2q_4+1}{w_4} \pm_4  \sqrt{2} R \frac{w_3}{w_4}\, \nn
\eeq 
Using that $w_3 = \frac{w_1+w_2}{2}$, $w_4 = \frac{w_1-w_2}{2}$, $q_3 = \frac{q_1 + q_2 - 1}{2}$ and $q_4 = \frac{q_1 - q_2 - 1}{2}$, after a few steps, we get $\mp_4=\pm_3=\pm_2 = \mp_1$ and then the Wilson lines are
\beq 
A_1 = \frac{2q_1}{w_1} \pm \frac{w_2}{w_1}\sqrt{2}R ,\quad
A_1 = \frac{2q_2}{w_2}\mp \frac{w_1}{w_2}\sqrt{2}R  ,\quad \\
A_1 =  \frac{2q_3+1}{w_3} \mp \frac{w_4}{w_3}\sqrt{2}R  ,\quad
A_1 =  \frac{2q_4+1}{w_4} \pm \frac{w_3}{w_4}\sqrt{2}R \label{wilsonlines1}
\eeq 
From here,
\beq 
(\sqrt{2}R)^{-1} = \mp( q_1 w_2 - q_2 w_1) \in \mathbb{Z} 
\eeq 
and then, after a few steps, we can prove that
\beq 
\frac{1}{\sqrt{2}R}, \quad {\mathbb{A}},\quad  \tfrac{R}{\sqrt{2}}\left (\tfrac12 \mathbb{A}^2+1 \right )\quad \in \mathbb{Z}\, ,
\eeq 
Defining integers $m = (\sqrt{2}R)^{-1}$ and $n = \mathbb{A}/\sqrt{2}$, all this type of enhancement points are given by 
\beq 
(R,A_1) = \left ( \frac{1}{m\sqrt{2}},\frac{n}{m}\right )
\quad \text{
such that} \
\frac{n^2 + 1}{2m}\in \mathbb{Z} \label{condicionfijos1}
\eeq 
%
%
%
and then
\beq 
R^{-2} =  2, 50, 338, 578, 1250, 1682, 2738, 3362, 5618, 7442, 8450, 10658, \dots 
\eeq 
These are all of the form $2C^2$ with $C$ an integer with prime divisors congruent to 1 mod 4. That is: $1, 5, 13, 17, 25, 29, 37, 41, 53, 61, 65, 73, 85, 89, 97, 101, 109, \dots$. Except for the $1$, these numbers are all Pythagorean primes or multiples of them.

We want to see if the $b$ lines considered here can be interposed with a $c$ line. 
$q_3$ and $q_4$ are suitable for curves $b$ with $w_3$ and $w_4$. For curves $c$ to coincide with them, we need $w_i$ even and $\tfrac{q_i(q_i+1)}{w_i}\in \mathbb{Z}$.
If one of the two curves $b$ has also a curve $c$ then we have an intersection between an $a$ and a $c$ curve. 
Analyzing all the possibilities, it can be shown that there are no $c$ curves that intersect with more than one other curve.

\subsection{Enhancements to $SU(2)\times SO(32)$ or $SU(2)\times E_8 \times E_8$ }

The equality $a_{w_1,q_1}(R) = b_{w_2,q_2}(R)$ arises for two type of radius
\beq 
R^{-2}=w_1^2 + 2w_2^2 \quad {\rm or} \quad
R^{-2}=2((w_1 - w_2)^2 + w_2^2)\, .
\eeq 
The second type gives $R^{-2}= w_1^2 + w_3^2$  if $w_2 = \tfrac{w_3 + w_1}{2}$,  which implies that there is an intersection with another  curve $a$ of winding $w_3$. Then, we  restrict to the first type, where $R^{-2}$ is odd for odd $w_1^2$. Thus the even $R^{-2}$ found in the previous section  cannot have additional curves $a$ or $b$ on the intersection.

For $R^{-2} = w_1^2 + 2 w_2^2$, the constraints are
\beq 
|2q_1 w_2 - (2q_2 + 1)w_1| = \sqrt{w_1^2 + 2w_2^2}\, , \quad
\tfrac{2q_1^2 - 1}{w_1} \in \mathbb{Z} \, , \quad
\tfrac{2q_2 (q_2 + 1)}{w_2} \in \mathbb{Z} 
\eeq 
The Wilson line can be written as
\beq 
A_1 = \tfrac{2q_1 \pm_1 2R w_2 }{w_1}\,  \quad {\rm or}\quad 
A_1 = \tfrac{2q_2 + 1 \pm_2 R w_1 }{w_2}\, ,
\eeq 
and equating them
leads to $\pm_2 = \mp_1$ and
\beq 
 R^{-1} = \mp \left ( 2q_1 w_2 - (2q_2 + 1) w_1 \right ) \, ,
\eeq 
implying that $R^{-1}$ is an odd number. After some algebra, we get
\beq 
\frac{1}{R}, \quad {\mathbb{A}},\quad  R\left (\tfrac12 \mathbb{A}^2+1 \right )\quad \in \mathbb{Z}\, ,
\eeq 
and then all this type of enhancement points satisfy
\beq 
(R,A_1) = \left ( \frac{1}{m} , \frac{2n}{m} \right )
\eeq 
for integer $m = R^{-1}$ and $n = \tfrac{R^{-1} A_1}{2}$, such that
\beq
\frac{2n^2 + 1}{m}\in \mathbb{Z}. \label{perd}
\eeq
%
We obtain
\beq 
R^{-1} = 3, 9, 11, 17, 19, 27, 33, 41, 43, 51, 57, 59, \dots
\eeq 
all integer numbers with prime divisors  congruent to 1 or 3 (mod 8). 

It is not hard to prove that  all the curves $b$ that intersect just one curve $a$ are superimposed by a curve $c$ (in the $\Gamma_8 \times \Gamma_8$ case).

\section{Other slices of moduli space}
\label{app:extrafigures}

Here we analyse two-dimensional slices of moduli space given by the radius and one parameter in the Wilson lines.
First we consider the $SO(32)$ theory compactified with Wilson lines of the form $A^I=\left( (A)_p,0_{16-p} \right)$.
We then show how the generalized Dynkin diagrams give us the points of enhanement located in the fundamental region (in the conventions of Appendix \ref{app:Dyndia}). Finally we invert the logic, and use the generalized Dynkin diagram for $\Gamma_8 \times \Gamma_8$ to find certain points of enhancement, and determine interesting slies of moduli space to explore. 

\subsection{Slices for the $SO(32)$  theory}

The results are summarized in the following figures, after which we present the calculations leading to them.


\begin{figure}[H] 
\begin{subfigure}{0.5\textwidth}
\centering
\includegraphics[width=.8\textwidth]{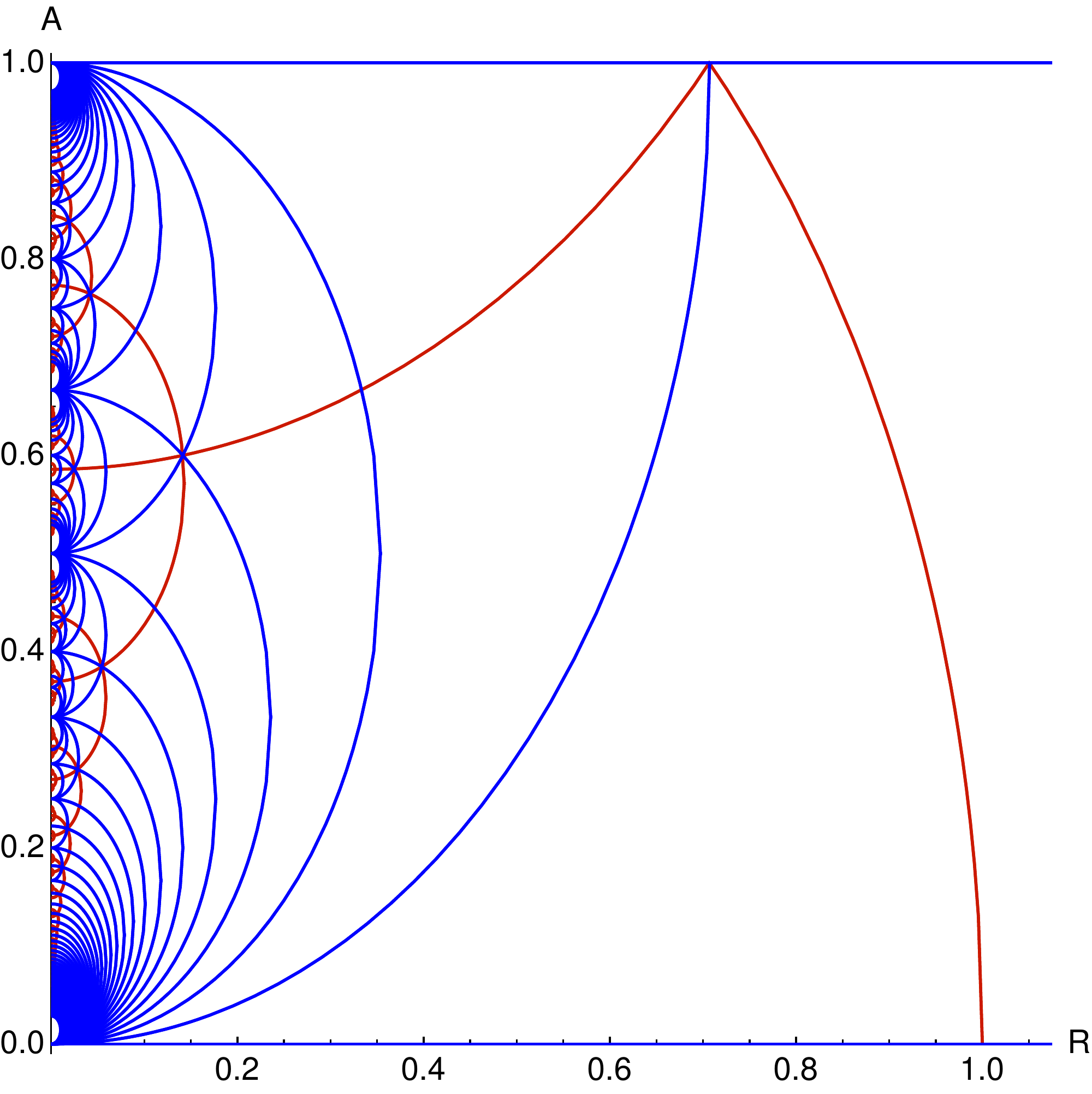}
\end{subfigure}
\begin{subfigure}{0.5\textwidth}
 \bleyenda
\cua{white}  &\quad  SO(30) \times U(1)^2
\\
\cua{rojo}  &\quad SU(2)\times SO(30) \times U(1)
\\
\cua{azul}  &\quad SO(32) \times U(1)
\\
\cua{azul}   +
\cua{rojo}   
&\quad SU(2) \times SO(32)
\\
\cua{azul}  +
\cua{azul}  +
\cua{rojo}  +\cua{rojo}  &\quad SO(34) 
 \eleyenda
\end{subfigure}
\caption{$SO(32)$ heterotic with Wilson line $A^I=\left( A ,0_{15} \right)$}
\label{fig:so32_n1}
\end{figure}
\begin{figure}[H] 
\begin{subfigure}{0.5\textwidth}
\centering
\includegraphics[width=.8\textwidth]{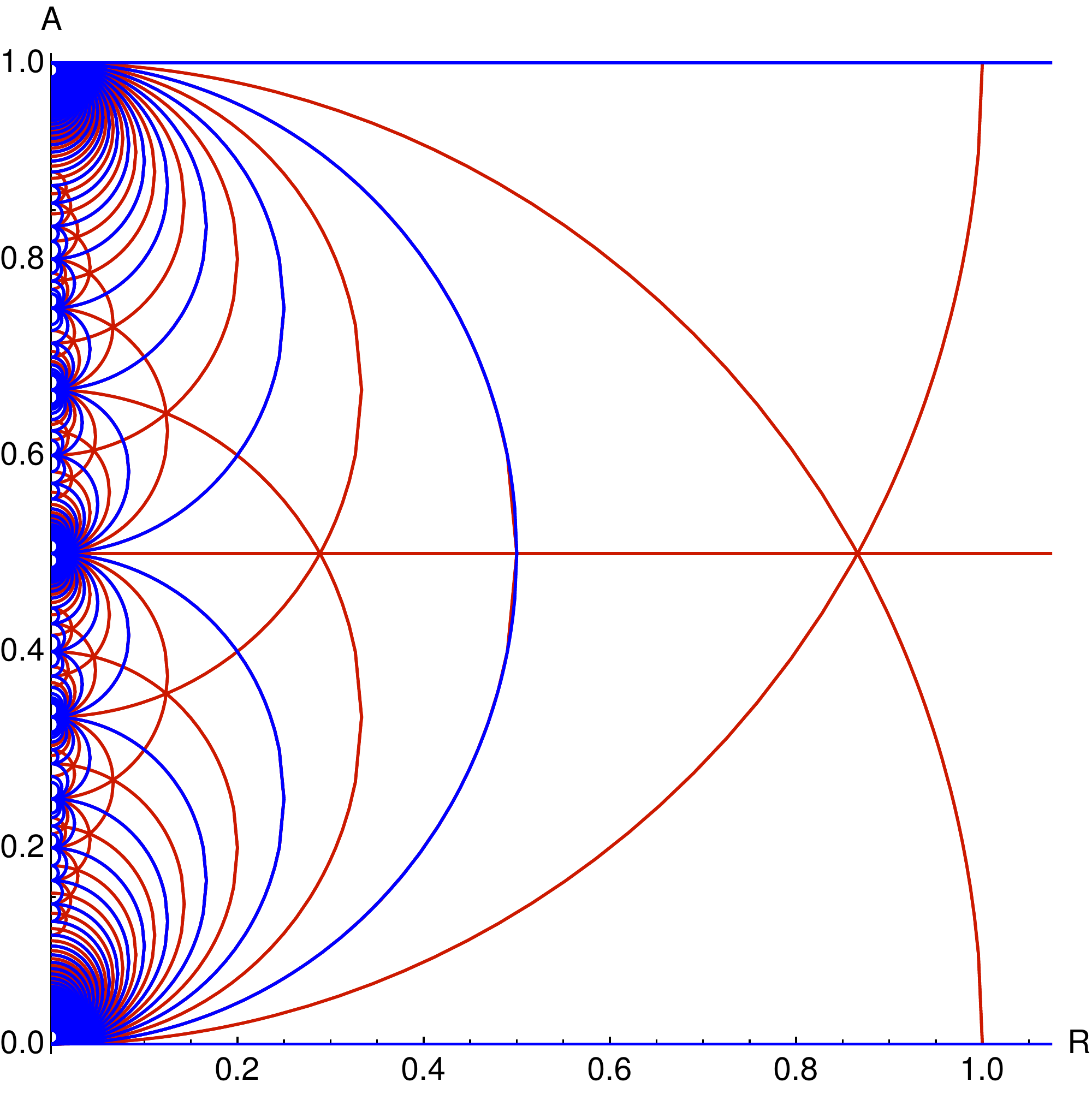}
\end{subfigure}
\begin{subfigure}{0.5\textwidth}
 \bleyenda
\cua{white}  &\quad  SU(2) \times SO(28)  \times U(1)^2
\\
\cua{rojo}  &\quad SU(2)\times SU(2) \times SO(28) \times U(1)
\\
\cua{azul}  &\quad SO(32)  \times U(1)
\\
\cua{azul}   +
\cua{rojo}   
&\quad SU(2) \times SO(32) 
\\
\cua{rojo}  +\cua{rojo}  +\cua{rojo}  &\quad SU(2) \times SU(3) \times SO(28)
 \eleyenda
\end{subfigure}
\caption{$SO(32)$ heterotic with Wilson line $A^I=\left( (A)_2 ,0_{14} \right)$}
\label{fig:so32_n2}
\end{figure}
\begin{figure}[H] 
\begin{subfigure}{0.5\textwidth}
\centering
\includegraphics[width=.8\textwidth]{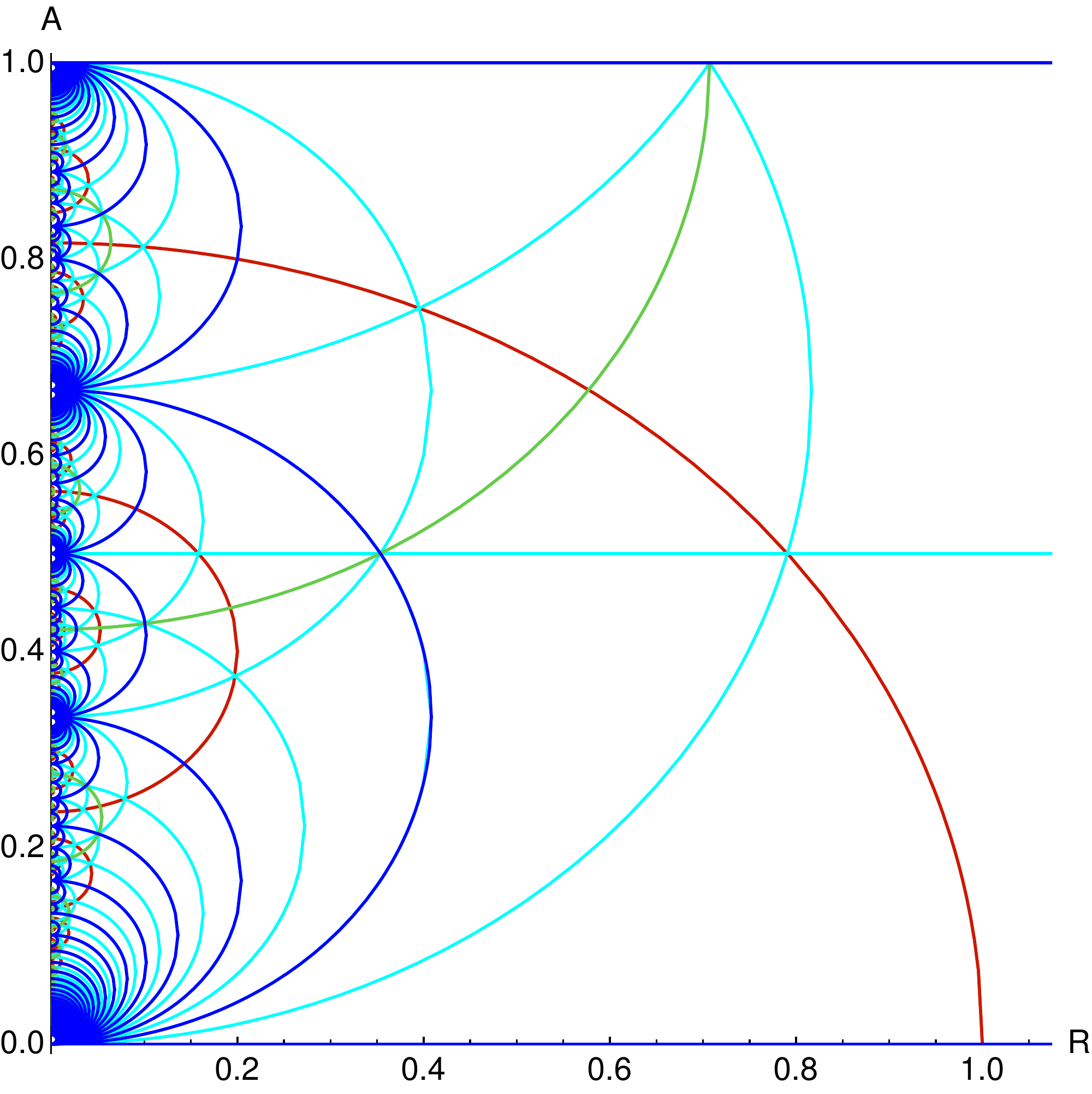}
\end{subfigure}
\begin{subfigure}{0.5\textwidth}
 \bleyenda
 \cua{white}  &\quad SU(3) \times SO(26) \times U(1)^2\\
\cua{rojo} &\quad   SU(2) \times SU(3) \times SO(26) \times U(1)  \\
\cua{cyan} &\quad  SO(6)\times SO(26)\times U(1)  \\
\cua{verde} &\quad SU(3) \times SO(28)\times U(1)  \\
\cua{azul} &\quad  SO(32)\times U(1)  \\
\cua{cyan}  + \cua{cyan}  + \cua{rojo}  &  \quad SU(5) \times SO(26) \\
\cua{cyan}  + \cua{cyan}  + \cua{azul}  + \cua{verde}  &  \quad SO(34) \\
\cua{rojo}  + \cua{verde}  &  \quad SU(2) \times SU(3) \times SO(28) \\
\cua{rojo}  + \cua{azul}  &  \quad SU(2) \times  SO(32) 
 \eleyenda
\end{subfigure}
\caption{$SO(32)$ heterotic with Wilson line $A^I=\left( (A)_3 ,0_{13} \right)$}
\label{fig:so32_n3}
\end{figure}

\begin{figure}[H] 
\begin{subfigure}{0.5\textwidth}
\centering
\includegraphics[width=.8\textwidth]{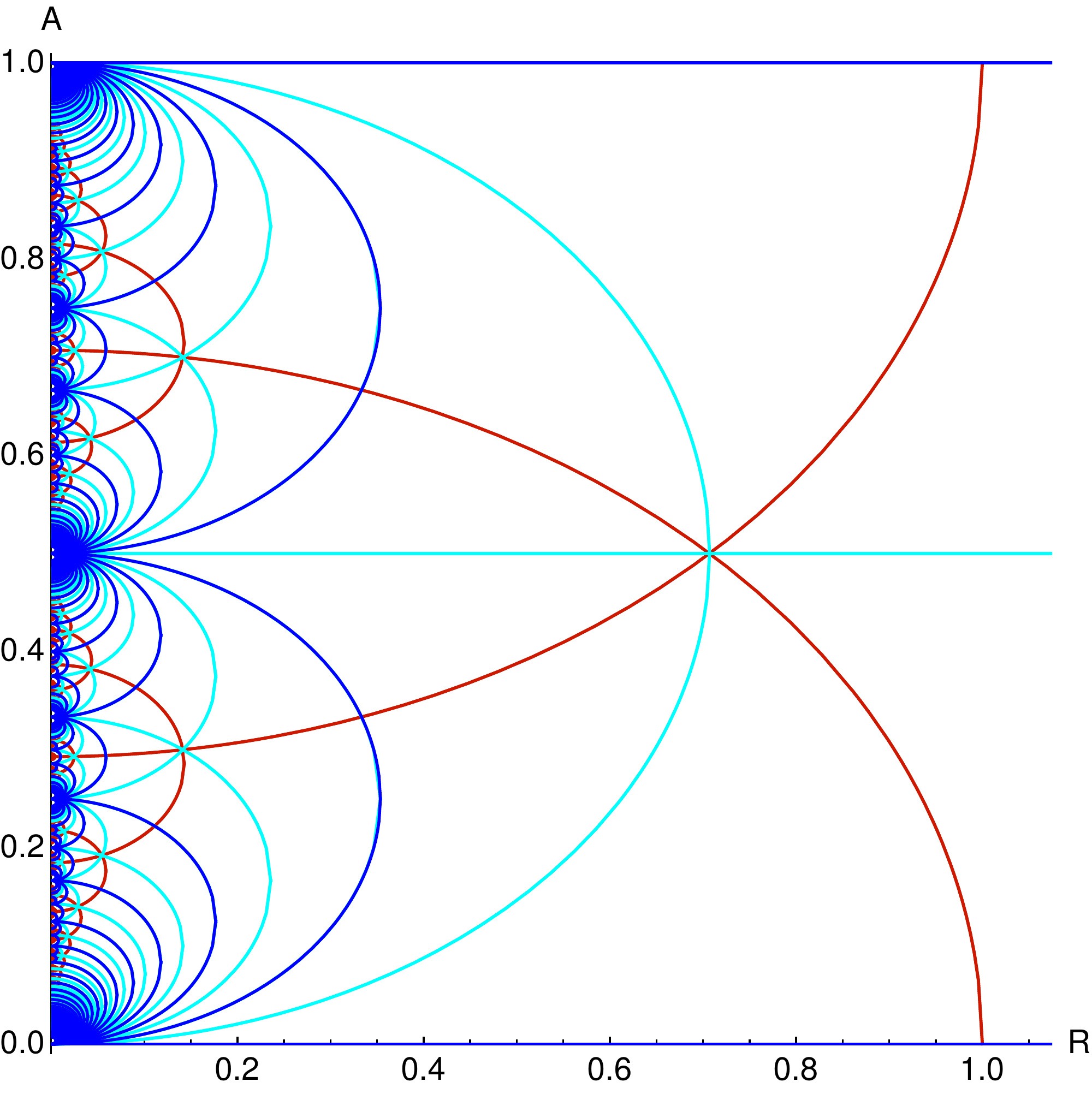}
\end{subfigure}
\begin{subfigure}{0.5\textwidth}
 \bleyenda
\cua{white}  &\quad SU(4) \times SO(24) \times U(1)^2 \\
\cua{rojo} &\quad   SU(2) \times SU(4) \times SO(24) \times U(1) \\
\cua{cyan} &\quad SO(8) \times SO(24) \times U(1)\\
\cua{azul} &\quad  SO(32) \times U(1)\\
\cua{cyan}  + \cua{cyan}  + \cua{rojo}  +\cua{rojo}  &  \quad SO(10) \times SO(24)\\
\cua{rojo}  + \cua{azul}  &  \quad SU(2) \times  SO(32) 
 \eleyenda
\end{subfigure}
\caption{$SO(32)$ heterotic with Wilson line $A^I=\left( (A)_4 ,0_{12} \right)$}
\label{fig:so32_n4}
\end{figure}

\begin{figure}[H] 
\begin{subfigure}{0.5\textwidth}
\centering
\includegraphics[width=.8\textwidth]{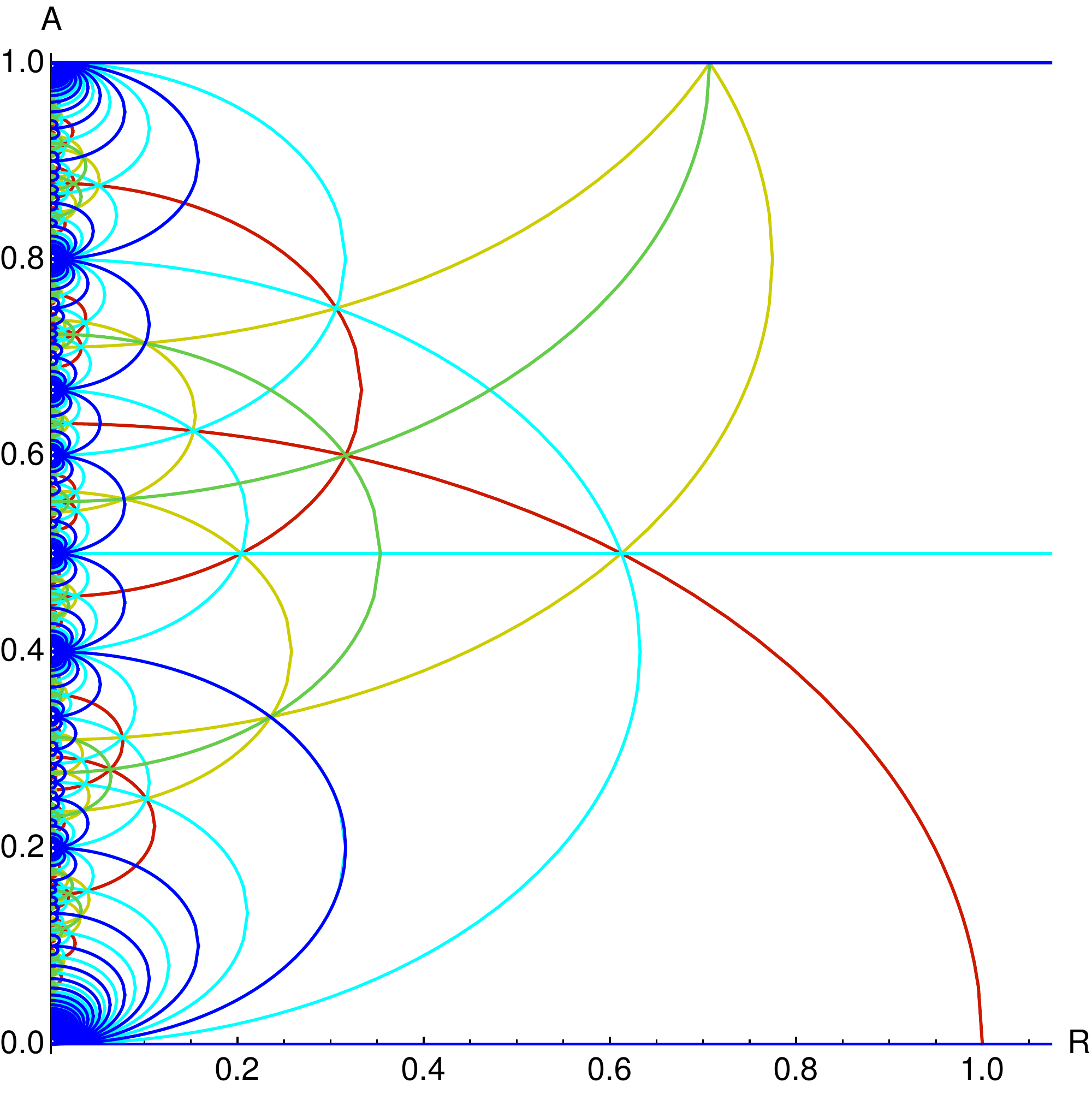}
 \bleyenda
\cua{white}  &\quad SU(5) \times SO(22) \times U(1)^2
 \eleyenda
\end{subfigure}
\begin{subfigure}{0.5\textwidth}
 \bleyenda
 \cua{rojo} &\quad   SU(2) \times SU(5) \times SO(22) \times U(1)\\
\cua{amarillo} &\quad  SU(6)\times SO(22)\times U(1) \\
\cua{cyan} &\quad SO(10) \times SO(22)\times U(1)\\
\cua{verde} &\quad SU(5) \times SO(24)\times U(1)\\
\cua{azul} &\quad  SO(32)\times U(1)\\
\cua{cyan}  + \cua{cyan}  + \cua{amarillo} +
\cua{rojo} &  \quad E_6 \times SO(22)\\
\cua{amarillo}  + \cua{amarillo}  + \cua{azul}  + \cua{verde}  &  \quad SO(34)\\
\cua{rojo}  +
\cua{rojo}  +
\cua{verde}  +\cua{verde}  &  \quad SU(5)\times SO(26)\\
\cua{cyan}  + \cua{verde}  &  \quad SO(10) \times  SO(24)\\
\cua{rojo}  + \cua{azul}  &  \quad SU(2) \times  SO(32)
 \eleyenda
\end{subfigure}
\caption{$SO(32)$ heterotic with Wilson line $A^I=\left( (A)_5 ,0_{11} \right)$}
\label{fig:so32_n5}
\end{figure}

\begin{figure}[H] 
\begin{subfigure}{0.5\textwidth}
\centering
\includegraphics[width=.8\textwidth]{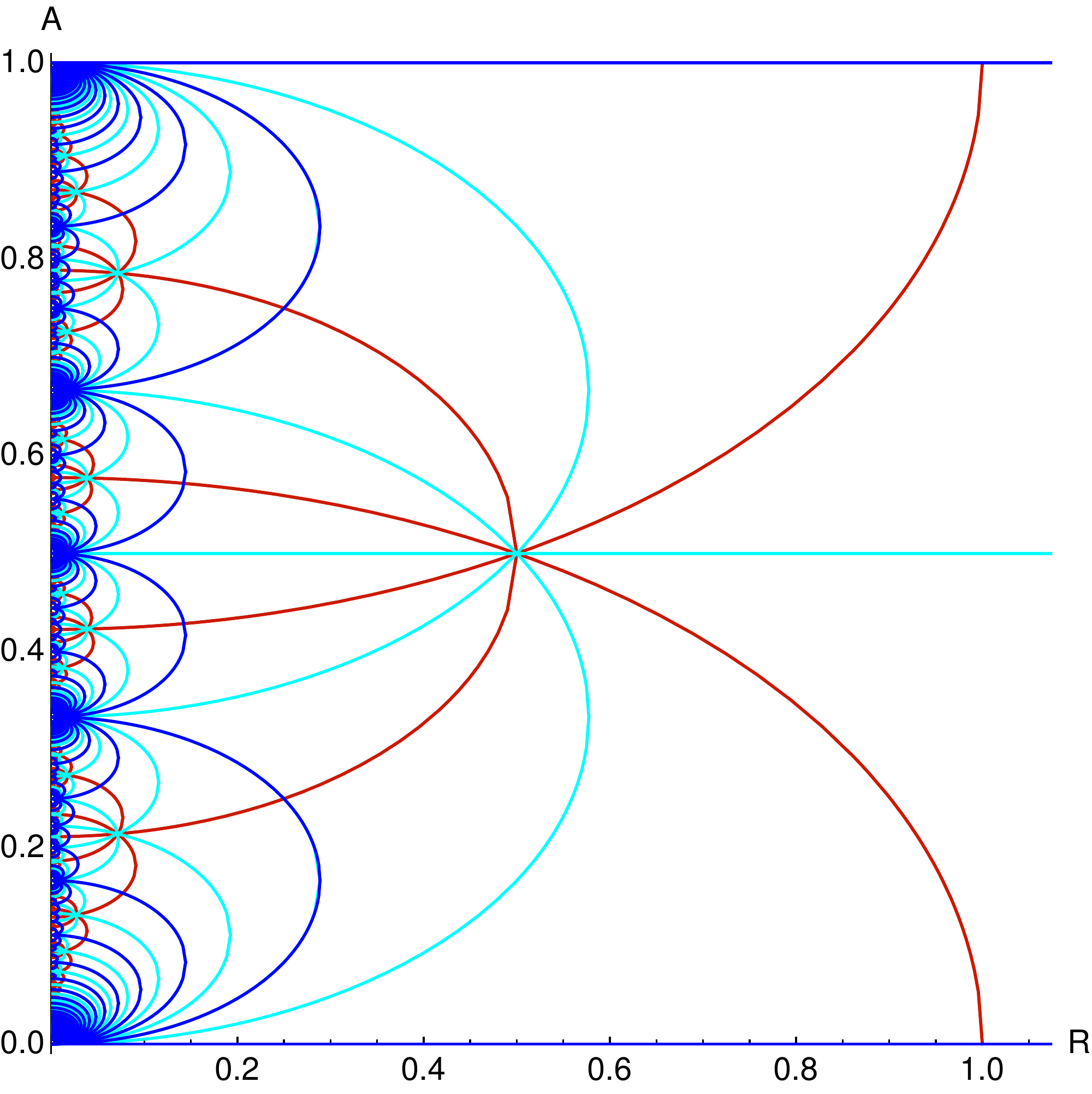}
\end{subfigure}
\begin{subfigure}{0.5\textwidth}
 \bleyenda
\cua{white}  &\quad SU(6) \times SO(20)\times U(1)^2 \\
\cua{rojo} &\quad   SU(2) \times SU(6) \times SO(20)\times U(1) \\
\cua{cyan} &\quad SO(12) \times SO(20)\times U(1) \\
\cua{azul} &\quad  SO(32) \times U(1) \\
\cua{rojo}  + \cua{azul}  &  \quad SU(2) \times  SO(32) \\
\cua{cyan}  + \cua{cyan}& +
\cua{cyan} + 
\cua{rojo} +
\cua{rojo} +
\cua{rojo}   \quad E_7 \times SO(20)  
 \eleyenda
\end{subfigure}
\caption{$SO(32)$ heterotic with Wilson line $A^I=\left( (A)_6 ,0_{10} \right)$}
\label{fig:so32_n6}
\end{figure}

\begin{figure}[H] 
\begin{subfigure}{0.5\textwidth}
\centering
\includegraphics[width=.8\textwidth]{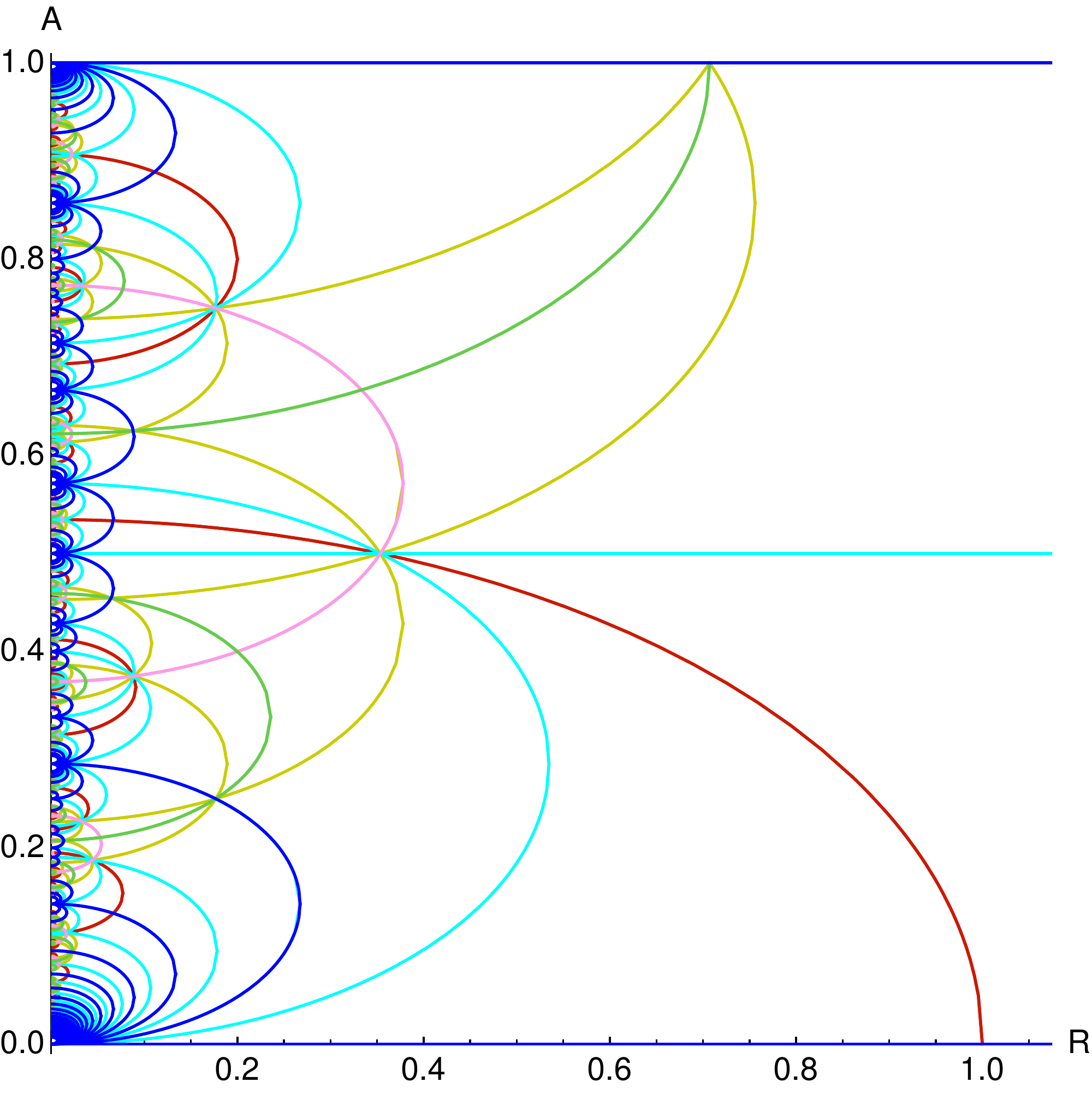}
 \bleyenda
\cua{white}  &\quad SU(7) \times SO(18)\times U(1)^2 
 \eleyenda
\end{subfigure}
\begin{subfigure}{0.5\textwidth}
 \bleyenda
 \cua{rojo} &\quad  SU(2) \times SU(7)\times SO(18)\times U(1)\\
\cua{azul} &\quad   SO(32) \times U(1)\\
\cua{verde} &\quad  SU(7)\times SO(20)\times U(1)  \\
\cua{cyan} &\quad SO(14) \times SO(18)\times U(1)  \\
\cua{amarillo} &\quad  SU(8)\times SO(18)\times U(1)\\
\cua{rosa} &\quad  E_7\times SO(18) \times U(1)\\
\cua{amarillo} +\cua{amarillo} +\cua{verde} +\cua{azul}&\quad SO(34)  
\\
\cua{azul} +\cua{rojo} &\quad SU(2) \times SO(32)\\
\cua{amarillo} +\cua{amarillo} +\cua{cyan} +\cua{cyan} +\cua{rojo} +\cua{rosa}&\quad E_8 \times SO(18) \\
\cua{rosa}+\cua{verde} 
&\quad E_7 \times SO(20)  
 \eleyenda
\end{subfigure}
\caption{$SO(32)$ heterotic with Wilson line $A^I=\left( (A)_7 ,0_{9} \right)$}
\label{fig:so32_n7}
\end{figure}

\begin{figure}[H] 
\begin{subfigure}{0.5\textwidth}
\centering
\includegraphics[width=.8\textwidth]{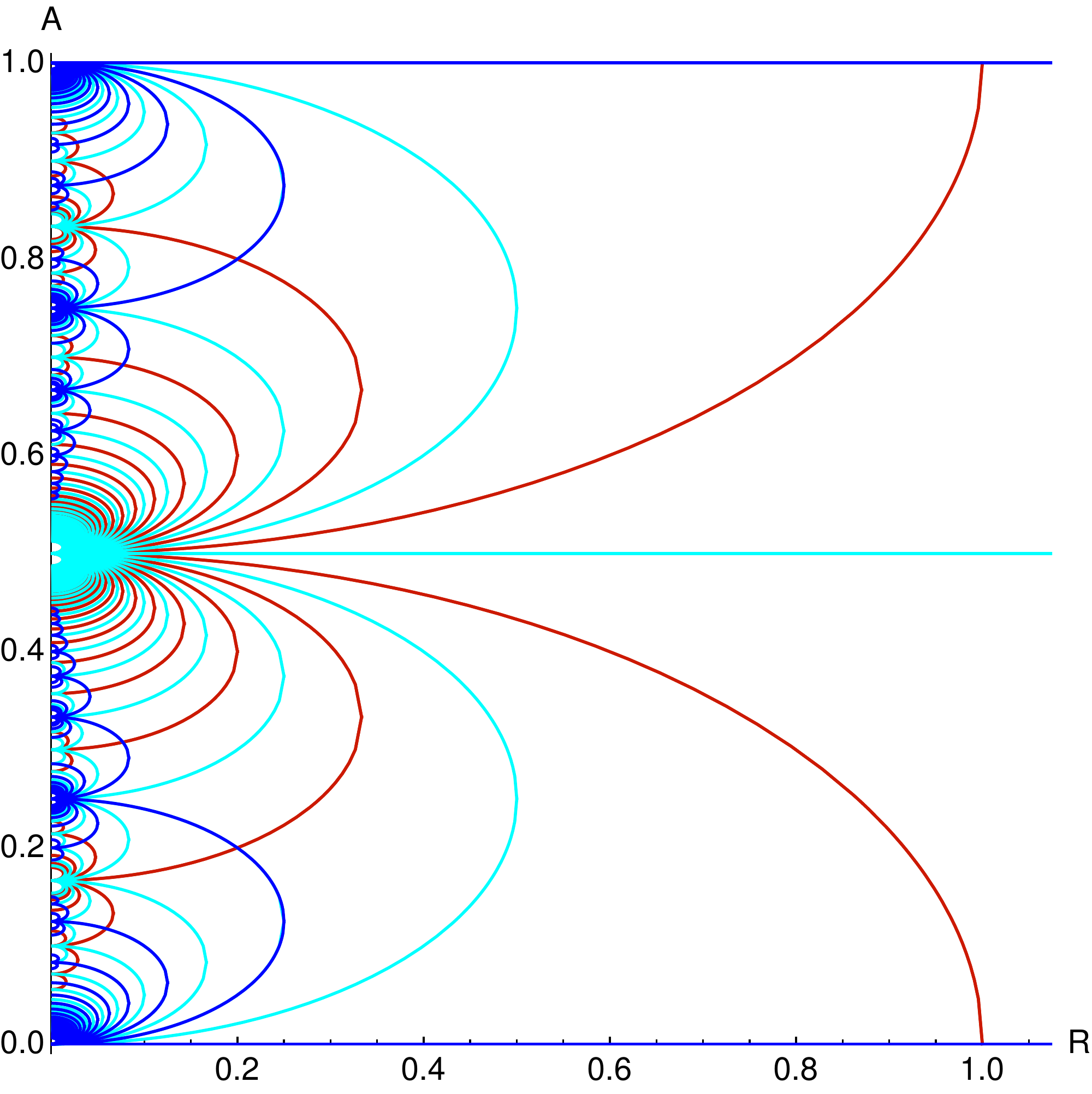}
\end{subfigure}
\begin{subfigure}{0.5\textwidth}
 \bleyenda
\cua{white}  &\quad SU(8) \times SO(16) \times U(1)^2 \\
\cua{rojo} &\quad   SU(2) \times SU(8) \times SO(16)\times U(1)  \\
\cua{cyan} &\quad  SO(16)\times SO(16)\times U(1)\\
\cua{blue} &\quad SO(32) \times U(1)\\
\cua{rojo}  + \cua{blue}  &  \quad SU(2) \times  SO(32)
 \eleyenda
\end{subfigure}
\caption{$SO(32)$ heterotic with Wilson line $A^I=\left( (A)_8 ,0_{8} \right)$}
\label{fig:so32_n8}
\end{figure}

\begin{figure}[H] 
\begin{subfigure}{0.5\textwidth}
\centering
\includegraphics[width=.8\textwidth]{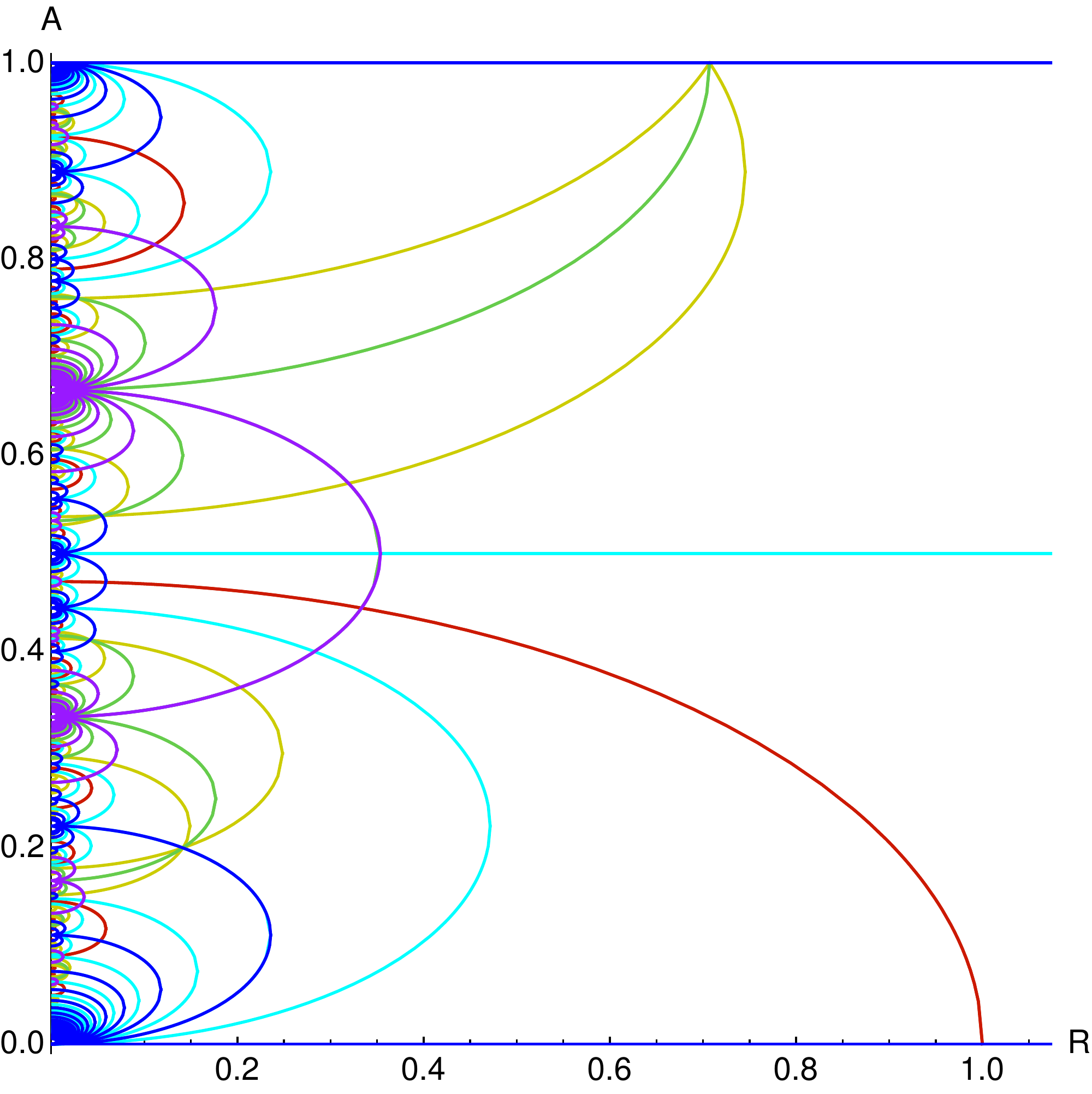}
 \bleyenda
\cua{white}  &\quad SU(9) \times SO(14) \times U(1)^2 \\
\cua{rojo} &\quad   SU(2) \times SU(9) \times U(1) \times SO(14)
 \eleyenda
\end{subfigure}
\begin{subfigure}{0.5\textwidth}
 \bleyenda
\cua{amarillo} &\quad SU(10) \times SO(14)\times U(1) \\
\cua{cyan} &\quad  SO(18) \times SO(14) \times U(1)\\
\cua{verde} &\quad  SU(9) \times SO(16) \times U(1)\\
\cua{azul} &\quad   SO(32)  \times U(1)\\
\cua{violeta} &\quad   SU(9) \times E_8\times U(1)\\
\cua{azul}  +\cua{verde} +\cua{amarillo} +\cua{amarillo} &  \quad SO(34)\\
\cua{violeta}  + \cua{cyan}  &  \quad SO(18) \times  E_8\\
\cua{violeta}  + \cua{amarillo}  &  \quad  SU(10) \times  E_8 \\
\cua{violeta}  + \cua{rojo}  &  \quad SU(2) \times SU(9) \times  E_8 \\
\cua{azul}  + \cua{rojo}  &  \quad SU(2) \times  SO(32) 
 \eleyenda
\end{subfigure}
\caption{$SO(32)$ heterotic with Wilson line $A^I=\left( (A)_9 ,0_{7} \right)$}
\label{fig:so32_n9}
\end{figure}

\begin{figure}[H] 
\begin{subfigure}{0.5\textwidth}
\centering
\includegraphics[width=.8\textwidth]{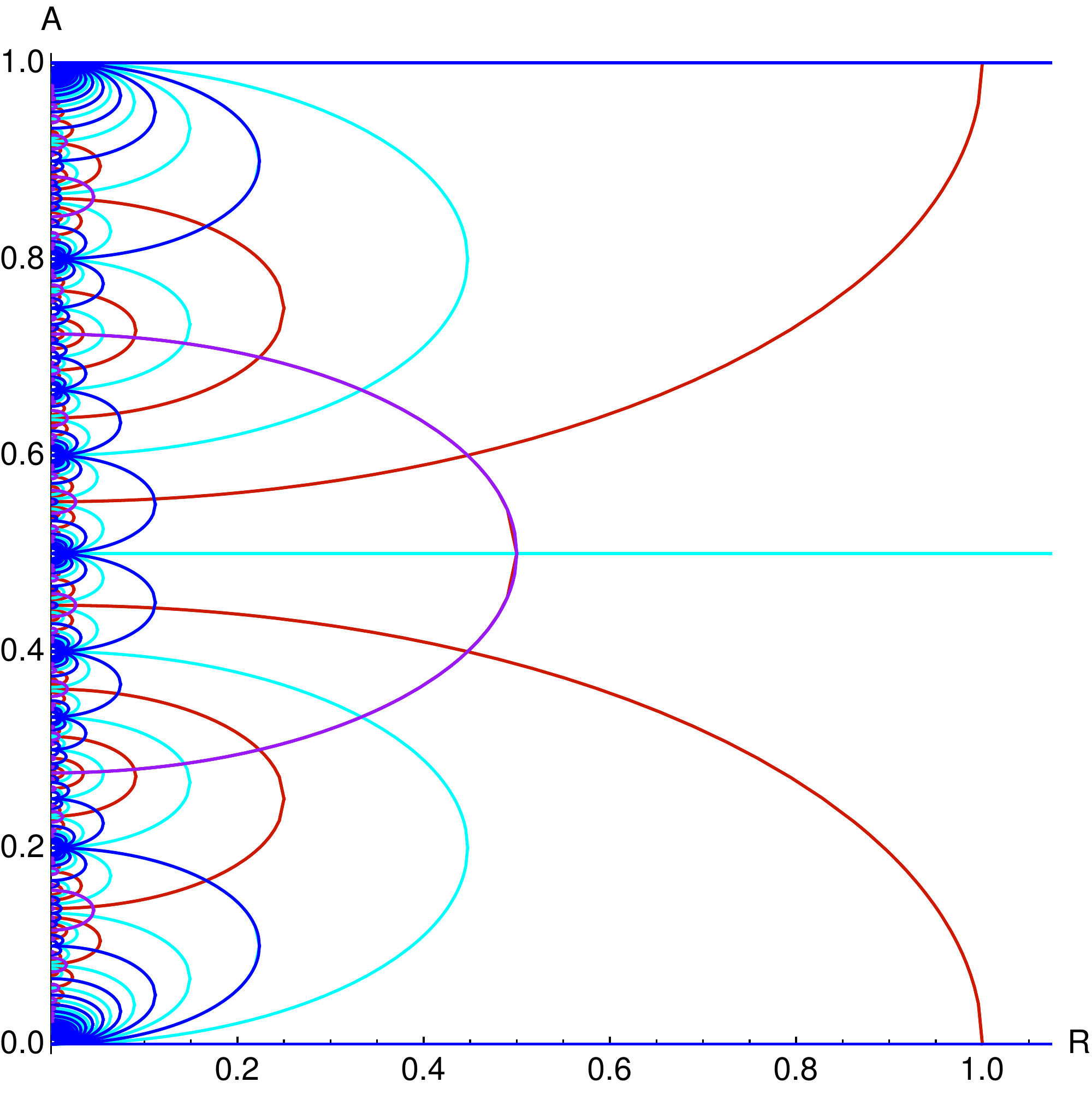}
\end{subfigure}
\begin{subfigure}{0.5\textwidth}
 \bleyenda
\cua{white}  &\quad SU(10) \times SO(12) \times U(1)^2 \\
\cua{rojo} &\quad   SU(2) \times SU(10) \times SO(12) \times U(1)\\
\cua{cyan} &\quad SO(20) \times SO(12) \times U(1)\\
\cua{blue} &\quad   SO(32) \times U(1) \\
\cua{violeta} &\quad  SU(10) \times E_7  \times U(1)\\
\cua{violeta}  + \cua{cyan}  &  \quad SO(20) \times  E_7\\
\cua{violeta}  + \cua{rojo}  &  \quad  SU(2) \times SU(10) \times  E_7 \\
\cua{blue}  + \cua{rojo}  &  \quad SU(2) \times  SO(32)
 \eleyenda
\end{subfigure}
\caption{$SO(32)$ heterotic with Wilson line $A^I=\left( (A)_{10} ,0_6 \right)$}
\label{fig:so32_n10}
\end{figure}

\begin{figure}[H] 
\begin{subfigure}{0.5\textwidth}
\centering
\includegraphics[width=.8\textwidth]{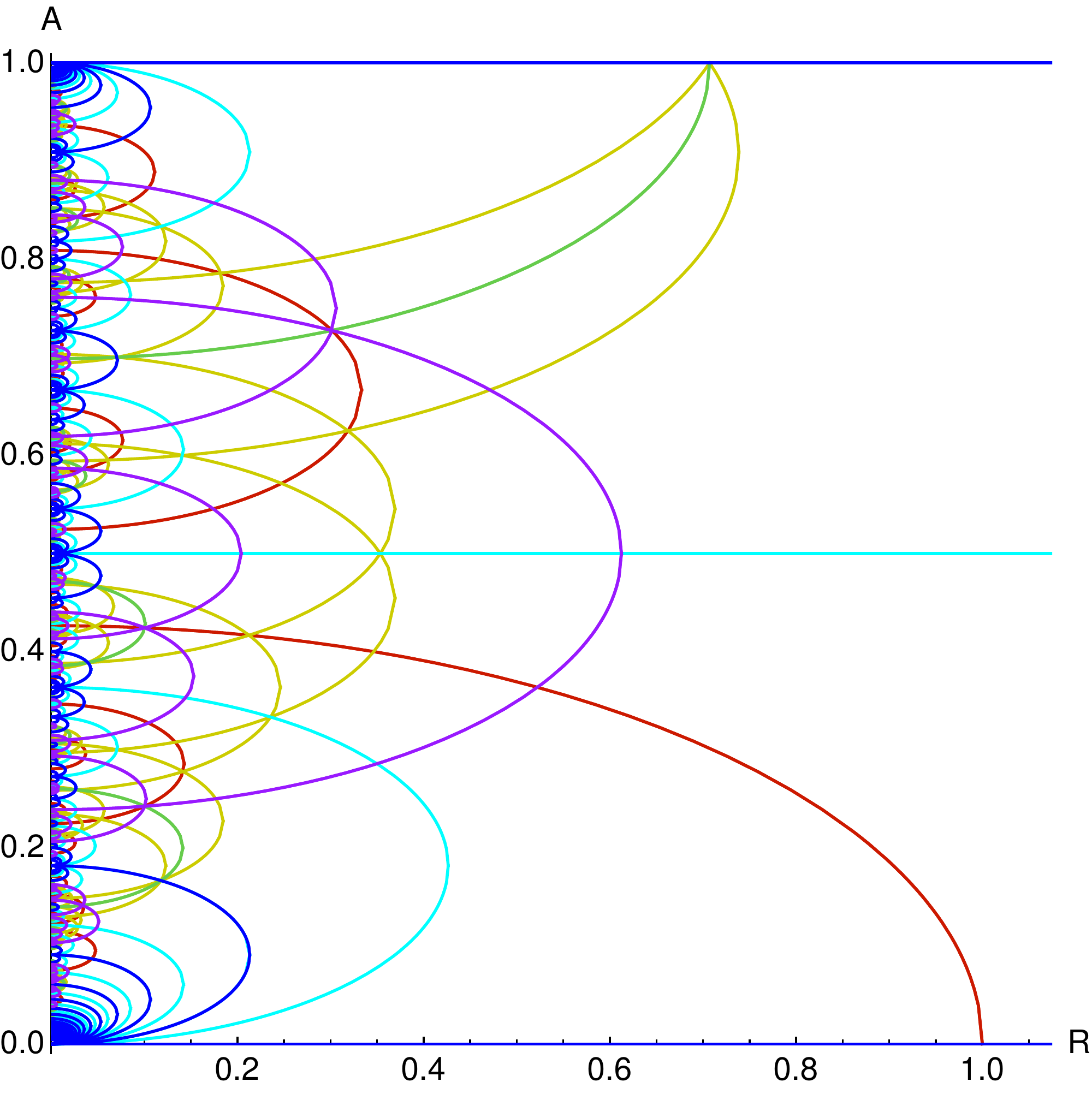}
 \bleyenda
\cua{white}  &\quad SU(11) \times SO(10) \times U(1)^2 \\
\cua{rojo} &\quad   SU(2) \times SU(11) \times SO(10)\times U(1) \\
\cua{amarillo} &\quad SU(12) \times SO(10)\times U(1) \\
\cua{cyan} &\quad  SO(22) \times SO(10) \times U(1)
 \eleyenda
\end{subfigure}
\begin{subfigure}{0.5\textwidth}
 \bleyenda
\cua{verde} &\quad  SU(11) \times SO(12) \times U(1)\\
\cua{azul} &\quad   SO(32) \times U(1)\\
\cua{violeta} &\quad   SU(11) \times E_6 \times U(1)\\
\cua{rojo} +\cua{amarillo} &  \quad SU(2) \times SU(12) \times SO(10)\times U(1)\\
\cua{azul}  +\cua{verde} +\cua{amarillo} +\cua{amarillo} 
&  \quad SO(34)\\
\cua{violeta}  +\cua{violeta} +\cua{verde} +\cua{rojo} 
&  \quad SU(11) \times E_7 
\\
\cua{cyan} +\cua{amarillo} +\cua{amarillo} 
&  \quad SO(24) \times SO(10)\\
\cua{rojo} +\cua{amarillo} +\cua{amarillo} 
&  \quad SU(13) \times SO(10) \\
\cua{violeta}  + \cua{cyan}  &  \quad SO(22) \times  E_6 
\\
\cua{violeta}  + \cua{amarillo}  &  \quad  SU(12) \times  E_6\\
\cua{violeta}  + \cua{rojo}  &  \quad SU(2) \times SU(11) \times  E_6 
\\
\cua{azul}  + \cua{rojo}  &  \quad SU(2) \times  SO(32)
 \eleyenda
\end{subfigure}
\caption{$SO(32)$ heterotic with Wilson line $A^I=\left( (A)_{11} ,0_5 \right)$}
\label{fig:so32_n11}
\end{figure}

\begin{figure}[H] 
\begin{subfigure}{0.5\textwidth}
\centering
\includegraphics[width=.8\textwidth]{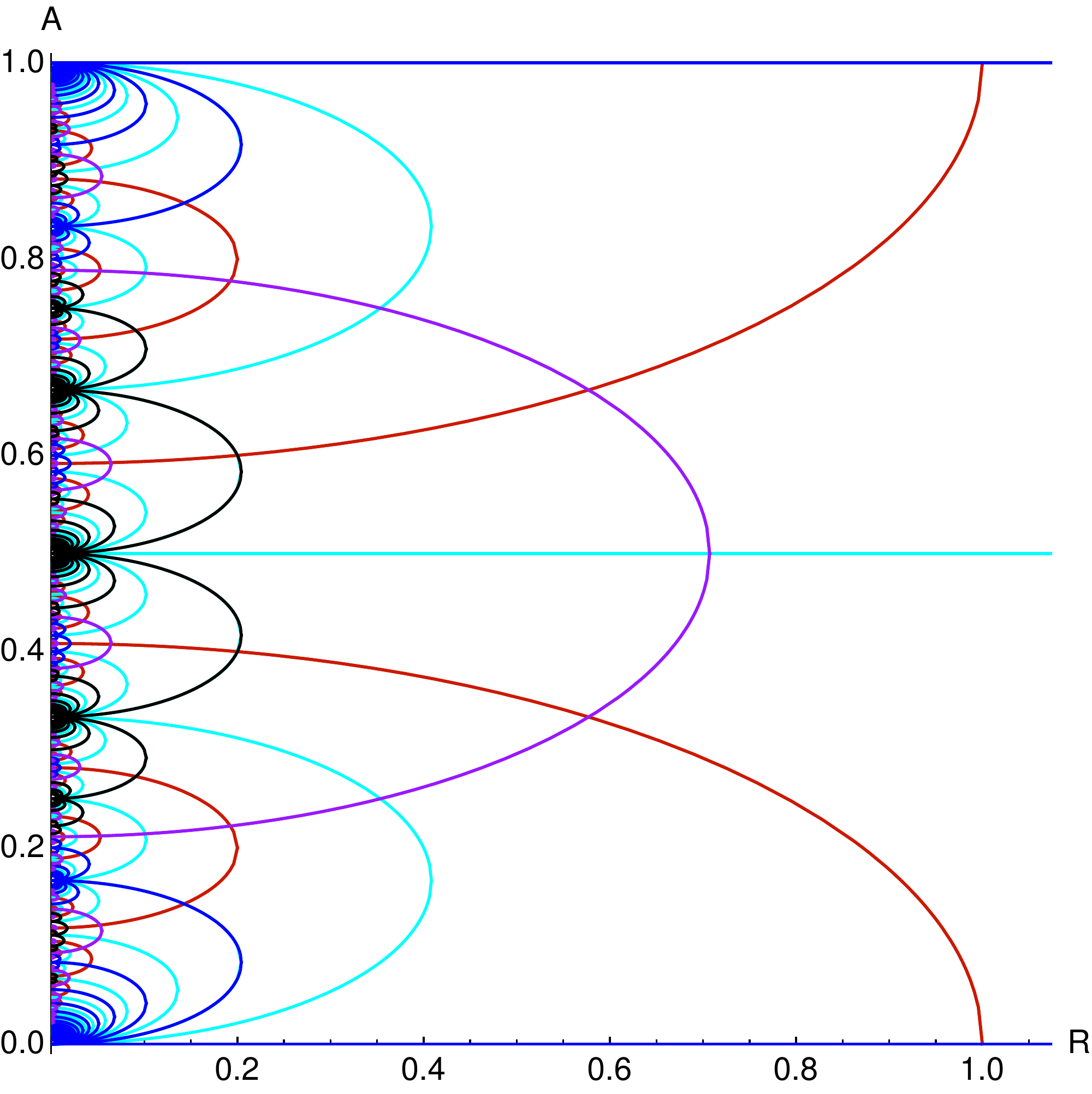}
\end{subfigure}
\begin{subfigure}{0.5\textwidth}
 \bleyenda
\cua{white}  &\quad SU(12) \times SO(8)  \times U(1)^2\\
\cua{rojo} &\quad   SU(2) \times SU(12) \times SO(8)\times U(1)\\
\cua{cyan} &\quad SO(24) \times SO(8) \times U(1)\\
\cuad{azul}{negro} &\quad SO(32)  \times U(1)\\
\cua{violeta} &\quad  SU(12) \times SO(10)\times U(1) \\
\cuad{azul}{negro}  + \cua{rojo}  &  \quad SU(2) \times SO(32) \\
\cua{violeta}  + \cua{rojo}  &  \quad SU(2) \times SU(12) \times  SO(10) \\
\cua{violeta}  + \cua{cyan}  &  \quad   SO(24)\times SO(10)
 \eleyenda
\end{subfigure}
\caption{$SO(32)$ heterotic with Wilson line $A^I=\left( (A)_{12} ,0_4 \right)$}
\label{fig:so32_n12}
\end{figure}

\begin{figure}[H] 
\begin{subfigure}{0.5\textwidth}
\centering
\includegraphics[width=.8\textwidth]{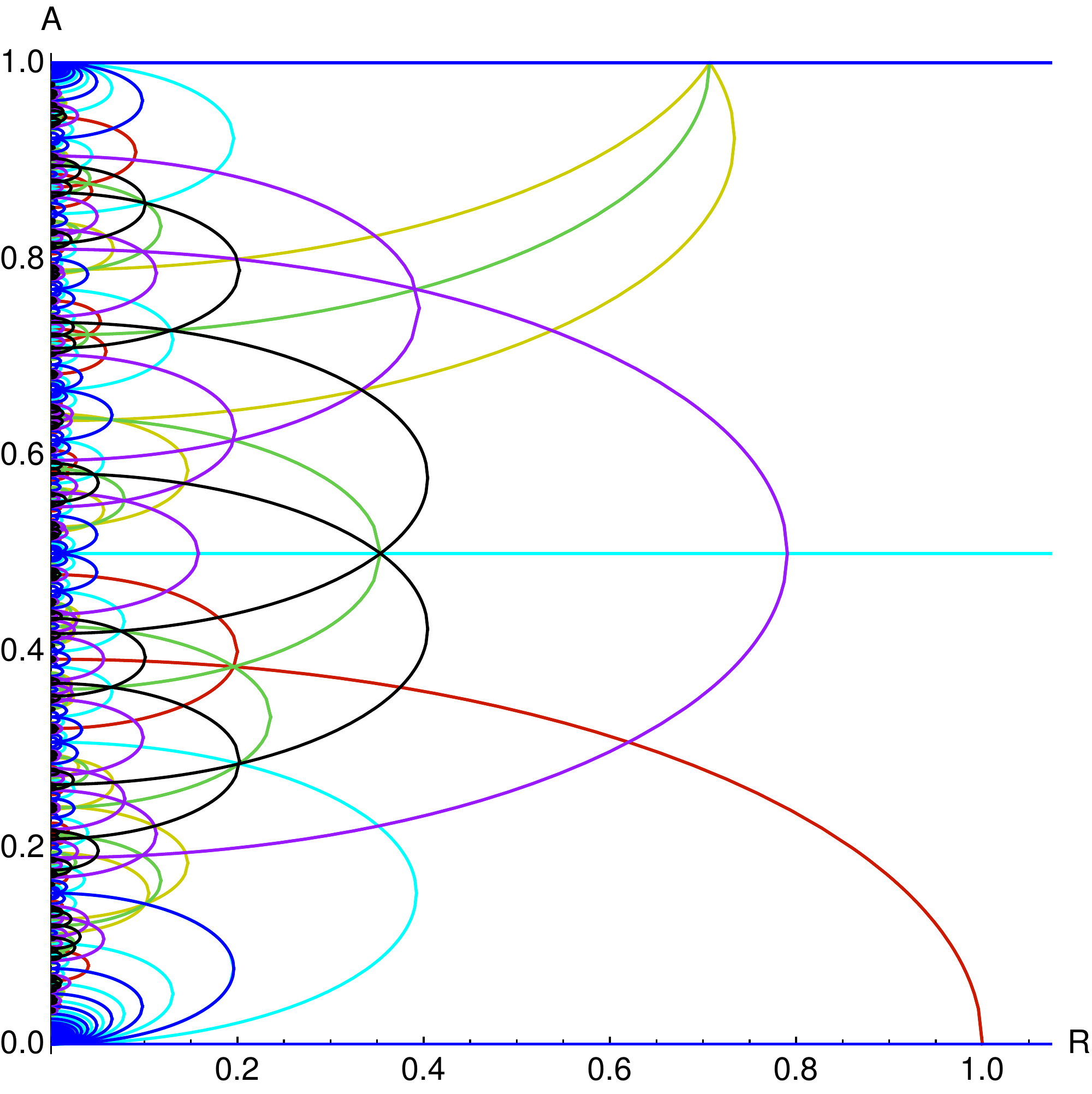}
 \bleyenda
\cua{white}  &\quad SU(13) \times SO(6)\times U(1)^2  \\
\cua{rojo} &\quad   SU(2) \times SU(13) \times SO(6) \times U(1)\\
\cua{amarillo} &\quad SU(14) \times SO(6) \times U(1)\\
\cua{cyan} &\quad  SO(26) \times SO(6) \times U(1)\\
\cua{verde} &\quad  SU(13) \times SO(8)\times U(1)
 \eleyenda
\end{subfigure}
\begin{subfigure}{0.5\textwidth}
 \bleyenda
\cua{azul} &\quad   SO(32) \times U(1)\\
\cua{violeta} &\quad   SU(13) \times SU(5)\times U(1) \\
\cua{negro} &\quad   SU(17) \times U(1)\\
\cua{azul}  +\cua{verde} +\cua{amarillo} +\cua{amarillo} 
&  \quad SO(34)  \\
\cua{violeta}  +\cua{violeta} +\cua{verde} 
&  \quad SU(13) \times SO(10)\\
\cua{negro} +\cua{negro} +\cua{verde} +\cua{cyan} 
&  \quad SO(28) \times SO(6) \\
\cua{rojo} +\cua{violeta} 
&  \quad SU(2) \times SU(13) \times SU(5)\\
\cua{rojo}  + \cua{rojo} +\cua{verde}  + \cua{verde} &  \quad SU(13) \times  SO(10) \\
\cua{rojo}  + \cua{azul}  &  \quad  SU(2) \times  SO(32)\\
\cua{negro}  + \cua{rojo}  &  \quad SU(2) \times SU(2) \times SU(17) \\
\cua{violeta}  + \cua{cyan}  &  \quad SO(26) \times  SU(5) \\
\cua{negro} +\cua{amarillo} +\cua{violeta} 
& \quad SU(18) \\ \cua{violeta} +\cua{amarillo} 
& \quad SU(14) \times SU(5)
 \eleyenda
\end{subfigure}
\caption{$SO(32)$ heterotic with Wilson line $A^I=\left( (A)_{13} ,0_3 \right)$}
\label{fig:so32_n13}
\end{figure}

\begin{figure}[H] 
\begin{subfigure}{0.5\textwidth}
\centering
\includegraphics[width=.8\textwidth]{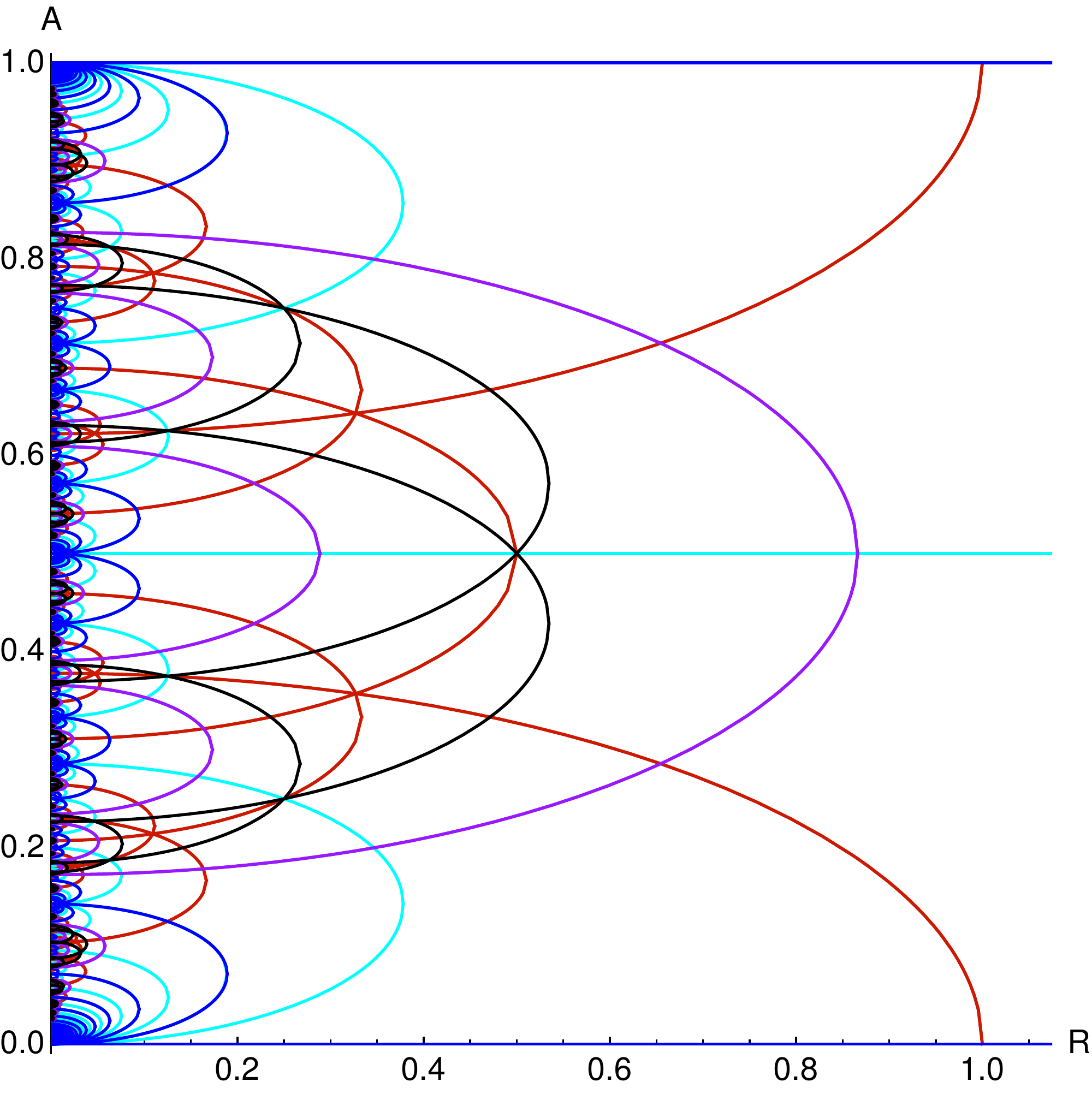}
 \bleyenda
\cua{white}  &\quad SU(14) \times SO(4)  \times U(1)^2\\
\cua{rojo} &\quad   SU(2) \times SU(14) \times SO(4)\times U(1)
 \eleyenda
\end{subfigure}
\begin{subfigure}{0.5\textwidth}
 \bleyenda
\cua{cyan} &\quad  SO(28) \times SO(4) \times U(1) \\
\cua{azul} &\quad   SO(32)\times U(1) \\
\cua{violeta} &\quad  SU(14) \times SU(3) \times SU(2) \times U(1)\\ 
\cua{negro} &\quad  SU(16) \times SU(2) 
\times U(1) \\
\cua{negro}  +\cua{negro} +\cua{cyan} +\cua{rojo} 
&  \quad SU(2) \times SO(32)  \\
\cua{rojo}  +\cua{rojo}  +\cua{rojo} 
&  \quad SU(3) \times SU(14) \times SO(4) \\
\cua{violeta} +\cua{cyan} 
&  \quad SO(28)  \times SU(3) \times SU(2)  
\\
\cua{rojo} +\cua{violeta} 
&  \quad SU(2) \times SU(14) \times SU(3) \times SO(4) \\
\cua{rojo}  + \cua{azul} &  \quad SU(2) \times  SO(32) \\
\cua{rojo}  + \cua{negro}  &  \quad  SU(2) \times  SU(16) \times SO(4)
 \eleyenda
\end{subfigure}
\caption{$SO(32)$ heterotic with Wilson line $A^I=\left( (A)_{14} ,0_2 \right)$}
\label{fig:so32_n14}
\end{figure}

\begin{figure}[H] 
\begin{subfigure}{0.5\textwidth}
\centering
\includegraphics[width=.8\textwidth]{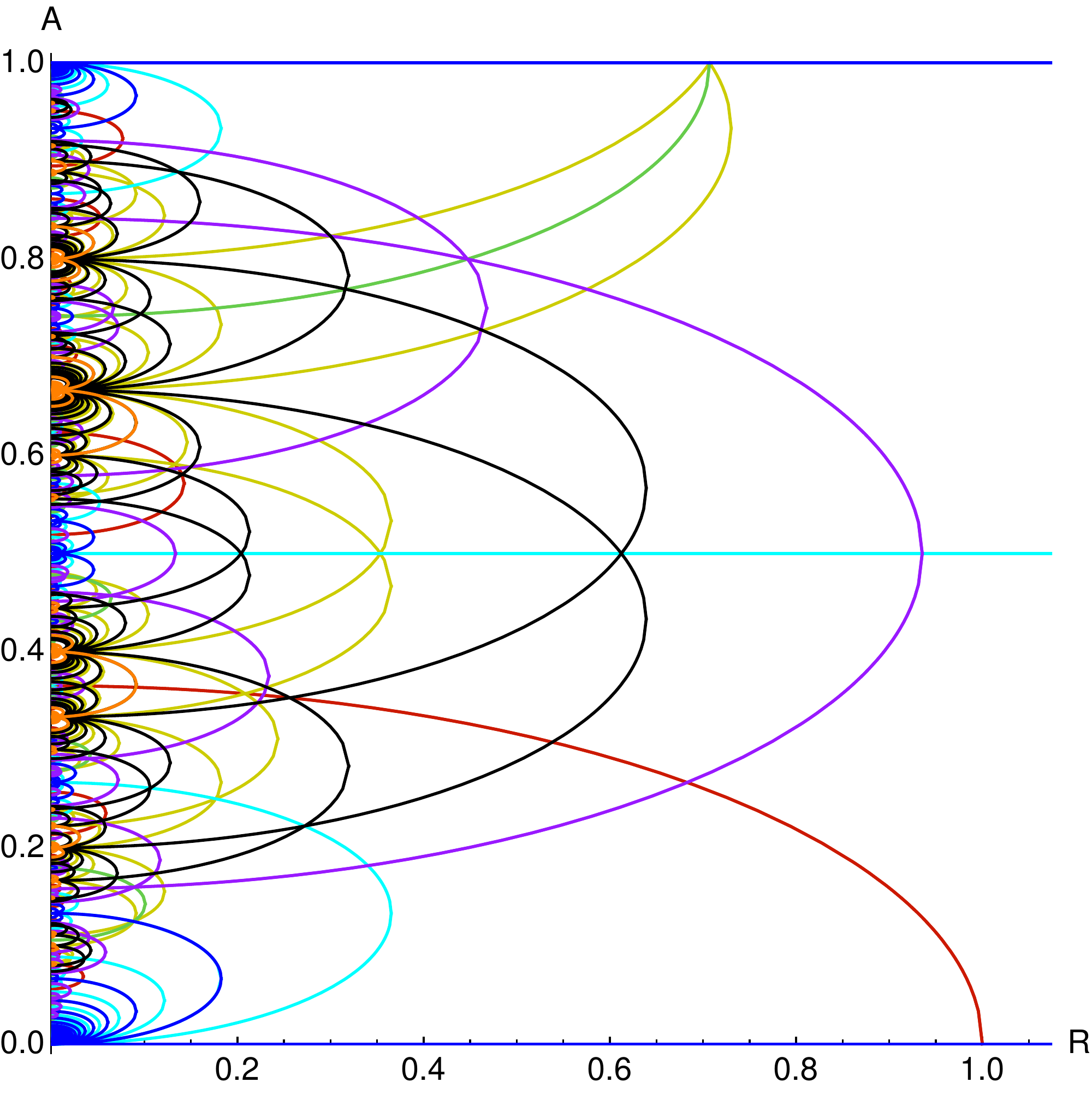}
 \bleyenda
\cua{white}  &\quad SU(15) \times U(1)^3 \\
\cuad{rojo}{violeta} &\quad   SU(2) \times SU(15) \times U(1)^2 \\
\cuad{amarillo}{negro} &\quad  SU(16) \times U(1)^2 \\
\cua{cyan} &\quad   SO(30) \times U(1)^2 
 \eleyenda
\end{subfigure}
\begin{subfigure}{0.5\textwidth}
 \bleyenda
\cua{verde} &\quad  SU(15) \times SO(4)\times U(1) \\ 
\cuad{azul}{naranja} &\quad  SO(32)\times U(1)\\
\cua{amarillo}  +\cua{amarillo} +\cua{verde} +\cua{azul}
&  \quad SO(34)
\\
\cuad{amarillo}{negro}  +\cuad{amarillo}{negro} +\cua{cyan} 
&  \quad   SO(32) \times U(1)   \\
\cua{violeta} +\cua{cyan} 
&  \quad SU(2) \times SO(30) \times U(1) \\
\cuad{amarillo}{negro} +\cuad{rojo}{violeta} 
&  \quad SU(2) \times SU(16) \times U(1)  \\
\cua{rojo}  + \cuad{azul}{naranja} &  \quad SU(2) \times  SO(32) \\
\cua{rojo}  + \cua{violeta}  &  \quad  SO(4) \times SU(15) \times U(1) \\
\cua{rojo}  + \cua{negro}  +\cua{negro} &  \quad   SU(17) \times U(1) \\
\cua{verde}  + \cua{negro}  +\cua{negro} &  \quad  SU(2) \times SU(17)\\
\cua{violeta}  + \cua{negro}  +\cua{amarillo} &  \quad  SO(4) \times SU(16)  \\
\cua{violeta}  + \cua{violeta}  +\cua{verde} &  \quad  SU(15) \times SU(3) \times SU(2)
 \eleyenda
\end{subfigure}
\caption{$SO(32)$ heterotic with Wilson line $A^I=\left( (A)_{15} ,0 \right)$}
\label{fig:so32_n15}
\end{figure}

\begin{figure}[H] 
\begin{subfigure}{0.5\textwidth}
\centering
\includegraphics[width=.8\textwidth]{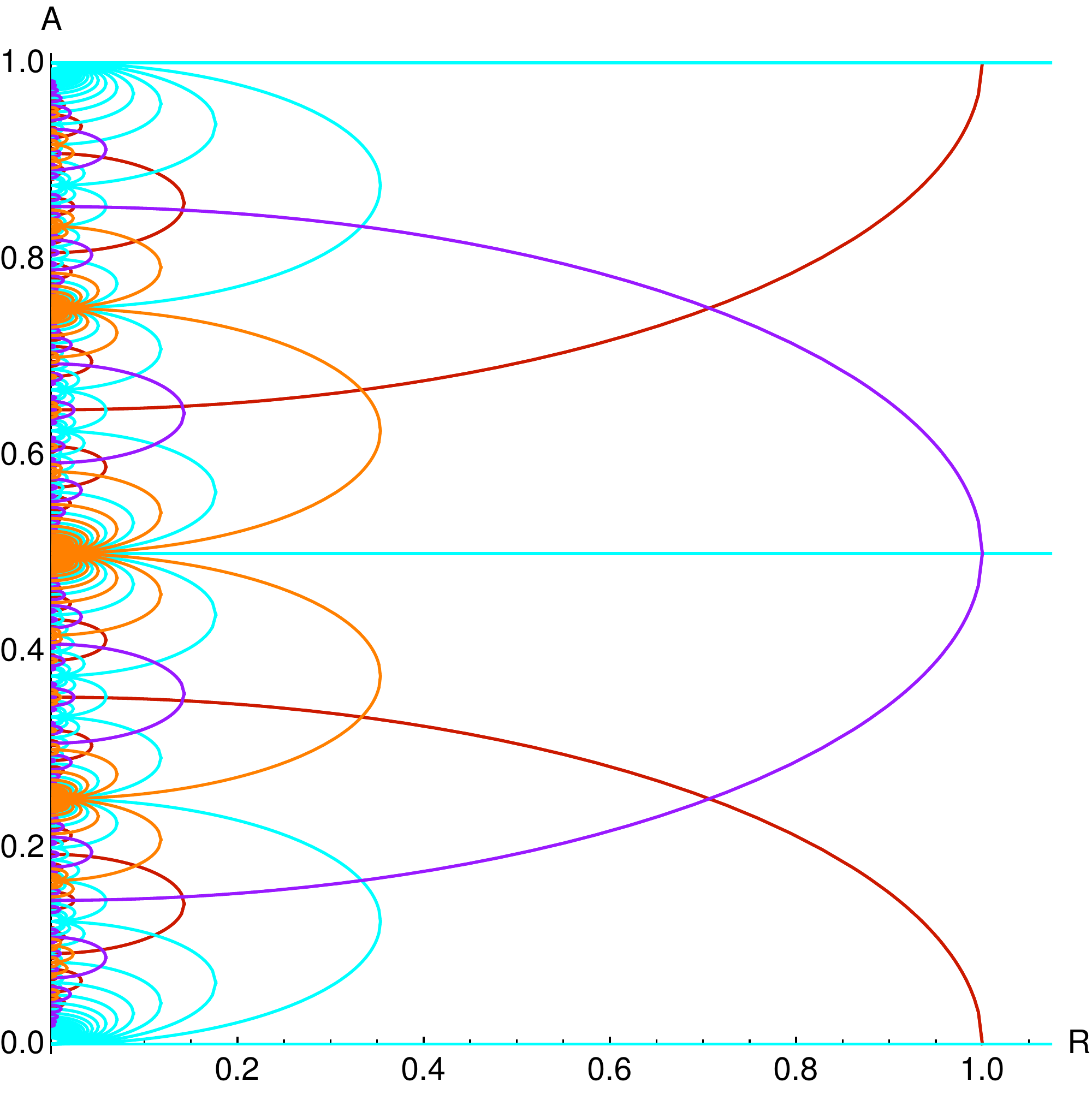}
\end{subfigure}
\begin{subfigure}{0.5\textwidth}
 \bleyenda
\cua{white}  &\quad SU(16)\times U(1)^2 \\
\cuad{rojo}{violeta} &\quad   SU(2) \times SU(16) \times U(1) \\
\cuad{cyan}{naranja} &\quad SO(32) \times U(1) \\
\cuad{rojo}{violeta}  + \cuad{cyan}{naranja}  &  \quad SU(2) \times  SO(32)\\
\cua{violeta}  + \cua{rojo}  &  \quad SO(4) \times  SU(16) \\
 \eleyenda
\end{subfigure}
\caption{$SO(32)$ heterotic with Wilson line $A^I=\left( (A)_{16} \right)$}
\label{fig:so32_n16}
\end{figure}

For Wilson lines of type $A^I = \left ( \left ( A \right )_{p}, 0_{16-p}\right )$ there are families of curves of enhancement parameterized by three integer numbers $\ax$, $\bx$ and $\dx$. Inside each family there are different curves corresponding to different winding numbers and different integer values for $q$. If $R$ is sufficiently small then $w$ can be arbitrarily large.
\bea
A_{w,\ax,\bx,\dx}(R) &=& \frac{p q + \ax-\tfrac{p \delta}{2} \pm \sqrt{\left(\ax-\tfrac{p\delta}{2}\right)^2  -  p \left(  |\ax|- \delta\ax +\bx + 4\delta -2  + 2w^2 R^2\right) }}{pw} \nn\\
&=& \frac{p q + \mu \pm \sqrt{\mu^2   -  p \left( \lambda + 2w^2 R^2\right) }}{pw} \, ,
\eea
where we defined:
\bea
\mu \ = \ \ax - \frac{p\delta}{2}\qquad{\rm and}\qquad
\lambda \ = \ |\ax|- \delta\ax +\bx + 4\delta -2 \, .\nn
\eea

The massless states associated with each family of curves are
\beq 
\pi =\left(\underline{(q \pm (1-\tfrac32\delta) )_{\ax},(q \pm \tfrac12\delta)_{p-{\ax}}},\underline{\bx \pm \tfrac12\delta,(\pm \tfrac12\delta)_{15-p}}\right)\, .
\eeq 

The possible values of the parameters are listed in the following table, with the colour we use to identify them on the figures and the corresponding gauge group.

\begin{center}
\begin{tabular}{|c | c | c | c |  c|} 
\hline
Colour & $\delta$& $\bx$ & $|\ax|$   
& Gauge group \\
\hline\hline 
\cua{white} & & & & $A_{p-1} \times  D_{16-p}$\\\hline
 \cua{rojo}  & $0$      & $0$  & $0$ &    $A_1 \times A_{p-1} \times D_{16-p}$
\\\hline
\cua{amarillo}  &
$0$ & $0$  & $1$ & $A_{p} \times D_{16-p}$ \\\hline
\cua{cyan}  &$0$ & $0$  & $2$ &$D_p \times  D_{16-p}$ \\\hline
\cua{rosa}  &$0$ & $0$  & $3$ & $E_p \times  D_{16-p}$  \\\hline
\cua{verde}  &$0$      & $1$  & $0$ &  $A_{p-1} \times  D_{17-p}$   \\\hline
\cua{azul}  &$0$ & $1$  & $1$ &  $D_{16}$ \\\hline
\cua{violeta}  &$1 $     & $0$  & $0$  & $A_{p-1} \times  E_{17-p}$  \\\hline
\cua{negro}  &$1$ & $0$  & $ 1$  & 
$D_{16}$ for $p=12$,\quad $A_{16}$ for $p=13$,\quad
$A_{15} \times A_1$ for $p=14$,\quad $A_{15} \times D_1$ for $p=15$
\\\hline
\cua{naranja}  &$1$ & $0$  & $ 2$   & $D_{16}$\\
\hline
\end{tabular} 
\end{center}

The number of states for each of these curves is given by
\beq 
2 \binom{p}{|\ax|} (32-2p)^{\bx}  2^{(15 - p+\delta_{p,16})\delta}\, .
\eeq

The allowed values for $q$ and $w$ are the ones that satisfy the quantization condition
\beq
\frac{pq^2 + 2\mu q  + \lambda }{2w} \in \mathbb{Z} \, .
\eeq

For  arbitrary $A$, we get the $3p^2 - 63p + 480$ roots of $U(1)^2 \times SU(p) \times SO(32-2p)$. If $A$ is half-integer we get the $4p^2 - 64p + 480$ roots of $U(1) \times SO(2p) \times SO(32-2p)$, so we can think of them as part of the family with $(\delta,\bx,\ax)=(0,0,2)$ and $w=0$ which give $p^2 - p$ additional states. For $p=2$ $(0,0,2)$ is equivalent to $(0,0,0)$. 

If $A_1$ is integer we get the $480$ roots of $SO(32) \times U(1)$, so we can think of them as part of the family with $(\delta,\bx,\ax)=(0,1,1)$ and $w=0$ superimposed with another one of the family with $(\delta,\bx,\ax)=(0,0,2)$ and $w=0$, which give $63p - 3p^2 = (64p - 4p^2) + (p^2 - p)$ additional states. 
For $p=16$ we only have the $(0,0,2)$.
We can classify some of the enhancements by the colours of the curves that intersect, we list them on the table below:

\begin{center}
\begin{tabular}{|c | c |} 
\hline
Colours &  Gauge group \\
\hline\hline 
$\cua{rojo}+\cuad{azul}{naranja}$ & $A_{1} \times  D_{16}$\\\hline
 $\cua{rojo}+\cua{violeta}$ & $A_1 \times A_{p-1} \times  E_{17-p}$\\\hline
 $\cua{cyan}+\cua{violeta}$ & $E_{p+1} \times  D_{16-p}$\\\hline
$\cua{violeta}+\cua{amarillo}$ & $A_{p} \times  E_{17-p}$\\\hline
$\cua{rojo}+\cua{rojo}+\cua{rojo}$ & $A_2 \times A_{p-1} \times  D_{16-p}$\\\hline
\end{tabular} 
\end{center}

\subsection{Relation to generalized Dynkin diagrams}

Here we show how some of the previous enhancement curves and points can be obtained from the generalized Dynkin diagram in \eqref{GDD}. 

For Wilson lines of the form $(0_{16-p},(A)_p)$ and at any radius, then the inequality $-A_2\le A_1$, as well as all the $A_i \le A_{i+1}$ inequalities are saturated except for $A_{16-p}=A_{17-p}$. This means that the gauge group is given by the generalized diagram with  all the nodes except for $16$, $18$, $19$ and $16-p$. Then the diagram that gives the enhancement symmetry is:
 \bea
\begin{dynkin} 
\dynkinline{1}{0}{2}{0};
\dynkinline{2}{0}{3}{0};  
\dynkinline{3}{0}{4}{0}; 
\dynkinline{4}{0}{5}{0};
\dynkindots{5}{0}{6}{0}; 
\dynkinline{6}{0}{7}{0};
\dynkinline{9}{0}{10}{0};
\dynkindots{10}{0}{11}{0};
\dynkinline{11}{0}{12}{0};
\dynkinline{12}{0}{13}{0};
\dynkinline{13}{0}{14}{0};
\dynkinline{14}{0}{15}{0};
\dynkinline{2}{0}{2}{1};
\foreach \x in {1,2,3,4,7,9,12,13,14,15}{\dynkindot{\x}{0}}
\dynkindot{2}{1} 
\foreach \x in {1,2,3,4,12,13,14,15}{\dynkinlabel{\x}{-1}{\x}}
\dynkinlabel{7}{-1}{15-p}
\dynkinlabel{9}{-1}{17-p}
\dynkinlabel{1}{1}{17}
 \end{dynkin}
 \eea 
which corresponds to the $A_{p-1} \times D_{16-p} (\times U(1)^2)=SU(p) \times SO(32-2p) (\times U(1)^2)$  at a generic value of $A$ and $R$.
Choosing particular values for them, we can saturate one or more inequalities associated to the missing nodes. 
To obtain the horizontal lines we have to pick an arbitrary $R$, which discards the nodes $18$ and $19$. To get the nodes $16-p$ or $16$ we have only one possibility: $A=0$ for the former, and $A=\tfrac12$ for the latter. We get, respectively:
\bea
\begin{dynkin} 
\dynkinline{1}{0}{2}{0};
\dynkinline{2}{0}{3}{0};  
\dynkinline{3}{0}{4}{0}; 
\dynkinline{4}{0}{5}{0};
\dynkindots{5}{0}{6}{0}; 
\dynkindots{6}{0}{7}{0};
\dynkinline{7}{0}{8}{0};
\dynkinline{8}{0}{9}{0}; 
\dynkindots{9}{0}{10}{0};
\dynkindots{10}{0}{11}{0};
\dynkinline{11}{0}{12}{0};
\dynkinline{12}{0}{13}{0};
\dynkinline{13}{0}{14}{0};
\dynkinline{14}{0}{15}{0};
\dynkinline{2}{0}{2}{1};
\foreach \x in {1,2,3,4,8,12,13,14,15}{\dynkindot{\x}{0}}
\dynkindot{2}{1} 
\foreach \x in {1,2,3,4,12,13,14,15}{\dynkinlabel{\x}{-1}{\x}}
\dynkinlabel{8}{-1}{16-p}
\dynkinlabel{1}{1}{17}
 \end{dynkin}\\
\begin{dynkin} 
\dynkinline{1}{0}{2}{0};
\dynkinline{2}{0}{3}{0};  
\dynkinline{3}{0}{4}{0}; 
\dynkinline{4}{0}{5}{0};
\dynkindots{5}{0}{6}{0}; 
\dynkinline{6}{0}{7}{0};
\dynkinline{9}{0}{10}{0};
\dynkindots{10}{0}{11}{0};
\dynkinline{11}{0}{12}{0};
\dynkinline{12}{0}{13}{0};
\dynkinline{13}{0}{14}{0};
\dynkinline{14}{0}{15}{0};
\dynkinline{14}{0}{14}{1};
\dynkinline{2}{0}{2}{1};
\foreach \x in {1,2,3,4,7,9,12,13,14,15}{\dynkindot{\x}{0}}
\dynkindot{14}{1}
\dynkindot{2}{1} 
\foreach \x in {1,2,3,4,12,13,14,15}{\dynkinlabel{\x}{-1}{\x}}
\dynkinlabel{7}{-1}{15-p}
\dynkinlabel{9}{-1}{17-p}
\dynkinlabel{15}{1}{16}
\dynkinlabel{1}{1}{17}
 \end{dynkin}
 \eea
and hence the gauge groups are $D_{16}=SO(32)$ ($\times U(1)$) and $D_{p} \times D_{16-p}=SO(2p)\times SO(32-2p)$ ($\times U(1)$) (blue and cyan lines).
Finally, choosing a specific value of $R$,  the inequality associated to the $18$th or  $19$th node (not both at the same time) can be saturated. This gives  maximal enhancements. In the $D_{16}$ case, the only possibility is to add the $18$th node, which gives  $A_1 \times D_{16}$ (intersection between a blue and a red curve):
\bea
\begin{dynkin} 
\dynkinline{1}{0}{2}{0};
\dynkinline{2}{0}{3}{0};  
\dynkinline{3}{0}{4}{0}; 
\dynkinline{4}{0}{5}{0};
\dynkindots{5}{0}{6}{0}; 
\dynkindots{6}{0}{7}{0};
\dynkinline{7}{0}{8}{0};
\dynkinline{8}{0}{9}{0}; 
\dynkindots{9}{0}{10}{0};
\dynkindots{10}{0}{11}{0};
\dynkinline{11}{0}{12}{0};
\dynkinline{12}{0}{13}{0};
\dynkinline{13}{0}{14}{0};
\dynkinline{14}{0}{15}{0};
\dynkinline{2}{0}{2}{1};
\foreach \x in {1,2,3,4,8,12,13,14,15}{\dynkindot{\x}{0}}
\dynkindot{2}{1} 
\dynkindot{14}{2}
\foreach \x in {1,2,3,4,12,13,14,15}{\dynkinlabel{\x}{-1}{\x}}
\dynkinlabel{8}{-1}{16-p}
\dynkinlabel{1}{1}{17}
\dynkinlabel{15}{2}{18}
 \end{dynkin}
 \eea
In the $D_p \times D_{16-p}$ case, one can add the $18$th or the $19$th node, depending on which  part of the diagram has less than $8$ nodes
\bea
\begin{dynkin} 
\dynkinline{1}{0}{2}{0};
\dynkinline{2}{0}{3}{0};  
\dynkinline{3}{0}{4}{0}; 
\dynkinline{4}{0}{5}{0};
\dynkindots{5}{0}{6}{0}; 
\dynkinline{6}{0}{7}{0};
\dynkinline{9}{0}{10}{0};
\dynkindots{10}{0}{11}{0};
\dynkinline{11}{0}{12}{0};
\dynkinline{12}{0}{13}{0};
\dynkinline{13}{0}{14}{0};
\dynkinline{14}{0}{15}{0};
\dynkinline{14}{0}{14}{1};
\dynkinline{2}{0}{2}{1};
\dynkinline{2}{1}{2}{2};
\foreach \x in {1,2,3,4,7,9,12,13,14,15}{\dynkindot{\x}{0}}
\dynkindot{14}{1}
\dynkindot{2}{1} 
\dynkindot{2}{2} 
\foreach \x in {1,2,3,4,12,13,14,15}{\dynkinlabel{\x}{-1}{\x}}
\dynkinlabel{7}{-1}{15-p}
\dynkinlabel{9}{-1}{17-p}
\dynkinlabel{15}{1}{16}
\dynkinlabel{1}{1}{17}
\dynkinlabel{1}{2}{19}
 \end{dynkin}\\
 \begin{dynkin} 
\dynkinline{1}{0}{2}{0};
\dynkinline{2}{0}{3}{0};  
\dynkinline{3}{0}{4}{0}; 
\dynkinline{4}{0}{5}{0};
\dynkindots{5}{0}{6}{0}; 
\dynkinline{6}{0}{7}{0};
\dynkinline{9}{0}{10}{0};
\dynkindots{10}{0}{11}{0};
\dynkinline{11}{0}{12}{0};
\dynkinline{12}{0}{13}{0};
\dynkinline{13}{0}{14}{0};
\dynkinline{14}{0}{15}{0};
\dynkinline{14}{0}{14}{1};
\dynkinline{2}{0}{2}{1};
\dynkinline{14}{1}{14}{2};
\foreach \x in {1,2,3,4,7,9,12,13,14,15}{\dynkindot{\x}{0}}
\dynkindot{14}{1}
\dynkindot{2}{1} 
\dynkindot{14}{2}
\foreach \x in {1,2,3,4,12,13,14,15}{\dynkinlabel{\x}{-1}{\x}}
\dynkinlabel{7}{-1}{15-p}
\dynkinlabel{9}{-1}{17-p}
\dynkinlabel{15}{1}{16}
\dynkinlabel{1}{1}{17}
\dynkinlabel{15}{2}{18}
 \end{dynkin}
 \eea
This accounts for $D_{p} \times E_{17-p}$ (intersection between a cyan and other curves) and $E_{p+1} \times D_{16-p}$ (intersection between a cyan and a purple curve).

For $R(A)$  (with arbitrary $A$) saturating the inequality associated to the $18$th node,  we obtain $A_1 \times A_{p-1} \times D_{16-p}$ (red curves):
 \bea
\begin{dynkin} 
\dynkinline{1}{0}{2}{0};
\dynkinline{2}{0}{3}{0};  
\dynkinline{3}{0}{4}{0}; 
\dynkinline{4}{0}{5}{0};
\dynkindots{5}{0}{6}{0}; 
\dynkinline{6}{0}{7}{0};
\dynkinline{9}{0}{10}{0};
\dynkindots{10}{0}{11}{0};
\dynkinline{11}{0}{12}{0};
\dynkinline{12}{0}{13}{0};
\dynkinline{13}{0}{14}{0};
\dynkinline{14}{0}{15}{0};
\dynkinline{2}{0}{2}{1};
\foreach \x in {1,2,3,4,7,9,12,13,14,15}{\dynkindot{\x}{0}}
\dynkindot{2}{1} 
\dynkindot{14}{2}
\foreach \x in {1,2,3,4,12,13,14,15}{\dynkinlabel{\x}{-1}{\x}}
\dynkinlabel{7}{-1}{15-p}
\dynkinlabel{9}{-1}{17-p}
\dynkinlabel{1}{1}{17}
\dynkinlabel{15}{2}{18}
 \end{dynkin}
 \eea 
 And in particular for  $A=\tfrac{4}{p}$, we have:
 \bea
\begin{dynkin} 
\dynkinline{1}{0}{2}{0};
\dynkinline{2}{0}{3}{0};  
\dynkinline{3}{0}{4}{0}; 
\dynkinline{4}{0}{5}{0};
\dynkindots{5}{0}{6}{0}; 
\dynkinline{6}{0}{7}{0};
\dynkinline{9}{0}{10}{0};
\dynkindots{10}{0}{11}{0};
\dynkinline{11}{0}{12}{0};
\dynkinline{12}{0}{13}{0};
\dynkinline{13}{0}{14}{0};
\dynkinline{14}{0}{15}{0};
\dynkinline{2}{0}{2}{1};
\dynkinline{2}{1}{2}{2};
\foreach \x in {1,2,3,4,7,9,12,13,14,15}{\dynkindot{\x}{0}}
\dynkindot{2}{1} 
\dynkindot{14}{2}
\dynkindot{2}{2} 
\foreach \x in {1,2,3,4,12,13,14,15}{\dynkinlabel{\x}{-1}{\x}}
\dynkinlabel{7}{-1}{15-p}
\dynkinlabel{9}{-1}{17-p}
\dynkinlabel{1}{1}{17}
\dynkinlabel{15}{2}{18}
\dynkinlabel{1}{2}{19}
 \end{dynkin}
 \eea 
 which gives the gauge group $A_1 \times E_{17-p} \times A_{p-1}$,  considered in section \ref{sec:SU2_E_SU} and  seen in the figures at the intersections between the red and purple curves.
 
 On the other hand,  choosing $R(A)$ so that it saturates the inequality associated to the $19$th node, we obtain $E_{17-p}\times A_{p-1}$ (purple curves): 
\bea
\begin{dynkin} 
\dynkinline{1}{0}{2}{0};
\dynkinline{2}{0}{3}{0};  
\dynkinline{3}{0}{4}{0}; 
\dynkinline{4}{0}{5}{0};
\dynkindots{5}{0}{6}{0}; 
\dynkinline{6}{0}{7}{0};
\dynkinline{9}{0}{10}{0};
\dynkindots{10}{0}{11}{0};
\dynkinline{11}{0}{12}{0};
\dynkinline{12}{0}{13}{0};
\dynkinline{13}{0}{14}{0};
\dynkinline{14}{0}{15}{0};
\dynkinline{2}{0}{2}{1};
\dynkinline{2}{1}{2}{2};
\foreach \x in {1,2,3,4,7,9,12,13,14,15}{\dynkindot{\x}{0}}
\dynkindot{2}{1} 
\dynkindot{2}{2} 
\foreach \x in {1,2,3,4,12,13,14,15}{\dynkinlabel{\x}{-1}{\x}}
\dynkinlabel{7}{-1}{15-p}
\dynkinlabel{9}{-1}{17-p}
\dynkinlabel{1}{1}{17}
\dynkinlabel{1}{2}{19}
 \end{dynkin}
 \eea 

Then we can colour the dots on the generalized Dynkin diagram depending on which curves saturate their inequality:
\bea  
\begin{dynkin} 
\dynkinline{1}{0}{2}{0};
\dynkinline{2}{0}{3}{0};  
\dynkinline{3}{0}{4}{0}; 
\dynkinline{4}{0}{5}{0};
\dynkindots{5}{0}{6}{0}; 
\dynkindots{6}{0}{7}{0};
\dynkinline{7}{0}{8}{0};
\dynkinline{8}{0}{9}{0}; 
\dynkindots{9}{0}{10}{0};
\dynkindots{10}{0}{11}{0};
\dynkinline{11}{0}{12}{0};
\dynkinline{12}{0}{13}{0};
\dynkinline{13}{0}{14}{0};
\dynkinline{14}{0}{15}{0};
\dynkinline{14}{0}{14}{1};
\dynkinline{2}{0}{2}{1};
\dynkinline{14}{1}{14}{2};
\dynkinline{2}{1}{2}{2};
\foreach \x in {1,2,3,4,12,13,14,15}{\dynkindot{\x}{0}}
\dynkindotc{8}{0}{azul}
\dynkindotc{14}{1}{cyan}
\dynkindot{2}{1} 
\dynkindotc{14}{2}{rojo}
\dynkindotc{2}{2}{violeta} 
\foreach \x in {1,2,3,4,12,13,14,15}{\dynkinlabel{\x}{-1}{\x}}
\dynkinlabel{8}{-1}{16-p}
\dynkinlabel{15}{1}{16}
\dynkinlabel{1}{1}{17}
\dynkinlabel{15}{2}{18}
\dynkinlabel{1}{2}{19}
 \end{dynkin}
 \eea
The enhancements corresponding to each  curve are obtained by removing all  the coloured nodes except the node with that colour. The intersection of curves give the group associated to the diagram obtained by keeping the nodes with the colours of the involved curves.

Something odd happens for $p=1$ 
\bea  
\begin{dynkin} 
\dynkinline{1}{0}{2}{0};
\dynkinline{2}{0}{3}{0};  
\dynkinline{3}{0}{4}{0}; 
\dynkinline{4}{0}{5}{0};
\dynkinline{5}{0}{6}{0}; 
\dynkinline{6}{0}{7}{0};
\dynkinline{7}{0}{8}{0};
\dynkinline{8}{0}{9}{0}; 
\dynkinline{9}{0}{10}{0};
\dynkinline{10}{0}{11}{0};
\dynkinline{11}{0}{12}{0};
\dynkinline{12}{0}{13}{0};
\dynkinline{13}{0}{14}{0};
\dynkinline{14}{0}{15}{0};
\dynkinline{14}{0}{14}{1};
\dynkinline{2}{0}{2}{1};
\dynkinline{14}{1}{14}{2};
\dynkinline{2}{1}{2}{2};
\foreach \x in {1,2,3,4,5,6,7,8,9,10,11,12,13,14}{\dynkindot{\x}{0}}
\dynkindotc{15}{0}{azul}
\dynkindotc{14}{1}{cyan}
\dynkindot{2}{1} 
\dynkindotc{14}{2}{rojo}
\dynkindotc{2}{2}{violeta} 
\foreach \x in {1,2,3,4,5,6,7,8,9,10,11,12,13,14,15}{\dynkinlabel{\x}{-1}{\x}}
\dynkinlabel{15}{1}{16}
\dynkinlabel{1}{1}{17}
\dynkinlabel{15}{2}{18}
\dynkinlabel{1}{2}{19}
 \end{dynkin}
 \eea
 For generic $A$ and $R$, this is $D_{15}$.
 For $A=1$ (cyan dot) we get $D_{16}$ and if we also take $R^2 = \tfrac12$ (red dot) we get $D_{17}$. If, on the other hand, we take $A=0$ (blue dot) then we get $D_{16}$ and if we also select $R^2 = 1$ (red dot) we get $A_1 \times D_{16}$. If we only take the appropriate $R$ to have the red dot, then we get $A_1 \times D_{15}$. To compare with  figure \ref{fig:so32_n1} we have to take into account that the cyan solutions are not well defined for $p<2$, and then we see them as blue curves.

 For $p=15$, the equation for the seventh node  no longer holds, and then we have:
\bea  
\begin{dynkin} 
\dynkinline{1}{0}{2}{0};
\dynkinline{2}{0}{3}{0};  
\dynkinline{3}{0}{4}{0}; 
\dynkinline{4}{0}{5}{0};
\dynkinline{5}{0}{6}{0}; 
\dynkinline{6}{0}{7}{0};
\dynkinline{7}{0}{8}{0};
\dynkinline{8}{0}{9}{0}; 
\dynkinline{9}{0}{10}{0};
\dynkinline{10}{0}{11}{0};
\dynkinline{11}{0}{12}{0};
\dynkinline{12}{0}{13}{0};
\dynkinline{13}{0}{14}{0};
\dynkinline{14}{0}{15}{0};
\dynkinline{14}{0}{14}{1};
\dynkinline{2}{0}{2}{1};
\dynkinline{14}{1}{14}{2};
\dynkinline{2}{1}{2}{2};
\foreach \x in {2,3,4,5,6,7,8,9,10,11,12,13,14,15}{\dynkindot{\x}{0}}
\dynkindotc{1}{0}{azul}
\dynkindotc{14}{1}{cyan}
\dynkindotc{2}{1}{azul} 
\dynkindotc{14}{2}{rojo}
\dynkindotc{2}{2}{violeta} 
\foreach \x in {1,2,3,4,5,6,7,8,9,10,11,12,13,14,15}{\dynkinlabel{\x}{-1}{\x}}
\dynkinlabel{15}{1}{16}
\dynkinlabel{1}{1}{17}
\dynkinlabel{15}{2}{18}
\dynkinlabel{1}{2}{19}
 \end{dynkin}
 \eea
 For generic $A$ and $R$ this is $A_{14}$. Selecting a specific $R$, we can turn on the red and/or the purple nodes to get $A_1 \times A_{14}$ or $D_2 \times A_{14}$. Selecting $A=\tfrac12$ (cyan dot) we obtain $D_{15}$ and for $A = 0$  both blue dots are turned on and we get $D_{16}$. Only choosing  $R=1$ (red dot) we get $A_1 \times D_{16}$.
 
 For $p=16$ we have a very different situation:
\bea  
\begin{dynkin} 
\dynkinline{1}{0}{2}{0};
\dynkinline{2}{0}{3}{0};  
\dynkinline{3}{0}{4}{0}; 
\dynkinline{4}{0}{5}{0};
\dynkinline{5}{0}{6}{0}; 
\dynkinline{6}{0}{7}{0};
\dynkinline{7}{0}{8}{0};
\dynkinline{8}{0}{9}{0}; 
\dynkinline{9}{0}{10}{0};
\dynkinline{10}{0}{11}{0};
\dynkinline{11}{0}{12}{0};
\dynkinline{12}{0}{13}{0};
\dynkinline{13}{0}{14}{0};
\dynkinline{14}{0}{15}{0};
\dynkinline{14}{0}{14}{1};
\dynkinline{2}{0}{2}{1};
\dynkinline{14}{1}{14}{2};
\dynkinline{2}{1}{2}{2};
\foreach \x in {1,2,3,4,5,6,7,8,9,10,11,12,13,14,15}{\dynkindot{\x}{0}}
\dynkindotc{14}{1}{cyan}
\dynkindotc{2}{1}{naranja} 
\dynkindotc{14}{2}{rojo}
\dynkindotc{2}{2}{violeta} 
\foreach \x in {1,2,3,4,5,6,7,8,9,10,11,12,13,14,15}{\dynkinlabel{\x}{-1}{\x}}
\dynkinlabel{15}{1}{16}
\dynkinlabel{1}{1}{17}
\dynkinlabel{15}{2}{18}
\dynkinlabel{1}{2}{19}
 \end{dynkin}
 \eea
 For  generic $A$ and $R$ this is $A_{15}$. Selecting a specific $R$,  the red and/or the purple nodes are turned on and we get $A_1 \times A_{15}$ or $D_2 \times A_{15}$. Selecting $A=\tfrac12$ (cyan dot) we obtain $D_{16}$ and for $A = 0$ (orange dot) we get $D_{16}$. Only choosing $R=1$ (red or purple dot) we get $A_1 \times D_{16}$.

The enhancements of the curves that correspond to the other colours cannot be obtained with this construction. On one hand we see from the figures that the Wilson lines that give these curves are not in the fundamental region in the conventions of Appendix \ref{app:Dyndia}. On the other hand, if this region is the fundamental region, it should contain all the possible enhancement groups, and as such all the curves with the different colours. However, it is easy to see that using this method, the Wilson lines in the fundamental region that give the missing enhancement groups are not of the form chosen, with $p$ equal components and the other zero. For example, to obtain the enhancement $A_p \times D_{16-p}$ corresponding to the yellow curves, we would need to replace the $15$th node with the $16$th one (and then add the 18th one), which requires $A_{16}=1-A_{15}$ which is not within the ansatz chosen for the Wilson lines.

\subsection{Slices for the $E_8 \times E_8$ theory}

We applied the method based on generalized Dynkin diagrams used to obtain the slices of moduli space containing desired enhanement groups, to the case of $\Gamma_8 \times \Gamma_8$. This forces us to consider now also Wilson lines where some of the components are of the form $1 - A$. The slices shown in the figures contain most of the enhacement groups discussed in the main text.

\begin{figure}[H]
\begin{subfigure}{0.5\textwidth}
\centering
\includegraphics[width=.8\textwidth]{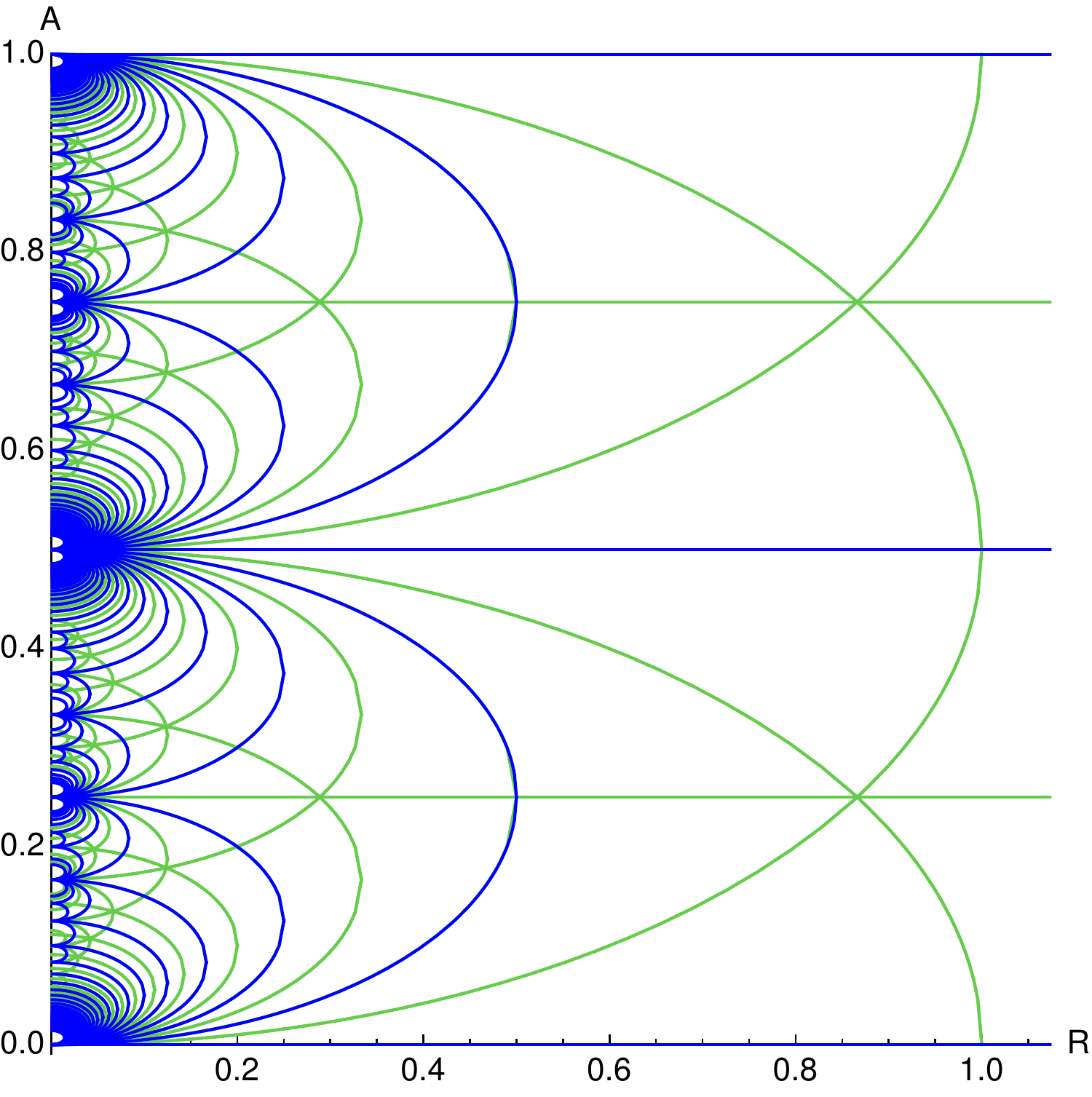}
\end{subfigure}
\begin{subfigure}{0.5\textwidth}
 \bleyenda
\cua{white}  &\quad E_7 \times E_8 \times U(1)^2 \\
\cua{verde} &\quad  SU(2) \times E_7 \times E_8 \times U(1) \\
\cua{azul} &\quad  E_8 \times E_8 \times U(1) \\
\cua{azul} + \cua{verde}  &  \quad SU(2)  \times  E_8 \times E_8 \\
\cua{verde} + \cua{verde}  + \cua{verde}  &  \quad SU(3) \times E_7 \times E_8
 \eleyenda
\end{subfigure}
\caption{$\Gamma_8 \times \Gamma_8$ heterotic with Wilson line $A^I=\left( (A)_{8},0_8 \right)$}
\label{fig:e8e8_n8_n0}
\end{figure}

\begin{figure}[H]
\begin{subfigure}{0.5\textwidth}
\centering
\includegraphics[width=.8\textwidth]{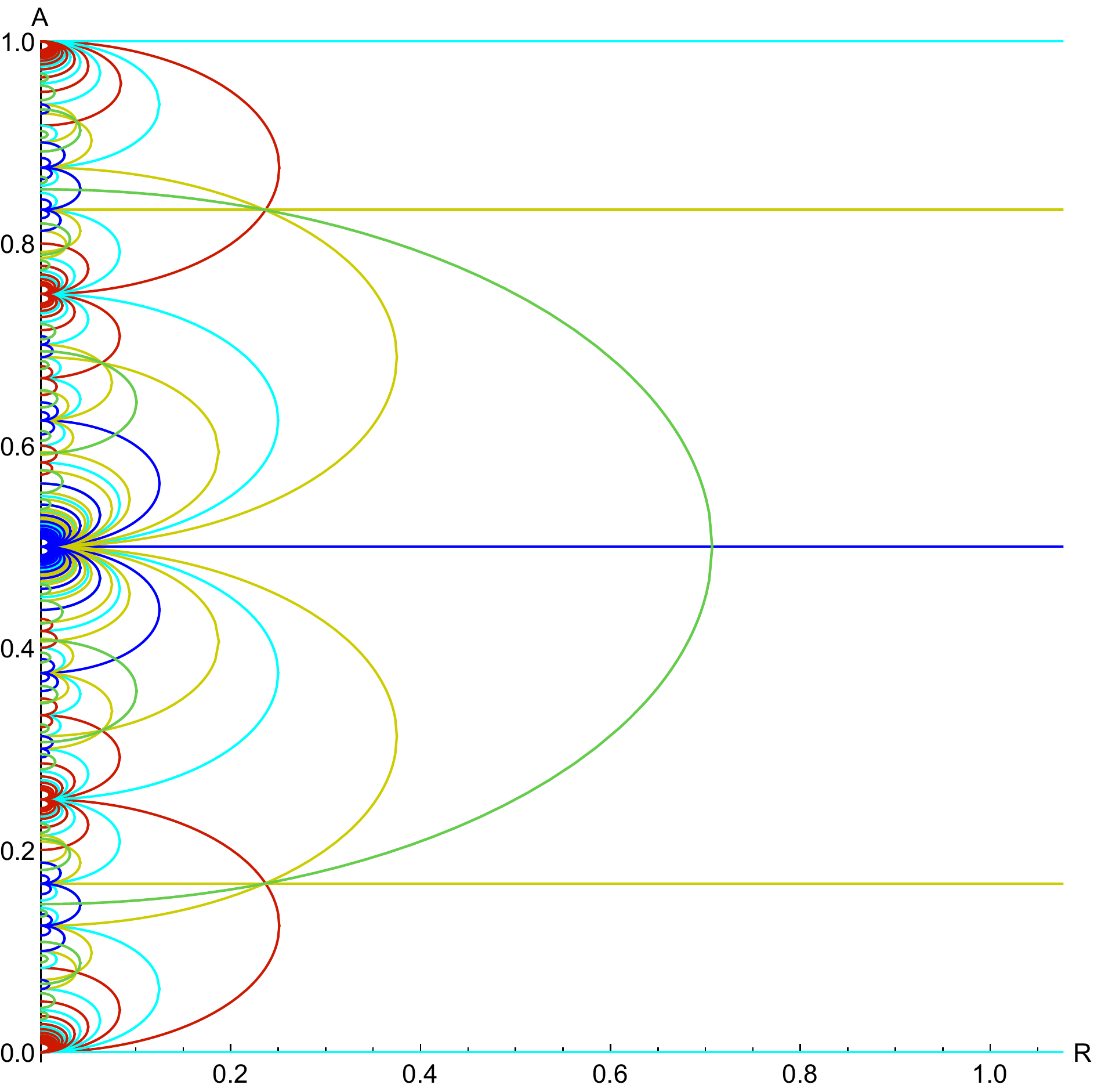}
\end{subfigure}
\begin{subfigure}{0.5\textwidth}
 \bleyenda
\cua{white}  &\quad SU(8) \times SO(16) \times U(1)^2 \\
\cua{rojo} &\quad  SO(32) \times U(1)\\
\cua{cyan} &\quad  SO(16) \times  SO(16) \times U(1) \\
\cua{amarillo} &\quad SU(9) \times SO(16)  \times U(1)\\
\cua{azul} &\quad  E_8 \times  SO(16) \times U(1)\\
\cua{verde} &\quad   SU(8)\times SO(18) \times U(1)\\
\cua{rojo} + \cua{amarillo}  + \cua{amarillo} + \cua{verde}  &  \quad SO(34) \\
\cua{azul} + \cua{verde}    &  \quad E_8 \times SO(18)
 \eleyenda
\end{subfigure}
\caption{$\Gamma_8 \times \Gamma_8$ heterotic with Wilson line $A^I=\left((A)_{7},1-A,1, 0_{7} \right)$}
\label{fig:e8e8_n0_n8b_11}
\end{figure}

\begin{figure}[H]
\begin{subfigure}{0.5\textwidth}
\centering
\includegraphics[width=.8\textwidth]{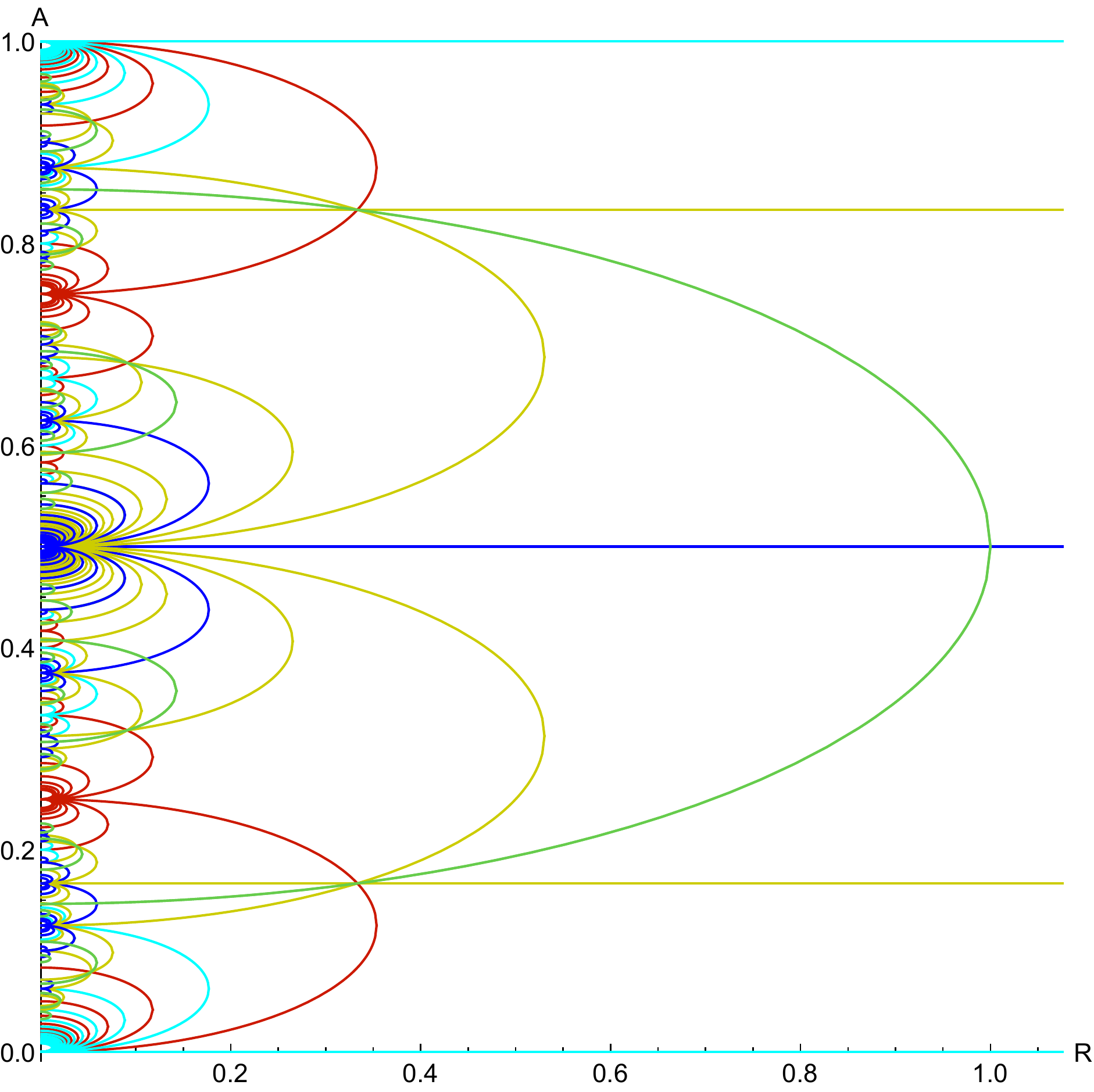}
\end{subfigure}
\begin{subfigure}{0.5\textwidth}
 \bleyenda
\cua{white}  &\quad SU(8) \times SU(8) \times U(1)^3 \\
\cua{rojo} &\quad  SU(16) \times U(1)^2\\
\cua{cyan} &\quad  SO(16) \times  SO(16) \times U(1) \\
\cua{amarillo} &\quad  SU(9) \times SU(9)  \times U(1)\\
\cua{azul} &\quad  E_8 \times  E_8 \times U(1)\\
\cua{verde} &\quad  SU(8) \times  SU(8) \times SU(2) \times U(1)\\
\cua{rojo} + \cua{amarillo}  + \cua{amarillo} + \cua{verde}  &  \quad SU(18) \\
\cua{azul} + \cua{verde}    &  \quad E_8 \times E_8 \times SU(2)
 \eleyenda
\end{subfigure}
\caption{$\Gamma_8 \times \Gamma_8$ heterotic with Wilson line $A^I=\left( (A)_{7},1-A,(A)_{7},1-A \right)$}
\label{fig:e8e8_n8b_n8b_11}
\end{figure}

\begin{figure}[H]
\begin{subfigure}{0.5\textwidth}
\centering
\includegraphics[width=.8\textwidth]{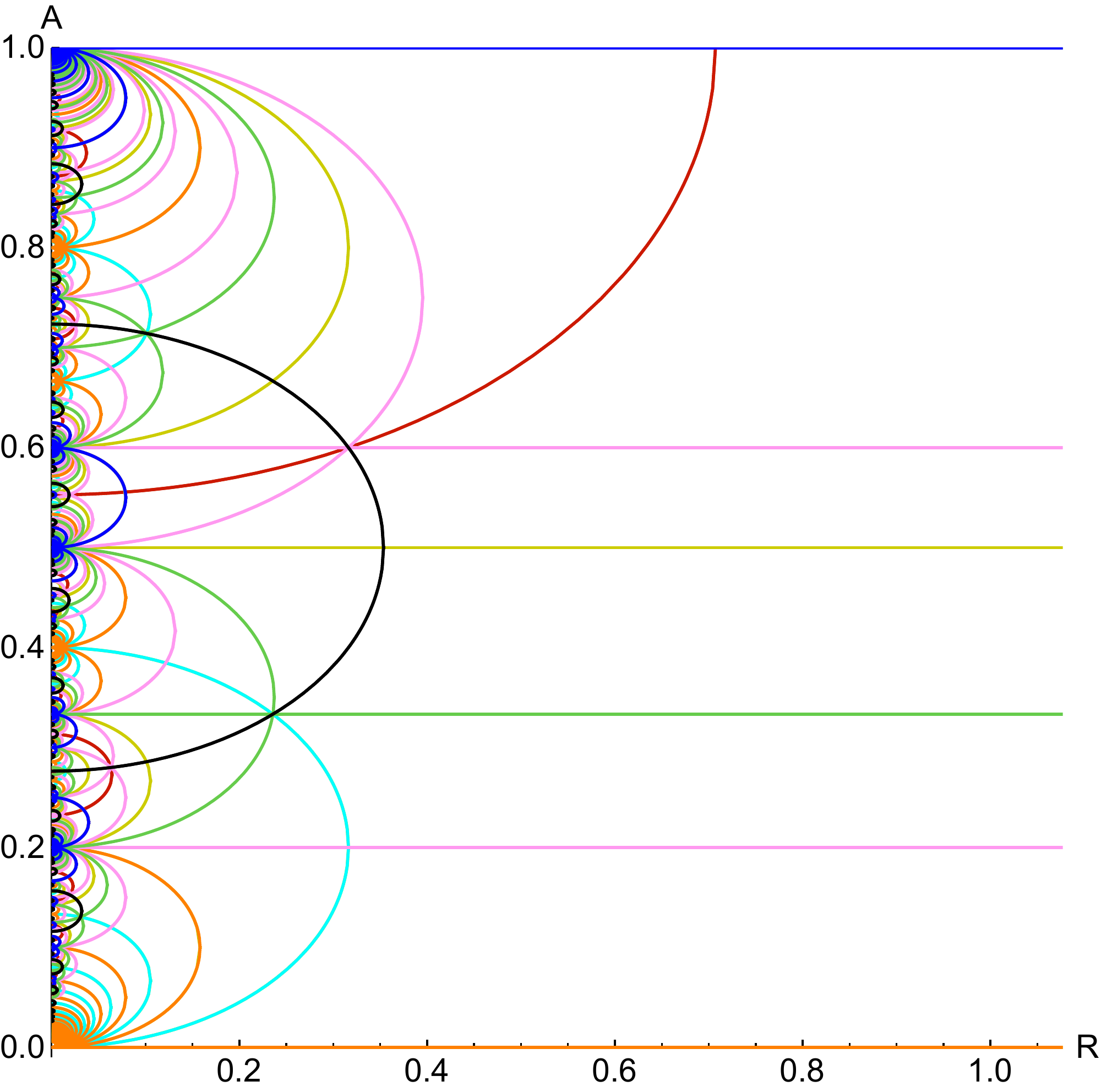}
 \bleyenda
\cua{white}  &\quad SO(16) \times SU(5) \times SO(6) \times U(1)^2 \\
\cua{rojo} &\quad  SO(18) \times SU(5) \times SO(6) \times U(1)
 \eleyenda
\end{subfigure}
\begin{subfigure}{0.5\textwidth}
 \bleyenda
\cua{amarillo} &\quad  SO(16) \times SO(10) \times SO(6) \times U(1)\\
\cua{cyan} &\quad  SO(26) \times  SO(6) \times U(1) \\
\cua{naranja} &\quad  SO(16) \times  SO(16) \times U(1) \\
\cua{rosa} &\quad  SO(16) \times  SU(5) \times SU(5) \\
\cua{verde} &\quad  SO(16) \times  SU(9)\\
\cua{azul} &\quad  SO(16) \times  E_8\\
\cua{negro} &\quad  SO(24) \times  SU(5)\\
\cua{rojo}  + \cua{azul}  &  \quad SO(18) \times  E_8 \\
\cua{amarillo}  + \cua{negro}  &  \quad SO(24) \times  SO(10) \\
\cua{cyan}  + \cua{rosa}  &  \quad 
SO(26) \times  SU(5) \\
\cua{rojo} + \cua{rosa}  + \cua{rosa} + \cua{negro}  &  \quad SO(26) \times  SU(5) \\
\cua{cyan} + \cua{verde}  + \cua{verde} + \cua{negro}  &  \quad SO(34)
 \eleyenda
\end{subfigure}
\caption{$\Gamma_8 \times \Gamma_8$ heterotic with Wilson line $A^I=\left( A-1, (A)_{4},0_{10},1 \right)$}
\label{fig:e8e8_n5_n0_11}
\end{figure}

\section{More on fixed points and dualities}
\label{app:B6}

Here we collect the details of the calculations involved in section \ref{sec:36}.
The transformations $O_{U}$ in \eqref{O} acting on a vector $Z$ of momentum, winding  and ``heterotic"  momenta, result in the transformed momenta 
  \beq
w'=& r \left ( n - \tfrac12|A|^2 w - A \cdot \pi \right )  \, ,\\
n'=&   \left ( \tfrac{1}{r} +  A' U A \right ) w 
+    A'      U \pi 
-r\tfrac{|A'|^2}{2}\left (  n  - \tfrac12|A|^2 w   - A \cdot \pi     \right )\, ,
 \\
\pi' =&  U \pi  +      U A w  - r A' \left (n -\tfrac12|A|^2 w - A \cdot \pi \right )\, .
\eeq 
Requiring these to be quantized 
leads to the conditions 
\bea \label{quantOfull}
&&r ,\, \frac{r|A|^2}{2} ,\, \frac{r|A'|^2}{2} \in \mathbb{Z}; \\
\frac{1}{r} + A' U A + r\frac{|A|^2}{2}\frac{|A'|^2}{2} \in \mathbb{Z} &&
\forall \,\,\pi \in \Gamma,\,\,\pi'\in\Gamma'  : \  \pi' U  \pi + r (\pi \cdot A) (\pi' \cdot A') \in \mathbb{Z} \, ;\nn\\ 
rA, \, rA' \in \Gamma \cap \Gamma' \, ; \qquad  &&
A'U + \frac{r|A'|^2}{2}A \in \Gamma\, ;\qquad
UA + \frac{r|A|^2}{2}A' \in \Gamma' \, \nn 
\eea
and $U \in O(16,\mathbb{Z})$.

We analyze these in more detail, depending whether the duality acts on the same theory or links two theories with different lattices $\Gamma$ and $\Gamma'$.

\subsection{$\Gamma \leftrightarrow \Gamma$}

For Wilson lines of the form \eqref{A1only}, and $U$ given by \eqref{Upm}, the quantization conditions \eqref{quantOfull} become
 \beq
  \tfrac{(Q \pm 1)^2}{r} \, , \,
  (Q \pm 1 )\sqrt{\tfrac{Q}{2r}} \, , \,  \sqrt{\tfrac{Qr}{2}}  \in \mathbb{Z} \, , \quad \text{and}\, \, \left\{ \begin{matrix} Q \in 2 \mathbb Z & \text{for} & U=\pm I \\ Q +1 \in 2 \mathbb Z & \text{for} & \, U=U_{\pm}
\end{matrix} \right.
\eeq
where $U_\pm$ are defined in \eqref{Upm},
\beq
Q \equiv  \frac{r A_1^2}{2} = \left. \frac12 \frac{A_1^2}{R^2} \right\vert_{\rm{fp}}=\left. \frac12 \mathbb A^2 \right\vert_{\rm fp}  \  
\eeq
and $\mathbb A$ is defined in \eqref{a}. Here we have used that for the fixed points $R=\Rsd$, one has $r = \Rsd^{-2}$ since $R'=\tfrac{1}{rR}=R$.

For $U=\pm I$ we define $p =   \tfrac{(Q \pm 1)^2}{r}$
and $q = Q/2$, then:
\beq 
p, \ \ q, \ \ r,\ \ \sqrt{pq}, \ \ \sqrt{qr} \  \in  \ \mathbb{Z} ;\quad  
\sqrt{pr} = 2q \pm 1\, .
\eeq 
Quotienting these equations, we see that $p, q$  $r$ can be written as 
\beq 
\sqrt{p} = t \sqrt{k} ,\quad
\sqrt{q} = n \sqrt{k},\quad
\sqrt{r} = m \sqrt{k} \, ,
\eeq 
with $t, n, m, k \in \mathbb{Z}$. 
Then
 \beq 
k^{-1}  =  \pm (tm - 2n^2) \in \mathbb{Z}\, ,
\eeq 
which implies $k=1$ and $t = \tfrac{2n^2 \pm 1}{m}$. Taking into account that $n = \sqrt{\tfrac{Q}{2}} = \tfrac{R^{-1}A_1}{2}$ and $m=\sqrt{r} = R^{-1}$ must be integers, the only condition is
\beq 
\frac{2n^2\pm 1}{m} \in \mathbb{Z}\, . \label{condiciondual1}
\eeq

For $U=U_\pm$, defining $p =   \tfrac{(Q \pm 1)^2}{2r}$ and $s = \tfrac{r}{2}$, 
\beq 
p, \  \ Q, \  \ s,\quad \sqrt{p Q}, \quad  \sqrt{Q s}  \in \mathbb{Z} \quad {\rm and} \quad
\sqrt{ps} = \tfrac{Q\pm1}{2}\, ,
\eeq 
where we have used  the fact that as $Q$, $2p$ and $\sqrt{pQ}$ are integers with $Q$ odd, implies that $p$ is also integer.
$\tfrac{Q+1}{2}$ is integer, then we have the same situation as in the first case. Analogously
\beq 
\sqrt{p} = t\sqrt{k},\quad
\sqrt{Q} = n\sqrt{k},\quad
\sqrt{s} = m\sqrt{k}
\eeq 
with $t,n, m, k \in \mathbb{Z}$. Then
\beq 
tmk = \tfrac{n^2 k \pm 1}{2}\,  \quad \implies \ \
k^{-1} = \pm(2tm -n^2) \in \mathbb{Z}\, ,
\eeq 
but this implies that $k=1$ and $t = \tfrac{n^2 \pm 1}{2m}$. Then, taking into account that $n=\sqrt{Q} = \tfrac{R^{-1}A_1}{\sqrt{2}}$ and $m=\sqrt{\tfrac{r}{2}}=\tfrac{R^{-1}}{\sqrt{2}}$, the only condition is:
\beq 
\frac{n^2 \pm 1}{2m} \in \mathbb{Z}  \label{condiciondual2}
\eeq 
The possible values of $m$ and $n$ that verify these conditions with the plus signs give the fixed points $(R,A_1)$ presented in Table 1. 

If we take $U=1$ then the quantization conditions \eqref{quantOfull} require $\pi \cdot \pi' + r (\pi \cdot A) (\pi' \cdot A')  \in \mathbb{Z} \,\, \left (\forall \,\,\pi,\pi' \in \Gamma \right)$. As $\pi$ and $\pi'$ belong to the same lattice, $\pi \cdot \pi' \in \mathbb{Z}$, and then $\frac{hh'}{r} \in \mathbb{Z}$, where $h= \pi \cdot rA$ and $h'= \pi' \cdot rA'$. Restricting to $r$ prime, this implies that either  $\pi \cdot A \in \mathbb{Z}$ or $\pi' \cdot A' \in \mathbb{Z}$.

 If $A$ does not satisfy this for any $\pi \in \Gamma$, then  $A' \in \Gamma$, and viceversa, i.e. either
$A \in \Gamma$ or $A' \in \Gamma$. But $ \frac{r|A'|^2}{2}\in\mathbb{Z}$, $A' + \frac{r|A'|^2}{2} A \in \Gamma$ and the reciprocal conditions imply $A, A' \in \Gamma$.

We just need to verify $\frac{1}{r} + A' \cdot A + r\frac{|A|^2}{2}\frac{|A'|^2}{2}  \in \mathbb{Z}$. But $A' \cdot A$ is integer and $|A|^2$, $|A'|^2$ are even, then we get: $\frac{1}{r}   \in \mathbb{Z}$, which is only possible for $r=1$. Then $1$ is the only non-composite possible value for $r$ when the duality does not change the lattice and $U=1$. 

\subsection{$\Gamma \leftrightarrow \Gamma'\neq \Gamma$ }

The quantization conditions \eqref{quantOfull} for the case where the dual lattice is not the original one become
\bea \label{quantcond2}
&&r ,\, \frac{r|A|^2}{2} ,\, \frac{r|A_U|^2}{2} \in \mathbb{Z}; \\
\frac{1}{r} + A_U \cdot A + r\frac{|A|^2}{2}\frac{|A_U|^2}{2} \in \mathbb{Z} &&
\forall \,\,\pi \in \Gamma,\,\,\pi_U\in\Gamma_U  : \  \pi_U \cdot \pi + r (\pi \cdot A) (\pi_U \cdot A_U) \in \mathbb{Z} \, ;\nn\\ 
rA, \, rA_U \in \Gamma \cap \Gamma_U \, ; \qquad  &&
A_U + \frac{r|A_U|^2}{2}A \in \Gamma\, ;\qquad
A + \frac{r|A|^2}{2}A_U \in \Gamma \nn
\eea
where $\Gamma_U$ is the lattice obtained by applying the transformation $U$ to all the elements of $\Gamma'$ and $A_U=A' U$. This proves the statements at the beginning of \ref{sec:gamaneqgamma'}. 

Restricting to the case $U=1$ we get the conditions
\bea \label{quantOfull_U_1}
&&r ,\, \frac{r|A|^2}{2} ,\, \frac{r|A'|^2}{2} \in \mathbb{Z}; \\
\frac{1}{r} + A' \cdot A + r\frac{|A|^2}{2}\frac{|A'|^2}{2} \in \mathbb{Z} &&
\forall \,\,\pi \in \Gamma,\,\,\pi'\in\Gamma'  : \  \pi \cdot \pi' + r (\pi \cdot A) (\pi' \cdot A') \in \mathbb{Z} \, ;\nn\\ 
rA, \, rA' \in \Gamma \cap \Gamma' \, ; \qquad  &&
A' + \frac{r|A'|^2}{2}A \in \Gamma\, ;\qquad
A + \frac{r|A|^2}{2}A' \in \Gamma' \, .\nn 
\eea

Given that $rA' \in \Gamma \cap \Gamma'$, then $k = \pi' \cdot rA'\in\mathbb Z$. We first analyze the  condition $ \pi \cdot \pi' + r (\pi \cdot A) (\pi' \cdot A') \in \mathbb{Z} $. Being both $h=r(\pi\cdot A)$ and $h'=r(\pi' \cdot A)$ integer, we get $\pi\cdot \pi' + \frac{hh'}{r}\in \mathbb{Z}$. For particular values of $\pi$ and  $\pi'$, one has:

If $\pi \in \Gamma \cap \Gamma'$ then $\pi \cdot \pi' \in \mathbb{Z}$ and $\frac{hh'}{r} \in \mathbb{Z}$. If we restrict again to non-composite values for $r$ then at least one of $h$ or $h'$ has to be divisible by $r$, and then $\pi \cdot A \in \mathbb{Z}$ or $\pi' \cdot A' \in \mathbb{Z}$.
If $A'$ is such that this does not hold for any $\pi' \in \Gamma'$, then $A$ must satisfy $\pi \cdot A \in \mathbb{Z}$ for all $\pi \in \Gamma \cap \Gamma'$, i.e. $A \in (\Gamma\cap\Gamma')^{*}$. 
In conclusion, if $A' \notin \Gamma'$, then $A \in (\Gamma\cap\Gamma')^{*}$, i.e. either one of $A$ or $A'$ has to be in one of those lattices.

Repeating this with $\pi' \in \Gamma \cap \Gamma'$, we get that  if $A \notin \Gamma$, then $A' \in (\Gamma\cap\Gamma')^{*}$.
But the additional restriction, $A' + \frac{r|A'|^2}{2}A \in \Gamma$,  necessarily gives $A'\in \Gamma$,  since $\frac{r|A'|^2}{2}\in \mathbb{Z}$ when $A \in \Gamma$. Analogously, when $A' \in \Gamma'$ we get  $A \in \Gamma'$.
Then the possible Wilson lines are
\beq
A, A' &\in \Gamma \, ,  \quad
A, A' &\in \Gamma' \, , \quad
A \in (\Gamma\cap\Gamma')^{*}\backslash \Gamma , &\,\, \quad A' \in (\Gamma\cap\Gamma')^{*}\backslash \Gamma',
\eeq
which implies $\pi \cdot A',\pi \cdot A \in \mathbb{Z} \ \forall\,\, \pi \in \Gamma \cap \Gamma'$.

The equation $A' + \frac{r|A'|^2}{2} A \in \Gamma$ is equivalent to $\pi \cdot A' + \frac{r|A'|^2}{2} (\pi \cdot A) \in \mathbb{Z}\,\, \forall\,\, \pi \in \Gamma$, but when  $\pi \in \Gamma \cap \Gamma'$ it holds  trivially. Then we only have to verify the following equations
\beq
\pi \cdot A' + \frac{r|A'|^2}{2} (\pi \cdot A) \in \mathbb{Z}\,\, \forall\,\, \pi \in \Gamma\backslash\Gamma',\\
\pi' \cdot A + \frac{r|A|^2}{2} (\pi' \cdot A') \in \mathbb{Z}\,\, \forall\,\, \pi' \in \Gamma'\backslash\Gamma \,\,.
\eeq

Depending on $\Gamma$ and $\Gamma'$, it is possible that when $\pi \in \Gamma \backslash \Gamma'$ and $ \pi' \in \Gamma' \backslash \Gamma$ (i.e. $\pi, \pi' \notin \Gamma \cap \Gamma'$), then  $\pi \cdot \pi' = \frac12 \,\, mod(1)$. Assuming one of these cases holds, the condition $\pi\cdot \pi' + \frac{hh'}{r}\in \mathbb{Z}$ turns into $\frac{hh'}{r} = \frac12 \,\, mod(1)$. That is, neither $h$ nor $h'$ must be divisible by $r$: $\pi \cdot A \notin \mathbb{Z}$ and $\pi' \cdot A' \notin \mathbb{Z}$.
These equations imply $A \notin \Gamma$ and $A' \notin \Gamma'$, and then the Wilson lines are
\beq
A \in (\Gamma\cap\Gamma')^{*}\backslash \Gamma\, , \quad
A' \in (\Gamma\cap\Gamma')^{*}\backslash \Gamma'\, .
\eeq
They can be split into two sets: $A, A' \in (\Gamma\cap\Gamma')^{*}\backslash (\Gamma \cup \Gamma')$ and $
A \in  \Gamma' \backslash \Gamma,
A' \in \Gamma \backslash \Gamma'$, where we used
 $\Omega \cap (\Omega \cap \Sigma)^{*} = \Omega$.


Now we analyze the condition $\pi \cdot A' + \frac{r|A'|^2}{2} (\pi \cdot A) \in  \mathbb{Z}\,\, \forall\,\, \pi \in \Gamma\backslash\Gamma'$. All the cases that we will study verify $\pi \cdot A =\frac12 \,mod(1) \,\, \forall\,\, \pi \in \Gamma\backslash\Gamma'$. Then the condition becomes
\beq
\pi \cdot A' + \frac{r|A'|^2}{4} \in\mathbb{Z} \,\forall\, \pi \in \Gamma \backslash \Gamma' \,\,.
\eeq
If $A' \in \Gamma$, then $\frac{r|A'|^2}{2}\in 2{\mathbb Z}$. Instead if $A' \notin \Gamma$, then  $\frac{r|A'|^2}{2}$ is odd.
Using the analogous equation for $A\in \Gamma'$,  $\frac{r|A|^2}{2}$ has to even and for $A\notin \Gamma'$, $\frac{r|A|^2}{2}$ odd.

Summarizing, the condition requires $\frac{r|A|^2}{2}$ even if $A \in \Gamma' \backslash \Gamma$ and odd if $A \in (\Gamma \cap \Gamma')^{*}\backslash (\Gamma \cup \Gamma')$ (and analogously for $A'$). If additionally $(\Gamma \cap \Gamma')^{*}\backslash(\Gamma \cup \Gamma')$ is an integer lattice (which always is in the cases of our interest) then  $\frac{r|A|^2}{2}$ is odd if  $r=2$.
Given that $r=2$ for $\Gamma \neq \Gamma'$, then $r=1 \iff \Gamma = \Gamma'$ (when restricting to non-composite values of $r$).

Another condition that must hold is $2A \in \Gamma \cap \Gamma'$. But this occurs trivially for the lattices that we consider. $2A$ will always be in the adjoint conjugacy class of $SO(16)\times SO(16)$, which is contained in all $\Gamma \cap \Gamma'$ we will study  (this could vary with other groups where, for instance, $(s)+(s)=(v)$...).

We now analyze the condition $\frac{1}{r} + A' \cdot A + r\frac{|A|^2}{2}\frac{|A'|^2}{2} \in \mathbb{Z}$, which can be rewritten as
\beq
\frac{1}{2}\left (1 + |A|^2 |A'|^2 \right ) + A' \cdot A +  \in \mathbb{Z}\, .
\eeq
If $A\cdot A'\in \mathbb{Z}$, then  $\frac{1}{2} (1+|A|^2|A'|^2) \in \mathbb{Z}$. This holds if both $|A|^2$ and $|A'|^2$ are odd, i.e. $A, A' \in (\Gamma \cap \Gamma')^{*}\backslash(\Gamma \cup \Gamma')$. The product of these Wilson lines verifies this as it is an integer lattice  by hypothesis.

If $A\cdot A' = \frac12 \, mod(1)$ then $\frac{1}{2}(1+|A|^2|A'|^2) = \frac12\, mod(1)$. This holds if at least one of the Wilson lines has even modulus squared, i.e. $A\in \Gamma'\backslash \Gamma$ and/or $A'\in \Gamma \backslash \Gamma'$. The product of these Wilson lines verifies this assuming the hypothesis holds: $\pi \cdot A =\frac12 \,mod(1) \,\, \forall\,\, \pi \in \Gamma\backslash\Gamma'$ and its dual.

Summarizing, if the following hypothesis hold
\bea
\pi \cdot \pi' = \frac12 \,\, mod(1) \,\,\forall\,\, \pi \in \Gamma \backslash \Gamma', &&\quad  \pi' \in \Gamma' \backslash \Gamma\, ,\quad
\pi \cdot A =\frac12 \,mod(1) \,\, \forall\,\, \pi \in \Gamma\backslash\Gamma'\, , \nn\\
\pi' \cdot A' =\frac12 \,mod(1) \,\, \forall\,\, \pi' \in \Gamma'\backslash\Gamma &&
(\Gamma \cap \Gamma')^{*}\backslash(\Gamma \cup \Gamma') \text{ is an integer lattice,}\nn
\eea
then the duality must have
\beq
r=&2 \, , \quad
A \in (\Gamma\cap&\Gamma')^{*}\backslash \Gamma\, ,
A' \in (\Gamma\cap&\Gamma')^{*}\backslash \Gamma'\nn
\eeq
and the following conditions must be satisfied:
\begin{itemize}
\item[-] If $A \in \Gamma\backslash \Gamma'$, then $|A|^2\in 2\mathbb Z$. 
\item[-] If $A' \in \Gamma'\backslash \Gamma$, then $|A'|^2\in 2\mathbb Z$. 
\item[-] If $A \in (\Gamma \cap \Gamma')^{*}\backslash(\Gamma \cup \Gamma')$,  then $|A|^2\in 2\mathbb Z+1$.
\item[-] $A' \in (\Gamma \cap \Gamma')^{*}\backslash(\Gamma \cup \Gamma')$ then $|A'|^2\in 2\mathbb Z+1$.
\end{itemize}
These can be replaced by the more restrictive conditions
\beq
\Gamma\backslash \Gamma' , \Gamma' \backslash \Gamma \, \  \ \text{  are even lattices}\, , \quad
(\Gamma \cap \Gamma')^{*}\backslash(\Gamma \cup \Gamma') \  \ \ \text{ is an  odd lattice.}\nn
\eeq

The hypothesis $\pi \cdot A =\frac12 \,mod(1) \,\, \forall\,\, \pi \in \Gamma\backslash\Gamma'$ and $
\pi' \cdot A' =\frac12 \,mod(1) \,\, \forall\,\, \pi' \in \Gamma'\backslash\Gamma$ can also be replaced by the more restrictive ones
\beq
\pi \cdot \pi' =\frac12 \,mod(1) \,\, \forall\,\, \pi \in \Gamma\backslash\Gamma', \,\, \pi' \in  (\Gamma\cap\Gamma')^{*}\backslash \Gamma = \left[\Gamma'\backslash\Gamma\right] \cup\left[ (\Gamma \cap \Gamma')^{*}\backslash(\Gamma \cup \Gamma')\right]\nn\\
\pi' \cdot \pi =\frac12 \,mod(1) \,\, \forall\,\, \pi' \in \Gamma'\backslash\Gamma, \, \,  \pi \in (\Gamma\cap\Gamma')^{*}\backslash \Gamma = \left[\Gamma\backslash\Gamma' \right]\cup \left[(\Gamma \cap \Gamma')^{*}\backslash(\Gamma \cup \Gamma')\right]\nn\\
\eeq
Then sufficient conditions for duality to exist  (and  only for $r=2$) are
\bea
\Gamma \backslash \Gamma',\,\Gamma' \backslash\Gamma \ \ \text{even  lattices }\, , \quad
(\Gamma \cap \Gamma')^{*}\backslash(\Gamma \cup \Gamma') \text{ odd lattice}\, ,\nn\\
\pi \cdot \pi' = \frac12 \, mod(1)\,\text{if $\pi$ and $\pi'$ belong to different lattices (from these three)} \label{hipotesis}
\eea

\subsection*{$SO(32) \leftrightarrow E_8\times E_8$}

The quantization conditions on the second line of \eqref{quantOfull} 
%
can be written as
\bea
r (\pi \cdot A) (\pi' \cdot A') \in \mathbb{Z} \text{ if } \pi \text{ or } \pi' \in   \Gamma_{(00),(ss)} \, ,\\
r (\pi \cdot A) (\pi' \cdot A')+ \tfrac12 \in \mathbb{Z}\text{ if } \pi \in \Gamma_{(vv),(cc)}\text{ and } \pi' \in   \Gamma_{(0s),(s0)} \, .
\eea
%
%
%
In the second situation, we get
\beq
\tfrac{1}{r} (\pi \cdot rA) (\pi' \cdot rA')+ \tfrac12  \in \mathbb{Z}\, , \quad
\tfrac{jl}{r} + \tfrac12  \in \mathbb{Z},\quad j, k\in \mathbb{Z}\, ,
\eeq
which imply that $r$ is even. We restrict to the simplest possibility, $r=2$, for which we get
\bea 
&& |A|^2 ,\, |A'|^2 \in \mathbb{Z}; \nn\\
\tfrac{1}{2}\left ( 1 + 2A' \cdot A + |A|^2 |A'|^2 \right ) \in \mathbb{Z} &&
\forall \,\,\pi \in \Gamma_{16},\,\,\pi'\in\Gamma_8 \times \Gamma_8  : \  \pi' \cdot \pi + 2 (\pi \cdot A) (\pi' \cdot A') \in \mathbb{Z} \, ;\nn\\ 
2A, \, 2A' \in \Gamma_{(00),(ss)} \, ; \qquad  &&
A' + |A'|^2 A \in \Gamma_{16}\, ;\qquad
A + |A|^2 A' \in \Gamma_8 \times \Gamma_8 \label{hipotesis}
\eea
The Wilson lines that satisfy \eqref{hipotesis} are
\beq
A \in (\Gamma\cap\Gamma')^{*}\backslash \Gamma = (0s),(s0),(vc),(cv)\\
A' \in (\Gamma\cap\Gamma')^{*}\backslash \Gamma' = (vv),(cc),(vc),(cv)
\eeq

\section{Three and four-point functions }
\label{app:3point}

For completeness, in this appendix we list the  scattering amplitudes of massless states of the (toroidally  compactified)  heterotic string that give the effective action \eqref{accionefectiva}. The results hold for arbitrary points of the moduli space, including enhanced and broken symmetry points, and differ only on the possible values taken by the indices and structure constants. Details of the calculations can be found in \cite{polchi,blumen,maha}.

We use the following expectation values 
\bea
\left < X^{\mu}(z)X^{\nu}(w) \right >&=&-\frac12 \eta^{\munu} ln(z-w)\, , \qquad \left < X^{\mu}(\bar z)X^{\nu}(\bar w) \right >=-\frac12 \eta^{\munu} ln(\bar z-\bar w)\, , \nn\\
\left < \bar\psi^{\mu}(\bar z)\bar\psi^{\nu}(\bar w) \right >&=&\frac{\eta^{\munu}}{\bar z - \bar w} \, , \quad \left < \phi(\bar z)\phi(\bar w) \right >=- ln(\bar z-\bar w)\, ,  \qquad 
\left < \bar\chi^m(\bar z)\bar\chi^n(\bar w) \right >= \frac{\delta^{mn}}{\bar z - \bar w}\, ,\nn\\
\left < Y_L^{\hat I}( z)Y_L^{\hat J}( w) \right >&=&- \delta^{\hat I\hat J}ln(z-w)\, , \quad\left < Y_R^{m}( \bar z)Y_R^{n}(\bar w) \right >=- \delta^{mn}ln(\bar z-\bar w)\, ,
\nn\\
\langle J^\Gamma(z_1) J^\Lambda(z_2)\rangle&=&\frac{\delta^{\Gamma\Lambda}}{z_{12}^2}\, , \qquad 
\langle J^\Gamma(z_1) J^\Lambda(z_2)J^\Omega(z_3\rangle=\frac{if^{\Gamma\Lambda\Omega}}{z_{12}z_{13}z_{23}}\,
 .\nn
\eea

\subsection{Three-point functions of massless states }

$\bullet$\textbf{ Three left vectors:}
\bea
A_{AAA}&=&
-\frac{i}{\sqrt{2}}C_{S^2}g_c^3  f^{\Gamma\Lambda\Omega}A_{\Gamma\mu}(k_1)A_{\Lambda\nu}(k_2)A_{\Omega\rho}(k_3) \left({k_1}^{\nu}\eta^{\mu\rho}  +{k_2}^{\rho}\eta^{\mu\nu}+k_3^{\mu}\eta^{\nu\rho}\right )\, \nn\\
&=&12\pi  g_c \sqrt{2} f^{\Gamma\Lambda\Omega}\partial_{\mu}A_{\Gamma\nu} A^{\mu}_{\Lambda} A^{\nu}_{\Omega} \, , \nn
\eea
where we used  $C_{S^2}=\frac{8\pi}{ g_c^2}$ from unitarity, and identified $k_1^{\mu}  {A_2}_{\gamma}^{\Lambda} {A_1}_{\rho}^{\Gamma}\rightarrow -i\partial^{\mu} {A_1}_{\rho}^{\Gamma} {A_2}_{\gamma}^{\Lambda}$.

$\bullet$\textbf{ Three tensors:}
\bea
A_{VVV}(k_1,\ex^{(1)},k_2,\ex^{(2)},k_3,\ex^{(3)})
&=&C_{S^2}g_c^3\frac{1}{2} \ex^{(1)}_{\mu\nu}(k_1)\ex^{(2)}_{\tau\sigma}(k_2)\ex^{(3)}_{\lambda\eta}(k_3) \left ( k_1^{\sigma}\eta^{\nu\eta}+ k_2^{\eta}\eta^{\nu\sigma}+k_3^{\nu}\eta^{\sigma\eta} \right )
\nn\\&&
\times\left ( \frac{1}{2} k_1^{\tau} k_2^{\lambda} k_3^{\mu} + k_1^{\tau}\eta^{\mu\lambda} +k_2^{\lambda}\eta^{\mu\tau}+ k_3^{\mu}\eta^{\tau\lambda}  \right ) 
\nn
\eea

$\bullet$\textbf{ Two  left vectors - one tensor}:
\bea
&&A^{VAA}(k_1, \ex_1, k_2, A_2, k_3, A_3)\nn\\
&& \ \ \ \ \ \ \ \ \ \ =4\pi g_c \Big(- k_2^{\mu} {\ex_1}_{\mu\nu} k_3^{\nu} A_{2\rho}^{\Gamma}A_{3\Gamma}^{\rho}  +k_2^{\mu}{\ex_1}_{\mu\nu}{A_2}^{\nu\Gamma}k_1^{\sigma}A_{3\Gamma\sigma} +k_3^{\mu}{\ex_1}_{\mu\nu}{A_3}^{\nu\Gamma}k_1^{\rho}A_{2\rho\Gamma}
\Big)\nn
\eea
Replacing $V=g, b$ or $D$, we get respectively
\beq
A^{gAA}(k_1, h_1, k_2, A_2, k_3, A_3)=4\pi g_c  g_{\mu\nu}\left ( \partial^{\mu}A_{\rho}^{\Gamma}\partial^{\nu}A^{\rho}_{\Gamma}- 2\partial^{\mu}A_{\rho}^{\Gamma}  \partial^{\rho}A^{\nu}_{\Gamma}\right ) \, ,\nn
\eeq
\beq
A^{bAA}(k_1, B_1, k_2, A_2, k_3, A_3)=-8\pi g_c  A_{\rho}^{\Gamma} \partial^{\mu}A^{\nu}_{\Gamma}\partial^{\rho} b_{\mu\nu} \nn
\eeq
or
\beq
A^{DAA}(k_1, D, k_2, A_2, k_3, A_3)
=4\pi g_c (k_3 A_2^{\Gamma})(k_2 A_{3\Gamma})=-\frac{4\pi g_c }{\sqrt{d-2}}D \partial^{\nu}A_{\mu}^{\Gamma} \partial^{\mu}A_{\Gamma\nu} \, . \nn
\eeq

$\bullet$\textbf{ Two  right vectors - one tensor}:
\beq
A^{V\bar A \bar A}(k_1, \ex_1, k_2, \bar A_2, k_3, \bar A_3)=4\pi g_c {\ex_1}_{\mu\nu}\bar{A_2}^m_{\rho}\bar{A_3}_{m\sigma} \left ( \frac{1}{2} k_1^{\rho} k_2^{\sigma} k_3^{\mu} +\eta^{\mu\rho}k_2^{\sigma}+ \eta^{\mu\sigma}k_1^{\rho} + \eta^{\rho\sigma}k_3^{\mu}  \right )k_3^{\nu}\nn
\eeq
which gives 
\beq
A^{g\bar A \bar A}
=4\pi g_c g_{\mu\nu}\left (\partial^{\mu}\bar A_{\rho} \cdot\partial^{\nu}\bar A^{\rho}
-   2 \partial^{\rho}\bar A^{\mu} \cdot\partial^{\nu}\bar A_{\rho}   +\frac{1}{2}   \partial^{\sigma}\partial^{\mu}\bar A_{\rho}\cdot \partial^{\rho}\partial^{\nu}\bar A_{\sigma}\right ) \nn
\eeq
\beq
A^{b\bar A \bar A}=
-8\pi g_c \partial^{\rho}B_{\mu\nu} \bar A_{\rho}  \cdot\partial^{\nu}\bar A^{\mu}=4\pi g_c   \partial^{\rho}B_{\mu\nu} \bar A_{\rho}\cdot \bar F^{\mu\nu} \nn
\eeq
 or
\beq
A^{D\bar A \bar A}=\frac{2\pi g_c}{\sqrt{d-2}} D \bar F^{\mu\nu}\cdot\bar F_{\mu\nu} \nn
\eeq

$\bullet$\textbf{ Two scalars - one left vector }:
\bea
A^{ASS}(k_1,A_1,k_2,S_2,k_3,S_3)&=&4 \pi g_c \sqrt{2} i f_{\Gamma\Lambda\Omega} k_2^{\mu} A_1^{\mu\Gamma}   S_2^{\Lambda m_2} S_3^{\Omega m_3}\delta_{m_2m_3}\nn\\
&=&4 \pi g_c \sqrt{2} f_{\Gamma\Lambda}{}^{\Omega}  A^{\mu\Gamma}  (\partial_{\mu} S^{\Lambda m}) S_{\Omega m}\nn
\eea


$\bullet$\textbf{ Two scalars - one tensor}:
\bea
A^{VSS}(k_1,\ex_1,k_2,S_2,k_3,S_3)&=&-4\pi g_c{\ex_1}_{\mu\nu} S_2^{\Gamma m_2} S_3^{\Lambda m_3} \kappa_{\Gamma \Lambda}  \delta_{m_2m_3}k_{2}^{\mu}k_3^{\nu} \nn\\
&=& -4\pi g_c S_2^{\Gamma m} S_{3\Gamma m} (k_2 \cdot \ex_1 \cdot k_3)\nn
\eea
This is only non-vanishing for $V=g$
\bea
A^{gSS}(k_1,\ex_1,k_2,S_2,k_3,S_3)&=& 4\pi g_c g_{\mu\nu}\partial^{\mu}S^{\Gamma m} \partial^{\nu}S_{\Gamma m}\, . \nn
\eea

$\bullet$\textbf{ One scalar - one  right vector - one  left vector}:
\bea
A^{SA\bar A}(k_1,S,k_2,A,k_3,\bar A)&=& -4\pi g_c S_{\Gamma m} k_1^{\mu}  A_{\mu}^{\Gamma} k_1^{\nu} \bar A^m_{\nu} = -4\pi g_c S_{\Gamma m}  \partial_{\mu}A_{\nu}^{\Gamma} \bar F^{m\mu\nu} \ \ \nn
\eea

\subsection{Four-point function of massless scalars }
We present some details of  this computation which, to our knowledge,  has not been previously published
\beq
\langle S_{(0)} S_{(0)} S_{(-1)} S_{(-1)}\rangle = S_{\Gamma m}S_{\Lambda n}S_{\Omega p}S_{\Delta q}\frac{|z_{34}|^{k_3 \cdot k_4}|z_{24}|^{k_2 \cdot k_4}|z_{14}|^{k_1 \cdot k_4}|z_{23}|^{k_2 \cdot k_3}|z_{13}|^{k_1 \cdot k_3}|z_{12}|^{k_1 \cdot k_2}}{\bar z_{34}} \\
\times   \left [-\frac{k_1 \cdot k_2}{2\bar z_{12}} \left ( \frac{\delta^{mn}\delta^{pq}}{\bar z_{12}\bar z_{34}}-\frac{\delta^{mp}\delta^{nq}}{\bar z_{13}\bar z_{24}}+\frac{\delta^{mq}\delta^{np}}{\bar z_{14}\bar z_{23}} \right )+ \frac{\delta^{mn}\delta^{pq}}{\bar z_{12}^2 \bar z_{34} }  \right ] \langle J_{\Gamma}(z_1)J_{\Lambda}(z_2)J_{\Omega}(z_3)J_{\Delta}(z_4)\rangle
\eeq
Using that
\bea
\langle J^{\Gamma}(z_1) J^{\Lambda}(z_2) J^{\Omega}(z_3) J^{\Delta}(z_4) \rangle &=& 
 \frac{\kappa^{\Gamma \Lambda}  \kappa^{\Omega \Delta}}{{ z_{34}}^2  z_{12}^2}+    \frac{\kappa^{\Gamma \Omega}\kappa^{\Lambda \Delta}}{ z_{13}^2  z_{24}^2}   +  \frac{\kappa^{\Gamma \Delta}\kappa^{\Lambda \Omega}}{ z_{14}^2  z_{23}^2}  \nn\\
&&-\frac{f^{\Gamma \Lambda \Pi} f ^{\Omega \Delta}{}_{\Pi}}{z_{12}   z_{23} z_{24}z_{34}}
+\frac{f^{\Gamma \Omega \Pi} f ^{\Lambda  \Delta}{}_{ \Pi}}{ z_{13}z_{23}z_{24}{ z_{34}}}   
 -\frac{f^{\Gamma \Delta \Pi} f ^{\Lambda \Omega}{}_{ \Pi}}{ z_{14} z_{23} z_{24}{ z_{34}}}   \nn\, ,
\eea
we get
\bea
&&A^{SSSS}(k_1,S_1,k_2,S_2,k_3,S_3,k_4,S_4)
\ = \ -  8\pi g_c^2 \frac{k_1 \cdot k_2}{2}
 S_{\Gamma_1 m}S_{\Gamma_2 n}S_{\Gamma_3 p}S_{\Gamma_4 q}  \nn \\ 
&& \ \ \ \ \ \ \ \ \ \ \ \ \ \ \ \ \ \ \ \ \ \ \ \ \ \times \int d^2 z_1 \frac{|z_{34}|^{k_3 \cdot k_4}|z_{24}|^{k_2 \cdot k_4}|z_{14}|^{k_1 \cdot k_4}|z_{23}|^{k_2 \cdot k_3}|z_{13}|^{k_1 \cdot k_3}|z_{12}|^{k_1 \cdot k_2}}{\bar z_{12}}\nn\\
&&  \ \ \ \ \ \ \ \ \ \ \ \ \ \ \ \ \ \ \ \ \ \ \ \ \ \ \ \ \ \ \ \ \ \ \ \times \left ( \frac{\delta^{mn}\delta^{pq} \bar z_{23}\bar z_{24}}{\bar z_{12}\bar z_{34}}\left (1 - \frac{2}{k_1 \cdot k_2} \right )-\frac{\delta^{mp}\delta^{nq}\bar z_{23}}{\bar z_{13}}+\frac{\delta^{mq}\delta^{np}\bar z_{24}}{\bar z_{14}} \right )\nn\\
&& \ \ \ \ \ \ \ \ \ \ \ \ \ \ \ \ \ \ \ \ \ \ \ \ \ \ \ \ \ \  \ \ \ \ \ \times \left (  \frac{\kappa^{\Gamma_1 \Gamma_2}  \kappa^{\Gamma_3 \Gamma_4}z_{23}z_{24}}{{ z_{34}}  z_{12}^2}  
-\frac{f^{\Gamma_1 \Gamma_2 \Pi} f ^{\Gamma_3 \Gamma_4}{}_{\Pi}}{z_{12}  }
- (2 \leftrightarrow 3) - (2 \leftrightarrow 4)
 \right ) \ \ \ \ \ \nn
 \eea

Taking $z_1=z, z_2 = 0$, $z_3 = 1$, $z_4 \rightarrow \infty$, the integral is
\bea
&&\lim_{x \to \infty} 
 \int d^2 z \frac{|1-\frac{z}{x}|^{k_1 \cdot k_4} |1 - z|^{k_1 \cdot k_3}|z|^{k_1 \cdot k_2}}{\bar z}\nn\\
&& \ \ \ \ \ \ \ \ \ \ \ \ \ \  \times  \left ( \delta^{mn}\delta^{pq} \frac{-1}{\bar z}\left (1 - \frac{2}{k_1 \cdot k_2} \right )-\delta^{mp}\delta^{nq}\frac{1}{1-\bar z}-\delta^{mq}\delta^{np}\frac{ 1}{\frac{\bar{z}}{x}- 1} \right )\nn \\
&&\ \ \ \ \ \ \ \ \ \ \ \ \ \  \times \left (  \frac{-\kappa^{\Gamma \Lambda}  \kappa^{\Omega \Delta}}{  z^2}-    \frac{\kappa^{\Gamma \Omega}\kappa^{\Lambda \Delta} }{( 1-z)^2  }   -  \frac{\kappa^{\Gamma \Delta}\kappa^{\Lambda \Omega}}{( 1-\frac{z}{x})^2 }         
-\frac{f^{\Gamma \Lambda \Pi} f ^{\Omega \Delta}{}_{\Pi}}{z  }
-\frac{f^{\Gamma \Omega \Pi} f ^{\Lambda  \Delta}{}_{ \Pi}}{ 1-z}   
 -\frac{f^{\Gamma \Delta \Pi} f ^{\Lambda \Omega}{}_{ \Pi}}{ z-x }   
 \right )\nn\\
&=& -2\pi (-1)^{\frac{-k_1(k_2+k_3)}{2}}\Big ( \frac{\delta^{mn}\delta^{pq}\left (1 - \frac{2}{k_1 \cdot k_2} \right )}{\Gamma(2-\frac{k_1k_2}{2})\Gamma(-\frac{k_1 k_3}{2})\Gamma(\frac{k_1 (k_2+k_3)}{2})}\nn\\
&&\ \ \ \ \ \ \ \ \ \ \ \ \ \  \ \ \ \ +\frac{\delta^{mp}\delta^{nq}}{\Gamma(1-\frac{k_1k_2}{2})\Gamma(1-\frac{k_1 k_3}{2})\Gamma(\frac{k_1 (k_2+k_3)}{2})}+\frac{\delta^{mq}\delta^{np}} {\Gamma(1-\frac{k_1k_2}{2})\Gamma(-\frac{k_1 k_3}{2})\Gamma(1+\frac{k_1 (k_2+k_3)}{2})}\Big )\nn \\ &&\ \ \ \ \ \ \ \ \ \ \ \ \ \ 
\times \  (-1)^{\frac{k_1(k_2+k_3)}{2}} \Big [ - \kappa^{\Gamma \Lambda}  \kappa^{\Omega \Delta}\Gamma\left(-1+\frac{k_1k_2}{2}\right)\Gamma\left(1+\frac{k_1 k_3}{2}\right)\Gamma\left(1-\frac{k_1 (k_2+k_3)}{2}\right)\nn\\
&&\ \ \ \ \ \ \ \ \ \ \ \ \ \  \ \ \ \ \ \ \ \ \ \ \ \ \ \ \ \ \ \ \ \ \ \ \ \  -    \kappa^{\Gamma \Omega}\kappa^{\Lambda \Delta}\Gamma\left (1+\frac{k_1k_2}{2}\right)\Gamma\left(-1+\frac{k_1 k_3}{2}\right)\Gamma\left(1-\frac{k_1 (k_2+k_3)}{2}\right)   \nn\\
&&\ \ \ \ \ \ \ \ \ \ \ \ \ \  \ \ \ \ \ \ \ \ \ \ \ \ \ \  \ \ \ \ \ \ \ \ \ \  -  {\kappa^{\Gamma \Delta}\kappa^{\Lambda \Omega}}\Gamma\left(1+\frac{k_1k_2}{2}\right)\Gamma\left(1+\frac{k_1 k_3}{2}\right)\Gamma\left(-1-\frac{k_1 (k_2+k_3)}{2}\right)          \nn\\
&&\ \ \ \ \ \ \ \ \ \ \ \ \ \ \ \ \ \ \ \ \ \   \ \ \ \ \ \ \ \ \ \ \ \  \ \ \ \ \  + \ 
f^{\Gamma \Lambda \Pi} f ^{\Omega \Delta}{}_{\Pi}\Gamma\left(\frac{k_1k_2}{2}\right) \Gamma\left(1+\frac{k_1 k_3}{2}\right)\Gamma\left(-\frac{k_1 (k_2+k_3)}{2}\right)\nn\\
&&\ \ \ \ \ \ \ \ \ \ \ \ \ \ \ \ \ \ \ \ \ \  \ \ \ \ \ \ \ \ \ \ \ \  \ \ \ \ \ + \ f^{\Gamma \Omega \Pi} f ^{\Lambda  \Delta}{}_{ \Pi}\Gamma\left(1+\frac{k_1k_2}{2}\right)\Gamma\left(\frac{k_1 k_3}{2}\right)\Gamma\left(-\frac{k_1 (k_2+k_3)}{2}\right)     
 \Big ]\nn
 \eea
 where we used
\beq
I(m,n,\ax,\bx)= \int d^2 z (1-z)^m z^n |z|^{2\ax} |1-z|^{2\bx} = 2\pi (-1)^{m+n} \times \\ \times \frac{\Gamma (1+n+\ax)\Gamma (1+m+\bx)\Gamma (-1-n-m-\ax-\bx)}{\Gamma (-\ax)\Gamma (-\bx)\Gamma (2+\ax+\bx)}\, .
\eeq

 In terms of Mandelstam variables $s = -2 k_1 \cdot k_2$, $t = -2 k_1 \cdot k_3$, $u = -2 k_1 \cdot k_4$ and 
 summing over all  cyclic orderings of the vertex operators to compensate for the fixing of $z_2$, $z_3$ and $z_4$, we get
\bea
A^{SSSS} &=& \frac{\pi^2}{12}  g_c^2 S_{\Gamma_1 m_1}S_{\Gamma_2 m_2}S_{\Gamma_3 m_3}S_{\Gamma_4 m_4} 
\frac{\Gamma(-s/4)\Gamma(-t/4)\Gamma(-u/4)}{\Gamma(1+s/4)\Gamma(1+t/4)\Gamma(1+u/4)} \nn\\
&& \ \ \ \ \ \ \ \ \ \ \ \ \ \ \ \ \ \ \  \ \ \ \   \times \ \Big (\delta^{m_1 m_2}\delta^{m_3 m_4}tu+\delta^{m_1 m_3}\delta^{m_2 m_4}su +\delta^{m_1 m_4}\delta^{m_2 m_3}st \Big )\nn\\
&& \ \ \ \ \ \ \ \ \ \ \ \ \ \ \ \ \ \ \  \ \ \ \  \times  \ \Big (  -3\frac{\kappa^{\Gamma_1 \Gamma_2}  \kappa^{\Gamma_3 \Gamma_4}tu}{(s+4)}-    3\frac{\kappa^{\Gamma_1 \Gamma_3}\kappa^{\Gamma_2 \Gamma_4}su}{(t+4)}    -  3\frac{{\kappa^{\Gamma_1 \Gamma_4}\kappa^{\Gamma_2 \Gamma_3}}st}{(u+4)}   \nn\\     
&& \ \ \ \ \ \ \ \ \ \ \ \ \ \ \ \ \ \ \  \ \ \    \  \ \ \ \ \ \ +\ tf^{\Gamma_1 \Gamma_2 \Pi} f ^{\Gamma_3 \Gamma_4}{}_{\Pi}
+sf^{\Gamma_1 \Gamma_3 \Pi} f ^{\Gamma_2  \Gamma_4}{}_{ \Pi}
+uf^{\Gamma_1 \Gamma_3 \Pi} f ^{\Gamma_4 \Gamma_2}{}_{\Pi}\nn   \\
&&
\ \ \ \ \ \ \ \ \ \ \ \ \ \ \ \ \ \ \  \ \ \ \ \ \ \ \ \ \ 
+\ tf^{\Gamma_1 \Gamma_4 \Pi} f ^{\Gamma_3  \Gamma_2}{}_{ \Pi}
+ sf^{\Gamma_1 \Gamma_4 \Pi} f ^{\Gamma_2 \Gamma_3}{}_{\Pi}
+ uf^{\Gamma_1 \Gamma_2 \Pi} f ^{\Gamma_4  \Gamma_3}{}_{ \Pi}
 \Big )\nn
\eea
Expanding on $s=t=u=0$ and using
\beq
\frac{\Gamma(-s/4)\Gamma(-t/4)\Gamma(-u/4)}{\Gamma(1+s/4)\Gamma(1+t/4)\Gamma(1+u/4)}=- \frac{64}{stu}-2\zeta (3) + \mathcal{O}(stu) \, ,
\eeq
we finally get
\bea A^{SSSS}&=&-\frac{16\pi^2}{3}  g_c^2 S_{\Gamma_1 m_1}S_{\Gamma_2 m_2}S_{\Gamma_3 m_3}S_{\Gamma_4 m_4} \nn\\
&&
 \times  \Big (\ \frac{t}s
 \delta^{m_1 m_2}\delta^{m_3 m_4}(f^{\Gamma_1 \Gamma_2 \Pi} f ^{\Gamma_3 \Gamma_4}{}_{\Pi}+ f^{\Gamma_1 \Gamma_4 \Pi} f ^{\Gamma_3  \Gamma_2}{}_{ \Pi})
 +\delta^{m_1 m_3}\delta^{m_2 m_4}f^{\Gamma_1 \Gamma_2 \Pi} f ^{\Gamma_3 \Gamma_4}{}_{\Pi} \nn\\
&&\ \ \ +\frac{t}u\delta^{m_1 m_4}\delta^{m_2 m_3}(f^{\Gamma_1 \Gamma_2 \Pi} f ^{\Gamma_3 \Gamma_4}{}_{\Pi}  
+f^{\Gamma_1 \Gamma_4 \Pi} f ^{\Gamma_3  \Gamma_2}{}_{ \Pi})      
+\delta^{m_1 m_2}\delta^{m_3 m_4}f^{\Gamma_1 \Gamma_3 \Pi} f ^{\Gamma_2  \Gamma_4}{}_{ \Pi}  \nn\\
&& \ \ \ 
 +\frac{s}t\delta^{m_1 m_3}\delta^{m_2 m_4} (f^{\Gamma_1 \Gamma_3 \Pi} f ^{\Gamma_2  \Gamma_4}{}_{ \Pi}+f^{\Gamma_1 \Gamma_4 \Pi} f ^{\Gamma_2 \Gamma_3}{}_{\Pi}) +\delta^{m_1 m_4}\delta^{m_2 m_3}f^{\Gamma_1 \Gamma_2 \Pi} f ^{\Gamma_4  \Gamma_3}{}_{ \Pi} 
  \nn\\
&&\ \ \  +\frac{u}t \delta^{m_1 m_3}\delta^{m_4 m_2}(f^{\Gamma_1 \Gamma_3 \Pi} f ^{\Gamma_4 \Gamma_2}{}_{\Pi}+ f^{\Gamma_1 \Gamma_2 \Pi} f ^{\Gamma_4  \Gamma_3}{}_{ \Pi})
 +\delta^{m_1 m_4}\delta^{m_3 m_2}f^{\Gamma_1 \Gamma_3 \Pi} f ^{\Gamma_4 \Gamma_2}{}_{\Pi}  \nn\\ &&\ \ \ 
 +\frac{u}s\delta^{m_1 m_2}\delta^{m_3 m_4}(f^{\Gamma_1 \Gamma_3 \Pi} f ^{\Gamma_4 \Gamma_2}{}_{\Pi} +f^{\Gamma_1 \Gamma_2 \Pi} f ^{\Gamma_4  \Gamma_3}{}_{ \Pi})       
 +\delta^{m_1 m_3}\delta^{m_4 m_2}f^{\Gamma_1 \Gamma_4 \Pi} f ^{\Gamma_3  \Gamma_2}{}_{ \Pi} \nn\\  &&\ \ \ 
+ \frac{s}u\delta^{m_1 m_4}\delta^{m_2 m_3}(f^{\Gamma_1 \Gamma_4 \Pi} f ^{\Gamma_2 \Gamma_3}{}_{\Pi}+f^{\Gamma_1 \Gamma_3 \Pi} f ^{\Gamma_2  \Gamma_4}{}_{ \Pi})
 +\delta^{m_1 m_2}\delta^{m_4 m_3}f^{\Gamma_1 \Gamma_4 \Pi} f ^{\Gamma_2 \Gamma_3}{}_{\Pi}  
       \Big ) \nn
\eea
which adds up to 
\beq 
A^{SSSS}=(4!)2\pi^2 g_c^2 S_{\Gamma m}S_{\Gamma'}{}^m S_{\Lambda n}S_{\Lambda'}{}^n f^{\Gamma \Lambda \Pi}f^{\Gamma' \Lambda'}{}_{\Pi} 
\label{MMMM} 
\eeq
when using
$\frac{t}s+\frac{s}t+\frac{u}t+\frac{t}u+\frac{s}u+\frac{u}s =-3$
and $S_{\Gamma m}S_{\Lambda}{}^m S_{\Gamma' n}S_{\Lambda'}{}^n f^{\Gamma \Lambda \Pi}f^{\Gamma' \Lambda'}{}_{\Pi}=0$.

\subsection{Three-point functions involving slightly massive states }
It is easy to see that the amplitudes of three massless right vectors or three massless scalars vanish at the enhancement points. However, in the neighborhood of these points,  the currents acquire dependence on ${\bf p}_R$ and then the amplitude of three scalars or that  of two left  and one right vectors get a non-vanishing value and  give extra terms in the effective action.
Here we compute the three point functions involving states that become massive when slightly moving away from the enhancement points, so that their masses are smaller than other massive string states which we are not considering.

$\bullet$\textbf{ One right vector - two massive left vectors}:
 \beq
 \frac{A^{\bar AA'A'}}{C_{S^2}g_c^3} =  \frac{\delta^{p_2 + p_3}}{\sqrt{2}}\bar A_{\mu m}  k_2^{\mu} {A'}^{p_2}_{\nu} {A'}^{p_3 \nu}   p_{2R}^m \nn
 \eeq
where we  used  $k_1+k_2+k_3=0$, $k_1^2=0$, $k_2^2=k_3^2=-m^2 = -2p_{2R}^2$,  $k_1 \cdot k_2 = k_1 \cdot k_3 = 0$ and $k_2 \cdot k_3 = 2p_{2R}^2$. This gives the term
\bea
\frac{-i}{\sqrt{2}} p_R^m \bar A_{\mu m}  {A'}^{-p \nu} \partial^{\mu} {A'}^{p}_{\nu}\nn
\eea
in the effective action.


$\bullet$\textbf{ One massless - two massive left vectors:}
\beq
\frac{A^{AA'A'}}{C_{S^2}g_c^3}=  \frac{\delta^{p_{2}+p_{3}}}{\sqrt{2}} p_{ 2L\hat I}   \left [ (A^{\hat I} \cdot k_2 )( {A'}^{p_2}  \cdot {A'}^{p_3}  ) + (k_{1} \cdot {A'}^{p_3}   )( A^{\hat I} \cdot {A'}^{p_2} )- (k_{1} \cdot {A'}^{p_2} )( A^{\hat I}\cdot {A'}^{p_3}  ) \right ] \nn
\eeq
giving in the effective action
\beq
-\frac{i}{\sqrt{2}}p_{\hat I L} \left [ {A'}^{-p \nu}A^{\hat I}_{\mu}\partial^{\mu}{A'}^{p}_{\nu}    + 2{A'}^{p \nu}  {A'}^{-p}_{\mu}\partial^{\mu}A^{\hat I}_{\nu}   \right ]\nn
\eeq

$\bullet$\textbf{ One massless tensor - two massive left vectors:}
\beq
\frac{A^{VA'A'}}{C_{S^2}g_c^3} 
=\frac12 \ex_{\mu\nu} \delta^{p_2+p_3}{A'}^{p_2}_{\mu_2} {A'}^{p_3}_{\mu_3} \Big ( -k_2^{\mu} k_3^{\nu} \eta^{\mu_2\mu_3} + k_1^{\mu_3}\eta^{\nu\mu_2}k_2^{\mu} + k_1^{\mu_2} \eta^{\nu\mu_3}k_3^{\mu} \Big )\nn
\eeq
giving in the effective action
\beq
 \frac{1}{2} \Big ( \ex_{\mu\nu} \partial^{\mu}{A'}^{p}_{\rho} \partial^{\nu}{A'}^{-p \rho}    -2\partial^{\rho}\ex_{\mu\nu} \partial^{\mu}{A'}^{p \nu} {A'}^{-p}_{\rho}  \Big )\nn
\eeq

$\bullet$\textbf{ One massless scalar - two massive left vectors:}
\beq
\frac{A^{SA'A'}}{C_{S^2}g_c^3} =
 \delta^{p_{2} + p_{3}}   S_{\hat I m} {A'}^{p_2}_{\mu_2} {A'}^{p_3}_{\mu_3} p^{m}_{2R} p^{\hat I}_{2L}  \eta^{\mu_2 \mu_3} = p^{\hat I}_{L}   S_{\hat I m} p^{m}_{R} {A'}^{p}_{\mu} {A'}^{-p \mu} \nn
\eeq

$\bullet$\textbf{ Three massive left vectors:}
\beq
\frac{A^{A'A'A'}}{C_{S^2}g_c^3} =\frac{-i\varepsilon(p_1,p_2)}{\sqrt2}{A'}^{p_1}_{\mu_1} {A'}^{p_2}_{\mu_2} {A'}^{-p_1-p_2}_{\mu_3}  \left (  k_2^{\mu_1}\eta^{\mu_2 \mu_3} + k_1^{\mu_3} \eta^{\mu_1\mu_2}+ k_3^{\mu_2} \eta^{\mu_1\mu_3} \right )\nn
\eeq
where 
we used $k_i \cdot {A'}^{p_i}= 0$ and
conservation of momentum implies $k_i \cdot k_j = -2p_{Ri} \cdot p_{Rj}$ and $p_{iL}^2 - p_{iR}^2 = 2 \longrightarrow p_{iL}\cdot p_{jL} - p_{iR}\cdot p_{jR} = -1$ if $i\neq j$. 

This gives in the effective action the term
\beq
\frac{3}{\sqrt{2}} \varepsilon(p_1,p_2) {A'}^{p_1}_{\nu}\partial^{\nu}{A'}^{p_2}_{\mu} {A'}^{-p_1-p_2 \mu} \nn\ .
\eeq

\section{Counting structure constants of $SO(32)$ and $E_8\times E_8$}
\label{app:X}
In this Appendix we count and compare the number of non-vanishinig  structure constants  of the $SO(32)$ and $E_8\times E_8$ algebras, which 
in the Weyl-Cartan basis are
\beq
f^{\alpha\beta}{}_{\gamma} = \delta^{\alpha+\beta}_{\gamma} f^{\alpha\beta}{}_{\alpha+\beta} + \delta^{\beta}_{\overline \alpha}\delta^A_{\gamma} f^{\underline\alpha\overline \alpha}{}_{A} 
\eeq
with $\overline\alpha = -\underline\alpha$. 

To calculate the number of combinations of  $\alpha, \beta$ indices giving non-vanishing structure constants of $SO(32)$ , it is convenient to denote the $480$  roots as
\beq
(i\pm,j\pm) =  (0_{i-1},\pm 1, 0_{j-i-1}, \pm 1, 0_{16-j})\, ,
\eeq
with $1 \leq i<j \leq 16$, and split them  in  subsets  $(+, +)$, $(-, -)$, $(+, -)$, $(-, +)$ of $120$ elements each. Then we have
\beq
|(+,+)+(+,+)|^2\geq 4 \ \ \ {\rm or}\ \
|(-,-)+(-,-)|^2\geq 4 \rightarrow \text{there are no  roots}\\
(i+,j+)+(k-,l-) = \begin{cases} 
						0 \text{ if } i=k,\, j=l \\
						(j+,l-) \text{ if } i=k,\, j<l \\
						(l-,j+) \text{ if } i=k,\, j>l \\
						(i+,k-) \text{ if } j=l,\, i<k \\
						(k-,i+) \text{ if } j=l,\, i>k \\																		  
						(i+,l-) \text{ if } j=k \\
						(k-,j+) \text{ if } i=l \\					
						\text{no roots} \text{ if } i\neq k \text{ and } j\neq l
						\end{cases}\\
\eeq
The number of pairs of roots $(i+,j+),(k-,l-)$ is:
\bea
\left\{\begin{matrix}120   & {\rm if}~ i=k, j=l\\  
\sum\limits_{i=1}^{15}(16-i)(16-i-1)=1120 &{\rm if}~ i=k, i<j, i<l\neq j\left\{\begin{matrix} 
560~{\rm with}~ j>l\\ 560~{\rm with} ~j<l\end{matrix}\right.\\
\\
\sum\limits_{j=1}^{15}(16-j)(16-j-1)=1120 &{\rm if}~ j=l, i\neq k\left\{\begin{matrix}
560~{\rm with}~ i>k \\ 560~{\rm with}~i<k\end{matrix}\right.\\
\sum\limits_{j=2}^{15}(j-1)(16-j)=560 &{\rm if}~j=k, i<j ~{\rm and }~j<l\\
560 &{\rm if}~ i=l
\end{matrix}\right.
\eea

Then there are
\beq
(+,+)+(-,-) = \begin{cases} 
						0 \longrightarrow 120 \\
						(+,-) \longrightarrow 3 \times 560 = 1680 \\
						(-,+) \longrightarrow 3 \times 560 = 1680 \\
						\end{cases}
\eeq
That is, $1680$ non-vanishing structure constants of type $f^{(+,+)(-,-)}{}_{(+,-)}$ and $1680$ of type $f^{(+,+)(-,-)}{}_{(-,+)}$.
And analogously, there are
$1680$ non-vanishing structure constants of type $f^{(-,-)(+,+)}{}_{(+,-)}$ and $1680$  of type $f^{(-,-)(+,+)}{}_{(-,+)}$.\\
\beq
(i+,j+)+(k+,l-) = \begin{cases} 
						(i+,k+) \text{ if } j=l,\, i<k \\
						(k+,i+) \text{ if} j=l,\, i>k \\	
						(k+,j+) \text{ if } i=l \\														  
						\text{ no  roots} \text{ if } i\neq l \neq j \text{ or } i=k\text{ ó } j=k
						\end{cases}\\
\eeq
The number of pairs  $(i+,j+),(k+,l-)$ with $j=l$, $i \neq k$ is $1120$, of which $560$ correspond to $i>k$ and $560$ to $i<k$. And for
 $i=l$ there are $560$ pairs. That is
\beq
(+,+)+(+,-) =	(+,+) \longrightarrow 3 \times 560 = 1680 ~{\rm non-vanishing} ~f^{(+,+)(+,-)}{}_{(+,+)}
\eeq
And analogously, there are
1680  non-vanishing structure constants of each of the types $f^{(+,-)(+,+)}{}_{(+,+)}, $
 $f^{(+,+)(-,+)}{}_{(+,+)},$ $
 f^{(-,+)(+,+)}{}_{(+,+)},$ $
f^{(-,-)(+,-)}{}_{(-,-)},$ $
 f^{(+,-)(-,-)}{}_{(-,-)},$ $
f^{(-,-)(-,+)}{}_{(-,-)},$ and $
f^{(-,+)(-,-)}{}_{(-,-)}.$

\beq
(i+,j-)+(k+,l-) = \begin{cases} 
						(i+,l-) \text{ if } j=k \\
						(k+,j-) \text{ if } i=l \\	
						\text{no  roots} \text{ if } i=k  \text{ or } j=l \text{ ó } i\neq l,\, j\neq k
						\end{cases}\\
\eeq
The number of pairs  $(i+,j-),(k+,l-)$  with $j=k$ is $560$ and with $i=l$ is also $560$.
Then we have
\beq
(+,-)+(+,-) =	(+,-) \longrightarrow 2 \times 560 = 1120 ~{\rm non-vanishing} ~f^{(+,-)(+,-)}{}_{(+,-)}
\eeq
And analogously,
$1120$  of the type $f^{(-,+)(-,+)}{}_{(-,+)}$.

\beq
(i+,j-)+(k-,l+) = \begin{cases} 
						0 \text{ if } i=k, j=l \\
						(j-,l+) \text{ if } i=k,\, j<l \\ 					
						(l+,j-) \text{ if } i=k,\, j>l \\
						(i+,k-) \text{ if } j=l,\, i<k \\
						(k-,i+) \text{ if } j=l,\, i>k \\
						\text{there are no roots} \text{ if } i=l  \text{ or } j=k \text{ or } i\neq k,\, j\neq l
						\end{cases}\\
\eeq
The number of pairs of roots  $(i+,j-),(k-,l+)$ verifying $i=k$, $j=l$ is $120$; 
with $i=k$, $j\neq l$ there are $1120$ of which $560$ correspond to $j>l$ and $560$ to $j<l$; and for
$j=l$, $i\neq k$ there are $1120$ of which $560$ correspond to $i>k$ and $560$ to $i<k$.
Then there are
\beq
(+,-)+(-,+) =	\begin{cases} 
						0 \longrightarrow 120 \\
						(+,-) \longrightarrow 2 \times 560 = 1120 \\
						(-,+) \longrightarrow 2 \times 560 = 1120 \\
						\end{cases}
\eeq
That is, $1120$ structure constants of type $f^{(+,-)(-,+)}{}_{(+,-)}$ and $1120$ of type $f^{(+,-)(-,+)}{}_{(-,+)}$. 
And analogously
$1120$ $f^{(-,+)(+,-)}{}_{(+,-)}$ and $1120$ $f^{(-,+)(+,-)}{}_{(-,+)}$. \\

Summarizing there are
$12\times 560 = 6720$ combinations giving $(+,+)$, $6720$ giving $(-,-)$, $6720$ giving  $(+,-)$ and  $6720$ giving $(-,+)$. That is  $26880$ non-vanishing structure  constants  $f^{\alpha\beta}{}_{\alpha+\beta}$, and
$4 \times 120 \times 16 = 7680$ non-vanishing structure   constants $f^{\alpha\bar \alpha}{}_{A}$.\\

In the case $E_8 \times E_8$, we denote the roots
\beq
(1;i\pm,j\pm) =  (0_{i-1},\pm 1, 0_{j-i-1}, \pm 1, 0_{16-j})\\
(2;i\pm,j\pm) =  (0_{8+i-1},\pm 1, 0_{j-i-1}, \pm 1, 0_{8-j})
\eeq
with $1 \leq i<j \leq 8$; and:
\beq
(1;s) =  	\left (\left(\pm \frac12 \right )_{8 \text{(even)}},0_8\right )\\
(2;s) =  	\left ( 0_8,\left(\pm \frac12 \right )_{8 \text{(even)}}\right )
\eeq
where $s$ can take $2^7$ values. It can be thought of as a binary number of $7$ digits (one depending on  the others because there must be an even number of $-$ signs).

Split the $480$ roots of $E_8 \times E_8$ into $8$ subsets of $28$ elements: $(1;+, +)$, $(1;-, -)$, $(1;+, -)$, $(1;-, +)$, $(2;+, +)$, $(2;-, -)$, $(2;+, -)$, $(2;-, +)$ and $2$ subsets $(1;s)$, $(2;s)$ of $128$ elements. 

If $|(1;\cdots)+(2;\cdots)|^2 =  4 ,
|(1;+,+)+(1;+,+)|^2\geq 4 , \ {\rm or} \
|(1;-,-)+(1;-,-)|^2\geq 4,\text{there are no roots}$.
\beq
(1;i+,j+)+(1;k-,l-) = \begin{cases} 
						0 \text{ if } i=k,\, j=l \\
						(1;j+,l-) \text{ if } i=k,\, j<l \\
						(1;l-,j+) \text{ if } i=k,\, j>l \\
						(1;i+,k-) \text{ if } j=l,\, i<k \\
						(1;k-,i+) \text{ if } j=l,\, i>k \\																		  
						(1;i+,l-) \text{ if } j=k \\
						(1;k-,j+) \text{ if } i=l \\					
						\text{there are no  roots} \text{ if } i\neq k \text{ and } j\neq l
						\end{cases}\\
\eeq
The number of pairs of roots $(1;i+,j+),(1;k-,l-)$ is
\bea
\left\{\begin{matrix}
28 & {\rm with}~ i=k, j=l \\
\sum\limits_{i=1}^{7}(8-i)(8-i-1) = 112 & {\rm with}~ i=k, j\neq l 
\left\{\begin{matrix}56~ {\rm with }~j>l\\ 56~{\rm with}~j<l\end{matrix}\right.\\
112
&{\rm with}~j=l, i\neq k \left\{\begin{matrix}56 &{\rm with }~i>k\\
56 &{\rm with }~i<k\end{matrix}\right.\\
\sum\limits_{j=2}^{7}(j-1)(8-j)= 56 &{\rm with}~j=k, \\
56&{\rm with}~ i=l\end{matrix}\right.
\eea
The second line counts the number of pairs $ j,l$  such that $ i<j $ and $ i<l\neq j $, and the fourth one, the number of pairs $i,l$  such that $ i<j  $ and $  j<l$.
Then we have
\beq
(1;+,+)+(1;-,-) = \begin{cases} 
						0 \longrightarrow 28 \\
						(+,-) \longrightarrow 3 \times 56 = 168 \\
						(-,+) \longrightarrow 3 \times 56 = 168 \\
						\end{cases}
\eeq
i.e. $168$ structure constants of type $f^{(1;+,+)(1;-,-)}{}_{(1;+,-)}$ and $168$ of type $f^{(1;+,+)(1;-,-)}{}_{(1;-,+)}$.
And analogously, there are
$168$ of type $f^{(1;-,-)(1;+,+)}{}_{(1;+,-)}$ and $168$ of type $f^{(1;-,-)(1;+,+)}{}_{(1;-,+)}$.

For the other  roots of the kind  $(1;\pm,\pm)$, the analysis is as in the $SO(32)$ case, but now the number of non-vanishing structure constants  is one tenth as before:
$12\times 56 = 672$ combinations giving  $(1;+,+)$, $672$ giving  $(1;-,-)$, $672$ giving  $(1;+,-)$ and  $672$ giving $(1;-,+)$.\\

We also have

\beq
(1;s)+(1;s)  = \begin{cases} 
						 0 \longrightarrow 2^7 = 128\\
						(1;+,+)\longrightarrow 28 \times 2^5 = 896\\
						(1;-,-) \longrightarrow 896\\
						(1;+,-) \longrightarrow 896\\
						(1;-,+) \longrightarrow 896\\																		  
						\end{cases}\\
\eeq
For the sum of two roots of the kind $(1;s)$ to give  $(1,1,0_{14})$ it is necessary that they are of the form $(+1/2,+1/2,r,s,t,u,v,w)$ and $(+1/2,+1/2,-r,-s,-t,-u,-v,-w)$. Then there are $2^5 =32$  possible choices of parameters $r, s, t, u, v$ ($w$ is not independent). Since there are $28$ roots of the kind $(1;+,+)$,  the number of non-vanishing structure constants  $f^{(1;s)(1;s)}{}_{(1;+,+)}$ is $32\times 28 = 896$.
And analogously there are $896$ $f^{(1;s)(1;s)}{}_{(1;+,+)}$, $896$ $f^{(1;s)(1;s)}{}_{(1;-,-)}$, $896$ $f^{(1;s)(1;s)}{}_{(1;+,-)}$ and $896$ $f^{(1;s)(1;s)}{}_{(1;-,+)}$, and
\beq
(1;s)+(1;+ +)  =  	(1;s)\longrightarrow 28 \times 2^5 = 896												
\eeq
To have $(1;s)+(1,1,0_{14})=(1;s)$, it is necessary that $(1;s)=(-1/2,-1/2,r,s,t,u,v,w)$. Then there are $2^5 =32$ possible choices of parameters $r, s, t, u, v$ ($w$ is not independent). Since there are $28$ roots of the kind $(1;+,+)$,  the number of non-vanishing  structure constants of the type $f^{(1;s)(1;+,+)}{}_{(1;s)}$ is $32\times 28 = 896$.
And analogously there are  $896$ structure constants of type $f^{(1;s)(1;-,-)}{}_{(1;s)}$, $896$ $f^{(1;s)(1;+,-)}{}_{(1;s)}$, $896$ $f^{(1;s)(1;-,+)}{}_{(1;s)}$, $896$ $f^{(1;+,+)(1;s)}{}_{(1;s)}$,$896$ $f^{(1;-,-)(1;s)}{}_{(1;s)}$, $896$ $f^{(1;+,-)(1;s)}{}_{(1;s)}$ and $896$ $f^{(1;-,+)(1;s)}{}_{(1;s)}$.
\footnote{Note that there are an even number of $-$ signs since the sign of two components is always modified. This agrees with the fact that the spinorial conjugation class  only changes to the conjugate one when adding a vector of the vectorial class.}

The same holds for the sum of two roots of  type $(2,\cdots)$, and
then there are a total of
$2\times (12\times 56 + 896 )= 3136$ combinations giving  $(+,+)$, $3136$ giving $(-,-)$, $3136$ giving $(+,-)$, $3136$ giving $(-,+)$ and $2\times 8 \times 896 = 14336$ giving $(1,s)$. That is $26880$ non-vanishing structure constants of  type $f^{\alpha\beta}{}_{\alpha+\beta}$. 

In addition, there are $2\times(4\times 28+128)\times16=7680$ structure constants 
of  type $f^{\alpha\bar \alpha}{}_{A}$.

In conclusion, the number of structure constants of type $f^{\alpha\beta}{}_{\alpha+\beta}$ is $26880$ 
and of type $f^{\underline\alpha\overline\alpha}{}_{A}$ is 7680, for both the $SO(32)$ and the $E_8 \times E_8$ groups.

\end{document}